\documentclass[10pt]{article}
\usepackage{amsfonts, amsmath, amssymb, amsthm}
\usepackage[english]{babel}
\usepackage{booktabs}
\usepackage{bm}
\usepackage{eucal}
\usepackage{enumerate}
\usepackage{float}
\usepackage{dsfont}
\usepackage{epsfig}
\usepackage {fontenc}
\usepackage[hmargin=1.6cm,vmargin=3cm]{geometry}
\usepackage{lscape}
\usepackage{mathrsfs}
\usepackage{natbib}
\usepackage{rotating}
\usepackage{verbatim}
\usepackage[table]{xcolor}
\usepackage{dirtytalk}
\usepackage{multirow}
\usepackage{mdframed}
\usepackage{subfigure}
\usepackage{adjustbox}
\usepackage{hyperref}
\usepackage{algorithm}
\usepackage{algpseudocode}
\date{}

\renewcommand{\thefootnote}{$\dagger$}

\begin{document}
\title{The underlap coefficient as a measure of \\a biomarker’s discriminatory ability}
\author{\textsc{Zhaoxi Zhang}, \textsc{Vanda~In\'acio}, \textsc{Miguel de Carvalho}, \\\textsc{for the Alzheimer's Disease Neuroimaging Initiative}}
\date{}
\maketitle 

\begin{abstract}
The first step in evaluating a potent ial diagnostic biomarker is to examine the variation in its values across different disease groups. In a three-class disease setting, the volume under the receiver operating characteristic surface and the three-class Youden index are commonly used summary measures of a biomarker’s discriminatory ability. However, these measures rely on a stochastic ordering assumption for the distributions of biomarker outcomes across the three groups. This assumption can be restrictive, particularly when covariates are involved, and its violation may lead to incorrect conclusions about a biomarker’s ability to distinguish between the three disease classes. Even when a stochastic ordering exists, the order may vary across different biomarkers in discovery studies involving dozens or even thousands of candidate biomarkers, complicating automated ranking.  To address these challenges and complement existing measures, we propose the underlap coefficient, a novel summary index of a biomarker's ability to distinguish between three (or more) disease groups, and study its properties.
Additionally, we introduce Bayesian nonparametric estimators for both the unconditional underlap coefficient and its covariate-specific counterpart. These estimators are broadly applicable to a wide range of biomarkers and populations. A simulation study reveals a good performance of the proposed estimators across a range of conceivable scenarios. We illustrate the proposed approach through an application to an Alzheimer’s disease (AD) dataset aimed to assess how four potential AD biomarkers distinguish between individuals with normal cognition, mild impairment, and dementia, and how and if age and gender impact this discriminatory ability.
\end{abstract}
\textsc{\footnotesize{key words: Alzheimer's disease; Bayesian nonparametrics; covariates; diagnostic biomarkers;  stochastic ordering; underlap coefficient.} }

\let\thefootnote\relax\footnotetext{Zhaoxi Zhang, School of Mathematics, University of Edinburgh, Scotland, UK (\textit{z.zhang-156@sms.ed.ac.uk}).Vanda In\'acio, School of Mathematics, University of Edinburgh, Scotland, UK (\textit{vanda.inacio@ed.ac.uk}).  Miguel de Carvalho,  School of Mathematics, University of Edinburgh, Scotland, UK  (\textit{miguel.decarvalho@ed.ac.uk}). Data used in preparation of this article were obtained from the Alzheimer's Disease Neuroimaging Initiative (ADNI) database (adni.loni.usc.edu). As such, the investigators within the ADNI contributed to the design and implementation of ADNI and/or provided data
	but did not participate in the analysis or writing of this report. A complete listing of ADNI
	investigators can be found at: \textcolor{blue}{\href{http://adni.loni.usc.edu/wp-content/uploads/how_to_apply/ADNI_Acknowledgement_List.pdf}{http://adni.loni.usc.edu/wp-content/uploads/how\_to\_apply/ADNI\_Acknowledgement\_List.pdf}}}

\section{\large{\textsf{INTRODUCTION}}}
Assessing the discriminatory ability of a continuous biomarker for diagnostic purposes is a common task in clinical research. The first fundamental step toward this goal is to evaluate how biomarker outcomes vary across different disease stages or groups. In this article, we investigate how potential biomarker of Alzheimer's disease (AD) simultaneously distinguish between individuals with normal cognition, mild cognitive impairment, and severe impairment or AD. As highlighted in \cite{Zhang2016}, suitable biomarkers that track the stages of AD progression could markedly accelerate drug development by providing earlier indications of drug efficacy. Thus, it is critical to develop proper statistical methods to evaluate how potential biomarkers can distinguish between the three aforementioned stages.

As a direct generalisation of receiver operating characteristic (ROC) curves, ROC surfaces have been developed to assess discriminatory ability in ordered three-class problems  \citep{nakas2004ordered}. In this setting, the area under the ROC curve and the Youden index extend to the volume under the ROC surface (VUS) and the three-class Youden index ($\text{YI}_3$), respectively \citep{nakas2014developments}. ROC surfaces, and consequently their summary indices, rely on a stochastic ordering assumption for the distributions of biomarker outcomes. When this assumption is strongly violated, using the VUS and $\text{YI}_3$ may be highly misleading. Such violations often reflect differences in scale, in addition to differences in location, for biomarker outcomes across the three groups. For many diseases or conditions, promising biomarkers may not follow a stochastic ordering. Even when the stochastic ordering holds, if the densities of biomarker outcomes cross more than twice, this may render both $\text{YI}_3$ and the VUS suboptimal as measures of discriminatory ability. As a result, when the VUS and  $\text{YI}_3$ are blindly applied to large panels of candidate biomarkers, those with good discriminatory ability but without a monotonic ordering may be overlooked. Moreover, even when a monotonic order exists but it is not the anticipated one (e.g.,  lower biomarker values indicate more severe disease stages), the biomarker may still be disregarded. While this issue can be manually corrected when dealing with a small number of biomarkers, it becomes impractical when evaluating dozens. 

The challenge of non-monotonic biomarkers has been quite extensively studied in the two-class disease setting (see, among many others, \citealp{Martinez2017,samawi2017notes,Carvalho2020, Bantis2021,Lindner2023}). However, the literature on the three-class setting, and beyond, remains sparse, and methods or adjustments from the two-class setting do not generalise easily to three (or more) classes. To the best of our knowledge, the only existing work is by \cite{yang2020non}, which proposes a non-monotonic transformation of  biomarker outcomes to improve VUS values. As acknowledged by the authors, the assumption of normally distributed biomarker outcomes is crucial for the methodology to work. We should also mention that, although not specifically tailored to non-monotonic biomarkers, \cite{Nakas2007} proposed  the construction of ROC graphs and a corresponding summary measure to assess the overlap of three groups under an umbrella ordering, i.e., for assessing the ability of a biomarker to differentiate one disease class from two others without requiring a specific ordering between the two latter classes.

In this article, we propose the underlap coefficient to address a gap in the literature and to complement existing summary measures of discriminatory ability for three or more disease classes.
The UNL is not based on any classification rule and directly quantifies differences in biomarker outcome distributions across three or more groups, without assuming a specific ordering of biomarker values among the groups.  As a result, it is widely applicable in the biomarker field and can be reliably used with panels of candidate biomarkers that include:  (i) biomarkers following the conventional monotonic ordering (where higher values correspond to more severe disease stages), (ii) biomarkers with a reverse ordering (where lower values indicate greater disease severity), and (iii) biomarkers exhibiting a non-monotonic ordering. The UNL is further geometrically appealing, has an intuitive interpretation as the `effective' number biomarker outcomes groups, and can be easily generalised to settings with more than three disease classes.  Furthermore, as discussed in Section \ref{underlap_section}, the UNL is computationally more efficient to estimate than existing ROC summary based measures, especially when the number of disease-classes increases.

We further introduce the covariate-specific underlap coefficient to account for patient heterogeneity in evaluating  a biomarker's ability to simultaneously distinguish between three or more classes. Ignoring such heterogeneity can lead to biased estimates of a biomarker's discriminatory ability \citep{Janes2008, Inacio2021}.  In the presence of covariates, assuming a specific ordering of biomarker outcomes across groups becomes an even more stringent assumption. 
Notably,  \cite{To2022} is the only study that adjusts for covariates in the case of more than two disease classes, specifically in the estimation of the VUS. 
Interestingly, in the conclusions section of this article, the following can be read: ``(...) in our paper methods are designed under the assumption that the monotone ordering
hypothesis holds for every value of the covariates. This assumption may not be satisfied in practice and its check might add, from a practical point of view, some complexity to the analysis.''

As a further contribution, we develop Bayesian nonparametric estimators for both the UNL and the covariate-specific UNL. At the core of both estimators lies a (dependent) Dirichlet process mixture of normal distributions, resulting in highly flexible methods that can adapt to intricate distributional features of the data. In the conditional setting, unlike the kernel-based approach of \cite{To2022}, our method can easily accommodate multiple covariates, whether continuous or categorical. Furthermore, we allow the entire conditional distribution to smoothly change with the covariates, rather than restricting the modeling to only the mean and scale, as in their approach. Consequently, our proposed estimators are broadly applicable to a wide range of biomarkers, making them particularly useful in analyses involving a large number of biomarkers. Furthermore, working within the Bayesian paradigm allows for both point and interval estimates of the UNL and the covariate-specific UNL to be obtained within a unified framework.

The rest of this paper is organised as follows. In Section 2, we introduce the underlap coefficient, examine its relationship with other summary measures of a biomarker's discriminatory ability, discuss its straightforward generalisation to settings with more than three disease classes, and introduce the covariate-specific underlap coefficient. In Section 3, we present our inferential framework for the underlap coefficient and its covariate-specific counterpart. In Section 4, we assess the performance of the unconditional and conditional estimators of the (covariate-specific) underlap coefficient using simulated data. In Section 5, we apply the underlap coefficient and its corresponding covariate-specific counterpart to evaluate the discriminatory ability of four biomarkers in distinguishing simultaneously between three different cognitive function stages in the context of Alzheimer's disease and examine whether and how this discriminatory ability is affected by age and gender. Finally, we offer concluding remarks in Section 6.

\section{\large{\textsf{THE UNDERLAP COEFFICIENT}}}
\label{underlap_section}
Let $Y$ be a univariate continuous random variable representing the biomarker outcome and let $D\in\{1,2,3\}$ be a ternary random variable indicating the true disease status. 
For example, in the context of our Alzheimer's disease  application, $D=1$ represents individuals with normal cognition, $D=2$ corresponds to individuals with mild cognitive impairment, and $D=3$ denotes individuals with severe dementia or Alzheimer's disease. We use the subscripts $1$, $2$, and $3$ to denote quantities conditional on $D=1$, $D=2$, and $D=3$, respectively. For example, $f_1 (F_1
)$, $f_2 (F_2)$, and $f_3 (F_3)$ denote the probability density (cumulative distribution) functions of $Y$ in groups $1$, $2$, and $3$. We further assume that the group to which each individual belongs is known without error, due to the existence of a gold standard biomarker or test. Our goal is to assess how the biomarker under investigation, which is potentially less invasive or costly, performs in distinguishing among the three groups, compared to the gold standard.

\subsection{The definition of the underlap coefficient}
We introduce the (three-class) underlap coefficient (UNL), an index that quantifies the degree of separation--or equivalently, overlap--between the distributions of biomarker outcomes in the three groups. It can therefore be regarded as a measure of the biomarker's ability to simultaneously distinguish among the three groups. Unlike the VUS and $\text{YI}_3$, the UNL does not rely on any form of stochastic ordering among the three distributions. The UNL is mathematically defined as
\begin{equation}
	\text{UNL}(f_1,f_2,f_3)=\int_{-\infty }^{+\infty} \max\left\{f_1(y),f_2(y),f_3(y)\right\}\text{d}y.
	\label{3_class_UNL_definition}
\end{equation}
The UNL ranges from one to three and can be intuitively interpreted as the `effective' number of distinct populations of biomarker outcomes among the three groups, analogous to the concept of effective sample size in Markov chain Monte Carlo methods. An underlap coefficient of three indicates complete separation of biomarker outcomes across the three groups, representing three distinct `effective' populations with no overlap. In this case, the biomarker perfectly discriminates among the groups. Conversely, at the other extreme, when the distributions of biomarker outcomes completely overlap, suggesting only one `effective' group, the underlap coefficient is one. This indicates that the biomarker cannot distinguish among the three (or even two) groups. UNL values between one and three represent varying degrees of separation (or, equivalently, overlap) among the distributions of biomarker outcomes in the three groups. Figure 1 in the Supplementary Material illustrates several examples. We note the case where two distributions of biomarker outcomes completely overlap but are entirely separated from the third, resulting in $\text{UNL} = 2$. This scenario represents two `effective' populations: one formed by the two identical distributions, which collectively contribute to a single `effective' population, and the other represented by the third, distinct `effective' population. The underlap coefficient is invariant under strictly increasing and differentiable transformations of $Y_1$, $Y_2$, and $Y_3$ (proof provided in Section A of the Supplementary Material). 

\subsection{On the relationship between the underlap and overlap coefficients}
\label{UNL_OVL_compare}
\begin{figure}[htbp]
	\centering
	\subfigure{
		\centering
		\includegraphics[width=0.38\textwidth]{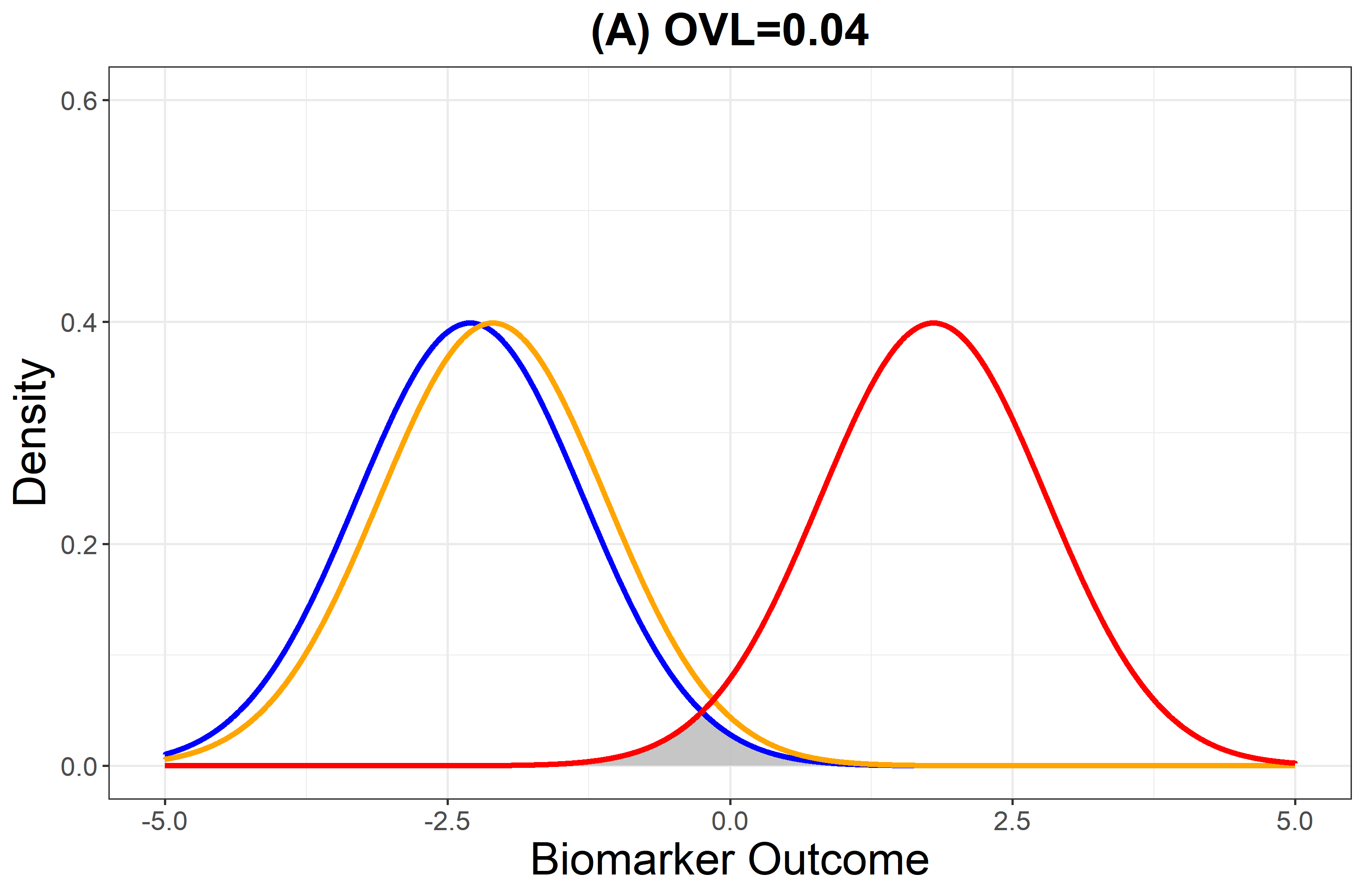}
	}
	\subfigure{
		\centering
		\includegraphics[width=0.38\textwidth]{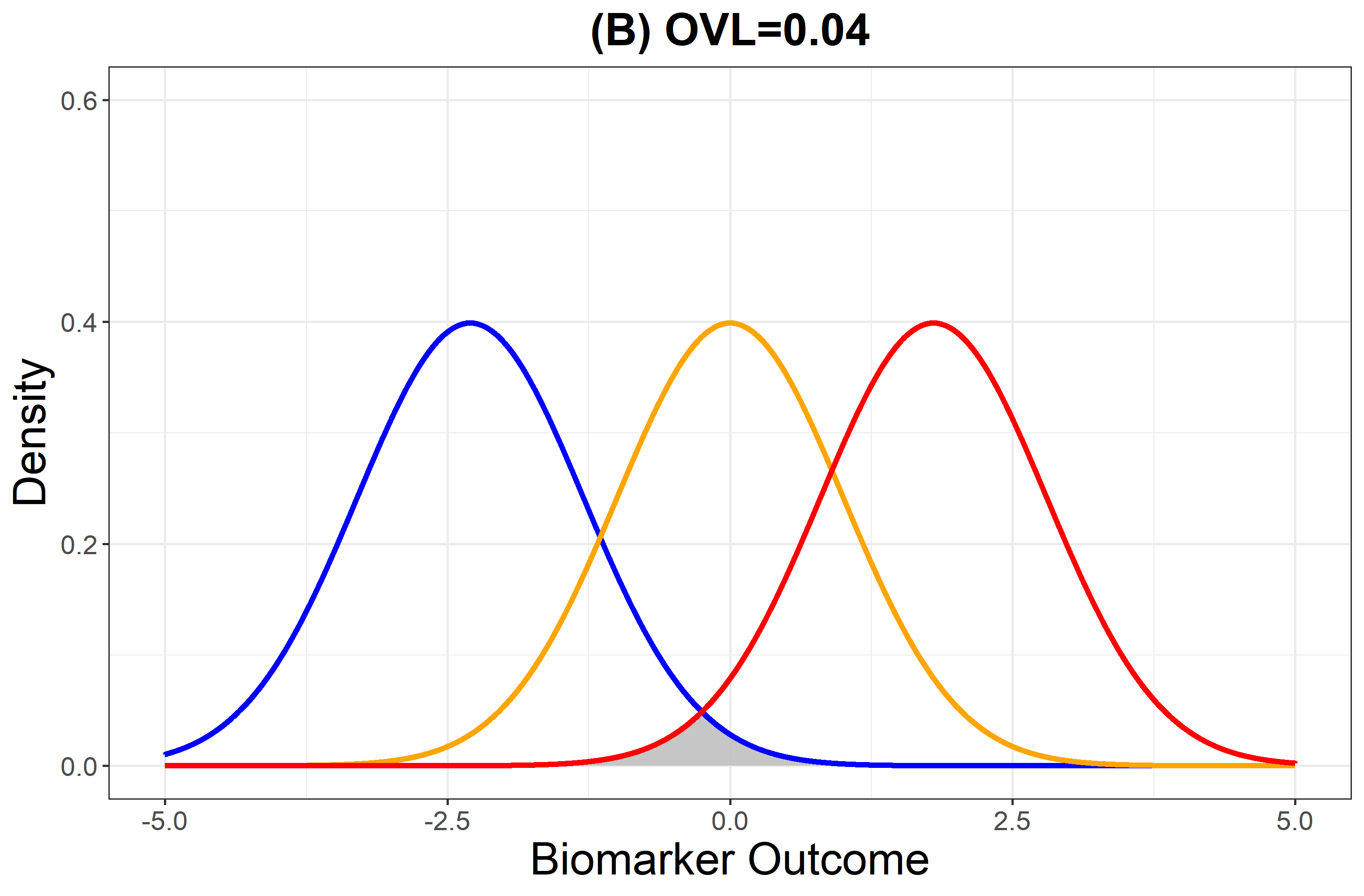}
	}
	\centering
	\caption{Example of the three-class overlap coefficient, as defined in \eqref{3_class_OVL_definition}, for normal distributions with two different degrees of separation between the biomarker outcome densities in groups 1 and 2. Blue lines: biomarker outcome density in group 1. Orange lines:  biomarker outcome density in group 2. Red lines: biomarker outcome density in group 3. Grey area: the three-class overlap coefficient. 
	}
	\label{3_overlap_example}
\end{figure}

The underlap coefficient is closely related to the two-class overlap coefficient, which is defined as the proportion of the area shared between two density functions and is mathematically expressed as
\begin{equation*}
	\text{OVL}(f_{d_1},f_{d_2})=\int_{-\infty }^{+\infty} \min\left\{f_{d_1}(y),f_{d_2},(y)\right\}\text{d}y,\quad d_1,d_2\in\{1,2,3\}.
	\label{2_class_OVL_definition}    
\end{equation*}
When developing a measure of how well a biomarker simultaneously discriminates between three disease groups, without relying on any stochastic ordering assumption, it is natural to extend the two-class overlap coefficient to the three-class case by considering
\begin{equation}
	\text{OVL}(f_1,f_2,f_3)=\int_{-\infty }^{+\infty} \min\left\{f_1(y),f_2(y),f_3(y)\right\}\text{d}y.
	\label{3_class_OVL_definition}    
\end{equation}
However, the three-class overlap coefficient defined in \eqref{3_class_OVL_definition} is not suitable for evaluating a biomarker's ability to simultaneously distinguish among three disease groups. To illustrate why, we consider the two hypothetical scenarios shown in Figure \ref{3_overlap_example}. Specifically, in Scenario A (left panel of Figure \ref{3_overlap_example}), the distributions of biomarker outcomes for groups 1 and 2 are nearly indistinguishable. Nevertheless, the three-class overlap is $0.04$, suggesting an almost perfect ability of the biomarker to simultaneously distinguish among the three groups. In contrast, in Scenario B (right panel of Figure \ref{3_overlap_example}), shifting the distribution of biomarker outcomes  to the right results in greater separation between the three distributions compared to Scenario A. However, the overlap coefficient remains $0.04$, as in Scenario A. By comparison, the underlap coefficient is $2.03$ for Scenario A and $2.38$ for Scenario B, reflecting the difference in the biomarker's discriminatory power across the two scenarios. We note that in these two scenarios, where a stochastic ordering of the biomarker outcome distributions holds, the VUS is $0.55$ in Scenario A and $0.84$ in Scenario B, while the three-class Youden index is $2.03$ for Scenario A and $2.38$ for Scenario B. 

\begin{figure}[!ht]\centering
	\includegraphics[width=7.5cm]{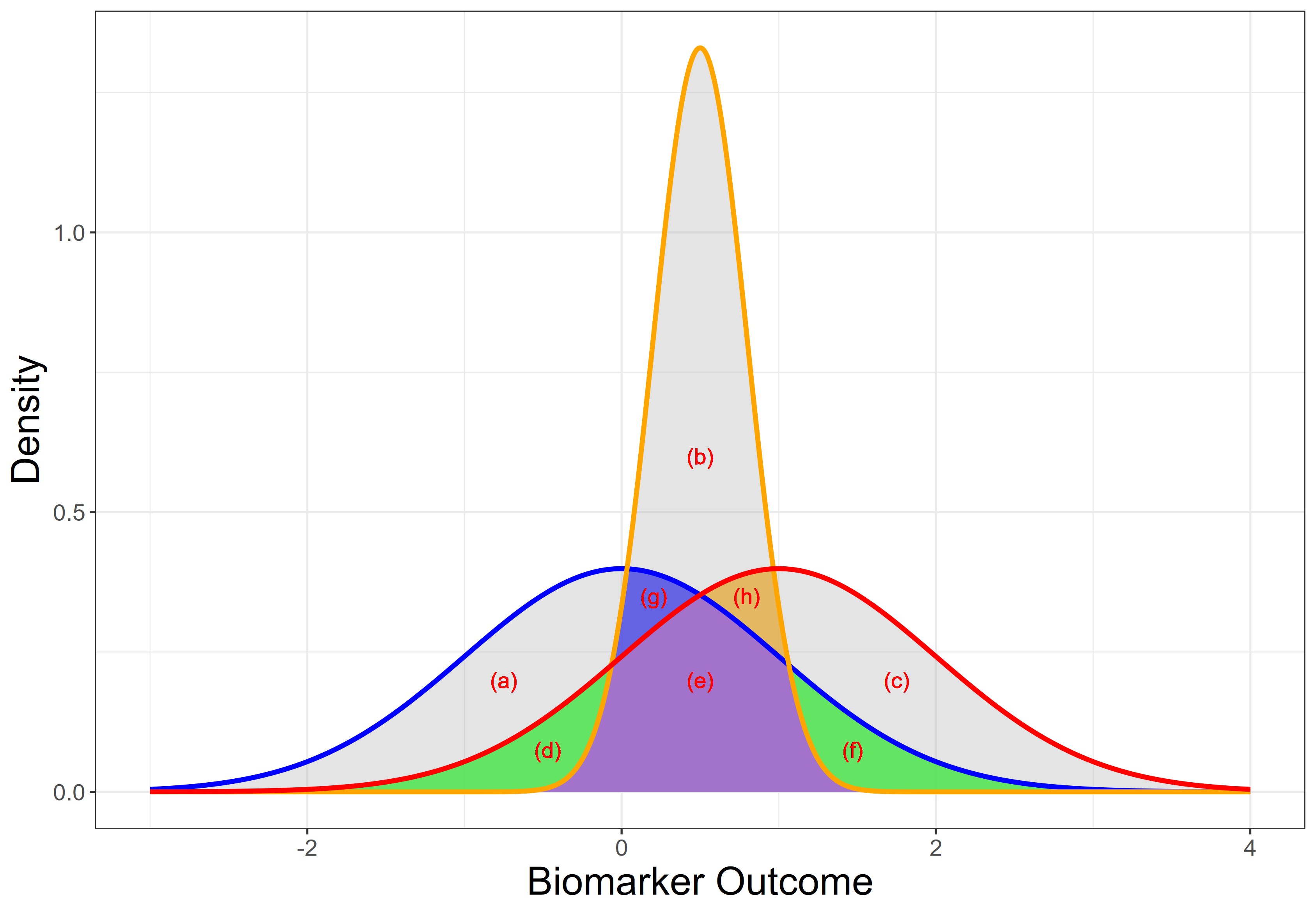}
	\caption{Graphical depiction of the underlap coefficient and its relationship to the overlap coefficient. $\text{OVL}(f_1,f_2,f_3)$: $\text{area}(e)$. $\text{OVL}(f_1,f_2)$: $\text{area}(g)+\text{area}(e)$. $\text{OVL}(f_2,f_3)$: $\text{area}(e)+\text{area}(h)$.\\$\text{OVL}(f_1,f_3)$: $\text{area}(d)+\text{area}(e)+\text{area}(f)$.\\$\text{UNL}(f_1,f_2,f_3)$: $\text{area}(a)+\text{area}(b)+\text{area}(c)+\text{area}(d)+\text{area}(e)+\text{area}(f)+\text{area}(g)+\text{area}(h)$. The blue, orange, and red lines denote the densities of biomarker outcomes in groups 1, 2, and 3, respectively.}
	\label{unl_example}
\end{figure}

While the underlap coefficient addresses the highlighted limitations of the overlap coefficient in the three-class setting, there remains a close relationship between the two. Specifically, the underlap coefficient can be expressed as a function of all three possible two-class overlap coefficients and  the three-class overlap coefficient, as follows
\begin{equation}
	\text{UNL}(f_1,f_2,f_3)= 3-\text{OVL}(f_1,f_2)-\text{OVL}(f_1,f_3)-\text{OVL}(f_2,f_3)+
	\text{OVL}(f_1,f_2,f_3).
	\label{UNL_OVL_relation}
\end{equation}
That is, as also graphically illustrated in Figure \ref{unl_example}, the UNL represents the cumulative area under the density functions of the three groups, accounting for overlapping regions only once. The derivation of the equality in \eqref{UNL_OVL_relation} is presented in Section A of the Supplementary Material.
It is worth noting that, in the two-class case, an analogous relationship also holds, i.e., 
\begin{equation*}
	\text{UNL}(f_{d_1}, f_{d_2}) = 2 - \text{OVL}(f_{d_1},f_{d_2}), \quad d_1,d_2\in\{1,2,3\}.
\end{equation*}
In this case, the UNL ranges from one to two, with a value of one indicating complete overlap of the two biomarker outcome distributions and a value of two indicating complete separation. However, in the two-class setting, the UNL does not offer any advantage over the OVL.

\subsection{On the relationship between the underlap coefficient and the three-class Youden index}
\label{unl_yi_section}
\begin{figure}[htbp]
	\centering
	\subfigure{
		\centering
		\includegraphics[width=0.38\textwidth]{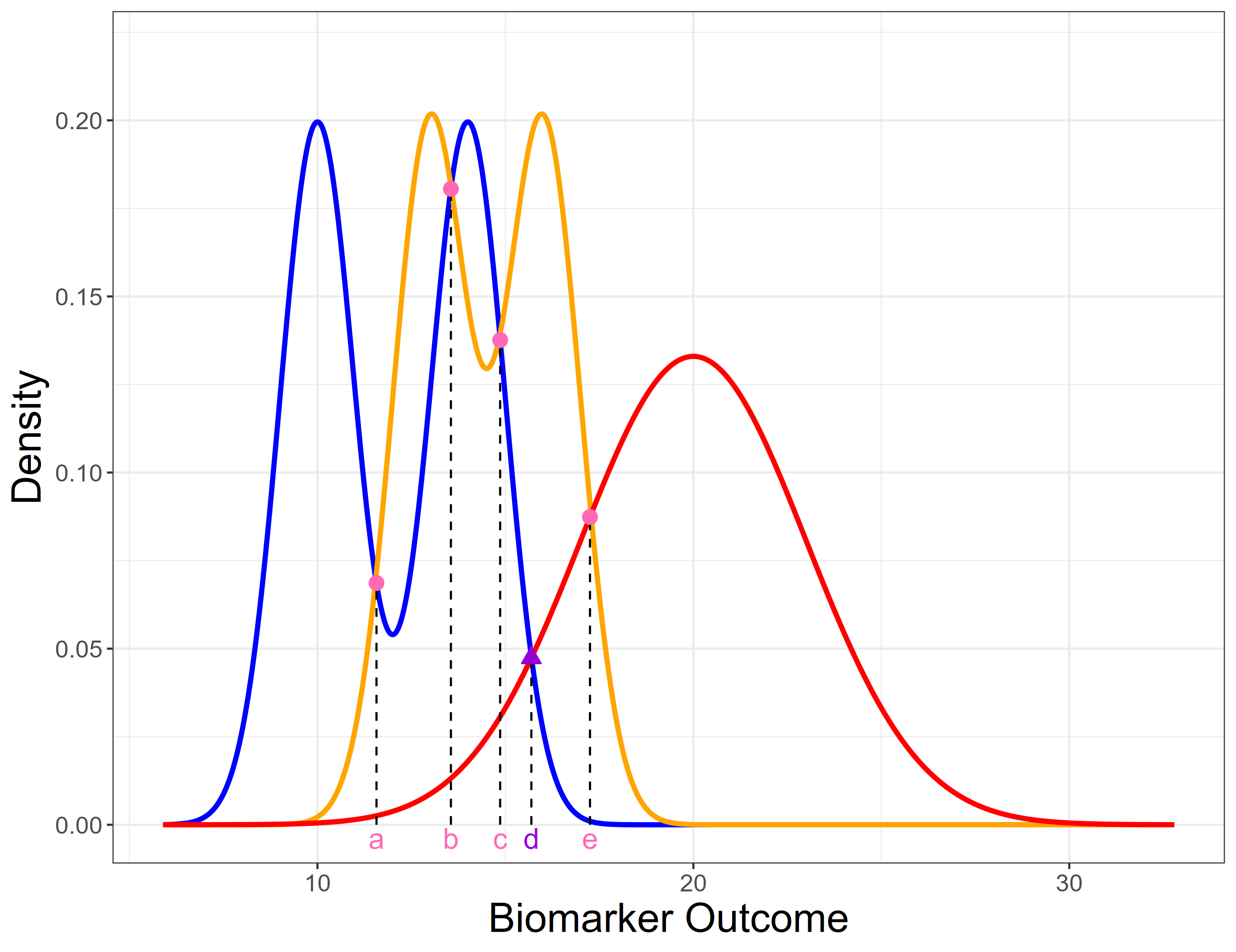}
	}
	\subfigure{
		\centering
		\includegraphics[width=0.38\textwidth]{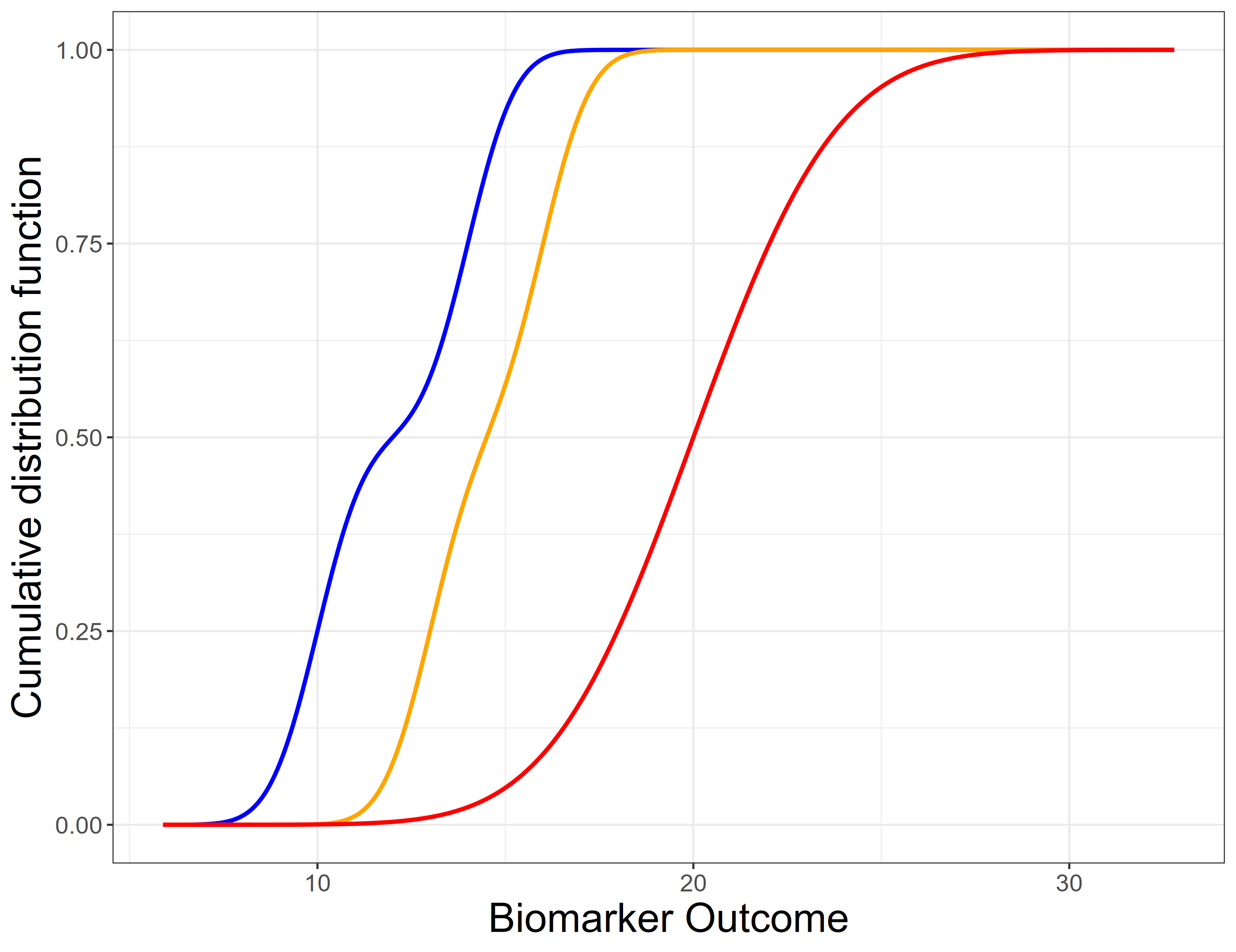}
	}
	\centering
	\caption{Left panel: Graphical illustration of inner and outer intersection points. The blue, orange, and red lines denote the densities of biomarker outcomes in groups 1, 2, and 3, respectively. Outer intersection points: $a$, $b$, $c$, and $e$ (pink points). Inner intersection point: $d$ (purple point). Right panel: cumulative distribution functions for this example. 
		The three-class Youden index for this example is 2.20, while the underlap coefficient is 2.25.}
	\label{inner_outer_plot}
\end{figure}

Without loss of generality, we assume that individuals in group 3 tend to have higher biomarker outcomes than those in group 2, who, in turn, tend to have higher outcomes than individuals in group 1.  This assumption implies a stochastic ordering of the biomarker outcome distributions across the three groups:
\begin{equation}\label{so}
	F_1(y)\geq F_2(y)\geq F_3(y),\quad \forall y \in \mathbb{R}.
\end{equation}
Let  $(c_1,c_2)\in \mathbb{R}^2$ be a pair of thresholds, with $c_1<c_2$. The probabilities of correct classification into each group can be written as
\begin{align*}
	p_1(c_1,c_2)&=\Pr(Y\le c_1\mid D=1)=F_1(c_1),\\
	p_2(c_1,c_2)&=\Pr(c_1<Y \le c_2\mid D=2)=F_2(c_2)-F_2(c_1),\\
	p_3(c_1,c_2)&=\Pr(Y>c_2\mid D=3)=1 - F_3(c_2).
\end{align*}
The three-class Youden index \citep{Nakas2010} can be calculated as follows:
\begin{align*}
	\text{YI}_3 &= \max_{c_1 < c_2} \{p_1(c_1,c_2) + p_2(c_1,c_2) + p_3(c_1,c_2)  \}\nonumber\\
	&= \max_{c_1 < c_2} \{ F_1(c_1)+F_2(c_2)-F_2(c_1)-F_3(c_2)+1\}.
\end{align*}
The three-class Youden index falls in the range $[1,3]$. When the densities of biomarker outcomes in the three groups completely overlap, $\text{YI}_3 = 1$, whereas $\text{YI}_3 = 3$ indicates perfect separation. If, in addition to the stochastic ordering in \eqref{so}, we assume that there is only one crossing point between $f_1$ and $f_2$, as well as between $f_2$ and $f_3$, then these crossing points correspond to the optimal cutoff points implied by the definition of the Youden index, and we may write
\begin{align}\label{unl_2_intersection}
	\text{YI}_3 &= F_1(c_1^\text{opt})+F_2(c_2^\text{opt})-F_2(c_1^\text{opt})-F_3(c_2^\text{opt})+1\\ \nonumber
	&=\int_{-\infty }^{c_1^\text{opt}} f_1(y)\text{d}y+\int_{c_1^\text{opt} }^{c_2^\text{opt}} f_2(y)\text{d}y+\int_{c_2^\text{opt}}^{+\infty } f_3(y)\text{d}y = \text{UNL} .
\end{align}
We note that the equivalence in \eqref{unl_2_intersection} holds even if $f_1$ and $f_3$ cross. However, this crossing must occur at what we define as an inner intersection or crossing point, as opposed to the intersections between $f_1$ and $f_2$, and between $f_2$ and $f_3$, which we term outer intersections. Specifically, a point $c \in \mathbb{R}$ is an outer intersection point if $f_{d_1}(c) = f_{d_2}(c) = M$ and $f_{d_3}(c)<M$. In contrast, $c$ is said to be an inner intersection point if $f_{d_1}(c) = f_{d_2}(c) = M$ and $f_{d_3}(c) >M$, where $d_1$, $d_2$, and $d_3$ each represent one of the three disease groups. Intuitively, outer intersection points can be interpreted as `change points' in the group with the highest density. Figure \ref{inner_outer_plot} illustrates these notions in a specific example. 

The equality in \eqref{unl_2_intersection} no longer holds if there are more than two outer intersections between the three densities. In such a case, the $K$ outer intersection points $\{c_i\}_{i=1}^{K}$ (with $K > 2$), divide the range of biomarker outcomes into $K+1$ intervals. Let the group with the highest density in the $i$th interval be denoted by $m_i$ ($m_i\in \{1,2,3\}$). The underlap coefficient can then be expressed as
\begin{align}
	\text{UNL} & = \int_{-\infty }^{c_1} f_{m_1}(y)\text{d}y+\int_{c_1}^{c_2} f_{m_2}(y)\text{d}y+\cdots+\int_{c_t}^{c_{t+1}}f_{m_{t+1}}(y)\text{d}y+\cdots+\int_{c_K}^{+\infty } f_{m_{K+1}}(y)\text{d}y
	\label{unl_multi_intersection}
\end{align} 
Even when $c_1^\text{opt}$ and $c_2^\text{opt}$ are both outer intersection points, the underlap coefficient in \eqref{unl_multi_intersection} evaluates a broader range of potential thresholds than the three-class Youden index in \eqref{unl_2_intersection}, thereby enabling a more accurate assessment. The stochastic ordering assumption is more likely to be violated when there are multiple outer intersections between the densities of the three groups, especially if one group's distribution has substantially higher variance than those of the other two. However, regardless of whether the stochastic ordering holds, when there are more than two outer intersections, the three-class Youden index will always be smaller than the underlap coefficient.
Situations leading to violations of the stochastic ordering also lead to $\text{UNL} > \text{YI}_3$.

Similar results were obtained by \cite{samawi2017notes} for the relationship between the overlap coefficient and the two-class Youden index.

\subsection{On the relationship between the underlap coefficient and the VUS}
The ROC surface is the three-dimensional plot in the unit cube depicting the probabilities of correct classification into each group as the
thresholds $c_1$ and $c_2$ vary, that is,
\begin{align*}
	&\{(p_1(c_1,c_2),p_2(c_1,c_2),p_3(c_1,c_2)) : (c_1,c_2)\in\mathbb{R}^2,\ c_1 < c_2\} \\
	&= \{(F_1(c_1), F_2(c_2) - F_2(c_1), 1 - F_3(c_2)): (c_1, c_2)\in\mathbb{R}^2,\ c_1 < c_2\}
\end{align*}
By writing $c_1=F_1^{-1}(p_1)$ and $c_2 = F_3^{-1}(1-p_3)$, we obtain the functional form of the ROC surface
\begin{equation*}
	\text{ROCS}(p_1,p_3) = \begin{cases}
		F_2(F_3^{-1}(1-p_3))-F_2(F_1^{-1}(p_1)),& \text{ if }   F_1^{-1}(p_1)<F_3^{-1}(1-p_3),\\
		0,& \text{otherwise},
	\end{cases}
\end{equation*}
where for notational simplicity, we have omitted the dependence of $p_1$ and $p_3$ on $c_1$ and $c_2$.
The VUS is defined as 
\begin{equation*}
	\text{VUS} = \int_{0}^{1} \int_{0}^{1} \text{ROCS}(p_1, p_3) \, \text{d}p_3 \, \text{d}p_1 = \Pr(Y_1 < Y_2 < Y_3).
	\label{VUS_definition}
\end{equation*}
The VUS takes the value $1/6$ when the three distributions completely overlap, and the value one when the three classes are perfectly discriminated in the correct order. 
Unlike the three-class Youden index, there is, to our knowledge, no equivalence between the UNL and the VUS under a stochastic ordering among the three groups. Nonetheless, we numerically explore the relationship between the UNL and the VUS in the proper trinormal setting, which is a special case of the popular trinomal model \citep{xiong2006measuring,noll2019}. In this setting, the biomarker outcomes for all three groups follow independent normal distributions with distinct means, denoted by $\mu_1$, $\mu_2$, and $\mu_3$, such that $\mu_1<\mu_2 <\mu_3$, and a common variance $\sigma^2$. In this case, the underlap coefficient can be written as
\begin{equation*}
	\text{UNL} =2\Phi\left(\frac{\mu_3-\mu_2}{2\sigma}\right)+2\Phi\left(\frac{\mu_2-\mu_1}{2\sigma}\right)-1,
\end{equation*}
whereas according to \cite{xiong2006measuring}, the VUS is given by
\begin{equation*}
	\text{VUS}=\int_{-\infty }^{+\infty } \Phi\left(y+\frac{\mu_2-\mu_1}{\sigma}\right) \Phi\left(-y+\frac{\mu_3-\mu_2}{\sigma}\right)\phi(y)\text{d}y, 
\end{equation*} 
where $\phi$ and $\Phi$ denote the density and distribution functions of the standard normal distribution, respectively.
As a concrete example, by fixing $\mu_2$ to a particular value ($7$ in this case, though any other value could have been used without loss of generality), and varying the group means $\mu_1$ and $\mu_3$ so that they move away from each other, we obtain the UNL and VUS  surfaces shown in Figure \ref{VUS_vs_UNL_3d}. Additionally, Figure 2 in the Supplementary Material shows three profile curves for specific $\mu_1$ values. As observed in these two figures, in this setting, where the VUS is a well-defined measure of discriminatory ability, the UNL and VUS follow a very similar trend, differing mainly in scale. Furthermore, the relationship between the VUS and the UNL in this simplified setting offers useful guidance for interpreting UNL values. For example, within this proper normal model, which ensures that the resulting ROC surface are monotonically increasing and concave (bowed downward), if $\text{VUS}<0.25$ is deemed clinically unacceptable, then $\text{UNL}<1.21$ may also be considered unacceptable. Understanding the relationship between the UNL and the VUS provides guidance on the expected UNL values for previously studied biomarkers based on their known VUS, particularly for those more accustomed to interpreting the VUS.

\begin{figure}[htbp]\centering
	\includegraphics[width=6.5cm]{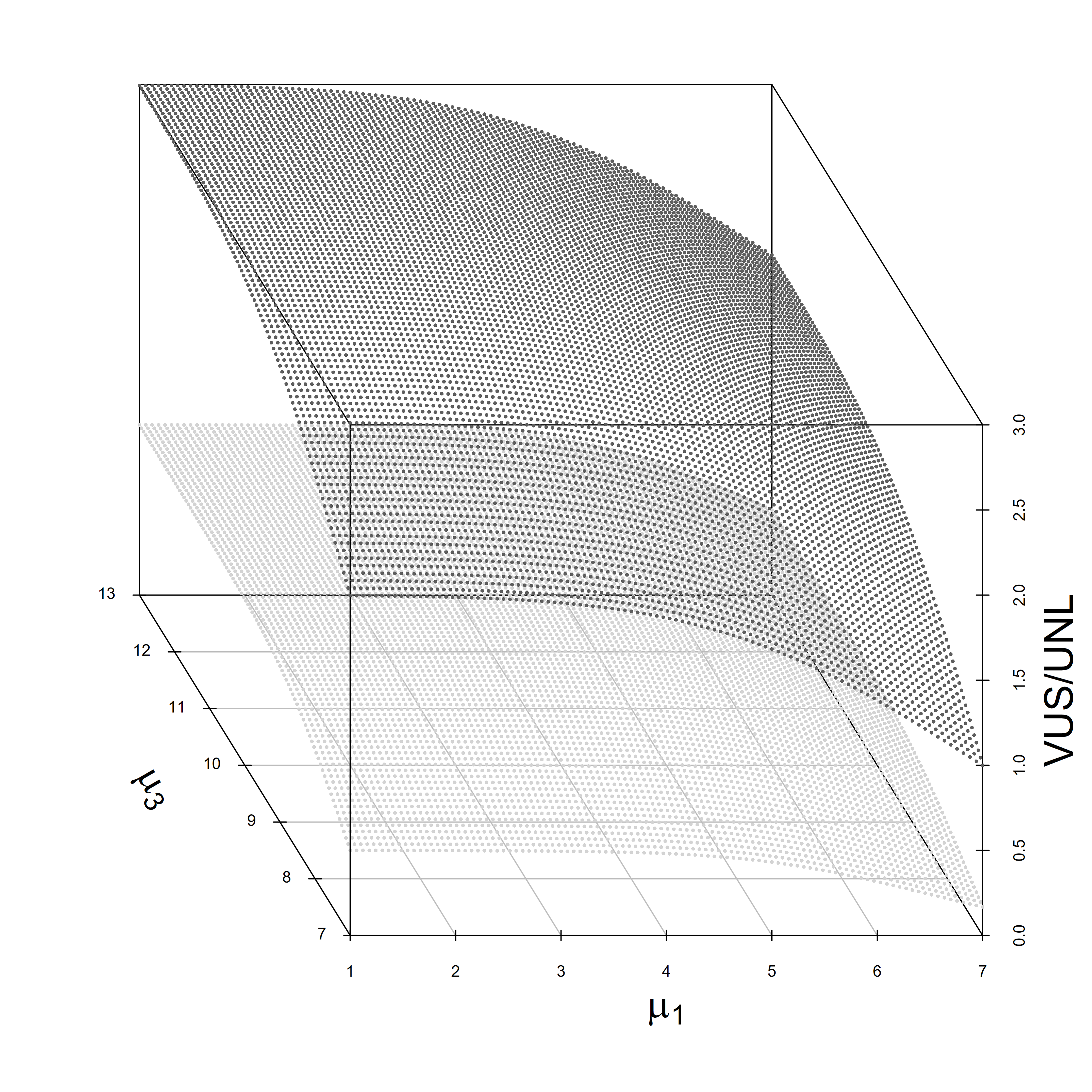}
	\caption{VUS and UNL surfaces based on three normal distributions for the biomarker outcome densities in each group, assuming a common variance. Here, $\mu_2=7$. Dark grey surface: UNL. Light grey surface: VUS.}
	\label{VUS_vs_UNL_3d}
\end{figure}

\subsection{The underlap coefficient when there are more than three disease classes}
\label{UNL_multiclass}
The underlap coefficient can be easily generalised to the multi-class setting. Let $H >3$ denote the number of disease groups, and let $f_1,\ldots,f_H$ represent the densities of the biomarker outcomes in each of these  $H$ groups. The mathematical definition of the multi-class UNL is as follows
\begin{equation*}
	\text{UNL}(f_1,\dots,f_H)=\int_{-\infty }^{+\infty} \max\left\{f_1(y),\ldots,f_H(y)\right\}\text{d}y.
	\label{multi_class_UNL_definition}
\end{equation*}
In this multi-class case, $\text{UNL} = 1$ when the densities of the biomarker outcomes completely overlap (no discriminatory ability and a single `effective' population of biomarker outcomes) and $\text{UNL} = H$ when the $H$ densities are completely separated from one another (perfect discriminatory ability and $H$ `effective' populations of biomarker outcomes).

The multi-class Youden index for the case where $H>3$ was introduced in \cite{nakas2013} and the volume under the surface  gives rise to the hypervolume under the ROC manifold  \citep{scurfield1996multiple,li2008roc}. As in the three-class case, both measures are constructed under the assumption of a stochastic ordering among the distributions of biomarker outcomes, an assumption that becomes increasingly stringent as the number of disease classes grows. In contrast, the multi-class UNL is free of any stochastic ordering assumption, the only such measure when $H>3$. Additionally, the computation of the UNL only requires, besides estimating the biomarkers' density functions in each group, computing a one-dimensional integral, regardless of the number of disease groups. The multi-class Youden index, on the other hand, involves solving $(H-1)$ optimization problems, while computing the VUS  requires  $(H-1)$-dimensional integration, which in turn demands sufficiently dense sampling in high-dimensional space for accurate approximation. The efficient computation of the hypervolume under the ROC manifold has indeed been the subject of recent research \citep{feng2023}.

\subsection{Covariate-specific underlap coefficient}
\label{sec:covariate_specific_UNL}
It is well established that subject-specific covariates can influence the distributions of biomarker outcomes across different groups, potentially leading to varying levels of discriminatory ability across subpopulations defined by covariate levels. Ignoring such heterogeneity and reporting a single UNL value may lead to misleading conclusions about a biomarker's performance. As highlighted in Section 1, the presence of covariates makes the stochastic ordering assumption even more stringent.

Let $\mathbf{X}_1$, $\mathbf{X}_2$, and $\mathbf{X}_3$ be vectors of covariates for each of the three groups. For simplicity, we assume that the same set of covariates is measured in each group. The covariate-specific underlap coefficient for a given covariate value, $\mathbf{x}$, is defined as follows
\begin{equation*}\label{UNL_inte_expression_covariate}
	\text{UNL}(f_1,f_2,f_3\mid \mathbf{x}) =\int_{-\infty }^{+\infty} \max\left\{f_1(y\mid\mathbf{x}),f_2(y\mid\mathbf{x}),f_3(y\mid\mathbf{x})\right\}\text{d}y,
\end{equation*}
where $f_d(y\mid \mathbf{x})$ denotes the conditional density of $Y_d$ given $\mathbf{X}_d=\mathbf{x}$, for $d\in\{1,2,3\}$. For each possible $\mathbf{x}$, we may obtain a different UNL value and, consequently, a potentially different level of discriminatory ability. By focusing on $\text{UNL}(f_1, f_2, f_3 \mid \mathbf{x})$, one can detect important interactions between covariates and disease stages, potentially revealing subgroups, defined by covariate values, in which a biomarker excels or fails to discriminate.

\section{\large{\textsf{INFERENTIAL FRAMEWORK}}}
\label{inference}
\subsection{Unconditional UNL estimator}
The definition of the UNL in Equation \eqref{3_class_UNL_definition} motivates a two-step modelling approach. The first step involves estimating the density functions of biomarker outcomes in each group, while the second step employs numerical integration.

\subsubsection{Step 1: Modelling $f_1$, $f_2$, and $f_3$}
Accurate estimation of the underlap coefficient requires  precise estimation of
the densities of biomarker outcomes in each group. In what follows, let $\{y_{1i}\}_{i=1}^{n_1}$, $\{y_{2j}\}_{j=1}^{n_2}$, and $\{y_{3k}\}_{k=1}^{n_3}$ represent three independent random samples of biomarker outcomes of sizes $n_1$, $n_2$, and $n_3$ from groups 1, 2, and 3, respectively, with
\begin{equation*}
	y_{11},\ldots ,y_{1n_1}|f_1 \overset{\text{iid}}\sim f_1, \quad y_{21},\ldots ,y_{2n_2}|f_2  \overset{\text{iid}}\sim f_2, \quad y_{31},\ldots ,y_{3n_3}|f_3  \overset{\text{iid}}\sim f_3.
\end{equation*}
The simplest approach would be to assume a normal distribution for $f_1$, $f_2$, and $f_3$, possibly after applying a transformation to the biomarker values (e.g., a logarithmic transformation). However, it may not be trivial to find a single transformation that makes the normality assumption plausible for all three groups. Furthermore, even if such a transformation is found for the biomarker data at hand, the transformed values may still exhibit nonstandard features, such as asymmetry or multiple modes. Mixtures of normal distributions can be used to represent a wide variety of density shapes. In particular, Dirichlet process mixtures (DPM) \citep[see, e.g.,][]{escobar1995} of normal distributions have been shown to accurately approximate any smooth density on the real line \citep{lo1984}. These mixture models have also been successfully applied to modelling biomarker outcome data in diagnostic studies (see, among others, \citealp{Erkanli2006,Hanson2008,Hwang2015, Chen2019,Inacio2022}).
Under a DPM of normal distributions model, the density function of biomarker outcomes in group 1 (the density functions of biomarker outcomes in groups 2 and 3 are analogously defined) can be written as
\begin{equation}
	f(y_{1i}\mid G_1) \equiv f_1(y_{1i}) = \int \phi(y_{1i}\mid\mu,\sigma^2) \text{d}G_1(\mu,\sigma^2),\quad G_1\sim \text{DP}(\alpha_1 ,G_{1}^*(\mu,\sigma^2)), 
	\label{dpm_definition1}
\end{equation}
where $\phi(y\mid \mu,\sigma^2)$ denotes the density function of a normal distribution evaluated at $y$ with mean $\mu$ and variance $\sigma^2$. The mixing distribution $G_1$ follows a Dirichlet process \citep{Ferguson1973} with centring distribution $\mathbb{E}[G_1(\mu,\sigma^2)] = G_1^{*}(\mu,\sigma^2)$ and precision parameter $\alpha_1 >0$. According to Sethuraman's stick-breaking representation \citep{Sethuraman1994}, $G_1$ can be expressed as an infinite weighted sum of point masses
\begin{align}
	G_1(\cdot )=\sum_{l=1}^{\infty}\omega_{1l}\delta_{(\mu_{1l},\sigma^2_{1l})}(\cdot),
	\label{dp_prior_definition}
\end{align}
where $\omega_{11}=v_{11}$, $\omega_{1l}=v_{1l}\prod_{m=1}^{l-1} (1-v_{1m})$, for $l\geq 2$, with $v_{1l} \overset{\text{iid}}\sim \text{Beta}(1,\alpha_1)$, for $l\geq 1$ and, independently, $(\mu_{1l},\sigma^2_{1l})\overset{\text{iid}} \sim G_1^*(\mu,\sigma^2)$. The density function in \eqref{dpm_definition1} can therefore be written as a countable mixture of normal densities
\begin{equation}\label{dpm_dens}
	f_{1}(y_{1i})= \sum_{l=1}^{\infty}\omega_{1l}\phi(y_{1i}\mid \mu_{1l},\sigma_{1l}^2).
\end{equation}
To facilitate posterior inference, a conditionally-conjugate centring distribution is specified as $G_1^{*} \equiv \text{N}(\mu\mid a_{\mu_{1}}, b^2_{\mu_1})\text{IG}(\sigma\mid a_{\sigma_{1}^{2}}, b_{\sigma_{1}^{2}})$, where $IG$ denotes the inverse gamma distribution. Under \eqref{dp_prior_definition}, the probabilities assigned to each component decrease rapidly with the index $l$ for typical choices of $\alpha_1$. Consequently, the model in \eqref{dpm_dens} can be reasonably approximated by a finite number of components. For this reason, posterior inference is  conducted using the blocked Gibbs sampler \citep{Ishwaran2001}, which truncates Sethuraman's representation in \eqref{dp_prior_definition} to a finite value, $L_1$, with $v_{1L_1} = 1$ to ensure that the mixture weights sum to one. The details of the Gibbs sampling scheme, although straightforward, are provided in Section B of the Supplementary Material for completeness.

\subsubsection{Step 2: Numerical integration}
Once Step 1 is completed and posterior realisations of the biomarker outcome density functions for each of the three groups are available, a posterior realisation of the underlap coefficient can be computed as follows
\begin{equation*}\label{unl_post}
	\text{UNL}^{(s)}(f_1,f_2,f_3)=\int_{-\infty }^{+\infty} \max\left\{f_1^{(s)}(y),f_2^{(s)}(y),f_3^{(s)}(y)\right\}\text{d}y,\quad f_{1}^{(s)}(y) = \sum_{l=1}^{L_1}\omega_{1l}^{(s)}\phi(y\mid \mu_{1l}^{(s)},(\sigma_{1l}^{2})^{(s)}),
\end{equation*}
for $s=1,\ldots,S$, where $S$ represents the number of Gibbs sampler iterations after burn-in, and where $f_{2}^{(s)}(y)$ and $f_{3}^{(s)}(y)$ are defined analogously to $f_{1}^{(s)}(y)$. The integral in \eqref{3_class_UNL_definition} is computed numerically using Simpson's rule, which is especially accurate for smooth integrand functions and employs an equally spaced grid of points that  should cover the range of biomarker outcomes across the three groups. A point estimate of the underlap coefficient can be obtained by taking the mean or median of the ensemble of posterior realisations $\{\text{UNL}^{(1)}(f_1,f_2,f_3), \ldots,\text{UNL}^{(S)}(f_1,f_2,f_3)\}$. A $100 (1-\alpha)\%$ credible interval can also be derived from the $100(\alpha/2)\%$ and $100(1-\alpha/2)\%$ percentiles of the same ensemble.

\subsection{Covariate-specific UNL estimator}
\label{section_conditional_UNL}
We now define $\{(\mathbf{x}_{1i},y_{1i})\}_{i=1}^{n_1}$, $\{(\mathbf{x}_{2j},y_{2j})\}_{j=1}^{n_2}$ and $\{(\mathbf{x}_{3k},y_{3k})\}_{k=1}^{n_3}$ as three independent random samples of covariates and biomarker outcomes, of sizes $n_1$, $n_2$ and $n_3$, drawn from groups 1, 2, and 3, respectively. Furthermore, we let $\mathbf{x}_{1i}=(x_{1i,1},\ldots,x_{1i,p})^{\prime}$, $\mathbf{x}_{2j}=(x_{2j,1},\ldots,x_{2j,p})^{\prime}$ and $\mathbf{x}_{3k}=(x_{3k,1},\ldots,x_{3k,p})^{\prime}$ denote $p$-dimensional vectors of covariates for all $i=1,\ldots,n_1$, $j=1,\ldots,n_2$, and $k=1,\ldots,n_3$. As in the unconditional case, the key to accurately estimating the covariate-specific underlap coefficient is to accurately estimate the conditional density in each group.

As a natural extension of model \eqref{dpm_definition1}, we consider a covariate-dependent Gaussian mixture model of the form
\begin{align*}
	f_1(y_{1i}\mid \mathbf{x}_{1i}) &= \int \phi(y_{1i}\mid \mu(\mathbf{x}_{1i}, \boldsymbol{\beta}),\sigma^2) \text{d}G_1,\mathbf{x}_{1i}(\boldsymbol{\beta},\sigma^2),\\
	G_1,\mathbf{x}_{1i}(\cdot) & = \sum_{l=1}^{\infty}\omega_{1l}(\mathbf{x}_{1i})\delta_{(\boldsymbol{\beta}_{1l},\sigma^2_{1l})}(\cdot), \quad (\boldsymbol{\beta}_{1l},\sigma^2_{1l})\overset{\text{iid}}\sim G_1^{*}(\boldsymbol{\beta},\sigma^2), 
\end{align*}
where $\omega_{11}(\mathbf{x}_{1i}) = v_{11}(\mathbf{x}_{1i})$ and $\omega_{1l}(\mathbf{x}_{1i}) = v_{1l}(\mathbf{x}_{1i})\prod_{m=1}^{l-1}\{1 - v_{1m}(\mathbf{x}_{1i})\}$, for $l \geq 2$, are covariate-dependent weights following a stick-breaking representation. We can therefore write the conditional density as
\begin{equation}\label{LSBP}
	f_1(y_{1i}\mid \mathbf{x}_{1i}) = \sum_{l=1}^{\infty}\omega_{1l}(\mathbf{x}_{1i})\phi(y_{1i}\mid \mu_1(\mathbf{x}_{1i},\boldsymbol{\beta}_{1l}),\sigma_{1l}^2).
\end{equation}
Our choice of a model in the form of \eqref{LSBP} is motivated by the fact that infinite mixture models with covariate-dependent stick-breaking weights possess desirable theoretical properties \citep{Barrientos2012,Pati2013} and offer a highly flexible tool for estimating conditional densities. To introduce covariate dependence in $v_{1l}(\mathbf{x}_{1i})$, we use a logit stick-breaking process \citep{Ren2011, rigon2021tractable}
\begin{equation}\label{lsbp_v}
	\text{logit}(v_{1l}(\mathbf{x}_{1i})) =  \gamma_{1l0} + g_{1l1}(x_{1i,1}) + \ldots + g_{1lp},(x_{1i,p}),\quad l\geq 1,
\end{equation}
where $g_{1lh}(x_{1i,h})$, $h=1,\ldots,p$, may encode the effect of a categorical covariate, a linear effect of a continuous covariate ($g_{1lh}(x_{1i,h}) = \gamma_{1lh}x_{1i,h}$), or a smooth nonlinear effect of a continuous covariate. In the latter case, the smooth function $g_{1lh}(\cdot)$ is approximated by a linear combination of cubic B-spline basis functions defined over a sequence of knots $\xi_{1h0}<\xi_{1h1} <\ldots< \xi_{1h K_{1h}} < \xi_{1h,K_{1h}+1}$, where $\xi_{1h0}$ and $\xi_{1h,K_{1h}+1}$  are boundary knots, while the remaining ones are interior knots. We write
\begin{equation*}
	g_{1lh}(x_{1i,h}) = \sum_{k=1}^{K_{1h}+3} B_{1hk}(x_{1i,h})\gamma_{1lhk} = \mathbf{B}_{1\boldsymbol{\xi}_{1h}}^{\prime}(x_{1i,h}) \boldsymbol{\gamma}_{1lh},
\end{equation*}
where $\mathbf{B}_{1\boldsymbol{\xi}_{1h}}(x_{1i,h}) = (B_{1h1}(x_{1i,h}),\ldots, B_{1h, K_{1h}+3}(x_{1i,h}))^{\prime}$ and $B_{1hk}(x)$ is the $k$th cubic B-spline basis function in group 1, evaluated at $x$, defined by the vector of knots $\boldsymbol{\xi}_{1h} = (\xi_{1h0},\ldots, \xi_{1h,K_{1h}+1})^{\prime}$. Lastly, $\boldsymbol{\gamma}_{1l1} = (\gamma_{1lh1},\ldots,\gamma_{1lh, K_{1h}+3})^{\prime}$ represents the corresponding coefficient vector. Generically, we can represent the logistic regression in \eqref{lsbp_v} as
\begin{equation*}
	\text{logit}(v_{1l}(\mathbf{x}_{1i})) = \mathbf{z}_{1i}^{\prime}\boldsymbol{\gamma}_{1l},\quad l\geq 1,
\end{equation*}
where $\mathbf{z}_{1i}$ is the design vector, with dimension $Q_1^{v}$, that collects all covariate values for individual $i$ in group 1 and may include dummy variables, continuous covariate values, and/or their cubic B-spline basis expansions. Similarly, $\boldsymbol{\gamma}_{1l}$ is the vector that collects all the corresponding parameters.

The mean of each Gaussian component can be written in a similar fashion, i.e., 
\begin{equation}\label{cmean_formulation}
	\mu_{1l}(\mathbf{x}_{1i}) = \beta_{1l0} + f_{1l1}(x_{1i,1}) + \ldots + f_{1lp},(x_{1i,p}),\quad l\geq 1,
\end{equation}
where again, each function $f_{1lh}(\cdot)$, $h=1,\ldots,p$, can encode the effects of categorical covariates, continuous covariates, and the B-spline expansions of continuous covariates and, generally, we may also write it as
\begin{equation*}
	\mu_{1l}(\mathbf{x}_{1i}) = \mathbf{u}_{1i}^{\prime}\boldsymbol{\beta}_{1l},\quad l\geq 1,
\end{equation*}
where $\mathbf{u}_{1i}$ and $\boldsymbol{\beta}_{1l}$ are vectors of dimension $Q_1^{\mu}$, defined similarly to $\mathbf{z}_{1i}$ and $\boldsymbol{\gamma}_{1l}$, respectively. To determine the functional form in which covariates enter the weights and  components' means, a model comparison criterion, such as the Watanabe-Akaike information criterion (WAIC) \citep{Watanabe2010, Gelman2014}, can be used. At this point, it is important to note that, according to our results reported in Section \ref{sim_study}, as well as those reported in \cite{Wade2025}, a linear specification for both the weights and  components' means is generally sufficient.

As in the unconditional case, we truncate the infinite mixture in \eqref{LSBP} at $L_1$ and therefore we let $v_{1L_1}(\mathbf{x}_{1i}) = 1$, for all $\mathbf{x}_{1i}$. Before proceeding, we note that although we have assumed an additive structure in \eqref{lsbp_v} and in \eqref{cmean_formulation}, the model can accommodate interactions between categorical covariates, continuous covariates, and between categorical and continuous covariates.  Based on Theorem 1 in \cite{rigon2021tractable}, the value $L_1$ need not be very large to accurately approximate the infinite mixture model. Furthermore, in the comparative study of \cite{Wade2025}, the truncated logit stick-breaking prior showed very good performance across a range of challenging scenarios. We complete the model specification with the following prior distributions
\begin{equation*}
	\boldsymbol{\gamma}_{1l} \sim \text{N}_{Q^{v}_1}(\mathbf{\mu}_{\boldsymbol{\gamma}_1},\boldsymbol{\Sigma}_{\boldsymbol{\gamma}_1}),\quad \boldsymbol{\beta}_{1l} \sim \text{N}_{Q^{\mu}_1}(\mathbf{\mu}_{\boldsymbol{\beta}_1},\boldsymbol{\Sigma}_{\boldsymbol{\beta}_1}),\quad \sigma^{2}_{1l}\sim\text{IG}(a_{\sigma^2_1}, b_{\sigma^2_1}),\quad l=1,\ldots, L_1.
\end{equation*}
It is worth noting that \cite{rigon2021tractable}, by exploring a reparametrisation of the covariate-dependent stick-breaking weights that relies on a set of sequential logistic regressions and by  leveraging a P\'olya-gamma data augmentation \citep{polson2013bayesian}, were able to derive the full conditional distribution of the weights parameters, $\{\boldsymbol{\gamma}_{1l}\}_{l=1}^{L_1}$, in closed form. Posterior inference is thus performed using a Gibbs sampler algorithm, as detailed in Section B of the Supplementary Material. 

Similarly to the unconditional case, for a given $\mathbf{x}$, a point estimate (or a credible interval) of the covariate-specific underlap coefficient can be obtained by computing the mean or median (or by considering the percentiles) over the ensemble of covariate-specific underlap coefficients $\{\text{UNL}^{(1)}(f_1,f_2,f_3\mid \mathbf{x}),\ldots, \text{UNL}^{(S)}(f_1,f_2,f_3\mid\mathbf{x})\}$, where
\begin{align*}
	\text{UNL}^{(s)}(f_1,f_2,f_3\mid \mathbf{x}) &=\int_{-\infty }^{+\infty} \max\left\{f_1^{(s)}(y\mid \mathbf{x}),f_2^{(s)}(y\mid \mathbf{x}),f_3^{(s)}(y \mid \mathbf{x})\right\}\text{d}y, \\
	f_1^{(s)}(y\mid \mathbf{x}) &= \sum_{l=1}^{L_1}\omega_{1l}^{(s)}(\mathbf{x})\phi(y\mid \mu_{1l}^{(s)}(\mathbf{x}),(\sigma_{1l}^{2})^{(s)}),
\end{align*}
for $s=1,\ldots, S$, where $f_2^{(s)}(y\mid \mathbf{x})$ and $f_3^{(s)}(y\mid \mathbf{x})$ are defined analogously to $f_1^{(s)}(y\mid \mathbf{x})$, and the integral is approximated numerically using Simpson's rule.

\section{\large{\textsf{SIMULATION STUDY}}}
\label{sim_study}
\subsection{Unconditional case}
\subsubsection{Simulation scenarios}
We considered the scenarios listed in Table \ref{tab:simu-scenario_uncon}. Figure 3 in the Supplementary Material shows the density functions of biomarker outcomes for each group in each scenario.
For each simulation scenario, we employed three distinct parameter configurations aimed at producing large, intermediate, and small values of the underlap coefficient. Scenario I corresponds to the case where the biomarker outcomes for the three groups follow normal distributions.
Scenario II corresponds to the case in which biomarker outcomes for all three groups follow skewed distributions. 
Finally, Scenario III involves mixtures of normal distributions for all three groups.
For each scenario and parameter configuration, we generated  a total of 100 datasets  using sample sizes of $(n_1,n_2,n_3)\in \{(100,100,100),(200,200,200),(500,500,500),(1000,1000,1000),(100,300,500)\}$. An unbalanced sample size was used, as this reflects what is commonly encountered in real-world settings.settings. We note that although a monotonic ordering of the biomarker outcome distributions is assumed in all scenarios, the main goal of this study is to assess the performance of the UNL estimator under different distributional assumptions for generating biomarker outcomes.

\renewcommand{\arraystretch}{1}
\begin{table}[!ht]
	\centering
	\begin{adjustbox}{width=0.95\textwidth,center}
		\begin{tabular}{c c c c c}
			\rowcolor{gray!100}
			Scenario &$Y_1$ & $Y_2$& $Y_3$&UNL\\
			
			I&$\text{N}(-3.25,1^2)$& $\text{N}(0,1^2)$&  $\text{N}(3.25,1^2)$& 2.792\\
			&$\text{N}(-1.3,1^2)$& $\text{N}(0,1^2)$&  $\text{N}(1.15,1^2)$& 1.919\\
			&$\text{N}(-0.2,1^2)$& $\text{N}(0,1^2)$&  $\text{N}(0.15,1^2)$& 1.139\\
			\rowcolor{gray!65}
			II&$\Gamma(3,1)$& $\text{SN}(6,2,5)$& $\text{SN}(8,2,5)$& 2.527\\
			\rowcolor{gray!85}
			&$\Gamma(3,1)$& $\text{SN}(2,2.5,5)$& $\text{SN}(4.25,2,5)$& 1.855\\
			\rowcolor{gray!65}
			&$\Gamma(1.5,1)$& $\text{SN}(0.1,2,5)$& $\text{SN}(0.25,2,5)$& 1.191\\
			III&$0.5\text{N}(-6,1^2)+0.5\text{N}(-3,1^2)$& $0.5\text{N}(0.5,1^2)+0.5\text{N}(3.25,1^2)$& $0.5\text{N}(3.5,1^2)+0.5\text{N}(6.25,1^2)$ &2.508\\
			&$0.5\text{N}(-2.25,1^2)+0.5\text{N}(0.5,1^2)$& $0.5\text{N}(2.75,1^2)+0.5\text{N}(5.5,1^2)$& $0.5\text{N}(3,1^2)+0.5\text{N}(5.75,1^2)$ &1.933\\
			&$0.5\text{N}(0.15,1^2)+0.5\text{N}(2.75,1^2)$& $0.5\text{N}(0.5,1^2)+0.5\text{N}(3,1^2)$& $0.5\text{N}(0.85,1^2)+0.5\text{N}(3.15,1^2)$ &1.143\\
			\hline
		\end{tabular}
	\end{adjustbox}
	\caption{Distributional assumptions for $Y_1$, $Y_2$ and $Y_3$ under Scenarios I, II, and III.}
	\label{tab:simu-scenario_uncon}
\end{table}

\subsubsection{Model specification}
To facilitate hyperparameter specification, biomarker outcomes were standardized so that the resulting mean is zero and the variance is one. We then transformed the data back to the original scale when calculating the relevant quantities. The hyperparameters were chosen as follows: \(a_{\mu_d} = 0\), \(b_{\mu_d}^2 = 10\), \(a_{\sigma^2_d} = 2\), and \(b_{\sigma^2_d} = 0.5\), for \(d \in \{1, 2, 3\}\). The rationale behind these hyperparameter values is as follows. Due to standardization of biomarker outcomes, we expect the mean of the components to be near zero, justifying the choice $a_{\mu_d} = 0$. The variance $b^2_{\mu_d}$ of the prior distribution for $\mu_{dl}$ determines where the samples values of $\mu_{dl}$ are likely to lie.  A value of 10 suggests a high probability that $\mu_{dl}$ lies within the interval $[-6, 6]$. Since the standardized biomarker outcomes have a variance of one, it is reasonable to expect the within-component variance to be smaller than the overall variance. Setting $a_{\sigma^2_d} = 2$ and $b_{\sigma^2_d} = 0.5$ leads to a prior distribution for $\sigma^2_{dl}$ with a finite mean of  $0.5$  and an infinite variance. We further set $\alpha_d = 1$, a commonly used default value that favours a small number of occupied mixture components relative to the sample size. Lastly, we set $L_d = 20$, allowing for a maximum of 20 mixture components. \cite{Ishwaran2000} showed the following distributional result for the tail of the stick-breaking weights: $\mathbb{E}(\sum_{l=L_{d}+1}^{\infty} \omega_{dl}) = \{\alpha_d/(\alpha_d+1)\}^{L_d}$. Thus,  setting $L_d=20$ and $\alpha_d=1$, results in an expected value approximately equal to zero.  Posterior inference is based on 5000 iterations, following a burn-in period in which the first 2000 iterations of the Gibbs sampler are discarded. For the numerical integration in Step 2, a grid of  $501$ points was used.

\subsubsection{Results}
\begin{figure}[htbp]
	\centering
	\subfigure{
		\centering
		\includegraphics[width=0.3\textwidth]{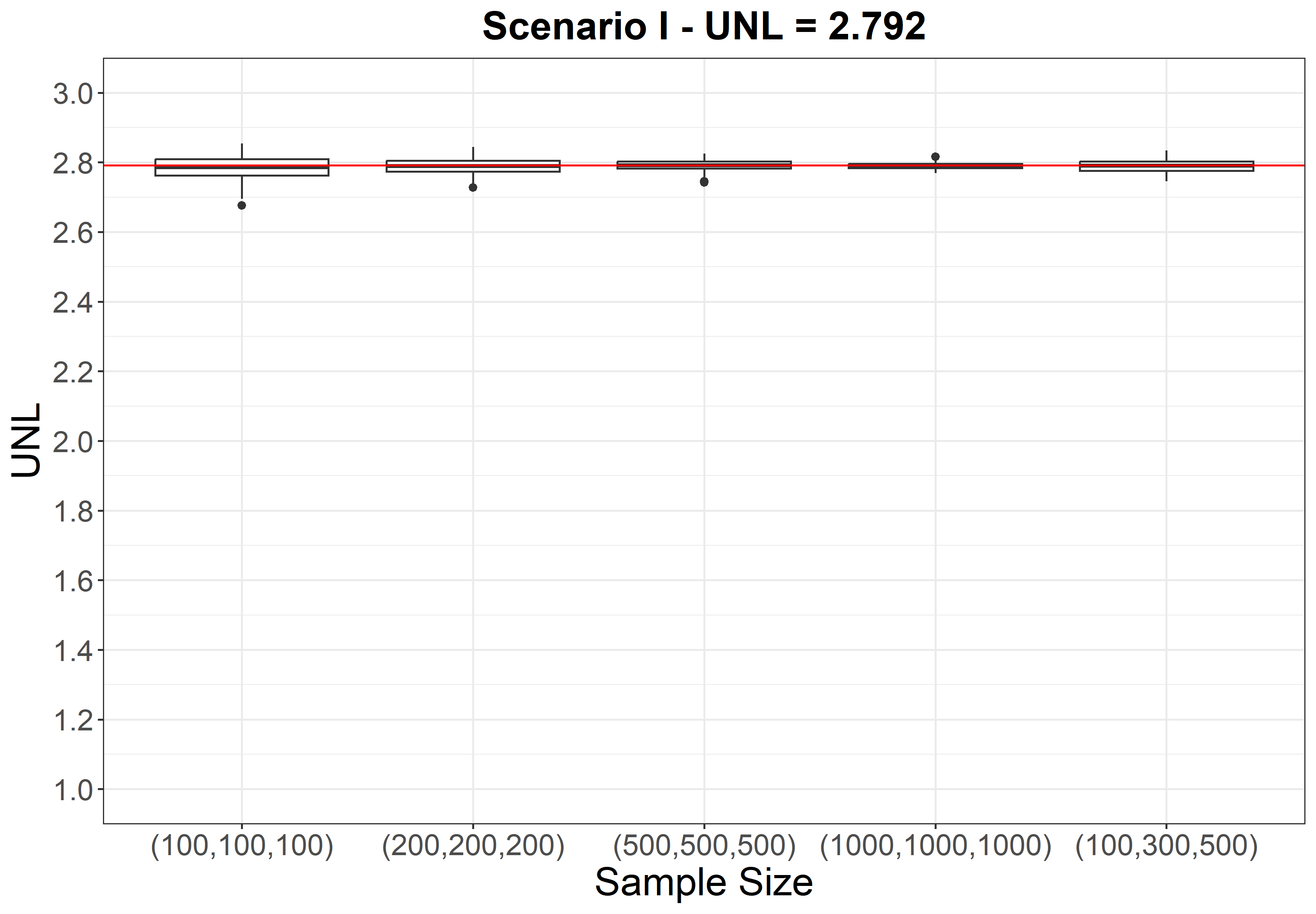}
	}
	\subfigure{
		\centering
		\includegraphics[width=0.3\textwidth]{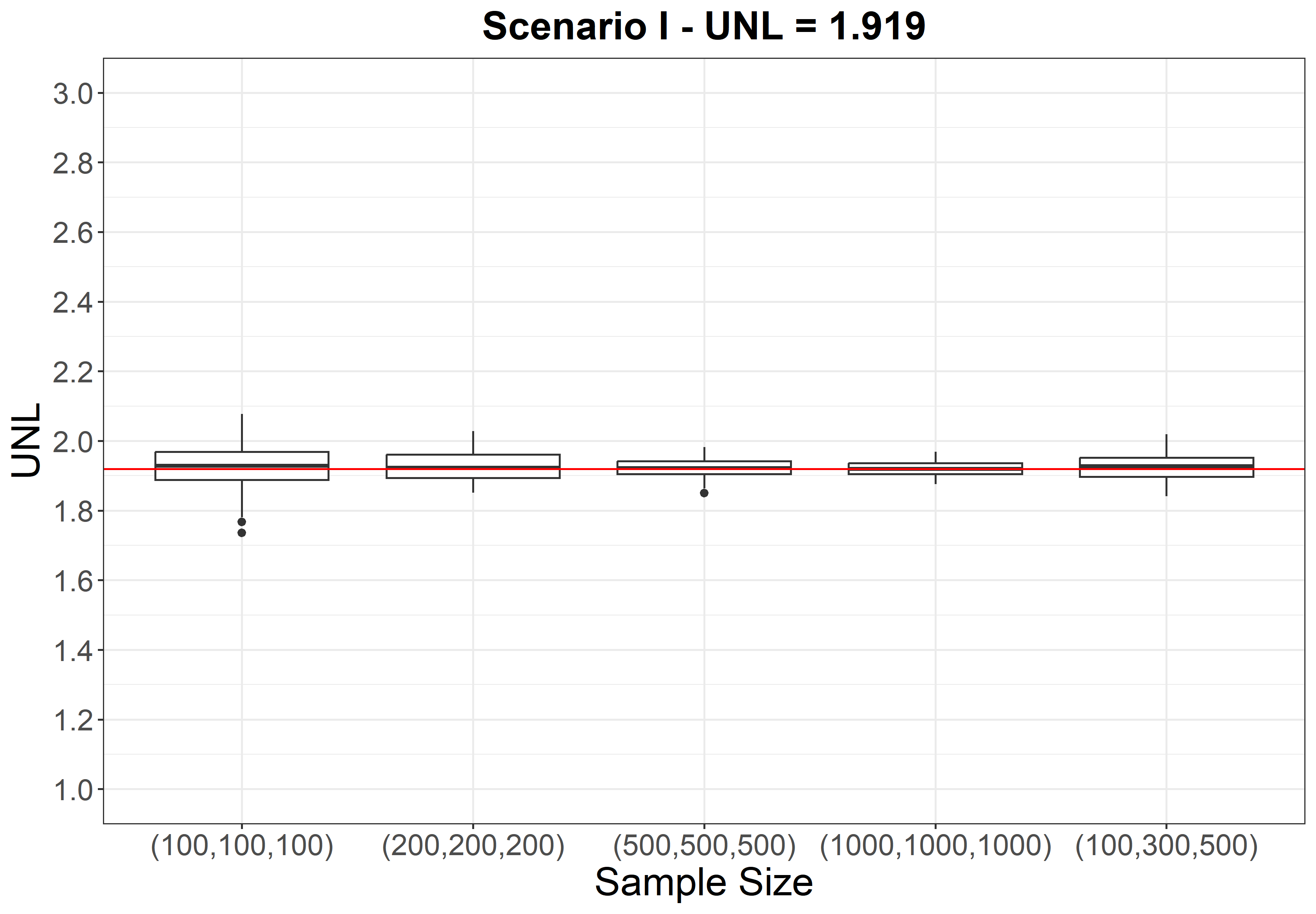}
	}
	\subfigure{
		\centering
		\includegraphics[width=0.3\textwidth]{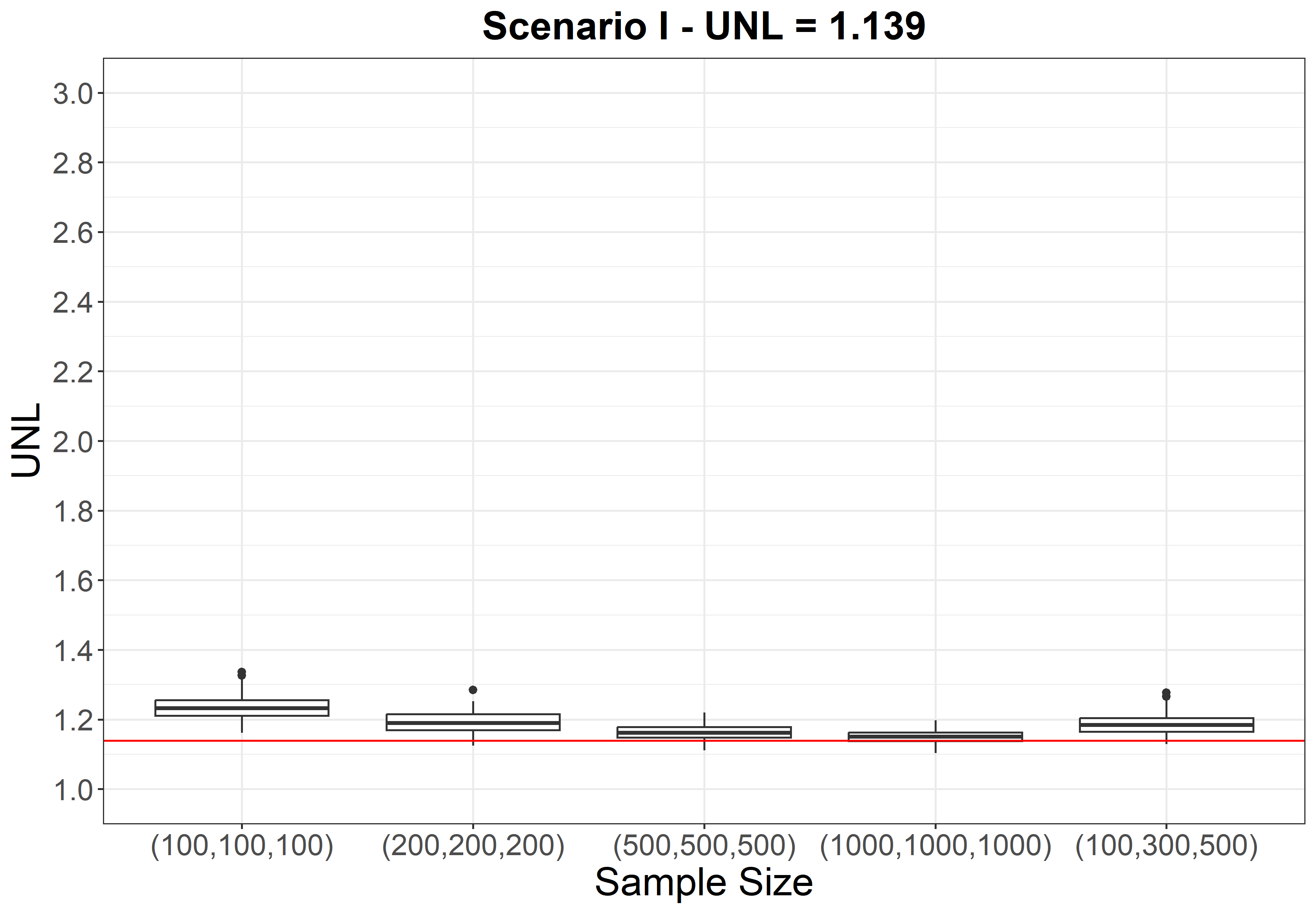}
	}
	\\
	\subfigure{
		\centering
		\includegraphics[width=0.3\textwidth]{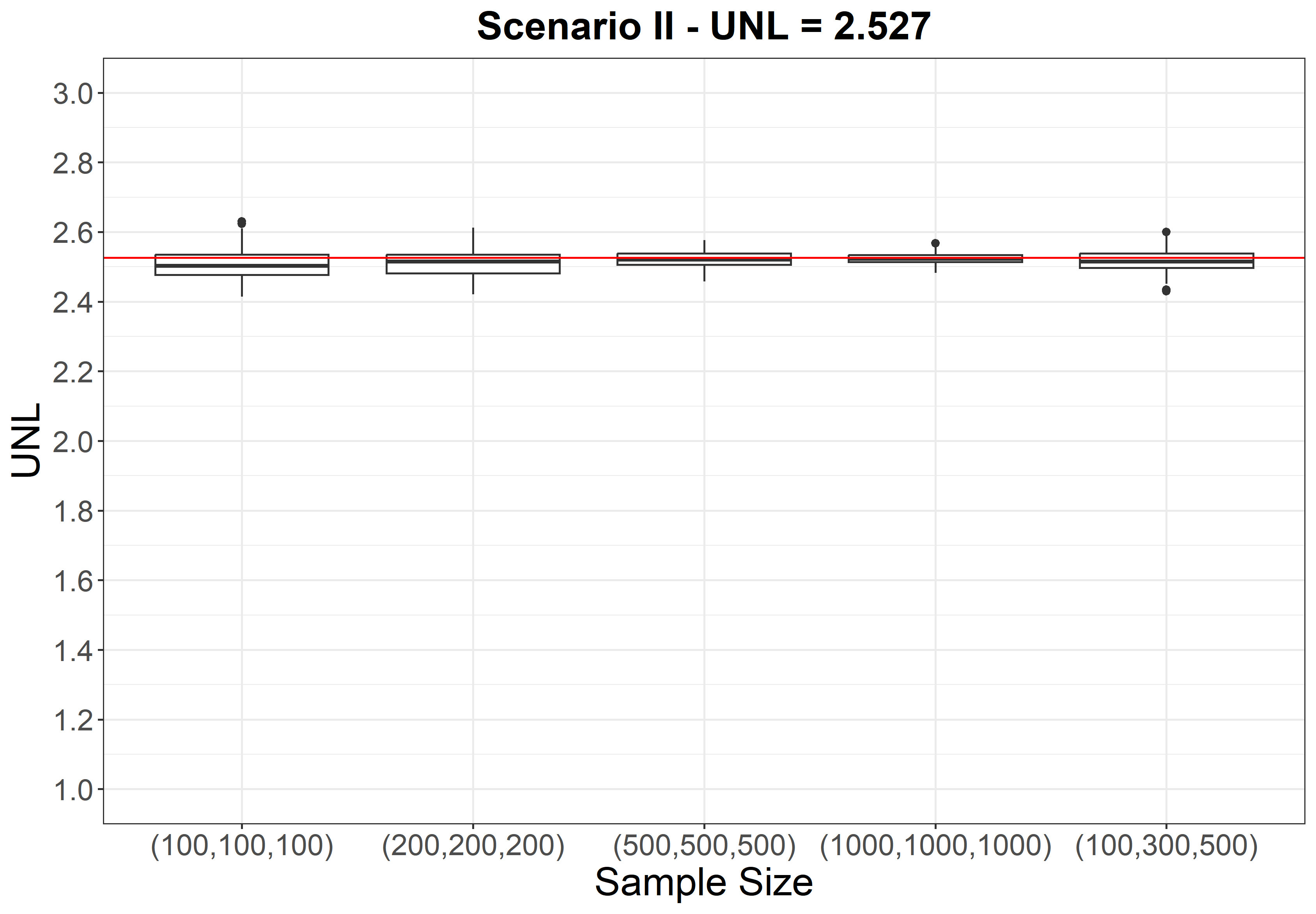}
	}
	\subfigure{
		\centering
		\includegraphics[width=0.3\textwidth]{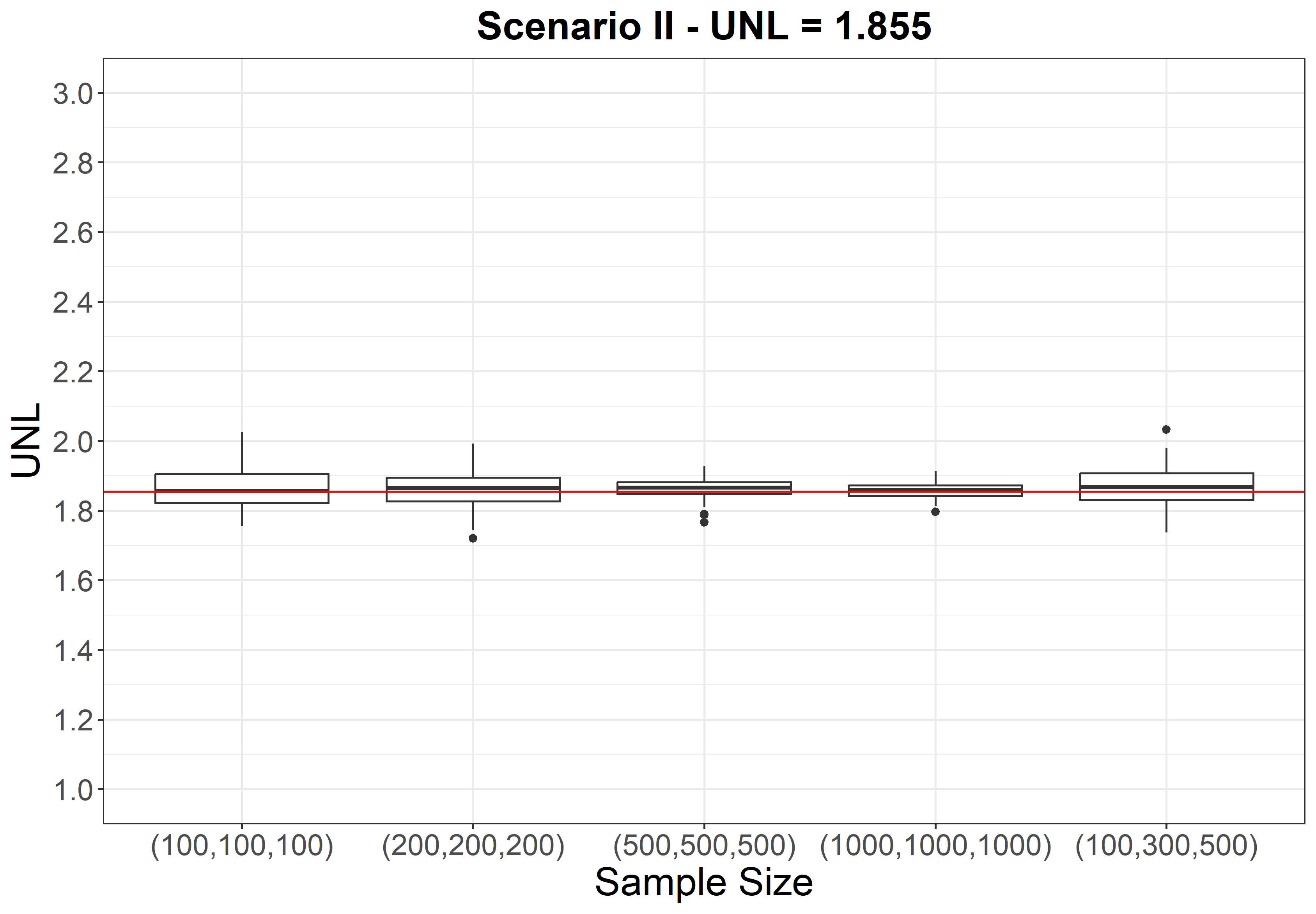}
	}
	\subfigure{
		\centering
		\includegraphics[width=0.3\textwidth]{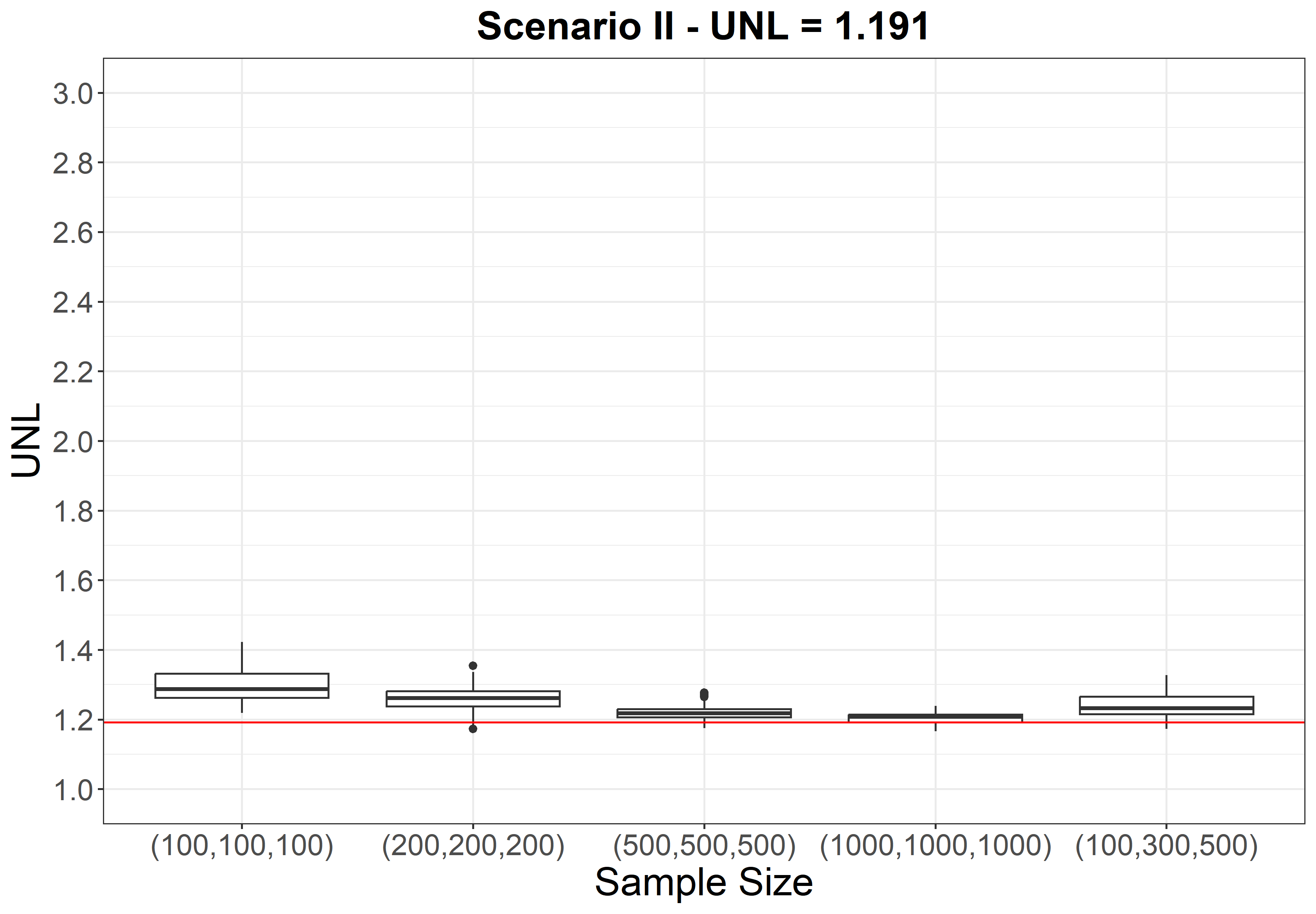}
	}
	\\
	\subfigure{
		\centering
		\includegraphics[width=0.3\textwidth]{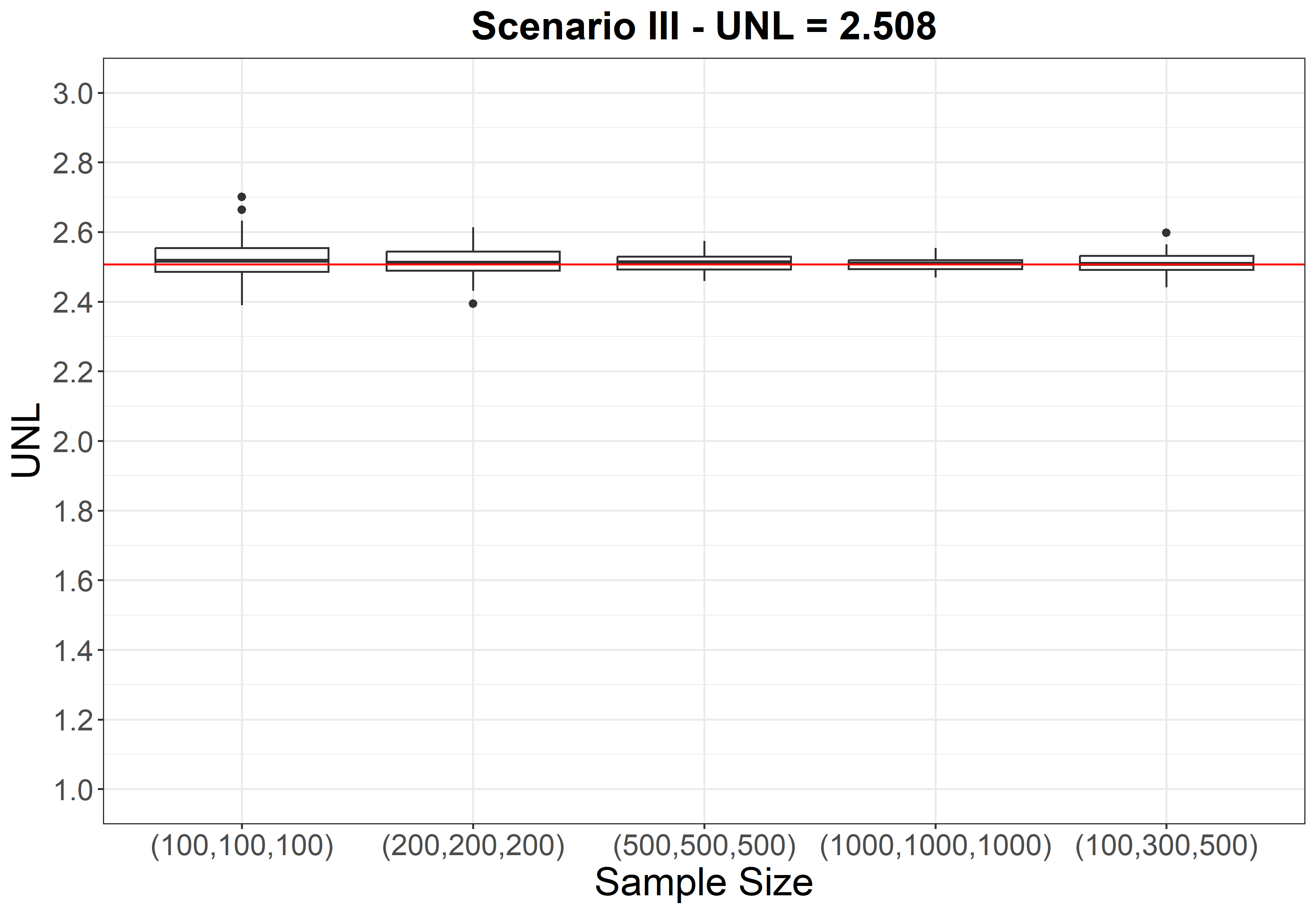}
	}
	\subfigure{
		\centering
		\includegraphics[width=0.3\textwidth]{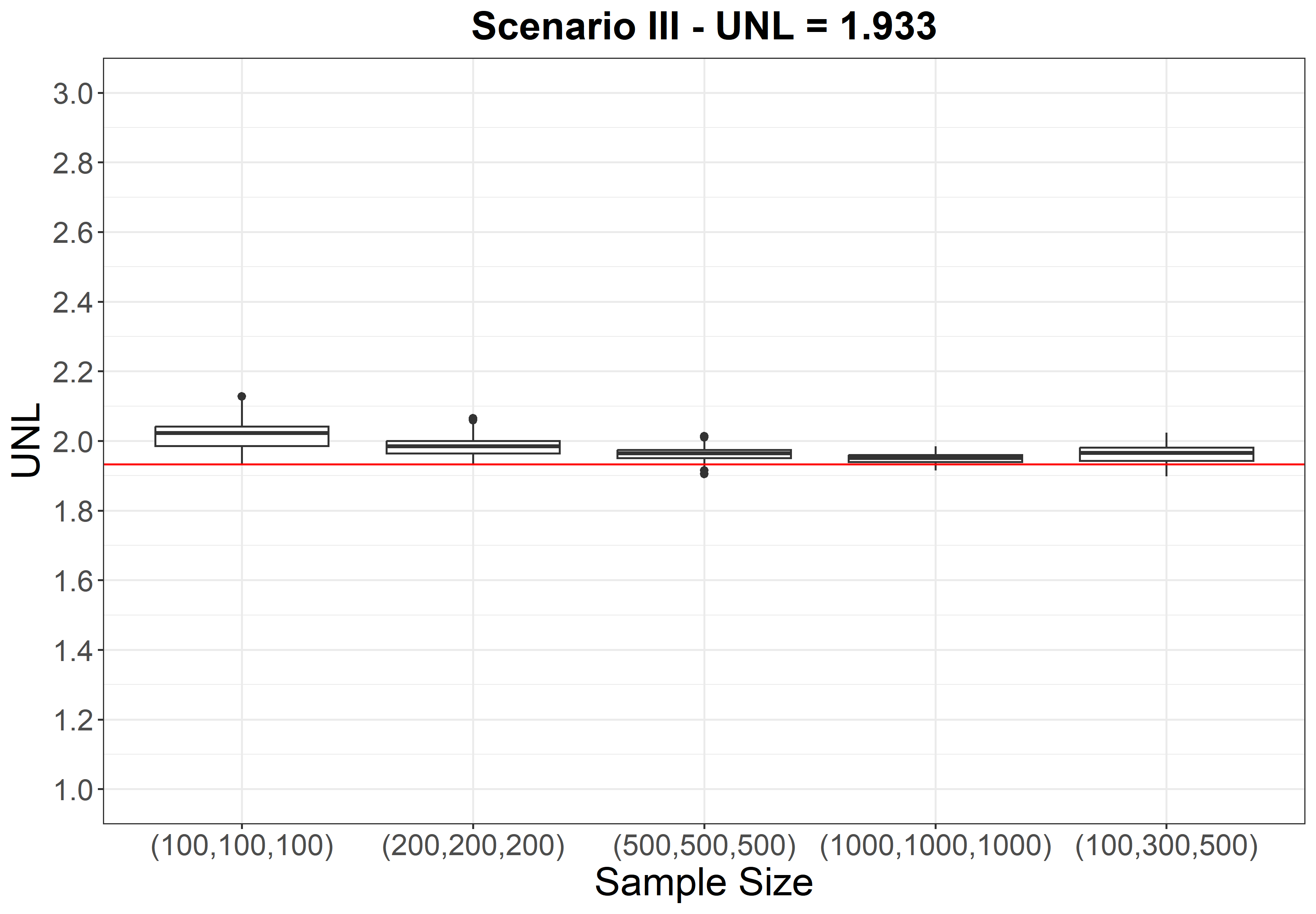}
	}
	\subfigure{
		\centering
		\includegraphics[width=0.3\textwidth]{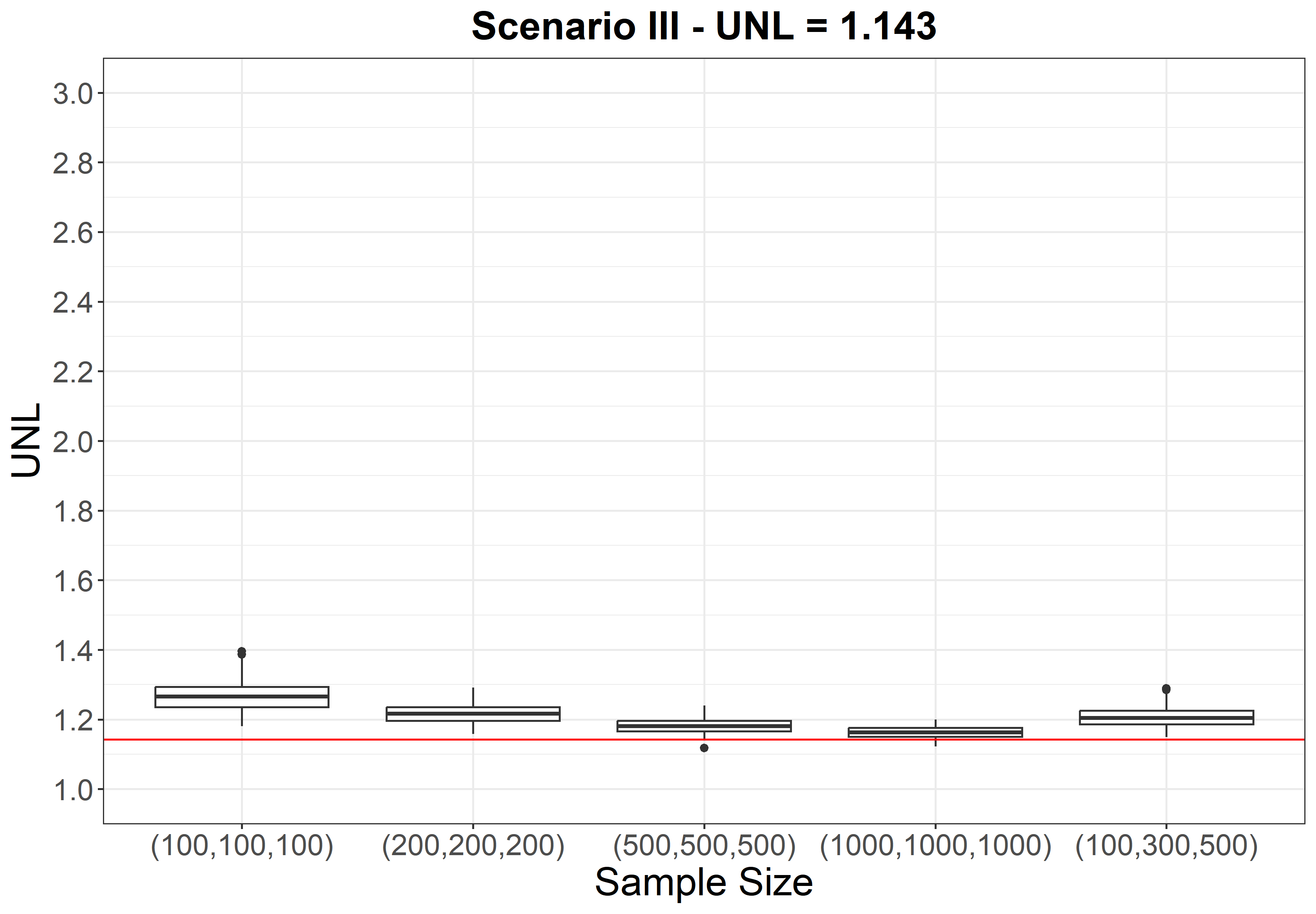}
	}
	\centering
	\caption{Boxplot of the estimates of the underlap coefficient (posterior median) across the 100 simulated datasets for different parameter configurations and sample sizes. The solid red line represents the true underlap coefficient.}
	\label{median_box_uncons}
\end{figure}

Figure \ref{median_box_uncons} presents the boxplots of the estimated underlap coefficients across the 100 datasets for all scenarios, distribution parameter configurations, and sample sizes considered. Our proposed DPM of normals based estimator generally provides unbiased estimates, except in  cases where the true underlap coefficent is very close to one or, in Scenario III, for the parameter configuration that produces a true underlap coefficient close to two, but where the densities of biomarker outcomes in groups 2 and 3 almost completely overlap. The bias in these cases, where there is substantial overlap between the densities of biomarker outcomes, is likely due to assuming an independent DPM prior in each group, with no borrowing of information. However, this bias diminishes as the sample size increases. Furthermore, the bias is never very large. For instance, the bottom right panel in Figure \ref{median_box_uncons} shows that even for the smaller sample size of $(n_1,n_2,n_3)=(100,100,100)$, the largest estimated underlap coefficient across the 100 datasets is about $1.4$, while the true value is $1.19$. Estimates of the underlap coefficient from the unbalanced sample size, $(n_1, n_2, n_3) = (100, 300, 500)$, are less accurate than those obtained from the balanced sample size $(n_1, n_2, n_3) = (500, 500, 500)$, as expected, but perform comparably to those obtained from the balanced sample size, $(n_1, n_2, n_3) = (200, 200, 200)$.

We also examined the empirical coverage probability of the $95\%$ credible intervals; the results are presented in Table 1 of the Supplementary Material. As observed, these probabilities are close to the nominal value, except when the underlap coefficient is close to one or, in Scenario II, when the underlap coefficient is close to two but the densities of biomarker outcomes in two of the groups have almost a completely overlap. This undercoverage is more pronounced for smaller sample sizes. Furthermore, Figure 4 in the Supplementary Material  shows the width of the $95\%$ credible intervals. As expected, the interval width decreases as sample size increases for balanced sample sizes. The interval width for the unbalanced sample size $(n_1, n_2, n_3) = (100, 300, 500)$ remains comparable to that of the balanced sample size $(n_1,n_2,n_3) = (200,200,200)$. This further reinforces our previous point: the undercoverage still observed in the largest sample size (in cases there is substantial overlap between biomarker outcomes densities) is due to the persistence of some bias, while the interval width tends to be smaller for such a large sample size.

\subsection{Covariate-specific case}
\subsubsection{Simulation scenarios and implementation details}
We consider three simulation scenarios, and for each scenario, we generate 100 datasets using the following sample sizes: $(n_1,n_2,n_3)\in \{(100,100,100),(200,200,200),(500,500,500),(1000,1000,1000),(100,300,500)\}$. In Scenario I, we consider a homoscedastic regression model with a linear covariate effect in the three groups
\begin{equation*}
	y_{1i}\mid x_{1i} \overset{\text{ind}}\sim \text{N}(0.25+x_{1i},1^2),\quad y_{2j}\mid x_{2j} \overset{\text{ind}} \sim \text{N}(1+1.5x_{2j},1.5^2),\quad y_{3k}\mid x_{3k} \overset{\text{ind}}\sim \text{N}(2.5+4x_{3k},1.75^2).    
\end{equation*}
Notably, while the effect of the covariate in the underlying regression models is linear, the resulting functional form of the covariate-specific underlap coefficient is nonlinear. In Scenario II, the covariate effect on the mean  of each regression model is nonlinear in  all three groups. Additionally, the variance in group 2 depends on the covariate in a nonlinear manner
\begin{align*}
	&y_{1i}\mid x_{1i}  \overset{\text{ind}}\sim \text{N}(-0.75+\sin(\pi x_{1i}+1.25),0.5^2),\quad y_{2j}\mid x_{2j}  \overset{\text{ind}}\sim \text{N}(0.75+\sin(\pi x_{2j}),(1.25+x_{2j}^2)^2), 
	\\
	& y_{3k}\mid x_{3k}  \overset{\text{ind}}\sim \text{N}(2.35+x_{3k}^2,1^2).
\end{align*}
Lastly, Scenario III corresponds to the most challenging setting. In addition to nonlinear covariate effects and heteroscedasticity, this scenario introduces skewness in the biomarker's distribution in one group, while in another group, a two-component mixture of normal distributions with covariate-dependent weights is used
\begin{align*}
	&y_{1i}\mid x_{1i}  \overset{\text{ind}}\sim  \text{N}(-0.75+\sin(\pi x_{1i}+1.25),1^2),\quad y_{2j}\mid x_{2j}  \overset{\text{ind}}\sim  \frac{e^{x_{2j}}}{1+e^{x_{2j}}}\text{N}(x_{2j},0.5^2)+\frac{1}{1+e^{x_{2j}}}N(x_{2j}^2,0.75^2), \\
	&y_{3k}\mid x_{3k}  \overset{\text{ind}}\sim  \text{Gamma}(3+x_{3k}^2,0.5+e^{x_{3k}}).
\end{align*}
In all three scenarios, $x_{1i},x_{2j},x_{3k} \sim \text{Unif}(-1,1)$, for $i=1,\dots,n_1$, $j=1,\dots,n_2$, and $k=1,\dots,n_3$.
For each simulated dataset, we fitted the Bayesian nonparametric model, based on a logit stick-breaking prior, introduced in Section \ref{section_conditional_UNL}. Posterior inference was obtained based on $5000$ iterations, following a burn-in period of $2000$ iterations. Analogous to the unconditional scenario, a grid of $501$ points was used for numerical integration using Simpson's rule. 

To determine how covariates affect the means and weights of the components, we expanded the strategy of \cite{rigon2021tractable} and \cite{Wade2025}. These authors modelled the covariate effect on each normal component’s mean linearly and used a basis expansion for the covariates effect on the weights.  Building on their approach, we considered either a linear (in the covariates) trend or a B-spline basis expansion (with the number of interior knots ranging from zero to four) for the weights or the component means. We guided our choice using the WAIC criterion and adopted a parsimonious strategy, selecting a more complex model over a simpler one (e.g., a linear trend over a B-splines basis expansion with one interior knot) only if the criterion value differs by more than 5 units. A similar strategy, though in a different context, was implemented by \cite{ariyo2020bayesian}. We emphasise that the B-splines basis expansion of the covariates was considered either on the weights or on the components' means, but not on both simultaneously. 

For Scenario I, a linear trend was selected for all three groups, both for the weights and the components’ means. In Scenario II, a linear trend was used in all groups for both the components'means and weights, except for group 1, where the covariate effect on the mean of each component was modelled using a B-spline basis with one interior knot, located at the median of the covariate sample. Following the notation in Section \ref{section_conditional_UNL}, and noting that in this case $p=1$, hence the index $h$ is dropped, we have  $\mu_{1l}(x_{1i}) = \mathbf{u}_{1i}^{\prime}\boldsymbol{\beta}_{1l}$,
where $\mathbf{u}_{1i}^{\prime} = (1, B_{11}(x_{1i}), B_{12}(x_{1i}), B_{13}(x_{1i}), B_{14}(x_{1i}))$, with $B_{1k}(x_{1i})$ being the $k$th B-spline basis function evaluated at $x_{1i}$ defined by the vector of knots $\boldsymbol{\xi}_1=(\xi_{10}, \xi_{11},\xi_{12})^{\prime}$, where $\xi_{10}$ and $\xi_{12}$ are boundary knots located at the minimum and maximum of the covariate values, respectively. Finally, in Scenario III, similar to the previous scenario, a linear trend was selected in all three groups for both the components' weights and means, except for group 1, where the covariate effect on the mean of each component was  modeled using a B-splines basis with no interior knots.
In this covariate setting, and to facilitate prior specification, both biomarker outcomes and covariates were standardised. We used
\begin{align*}
	\boldsymbol{\mu}_{\boldsymbol{\gamma}_d} = \mathbf{0}_{2},\quad \boldsymbol{\Sigma}_{\boldsymbol{\gamma}_d} =10\mathbf{I}_{2} ,\quad\boldsymbol{\mu}_{\boldsymbol{\beta}_d} = \mathbf{0}_{Q_d^{u}},\quad \boldsymbol{\Sigma}_{\boldsymbol{\beta}_d} = 10 \mathbf{I}_{Q_d^{u}},\quad a_{\sigma^2_{d}} = 2,\quad b_{\sigma^2_{d}} = 0.5, \quad L_d=20,\quad d\in\{1,2,3\},
\end{align*}
where, as the previous description makes clear, $Q_d^{u}$, the length of the design vector for the components' means in group $d$, differs for each scenario. However, for notational simplicity, we do not index it by the scenario.

\subsubsection{Results}
\begin{figure}[htpb]
	\centering
	\subfigure{
		\includegraphics[width=0.3\linewidth]{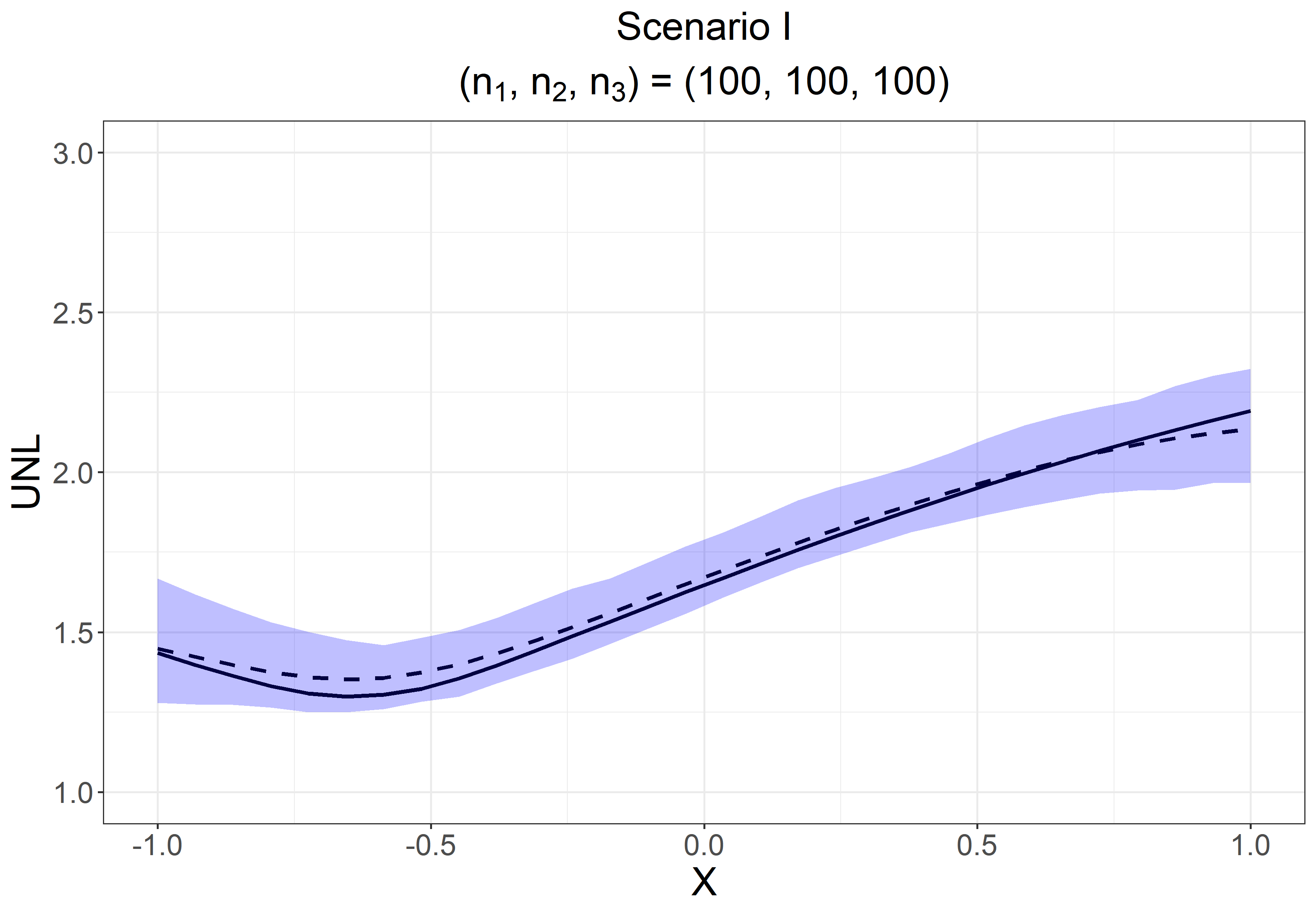}
	}
	\subfigure{
		\includegraphics[width=0.3\linewidth]{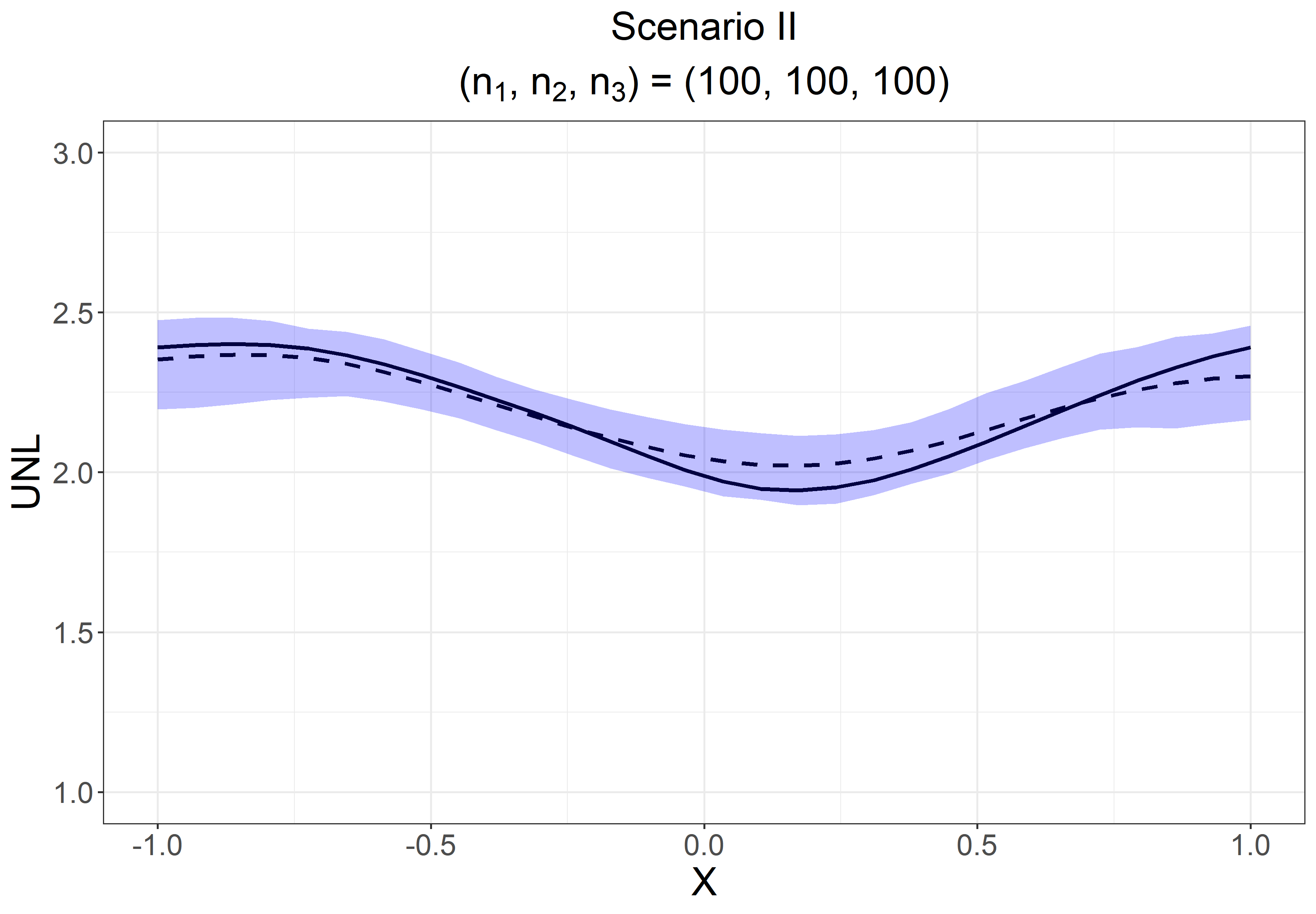}
	}
	\subfigure{
		\includegraphics[width=0.3\linewidth]{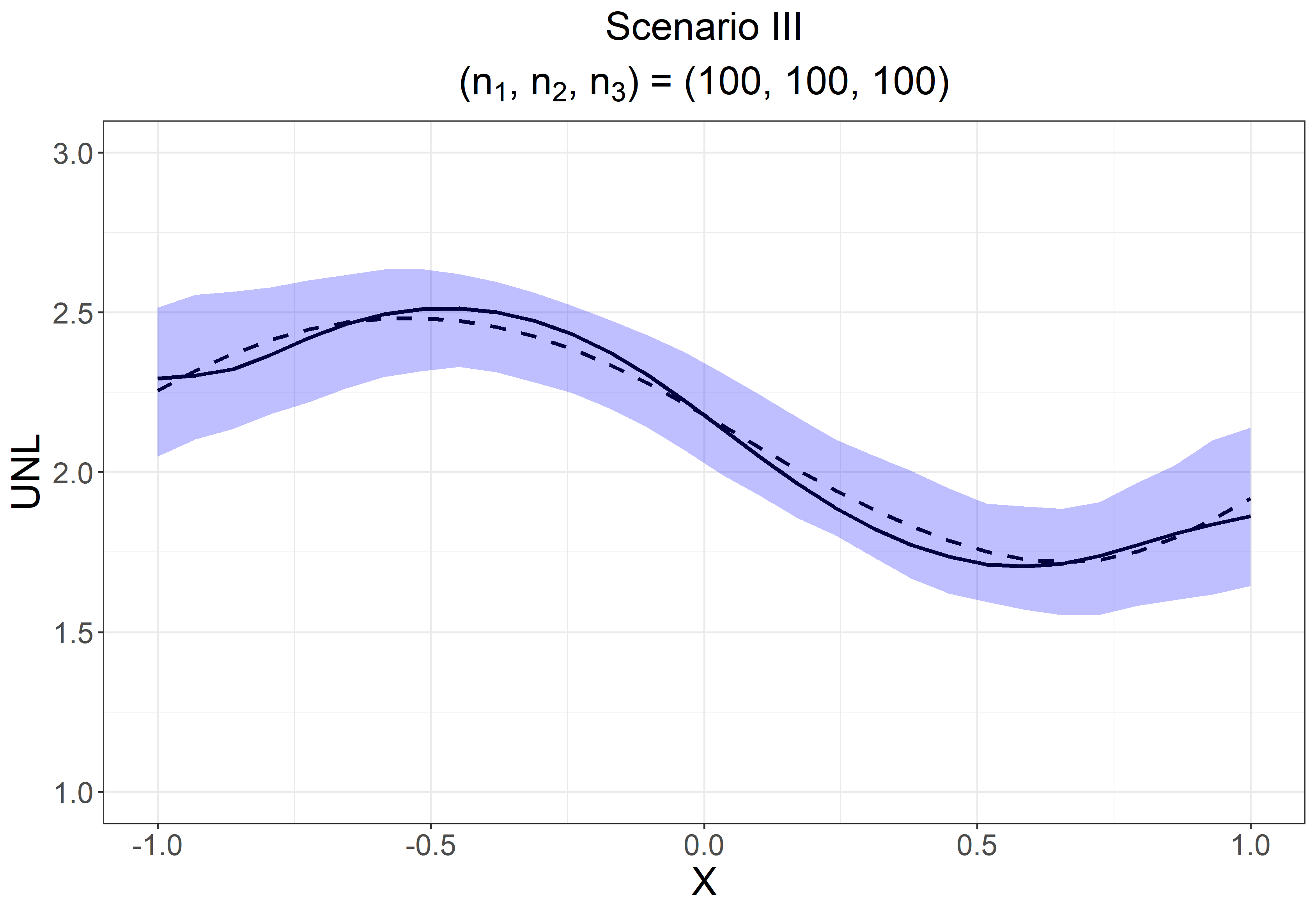}
	}
	\\
	\subfigure{
		\includegraphics[width=0.3\linewidth]{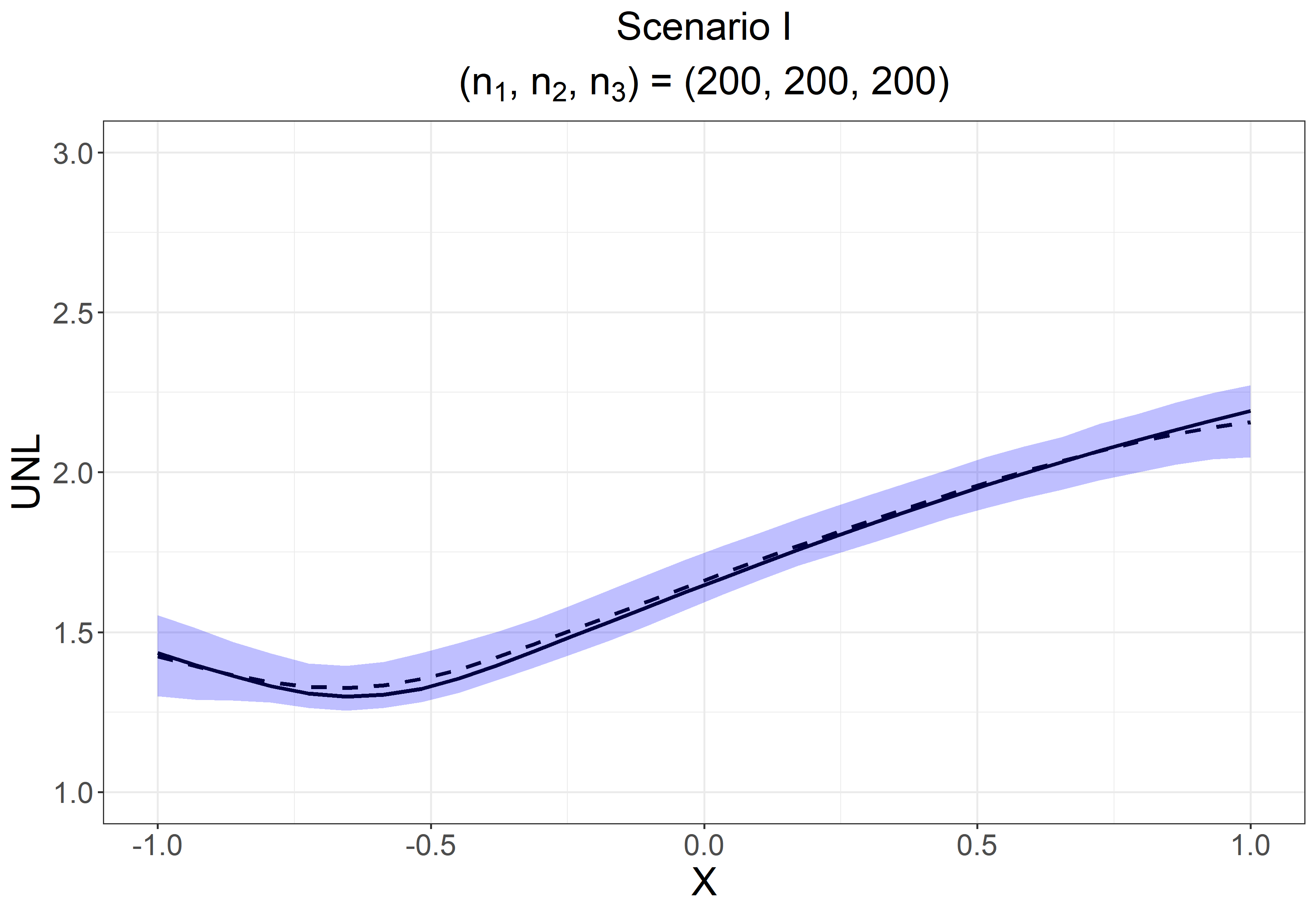}
	}
	\subfigure{
		\includegraphics[width=0.3\linewidth]{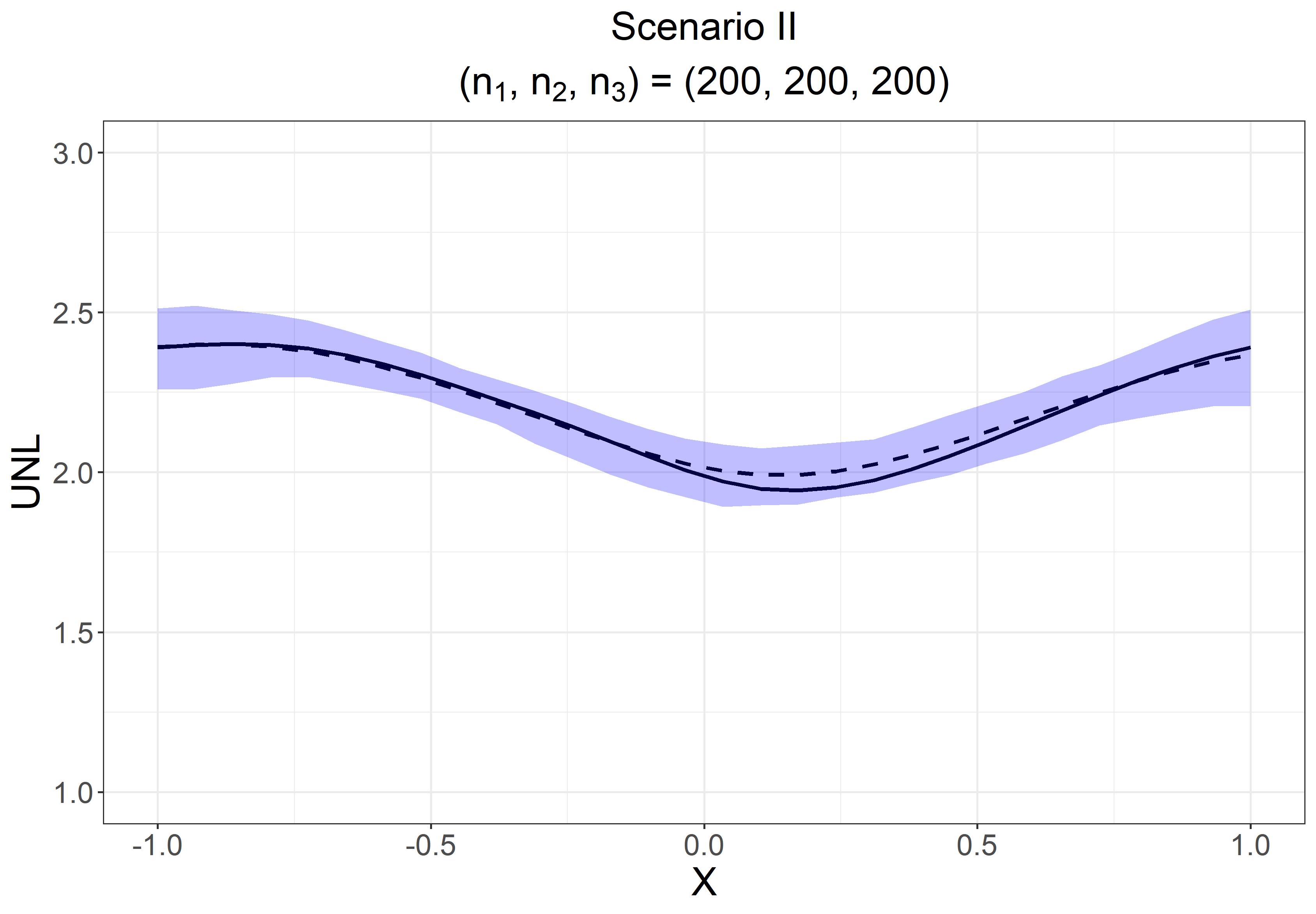}
	}
	\subfigure{
		\includegraphics[width=0.3\linewidth]{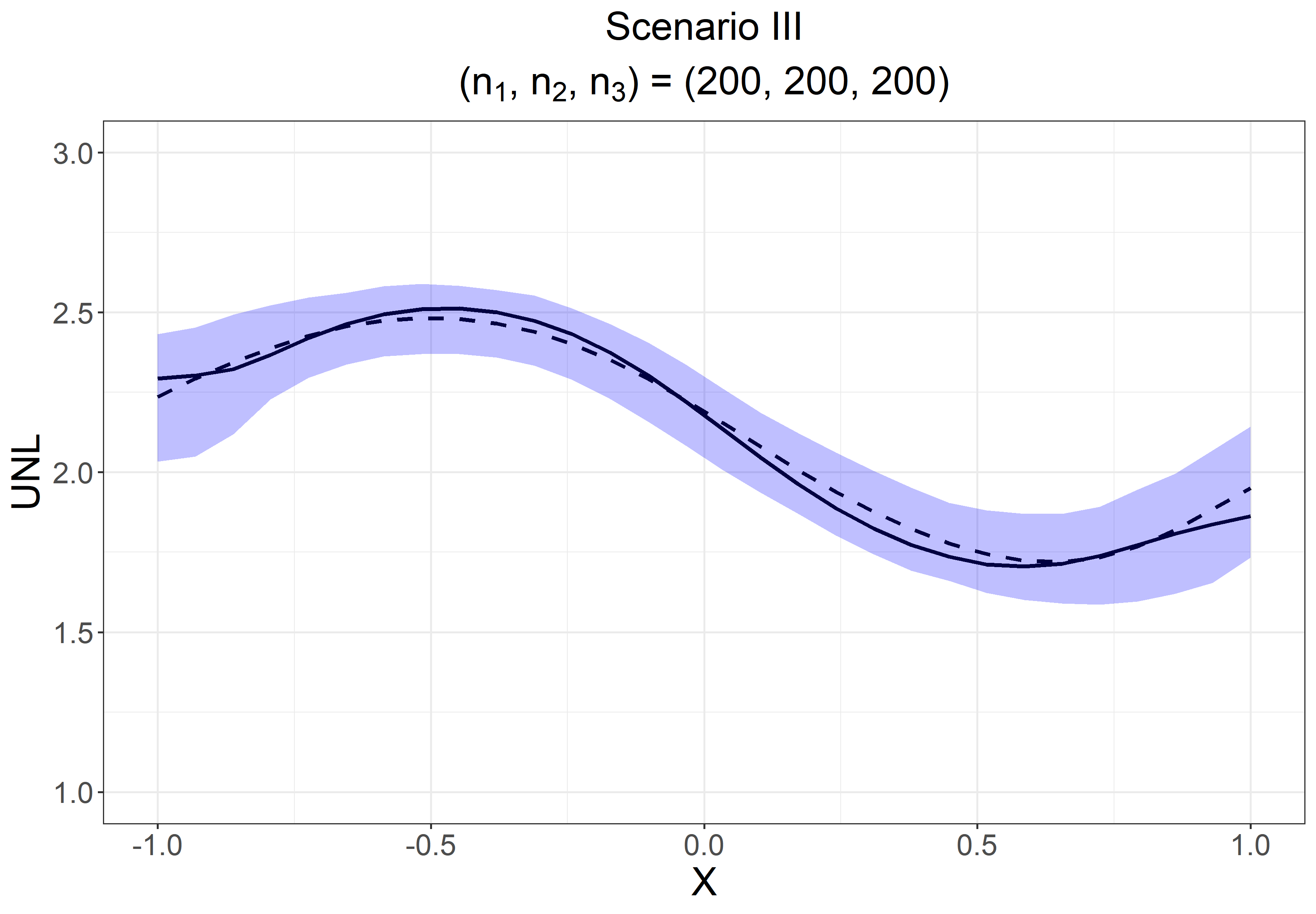}
	}
	\\
	\subfigure{
		\includegraphics[width=0.3\linewidth]{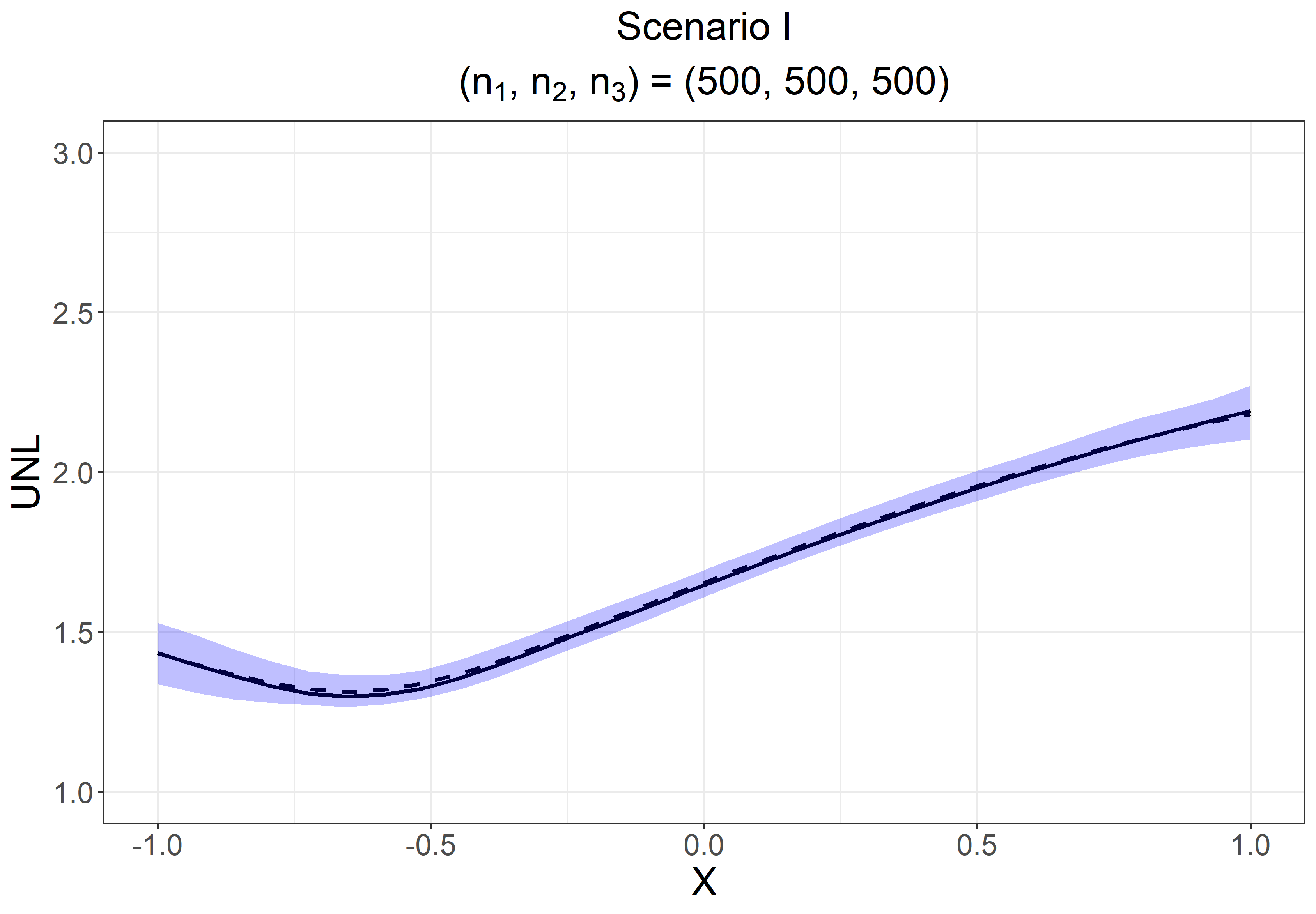}
	}
	\subfigure{
		\includegraphics[width=0.3\linewidth]{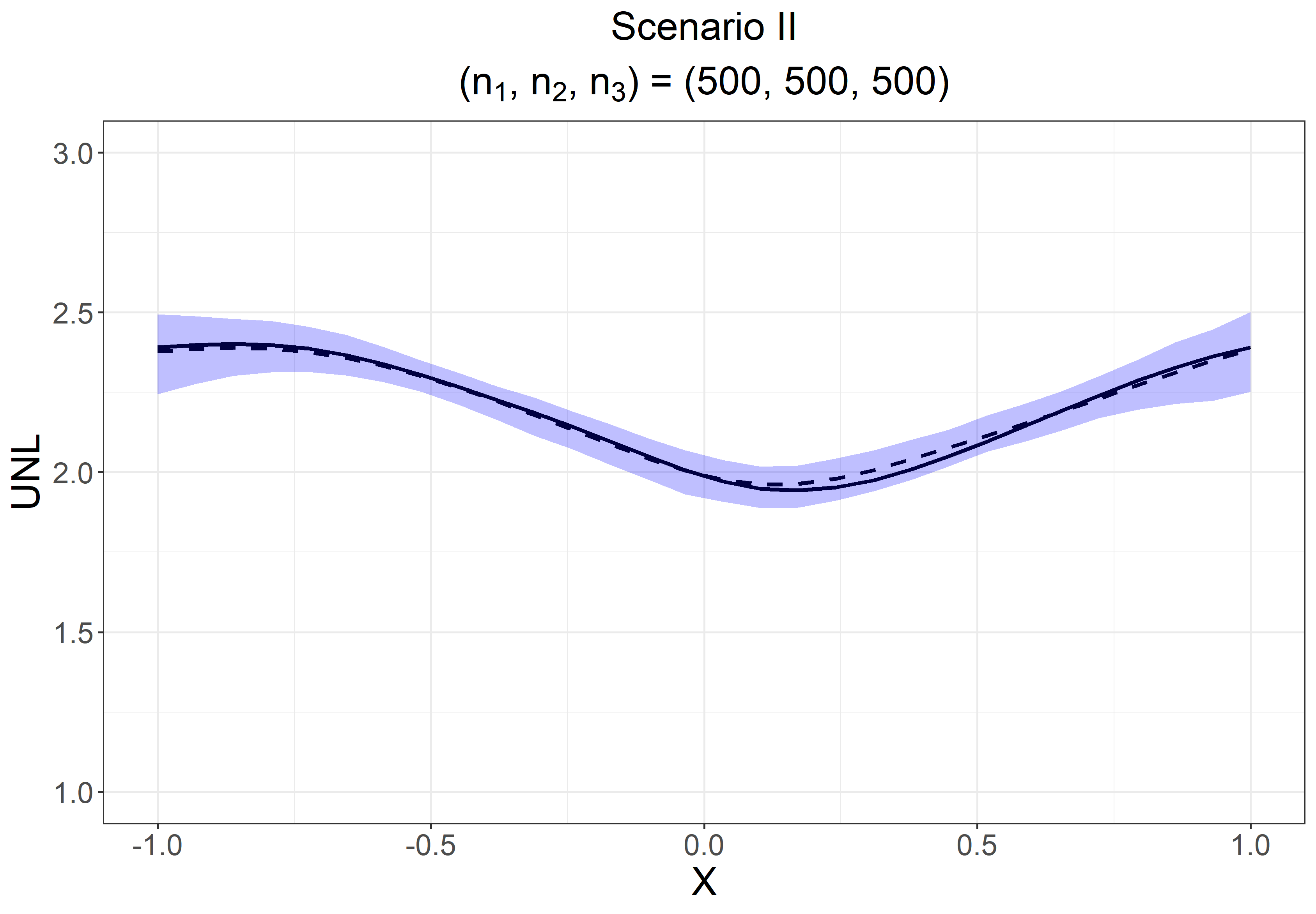}
	}
	\subfigure{
		\includegraphics[width=0.3\linewidth]{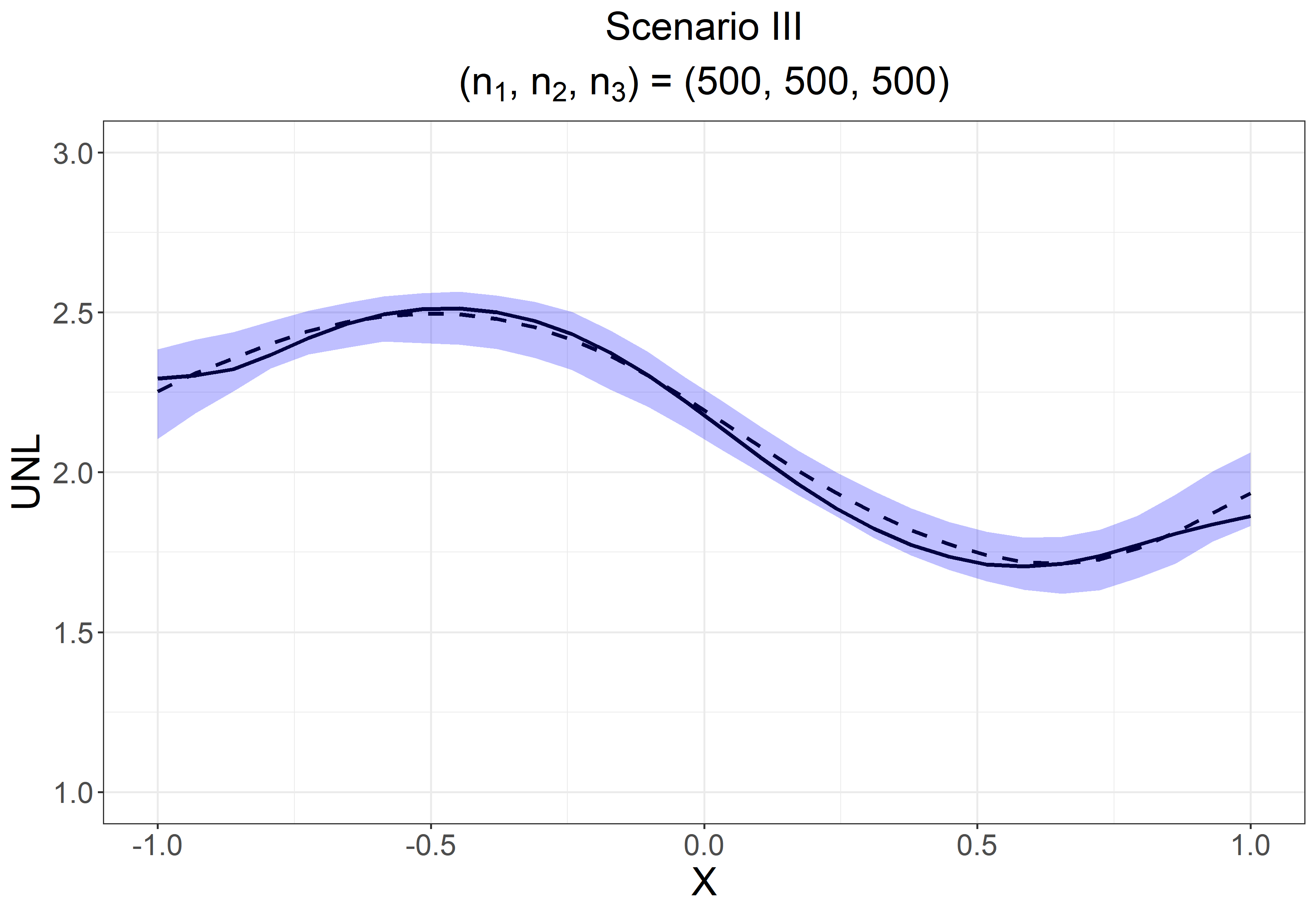}
	}
	\\
	\subfigure{
		\includegraphics[width=0.3\linewidth]{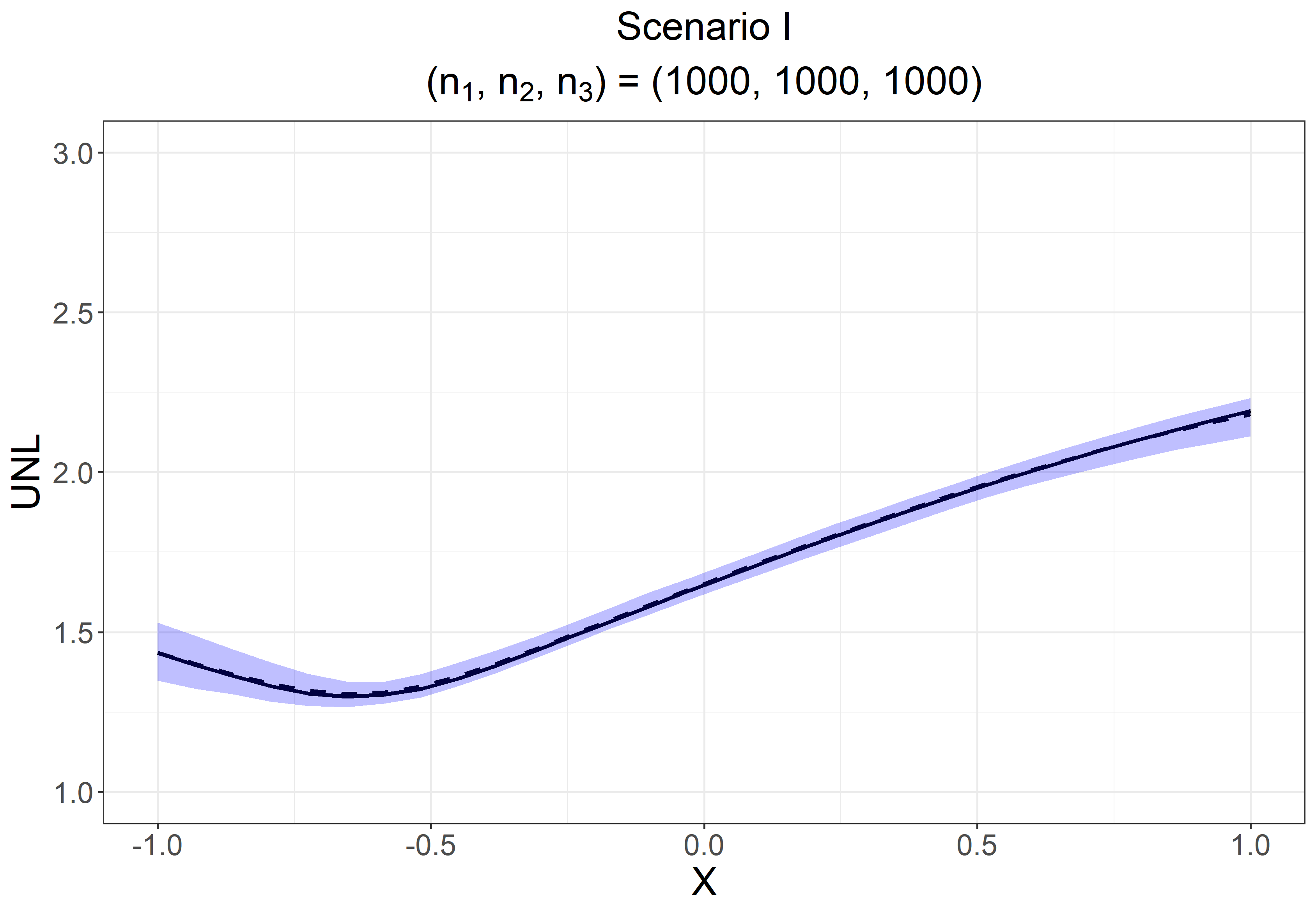}
	}
	\subfigure{
		\includegraphics[width=0.3\linewidth]{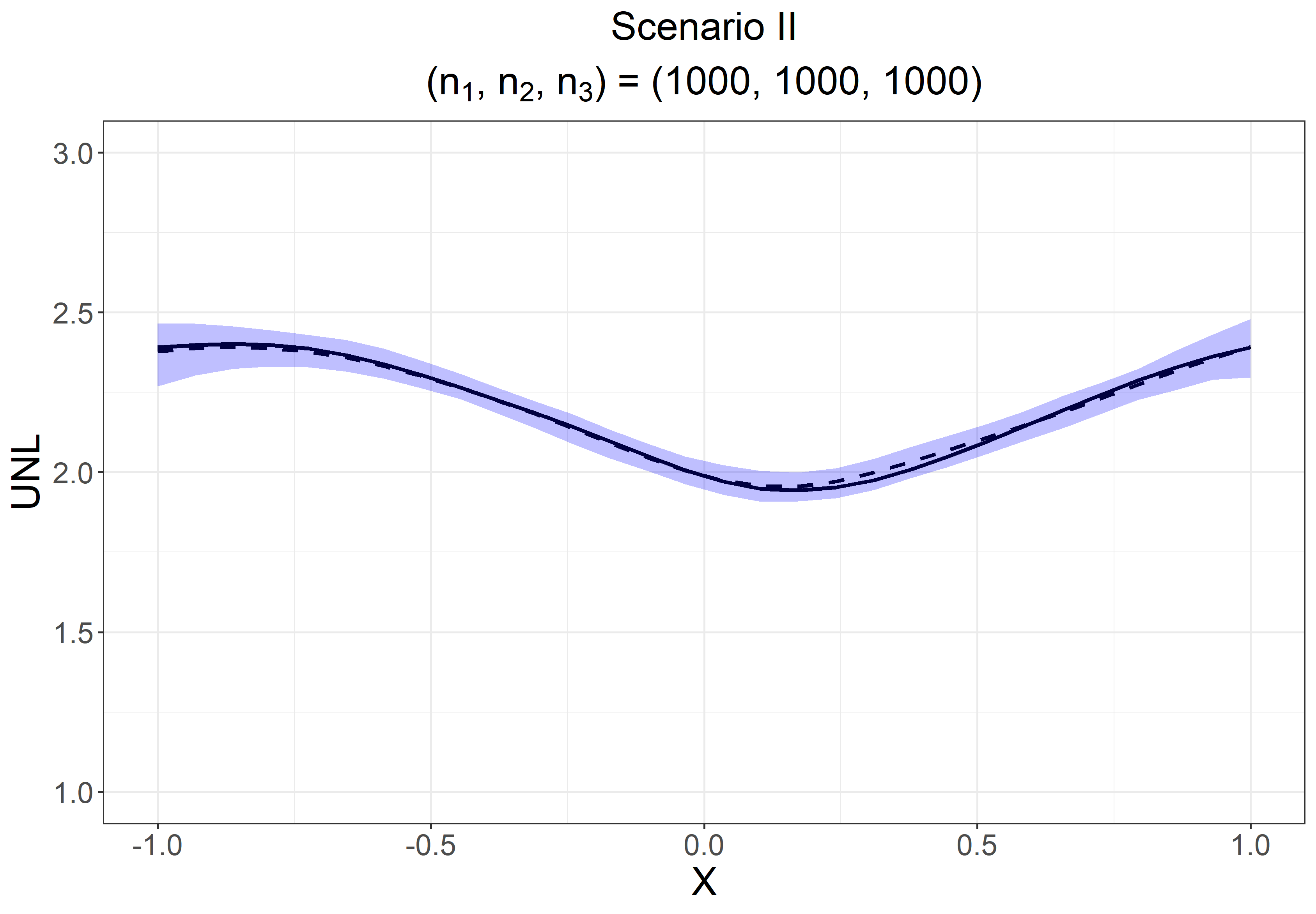}
	}
	\subfigure{
		\includegraphics[width=0.3\linewidth]{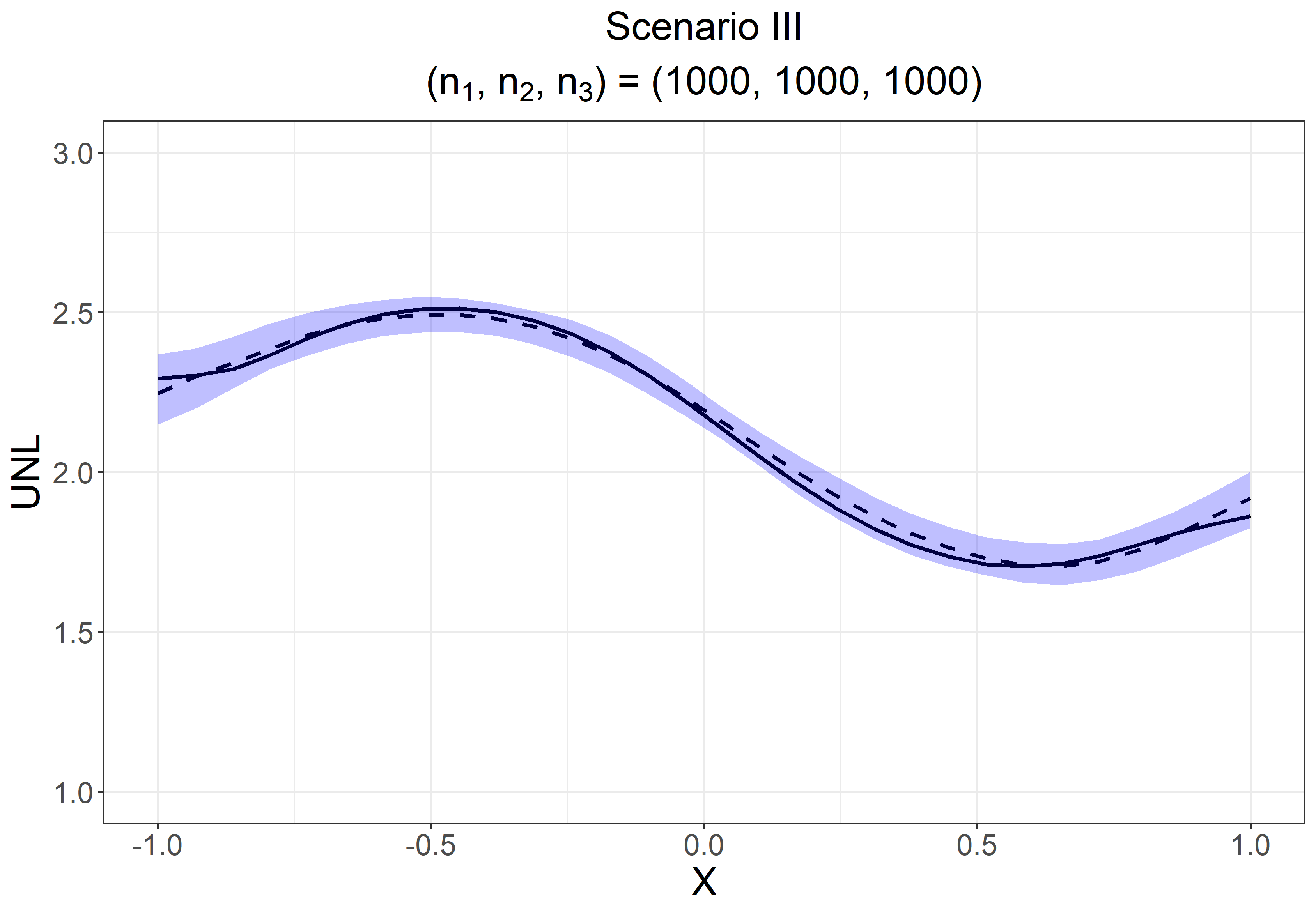}
	}
	\\
	\subfigure{
		\includegraphics[width=0.3\linewidth]{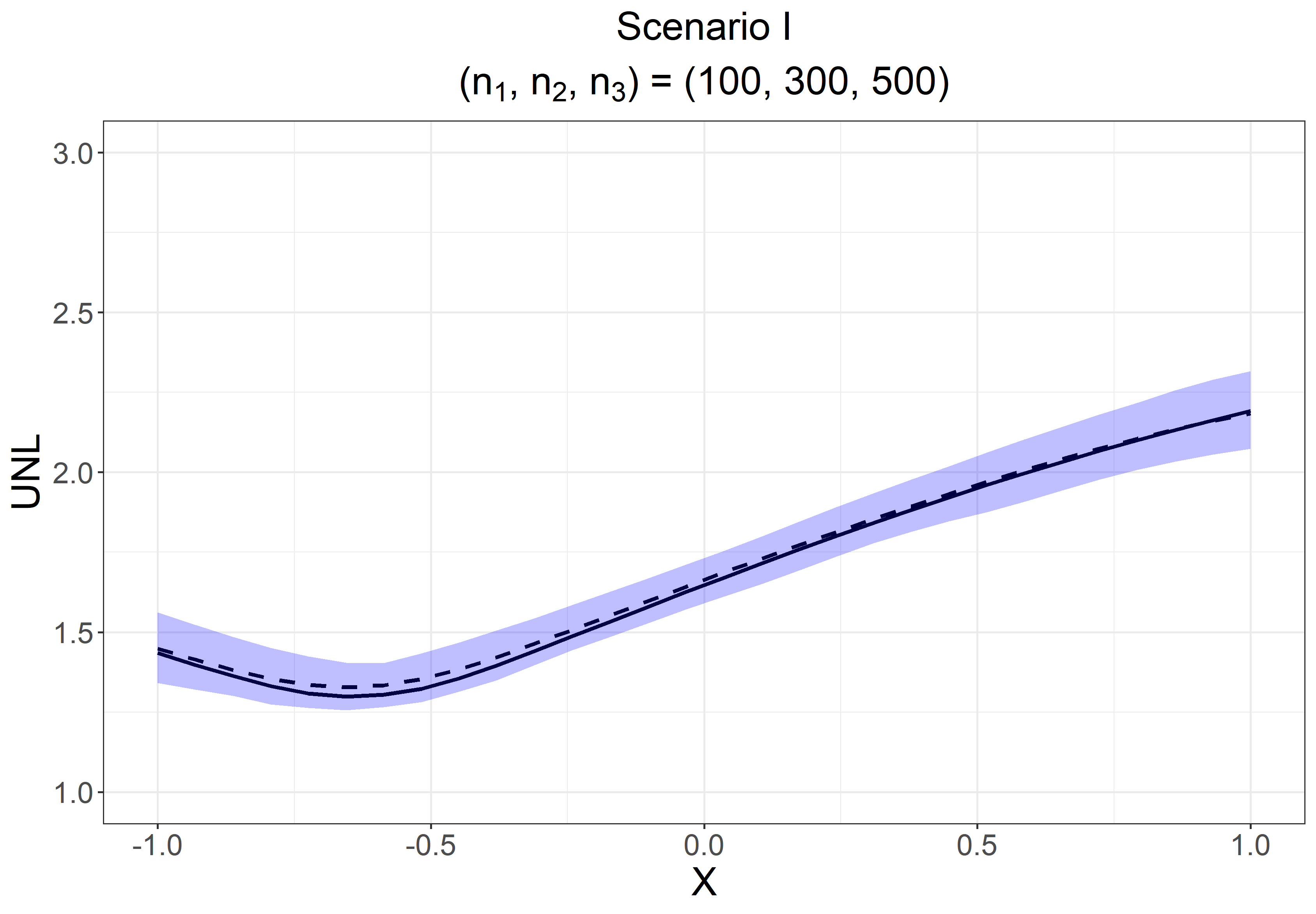}
	}
	\subfigure{
		\includegraphics[width=0.3\linewidth]{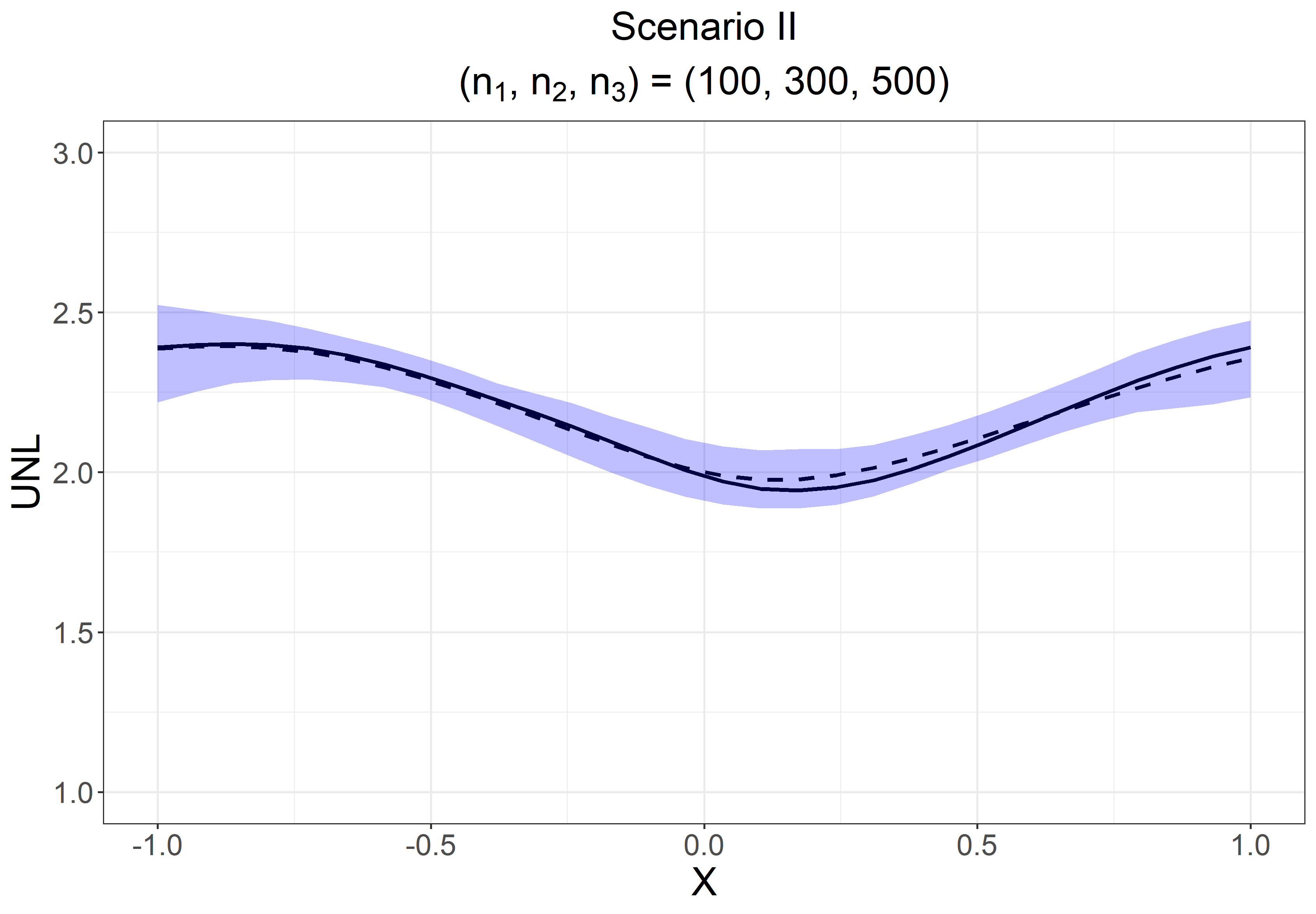}
	}
	\subfigure{
		\includegraphics[width=0.3\linewidth]{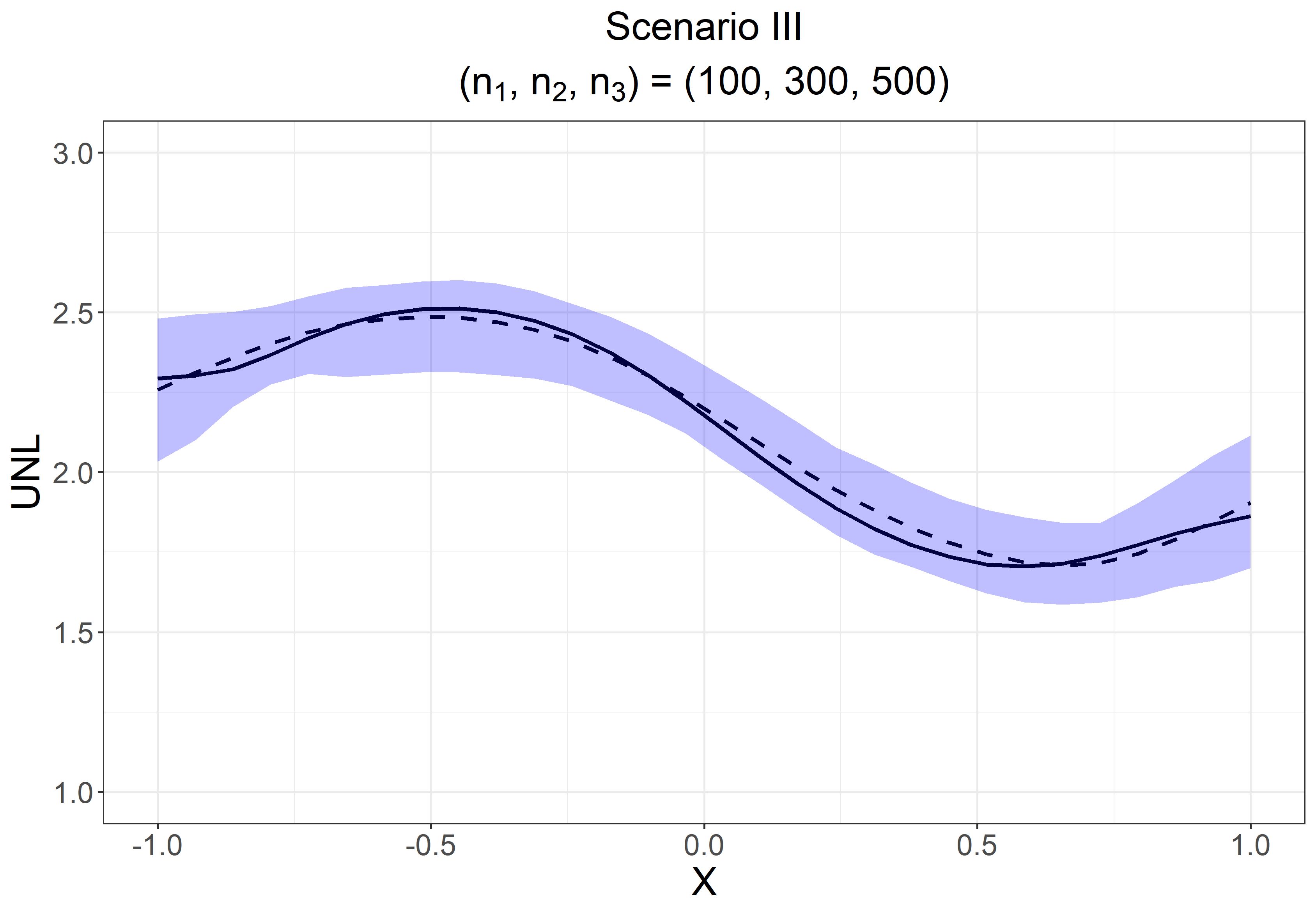}
	}
	\caption{True conditional underlap coefficient (solid black line) and average value across the 100 simulated datasets (dashed line) of the posterior median of the covariate-specific underlap coefficient. The shaded areas are bands constructed using the 2.5\% and 97.5\% pointwise quantiles across of these 100 posterior medians.}
	\label{plot_con_mean_quantile_kernel}
\end{figure}

In Figure \ref{plot_con_mean_quantile_kernel}, for each scenario and sample size considered, we present the average of the posterior medians computed across the 100 simulated datasets. Additionally, the 2.5\% and 97.5\% pointwise simulation quantiles of these posterior medians are also displayed. As observed, the Bayesian nonparametric estimator successfully recovers the true functional form of the covariate-specific underlap coefficient across all three scenarios. There is a small amount of bias in Scenarios II and III, which diminishes as the sample size increases.
The frequentist coverage probability of the 95\% credible intervals was also investigated and the results are presented in Figure 5 of the Supplementary Material. Examination of this figure reveals that in Scenarios I, for all sample sizes and covariate values considered, the empirical coverage probabilities are close to the nominal value of $0.95$. Conversely, in Scenarios II and III, the empirical coverage probability is moderately lower than $0.95$ for some covariate values. This is not surprising, as the covariate values associated with diminished coverage probabilities coincide with those exhibiting bias in the estimations. As the sample size increases, the bias correspondingly decreases; however, the pointwise  $95\%$  credible intervals also become narrower, causing them to fail to encompass the nominal UNL value.

\section{\large{\textsf{APPLICATION TO  ALZHEIMER'S DISEASE }}}
Alzheimer's disease (AD) is an irreversible brain disorder that progressively destroys memory and thinking abilities. Due to its increasing prevalence, it also constitutes a major public health concern. According to
the World Health Organization, AD is the most common form of dementia and may contribute to 60--70\% of
cases. We applied our methods to data derived from the Alzheimer’s Disease Neuroimaging Initiative (ADNI). ADNI was launched in 2004 as a public-private partnership with the goal of developing and identifying biomarkers for the early detection and tracking of Alzheimer’s disease. The data used in our analysis were obtained from the ADNIMERGE \texttt{R} package  \citep{adni_Data}. Study participants were classified into the following three disease stages based on  neuropsychological tests: cognitively normal (CN) (normal aging, including individuals with subjective memory complaints), mild cognitive impairment (MCI) (including individuals with early and late MCI), and severe dementia or AD. We consider the following four biomarkers measured at baseline: cerebrospinal fluid (CSF) Abeta, CSF Tau, CSF phosphorylated tau (pTau), and the hypometabolic convergence index (HCI).  
In our analysis, we aim to evaluate the individual ability of each of these four biomarkers to distinguish between the three disease stages: CN, MCI, and AD. Furthermore, it is crucial to understand if and how this discriminatory ability changes with patient characteristics, namely age and gender. 

We analyse data on 1007 subjects with complete data on disease stage, age, gender, HCI, CSF Abeta, CSF Tau, and CSF pTau. Among the 1007 subjects, 307 are classified in the CN stage, 520 in the MCI group, and 180 in the AD stage. Some biomarker values were subject to a limit of detection. Specifically, CSF Abeta values greater than 1700 pg/mL were not recorded exactly and were instead reported as 1700, while those below 200 pg/mL were recorded as 200. Similarly, CSF tau values exceeding 1300 pg/ml were capped at 1300, and values below 80 pg/mL were recorded as 80. For pTau, outcomes greater than 120 pg/ml were recorded as 120, and those below 8 pg/mL were recorded as 8. For HCI, CSF Tau, and CSF pTau, larger biomarker values indicate worse disease staging. In contrast, for the CSF Abeta biomarker, the order is reversed, that is, smaller biomarker  values are associated with worse disease staging.

We begin with an exploratory data analysis, and Figure 6 in the Supplementary Material presents scatterplots of age against each of the four biomarkers across the three disease groups, stratified by gender. Although not highly pronounced, some variability is observed with respect to both age and gender. The scatterplots do not suggest an interaction between age and gender in any combination of disease group and biomarker. We therefore fitted the logit stick-breaking process prior described in Section \ref{inference}, with the effects of age and gender modelled additively. To determine how the age effect should be modelled, we followed the same strategy outlined in Section \ref{sim_study}. For all combinations of disease groups and biomarkers, except for one, a linear effect of age was used when modelling both the components' weights and means. In the CN group for  the Abeta biomarker, the age effect was modelled using a B-spline basis with no interior knots. Posterior inference is based on 5000 iterations after discarding a burn-in period of 5000 iterations. The same prior information as in Section \ref{sim_study} was used. Traceplots and Geweke statistics do not suggest a lack of convergence in the chains for the conditional density, the identifiable quantity of interest here (results not shown). The effective sample sizes of the conditional density chains are reasonably large, enabling the computation of low and high quantiles with sufficient precision (results not shown). To assess model fit for each biomarker within each disease group, we conducted posterior predictive checks. Specifically, 5000 replicate datasets were generated from the posterior predictive distribution for each disease group  and biomarker and compared to the corresponding biomarker values in each group using specific test quantities. We follow \cite{Gabry2019}, who suggest choosing statistics that are `orthogonal' to the model parameters. Since we  use a location-scale mixture of normal distributions model, we investigate how well the posterior predictive distribution captures skewness and kurtosis. From Figures 7 and 8 in the Supplementary Material, we see that our model does an excellent job of capturing both quantities. Furthermore, Figure 9 in the Supplementary Material presents the kernel density estimates of the 5000 replicate datasets for each disease group and biomarker, compared to the kernel density estimate of each biomarker outcomes in each group. It is evident that, for each disease group and biomarker, our model can generate simulated data that closely resembles the observed biomarker levels.

\begin{figure}[htpb]
	\centering
	\subfigure{
		\includegraphics[width=0.48\linewidth]{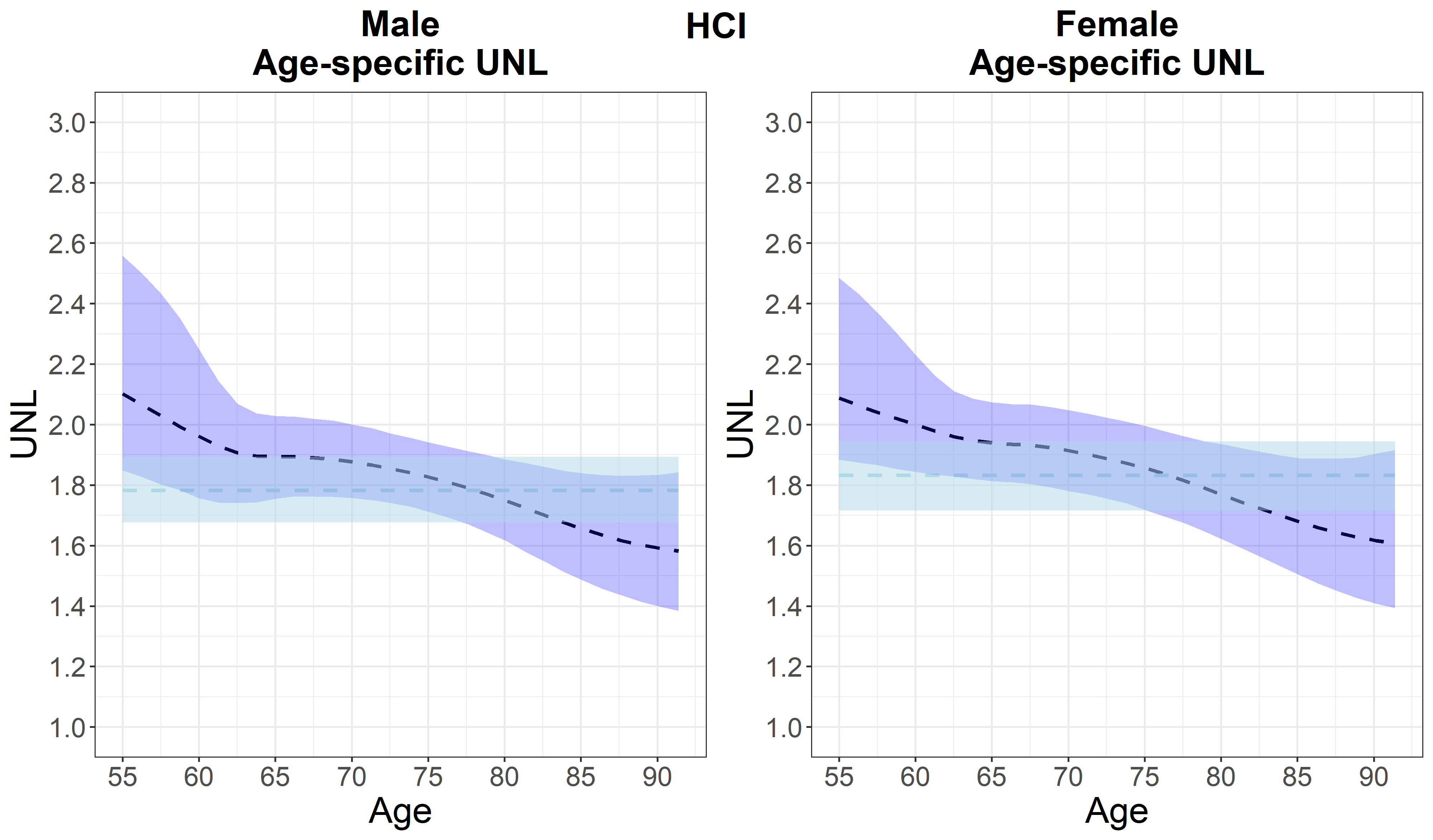}
	}
	\subfigure{
		\includegraphics[width=0.48\linewidth]{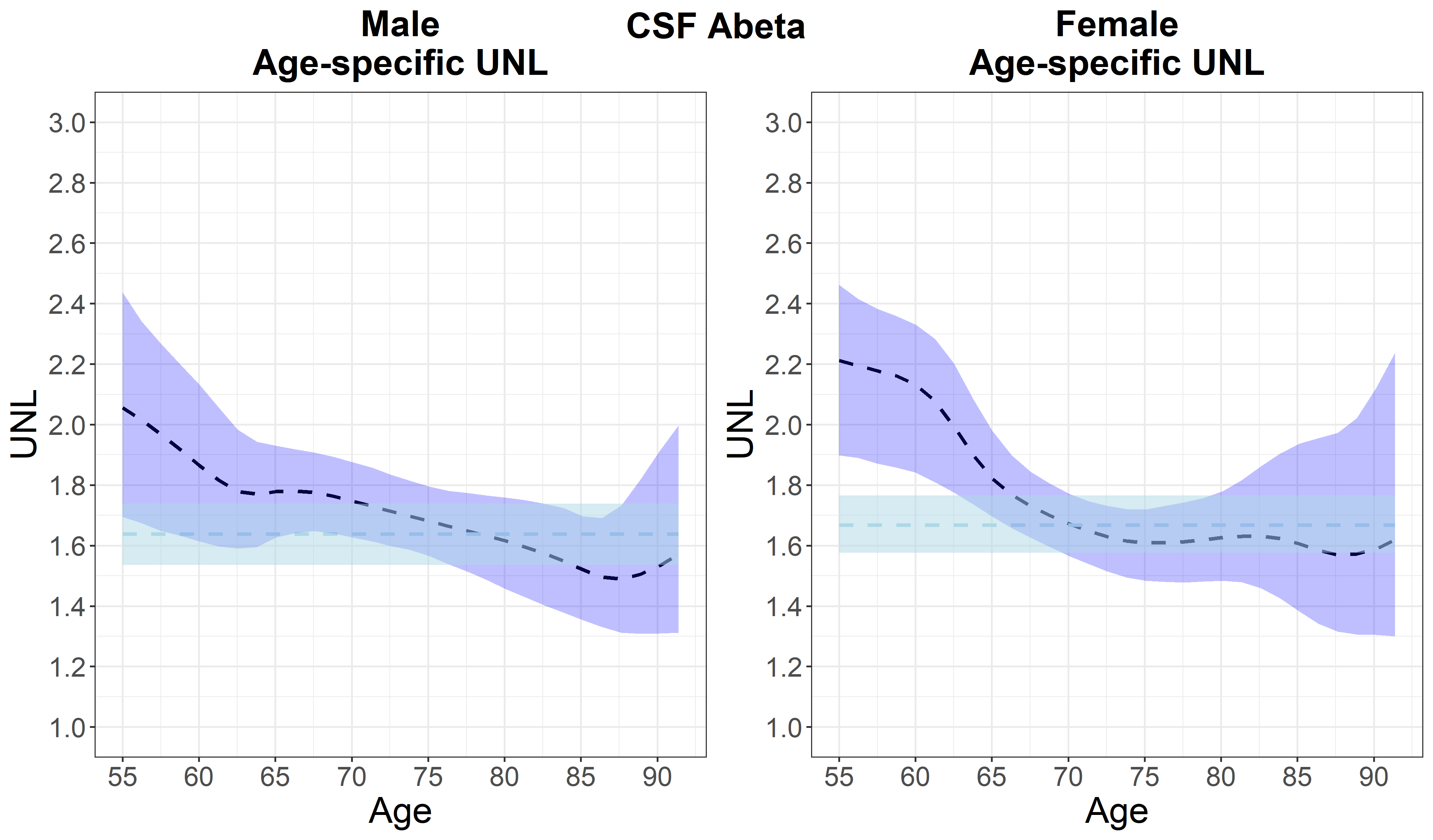}
	}
	\\
	\subfigure{
		\includegraphics[width=0.48\linewidth]{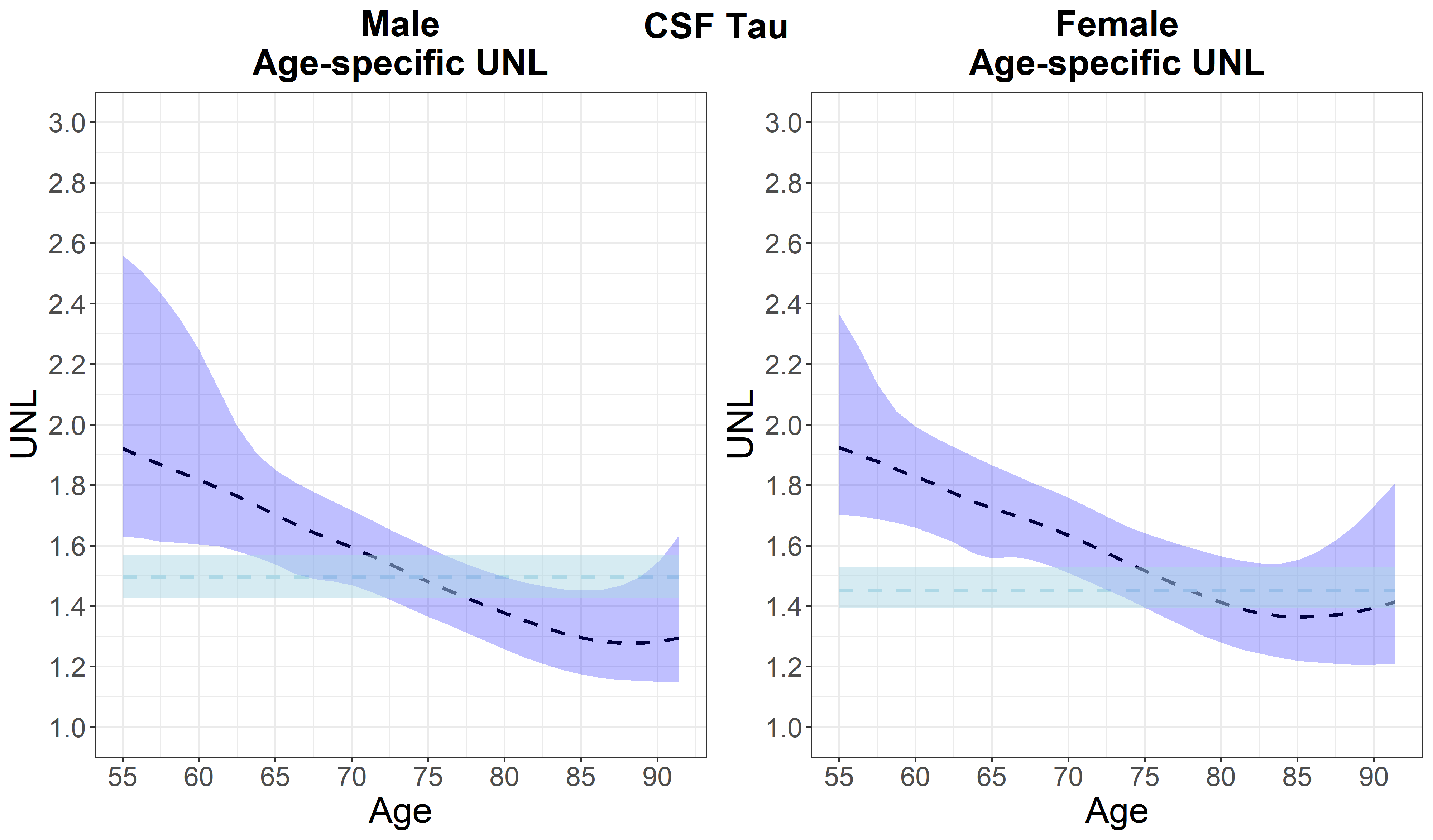}
	}
	\subfigure{
		\includegraphics[width=0.48\linewidth]{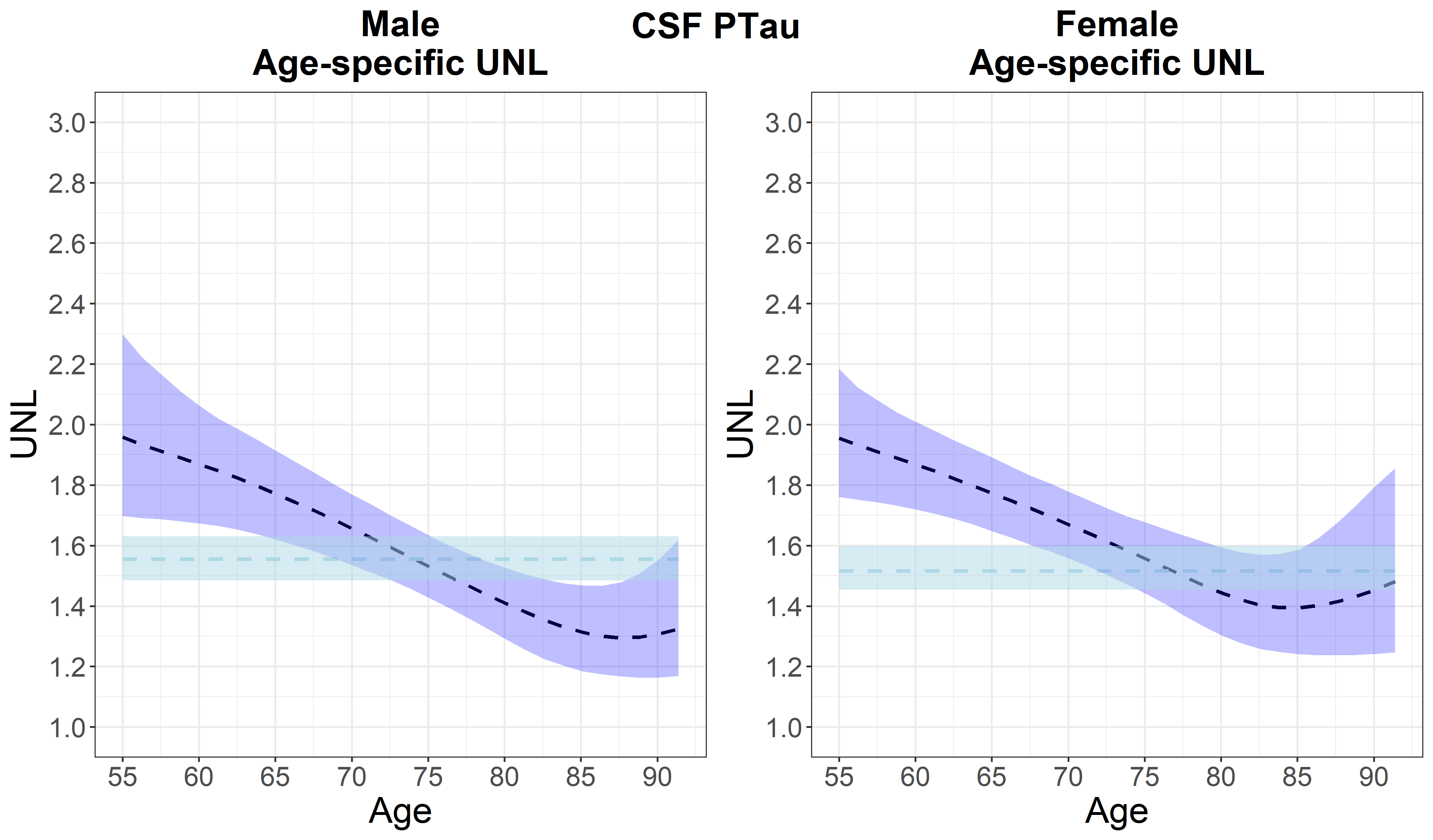}
	}
	\caption{Black dashed line and dark blue ribbon: posterior median and 95\% pointwise credible band for the age-and-gender-specific underlap coefficient for each of the four biomarkers. Light blue dashed line and ribbon: posterior median and 95\% credible interval for the underlap coefficient for the each of four biomarkers, ignoring the age effect.}
	\label{unl_age_gender_specific}
\end{figure}

In Figure \ref{unl_age_gender_specific}, we present the estimated age-and-gender-specific UNL for each of the four biomarkers, for ages between 55 and 91 years old. We observe that the ability to simultaneously distinguish between CN, MCI, and AD stages decreases with age for both genders and across all four biomarkers. We also notice that the discriminatory ability of the biomarkers seems to be almost identical for both males and females. Regarding the age effect, we have also computed the gender-specific UNL while ignoring the age effect. As it can be observed in Figure  \ref{unl_age_gender_specific}, the 95\% credible band of the gender-specific UNL falls outside the 95\% credible band of the gender-and-age-specific UNL, for a substantial range of ages in both the CSF Tau and pTau biomarkers. This highlights the importance of including age as a covariate: neglecting it would lead, in this case, to underestimating biomarker performance in younger individuals and overestimating it in older individuals. This figure also suggests that the HCI and Abeta biomarkers are better in simultaneously distinguishing among the three groups than the Tau and pTau biomarkers. In a head-to-head comparison, the HCI biomarker appears to outperform  Abeta across all ages for men. In contrast, Abeta is slightly superior to HCI for women between approximately 55 and 63 years of age, though uncertainty is high in this range. Nonetheless, both biomarkers appear to have  very similar discriminatory ability.  Furthermore, Figure 10 in the Supplementary Material shed further light on this issue. In this figure, we compute the posterior probability that the HCI biomarker outperforms each of the other three biomarkers, as measured by its UNL value, across the same grid of ages considered previously and for both genders. As observed, the probability that the HCI UNL value exceeds that of the Tau biomarker is always above 0.7 for all ages and both genders. The same conclusion holds for the comparison between  HCI and pTau. In turn, the posterior probability that the HCI UNL value exceeds that of the Abeta biomarker is above $0.6$ for most ages for males, while in females, it is only above $0.5$ for ages  over approximately 63 years old. 

The discriminatory ability, as mentioned previously, decreases with age and is only moderate. For instance, for the HCI biomarker, it ranges from roughly $2.1$ to $1.6$, whereas for the CSF pTau in males, it ranges from about $1.95$ to $1.3$. To better understand how these values translate into the separation between the densities of biomarker outcomes across the three disease stages, Figures 11--14 in the Supplementary Material show the estimated densities of biomarker outcomes for each biomarker at six specific age levels and for both genders. As can be seen, the densities of biomarker outcomes for all biomarkers show substantial overlap between the CN and MCI stages, regardless of age and gender. These figures also highlight the need for a flexible model for the conditional distribution of the biomarkers.

Lastly, Figure 15 in the Supplementary Material shows the estimated cumulative distribution function for the HCI, CSF Abeta, and CSF Tau biomarkers for specific ages and genders. For both the HCI and CSF Tau biomarkers, an assumed stochastic ordering would require that the estimated distribution function of the CN group lies above that of the MCI group for all biomarker values, which, in turn, should be above that of the AD group at all biomarker levels. For the CSF Abeta biomarker, the ordering is reversed. For all three cases presented in the figure, the stochastic ordering is violated to some extent. However, we recognise that uncertainty is quite high, and these violations tend to occur at ages near the boundaries of the age grid considered, where data are scarce. Additionally, in Figure 16 of the Supplementary Material we show the estimated three-class Youden index against the estimated UNL. We note that the estimates of the cumulative distribution functions used to estimate the three-class Youden index are the same as those used to estimate the UNL. We observe that both estimated indices largely coincide for both males and females and across most of the ages considered. However, some discrepancies exist, especially for the youngest and oldest ages considered, where, as mentioned, data are scarce. We also show in Figure 17 in the Supplementary Material the estimated three-class Youden index that one would obtain if the incorrect order among the biomarker values in the three groups were assumed. We can observe how misleading the results are. Although in this case we have only considered four biomarkers, making it easy to determine an order for each biomarker, this approach becomes impractical when many more biomarkers are available. Assuming an incorrect order may lead to overlooking biomarkers with strong discriminatory ability between the three groups.

\section{\large{\textsf{CONCLUDING REMARKS}}}
We have introduced the underlap coefficient as a measure of a biomarker's discriminatory ability in a three-class disease setting. Besides its graphical interpretation, the UNL also has an intuitive interpretation in terms of the number of `effective' populations of biomarker outcomes, and can be straightforwardly generalised to cases with more than three-disease classes. In contrast to traditional ROC-based summary measures, such as the VUS and the three-class Youden index, the UNL does not require an assumed monotonic ordering of disease stages. The UNL can therefore be broadly applied in the biomarker field, as it may be safely applied to any potential panel containing: (i) biomarkers that follow the usual monotonic ordering (i.e., higher biomarker outcomes are associated with worse disease staging), (ii) biomarkers following a reverse ordering (i.e., lower biomarker outcomes indicate more severe disease stages), (iii) biomarkers with non-monotonic ordering. It is important to emphasise that UNL is designed to be used for evaluating a biomarker's discriminatory potential in distinguishing between the three diseae stages during its discovery phase, rather than for determining the classification thresholds to be used in clinical practice. This is because the UNL is independent of any threshold-based classification rule. In some sense, the VUS shares this `limitation' as well, as it serves as a summary measure of discriminatory ability across all thresholds. 
Recognising the importance of handling subject heterogeneity, we further introduced the  covariate-specific UNL, which allows for determining the optimal and suboptimal populations, as defined by the covariate values, where the biomarker performs optimally or suboptimally.

To ensure broad applicability, we proposed a Bayesian nonparametric inferential framework, based on a (dependent) Dirichlet process mixture of normal distributions for both the unconditional UNL and its covariate-specific counterpart. An extensive simulation study, involving different distributional assumptions for biomarker outcomes in each class as well as different functional effects of covariates, demonstrates that our methods perform very well in terms of both point estimation and coverage probability. In the unconditional setting, we observed some bias, particularly for the smaller sample sizes, in cases where the UNL is very close to one, regardless of the distributional assumption for the biomarker outcomes. Note that in these cases, there is almost complete overlap between the densities of biomarker outcomes across the three groups, and our modelling framework is based on modelling these groups independently. A possible remedy is to introduce a hierarchical structure that allows borrowing information across the three groups. This should also be advantageous in the covariate setting, particularly for smaller sample sizes and in regions of the covariate space where data are relatively sparse. This constitutes a future avenue of research.

\section*{Acknowledgments}
The authors thank Finn Lindgren, Mar\'ia Xos\'e Rodr\'iguez-\'Alvarez, and Sara Wade for helpful discussions.  Data collection and sharing for the Alzheimer's Disease Neuroimaging Initiative (ADNI) is
funded by the National Institute on Aging (National Institutes of Health Grant
U19AG024904). The grantee organization is the Northern California Institute for Research
and Education. In the past, ADNI has also received funding from the National Institute of
Biomedical Imaging and Bioengineering, the Canadian Institutes of Health Research, and
private sector contributions through the Foundation for the National Institutes of Health
(FNIH) including generous contributions from the following: AbbVie, Alzheimer’s Association;
Alzheimer’s Drug Discovery Foundation; Araclon Biotech; BioClinica, Inc.; Biogen; BristolMyers Squibb Company; CereSpir, Inc.; Cogstate; Eisai Inc.; Elan Pharmaceuticals, Inc.; Eli
Lilly and Company; EuroImmun; F. Hoffmann-La Roche Ltd and its affiliated company
Genentech, Inc.; Fujirebio; GE Healthcare; IXICO Ltd.; Janssen Alzheimer Immunotherapy
Research \& Development, LLC.; Johnson \& Johnson Pharmaceutical Research \&
Development LLC.; Lumosity; Lundbeck; Merck \& Co., Inc.; Meso Scale Diagnostics, LLC.;
NeuroRx Research; Neurotrack Technologies; Novartis Pharmaceuticals Corporation; Pfizer
Inc.; Piramal Imaging; Servier; Takeda Pharmaceutical Company; and Transition
Therapeutics.

\bibliographystyle{hapalike}
\bibliography{references}

\newpage
\section*{\large{\textsf{SUPPLEMENTARY MATERIAL}}}
\setcounter{figure}{0}  
\subsection*{Further properties of the underlap coefficient and additional figures}

\subsubsection*{Invariance property of the UNL}
We follow the proof in \cite{schmid2006nonparametric} in the context of the two-class overlap coefficient. 
Let $\psi$ be a strictly increasing and differentiable transformation defined on the support of $Y_1$, $Y_2$, and $Y_3$, and let $u=\psi(y)$. Then
\begin{equation*}
	\frac{f_1(\psi^{-1}(u))}{{\psi}'(\psi^{-1}(u))},\quad  \frac{f_2(\psi^{-1}(u))}{{\psi}'(\psi^{-1}(u))},\quad \frac{f_3(\psi^{-1}(u))}{{\psi}'(\psi^{-1}(u))},
\end{equation*}
\noindent
are the densities of $\psi(Y_1)$, $\psi(Y_2)$, and $\psi(Y_3)$, respectively.

\noindent The underlap of $\psi(Y_1)$, $\psi(Y_2)$, and $\psi(Y_3)$ is given by
\begin{align*}
	\text{UNL}(\psi(Y_1),\psi(Y_2),\psi(Y_3))=&\int_{-\infty}^{\infty}\max\left\{\frac{f_1(\psi^{-1}(u))}{{\psi}'(\psi^{-1}(u))},\frac{f_2(\psi^{-1}(u))}{{\psi}'(\psi^{-1}(u))},\frac{f_3(\psi^{-1}(u))}{{\psi}'(\psi^{-1}(u))}\right\}\text{d}u \\
	=&\int_{-\infty}^{\infty}\max\{f_1(\psi^{-1}(u)),f_2(\psi^{-1}(u)),f_3(\psi^{-1}(u))\}\frac{1}{{\psi}'(\psi^{-1}(u))}\text{d}u \\
	=&\int_{-\infty }^{+\infty} \max\left\{f_1(y),f_2(y),f_3(y)\right\}\text{d}y\\
	= & \text{ UNL}(Y_1,Y_2,Y_3)\equiv \text{UNL}(f_1,f_2,f_3).
\end{align*}
\noindent
Hence, the underlap coefficient is invariant with respect to strictly increasing and differentiable transformations of $Y_1$, $Y_2$ and $Y_3$.

\subsubsection*{Relationship between the underlap and overlap coefficients}
The three-class UNL can be rewritten as
	\begin{align}
		\text{UNL}(f_1,f_2,f_3)=&\int_{-\infty }^{+\infty} \max\left\{f_1(y),f_2(y),f_3(y)\right\}\text{d}y\nonumber\\
		=& \int_{-\infty }^{+\infty}\text{I}\left(f_1(y)>\max\{f_2(y),f_3(y)\}\right)f_1(y)\text{d}y+\int_{-\infty }^{+\infty}\text{I}\left(f_2(y)>\max\{f_1(y),f_3(y)\}\right)f_2(y)\text{d}y\nonumber\\
		+&\int_{-\infty }^{+\infty}\text{I}\left(f_3(y)>\max\{f_1(y),f_2(y)\}\right)f_3(y)\text{d}y.
		\label{UNL_relation_OVL_detail1}
	\end{align}
\noindent
We denote the regions
	\begin{align*}
		&\{y: f_1(y)>\max\{f_2(y),f_3(y)\}\},\\
		&\{y: f_2(y)>\max\{f_1(y),f_3(y)\}\}, \\
		&\{y: f_3(y)>\max\{f_1(y),f_2(y)\}\},
	\end{align*}
by $A_1$, $A_2$, and $A_3$, respectively. We further denote the complements of $A_1$, $A_2$, and $A_3$ by $A_1^c$, $A_2^c$, and $A_3^c$, respectively. These are, respectively, expressed as
	\begin{align*}
		&\{y: f_1(y)\le f_2(y)\}\cup\{y: f_1(y)\le f_3(y)\},\\ 
		&\{y: f_2(y)\le f_1(y)\}\cup\{y: f_2(y)\le f_3(y)\},\\ 
		&\{y: f_3(y)\le f_1(y)\}\cup\{y: f_3(y)\le f_2(y)\}.
	\end{align*}
\noindent
Using this terminology, we can further rewrite (\ref{UNL_relation_OVL_detail1}) as
\begin{align*}
	\text{UNL}(f_1,f_2,f_3)
	& = \int_{A_1 }f_1(y)\text{d} y+\int_{A_2}f_2(y)dy+\int_{A_3}f_3(y)\text{d} y\\
	&=\left(1-\int_{A_1^c }f_1(y)\text{d}y\right)+\left(1-\int_{A_2^c}f_2(y)\text{d}y\right)+\left(1-\int_{A_3^c}f_3(y)\text{d}y\right)\\
	&=3-\int_{A_1^c }f_1(y)\text{d}y -\int_{A_2^c}f_2(y)\text{d}y-\int_{A_3^c}f_3(y)\text{d}y\\
	& = 3 - \left(\int_{\{f_1(y)\le f_2(y)\}}f_1(y)\text{d}y + \int_{\{f_1(y)\le f_3(y)\}}f_1(y)\text{d}y - \int_{\{f_1(y)\le \min \{f_2(y),f_3(y)\}\}}f_1(y)\text{d}y \right)\\
	&~~~~~ - \left(\int_{\{f_2(y)\le f_1(y)\}}f_2(y)\text{d}y + \int_{\{f_2(y)\le f_3(y)\}}f_2(y)\text{d}y - \int_{\{f_2(y)\le \min \{f_1(y),f_3(y)\}\}}f_2(y)\text{d}y \right)\\
	&~~~~~ - \left(\int_{\{f_3(y)\le f_1(y)\}}f_3(y)\text{d}y + \int_{\{f_3(y)\le f_2(y)\}}f_3(y)\text{d}y - \int_{\{f_3(y)\le \min \{f_1(y),f_2(y)\}\}}f_3(y)\text{d}y \right)\\
	& = 3-\left(\int_{\{f_1(y)\le f_2(y)\}}f_1(y)\text{d}y+\int_{\{f_2(y)\le f_1(y)\}}f_2(y)\text{d}y\right) \\
	&~~~~~-\left(\int_{\{f_2(y)\le f_3(y)\}}f_2(y)\text{d}y+\int_{\{f_3(y)\le f_2(y)\}}f_3(y)\text{d}y\right) \\ 
	&~~~~~-\left(\int_{\{f_1(y)\le f_3(y)\}}f_1(y)\text{d}y+\int_{\{f_3(y)\le f_1(y)\}}f_3(y)\text{d}y\right) \\ 
	&~~~~~+\int_{\{f_1(y)\le \min \{f_2(y),f_3(y)\}\}}f_1(y)\text{d}y\\
	&~~~~~ + \int_{\{f_2(y)\le \min \{f_1(y),f_3(y)\}\}}f_2(y)\text{d}y \\
	& ~~~~~+\int_{\{f_3(y)\le \min \{f_1(y),f_2(y)\}\}}f_3(y)\text{d}y \\
	& =3-\int_{-\infty }^{+\infty} \min\left\{f_1(y),f_2(y)\right\}\text{d}y
	-\int_{-\infty }^{+\infty} \min\left\{f_2(y),f_3(y)\right\}\text{d}y
	-\int_{-\infty }^{+\infty} \min\left\{f_1(y),f_3(y)\right\}\text{d}y  \\ 
	&~~~~~ +\int_{-\infty }^{+\infty} \min\left\{f_1(y),f_2(y),f_3(y)\right\}\text{d}y \\
	& =3-\text{OVL}(f_1,f_2)-\text{OVL}(f_2,f_3)-\text{OVL}(f_1,f_3)+\text{OVL}(f_1,f_2,f_3).
\end{align*}
We have thus established the connection between the underlap and the two- and three-class overlap coefficients.

\clearpage

\subsection*{Additional figures}
\begin{figure}[H]
	\centering
	\subfigure{
		\centering
		\includegraphics[width=0.28\textwidth]{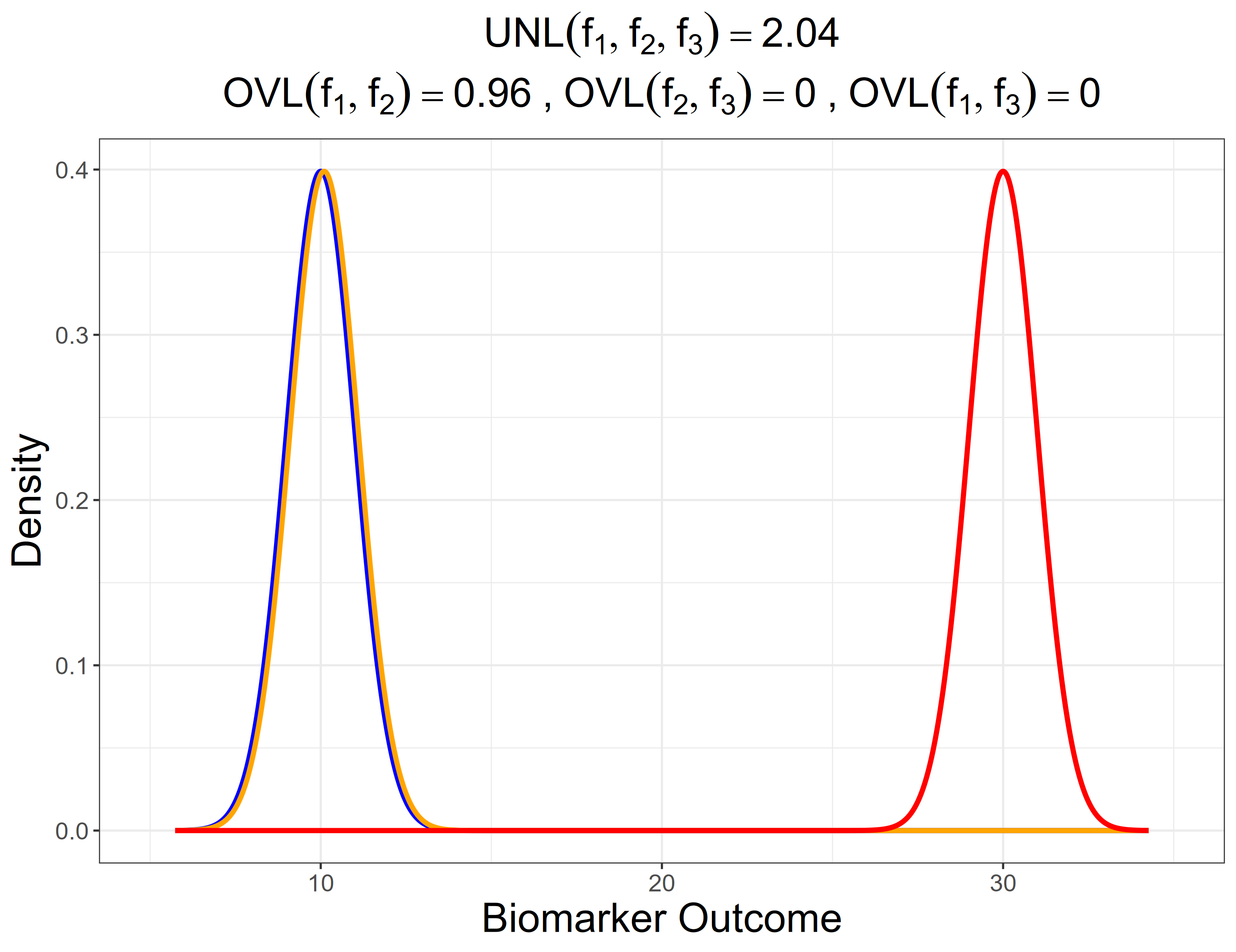}
	}
	\subfigure{
		\centering
		\includegraphics[width=0.28\textwidth]{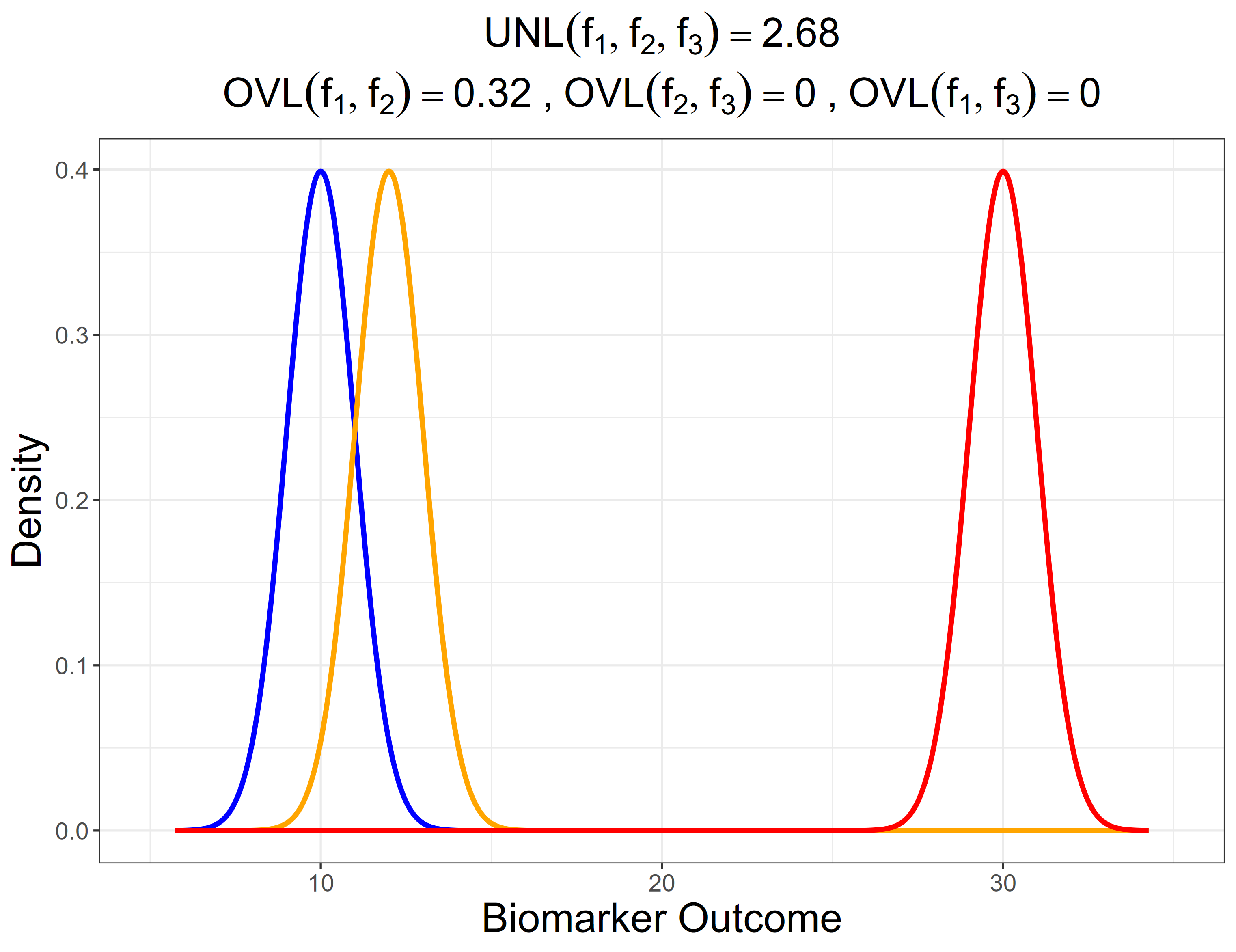}
	}
	\subfigure{
		\centering
		\includegraphics[width=0.28\textwidth]{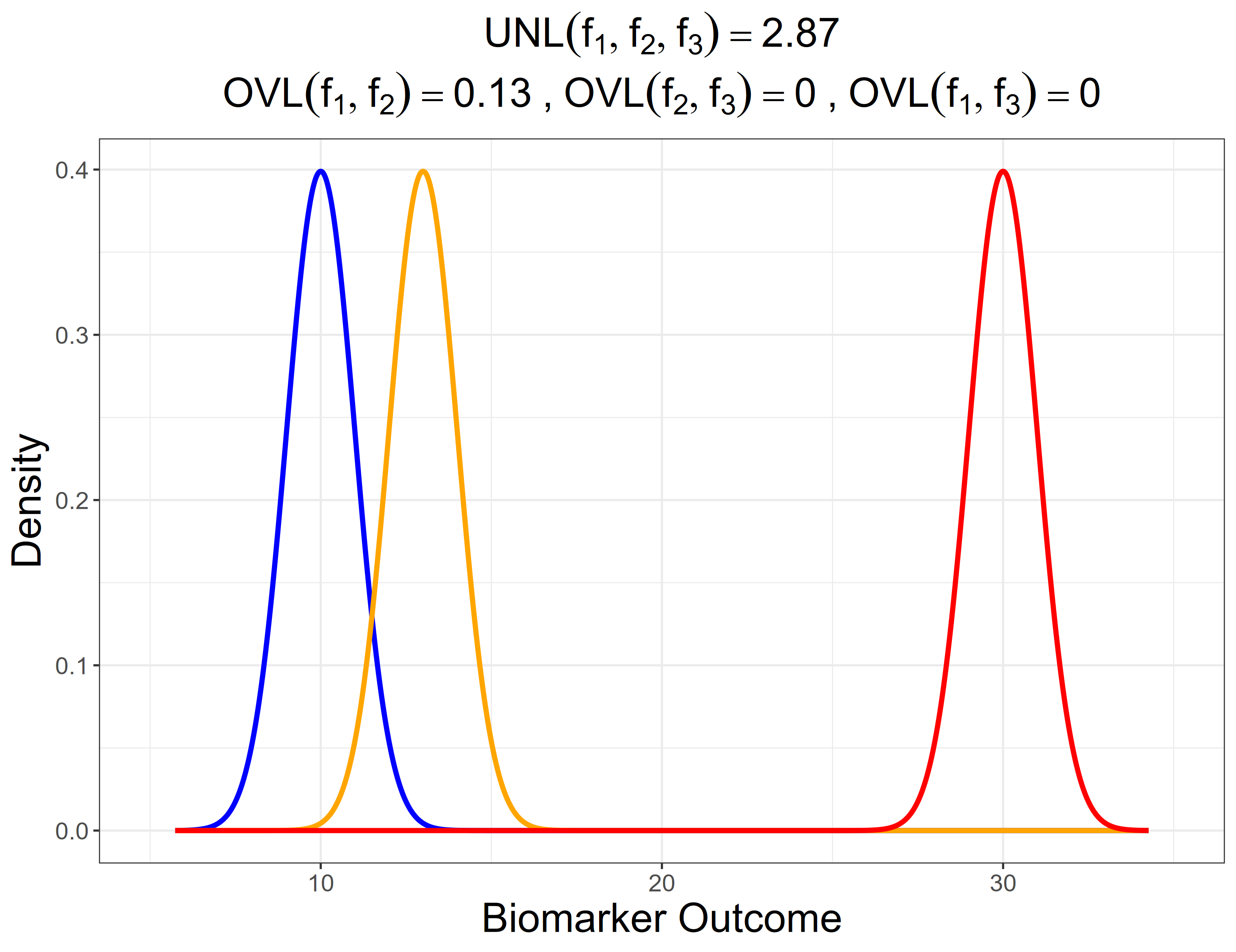}
	}
	\\
	\subfigure{
		\centering
		\includegraphics[width=0.28\textwidth]{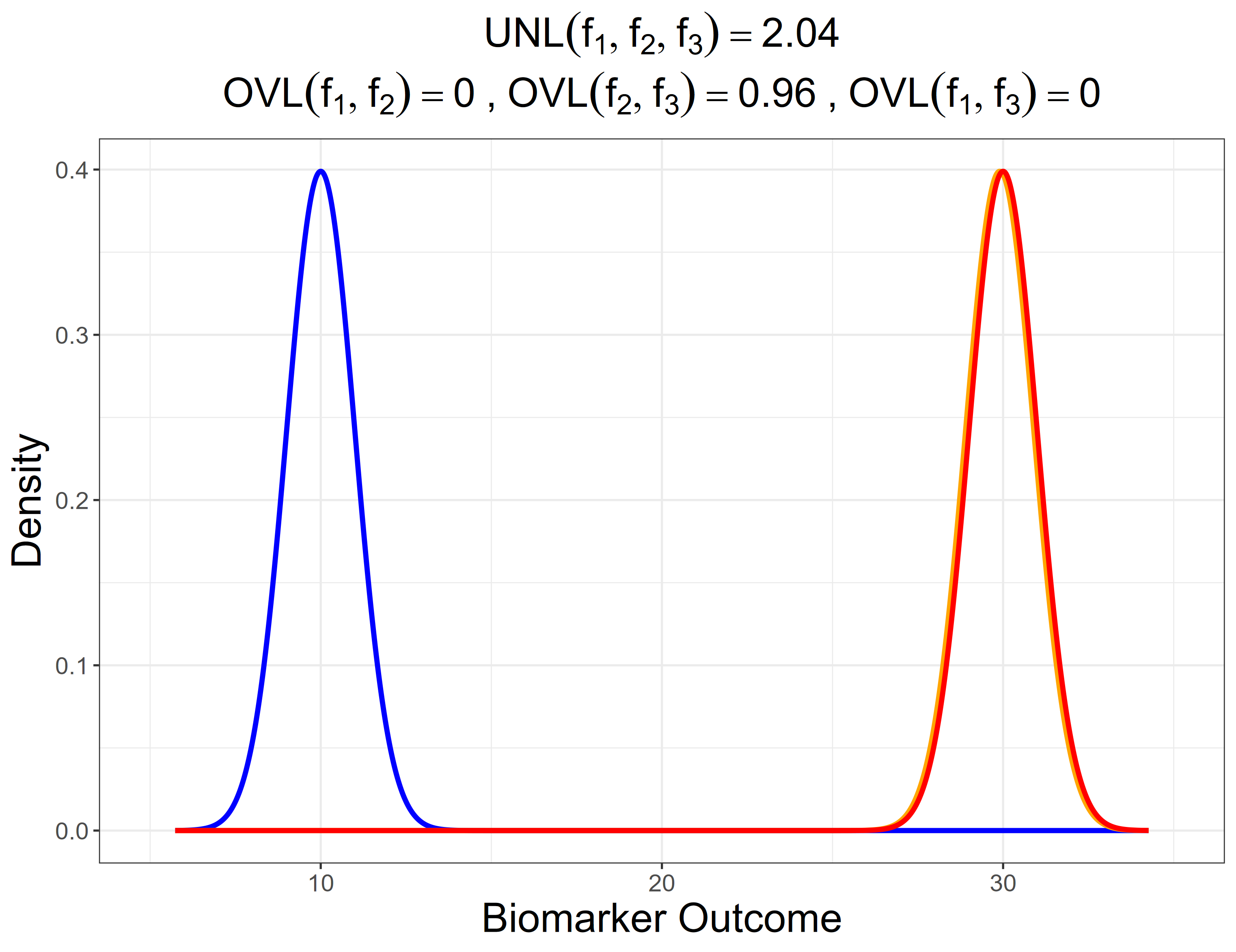}
	}
	\subfigure{
		\centering
		\includegraphics[width=0.28\textwidth]{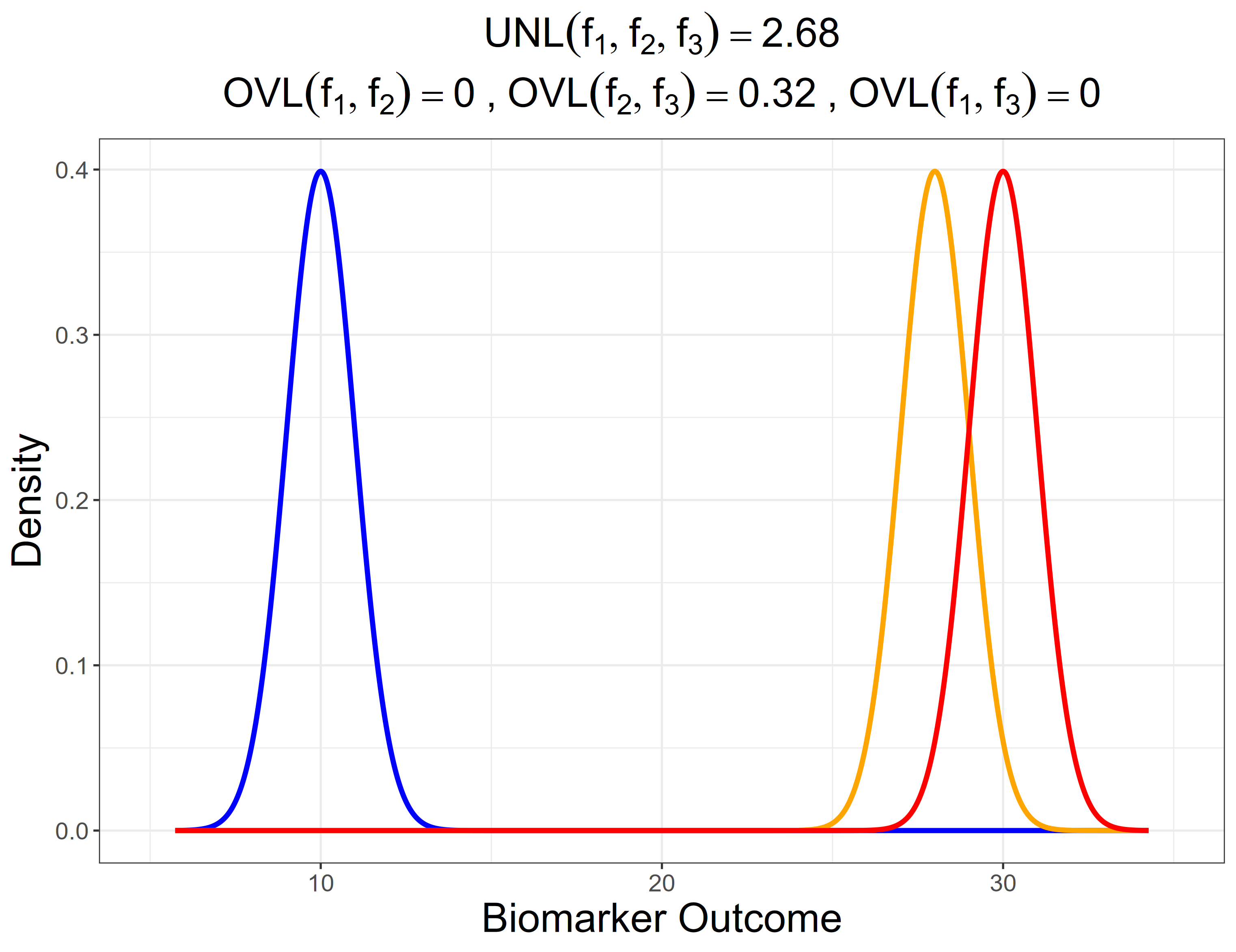}
	}
	\subfigure{
		\centering
		\includegraphics[width=0.28\textwidth]{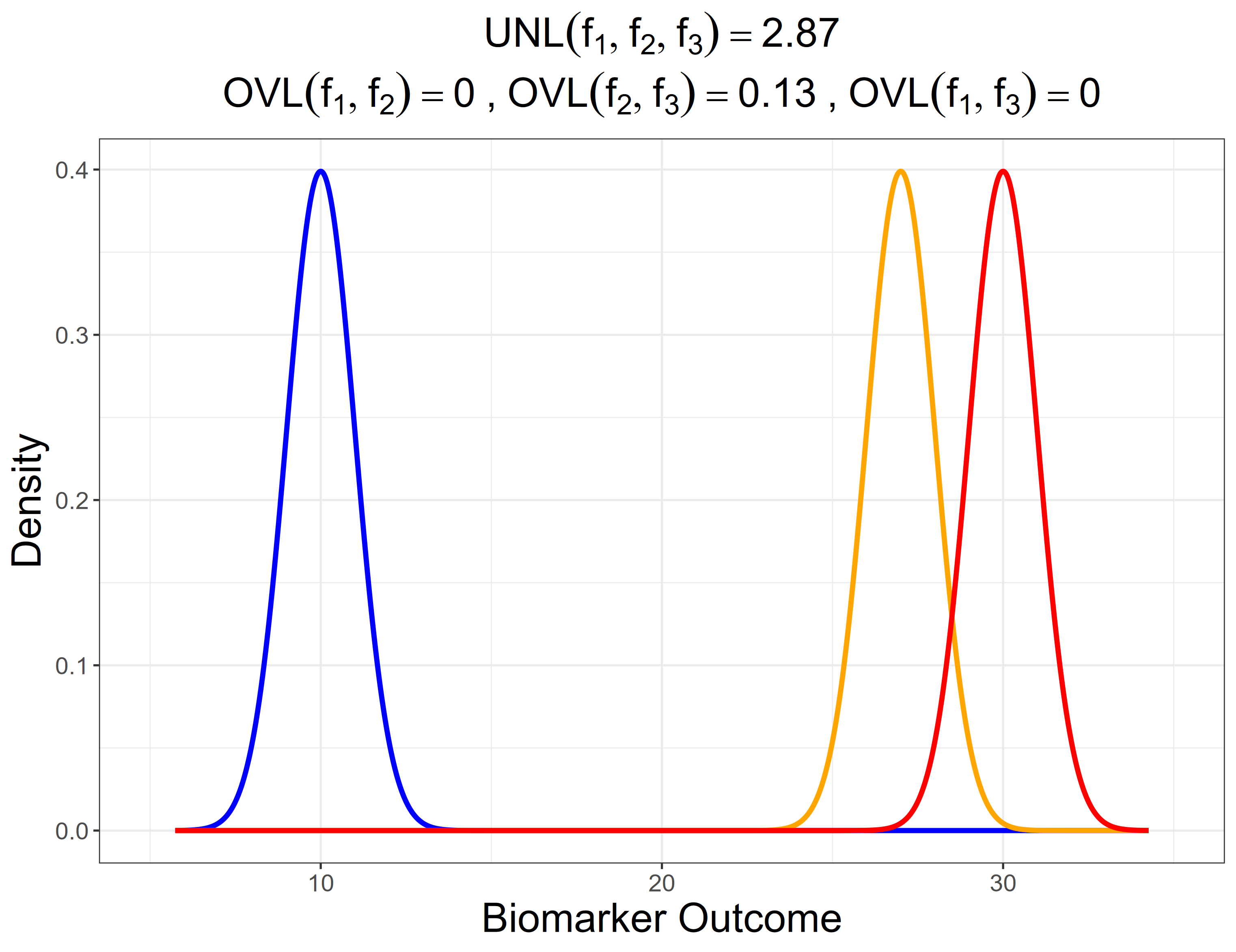}
	}
	\\
	\subfigure{
		\centering
		\includegraphics[width=0.28\textwidth]{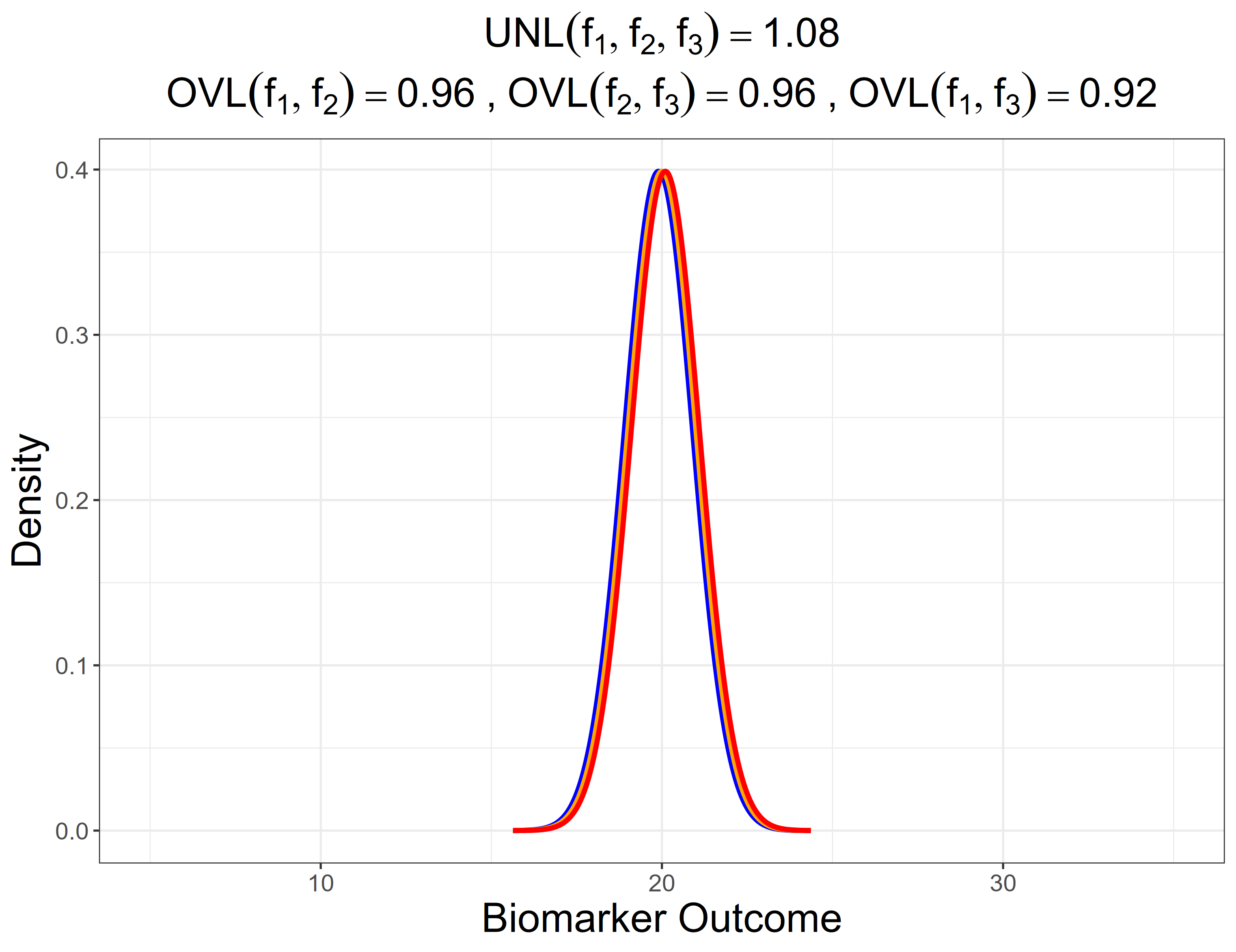}
	}
	\subfigure{
		\centering
		\includegraphics[width=0.28\textwidth]{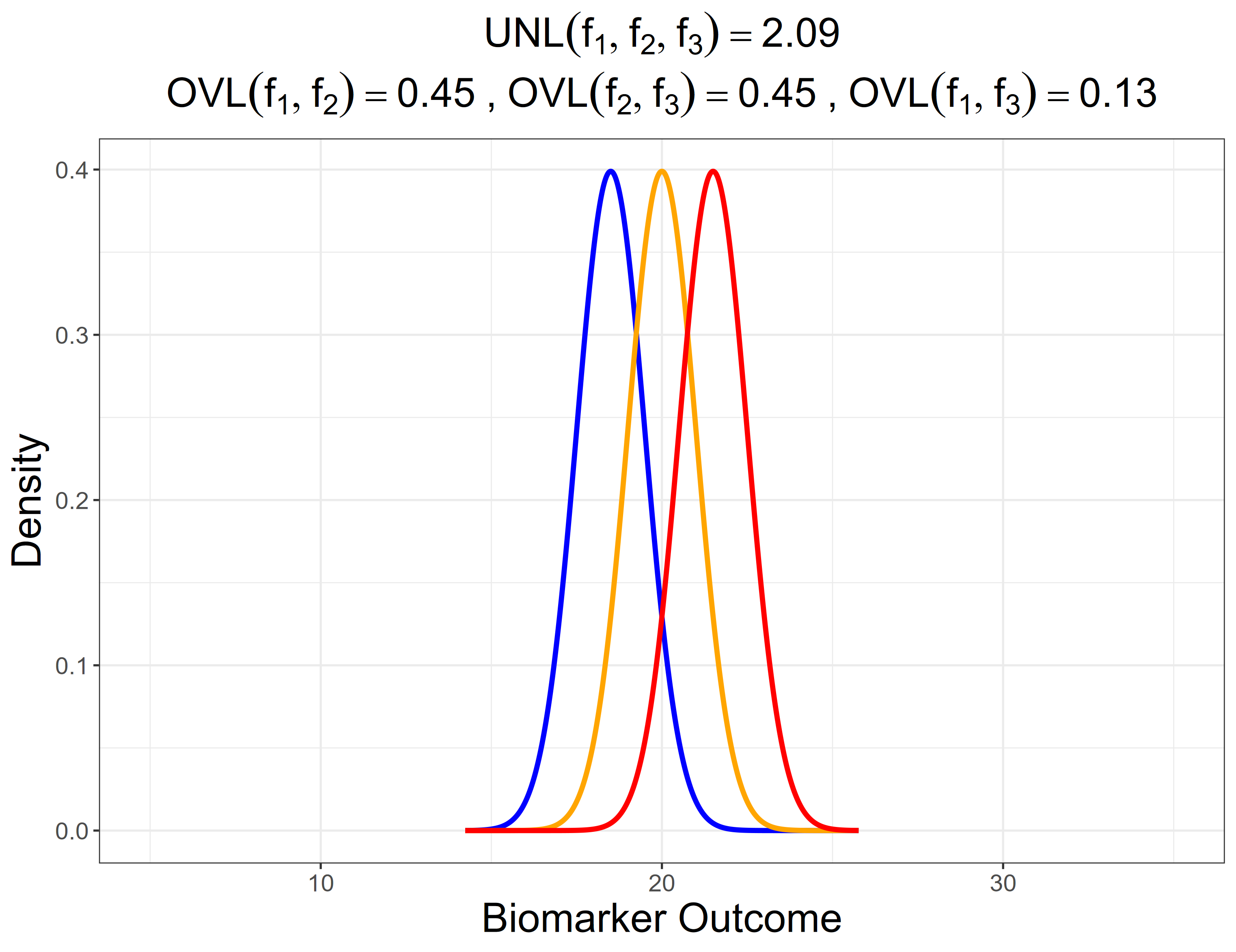}
	}
	\subfigure{
		\centering
		\includegraphics[width=0.28\textwidth]{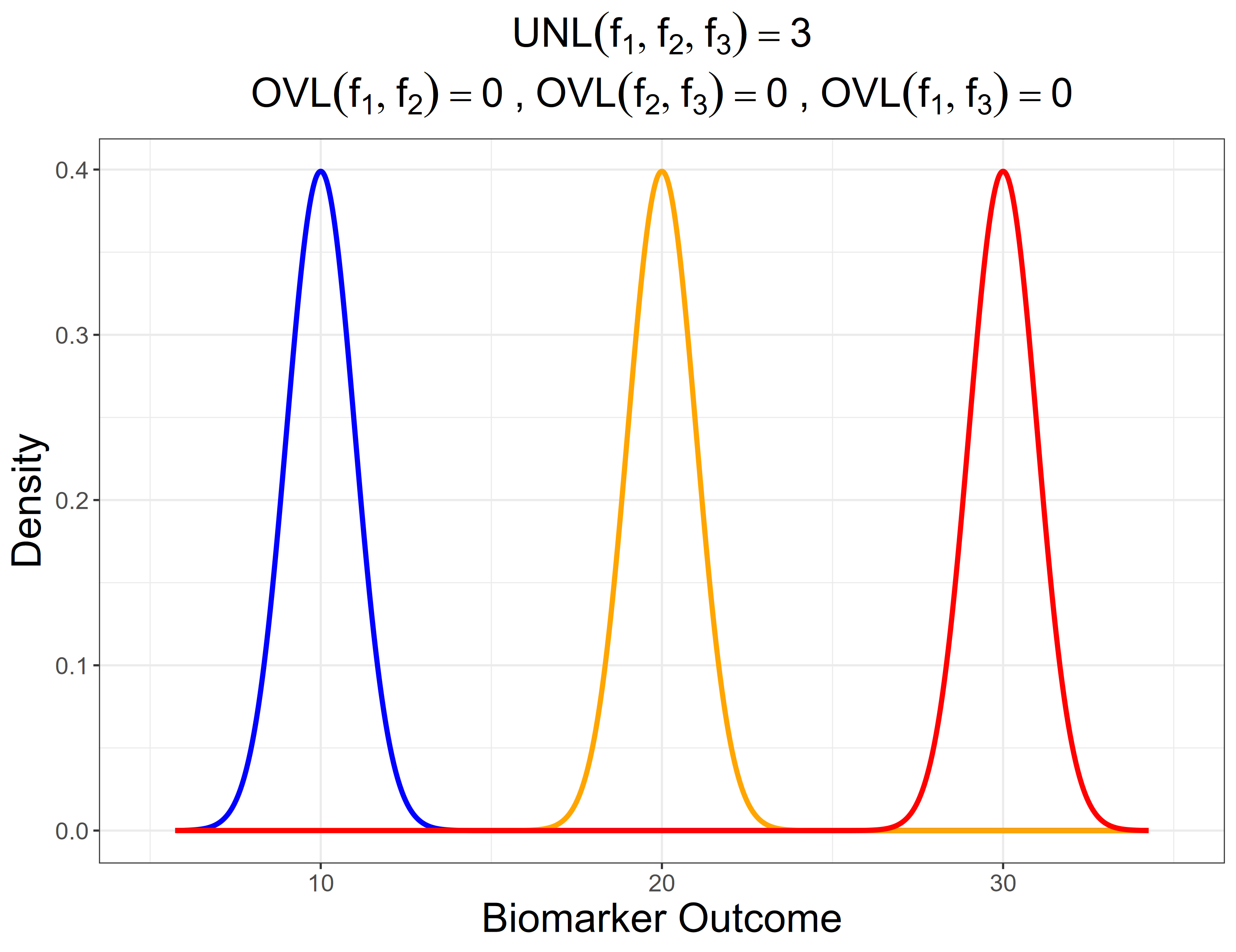}
	}
	\centering
	\caption{Examples of different degrees of separation across the three groups. The blue, orange, and red lines represent the densities of biomarker outcomes in groups 1, 2, and 3, respectively. }
	\label{separation_example_plots}
\end{figure}

\begin{figure}[H]
	\centering
	\subfigure{
		\centering
		\includegraphics[width=0.3\textwidth]{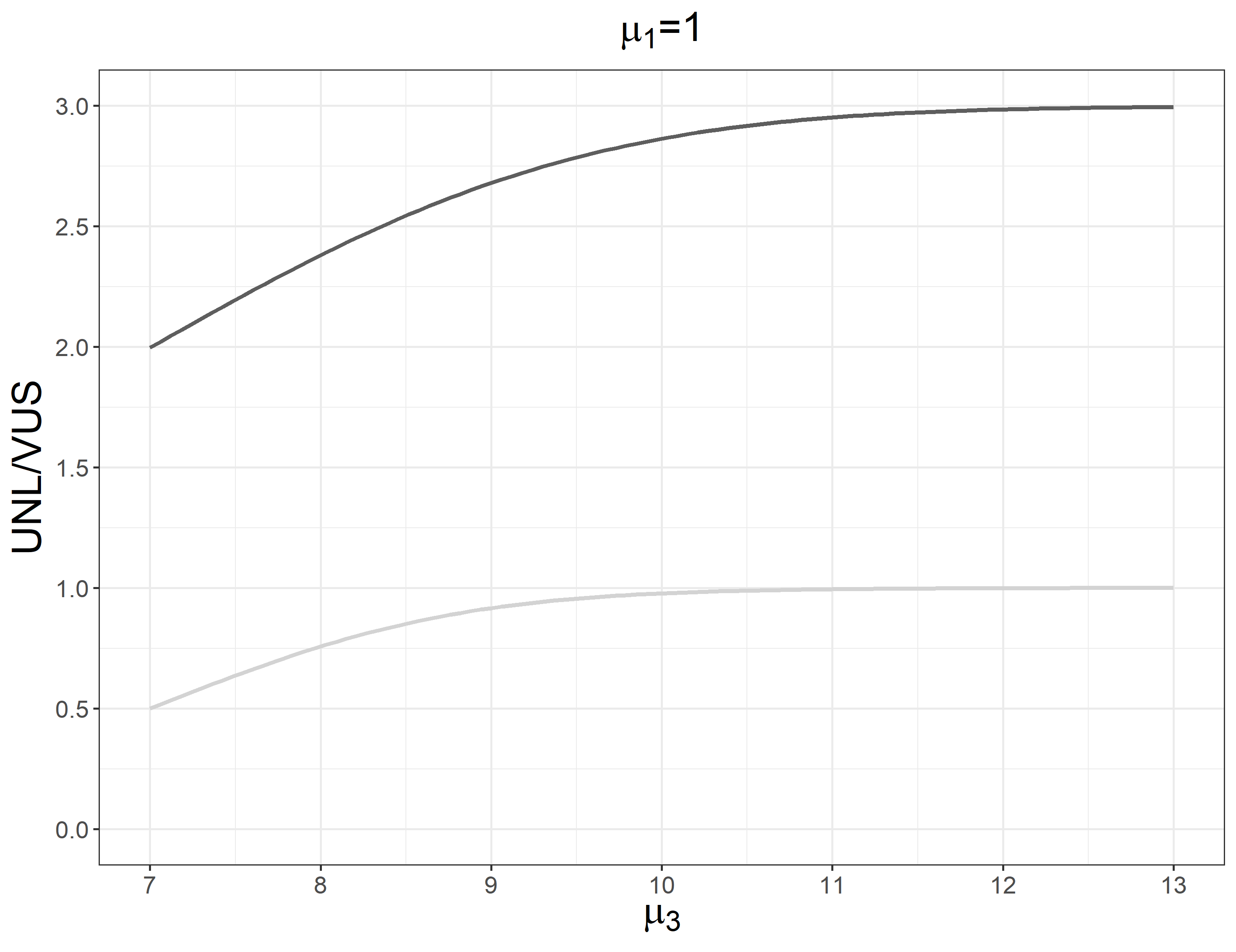}
	}
	\subfigure{
		\centering
		\includegraphics[width=0.3\textwidth]{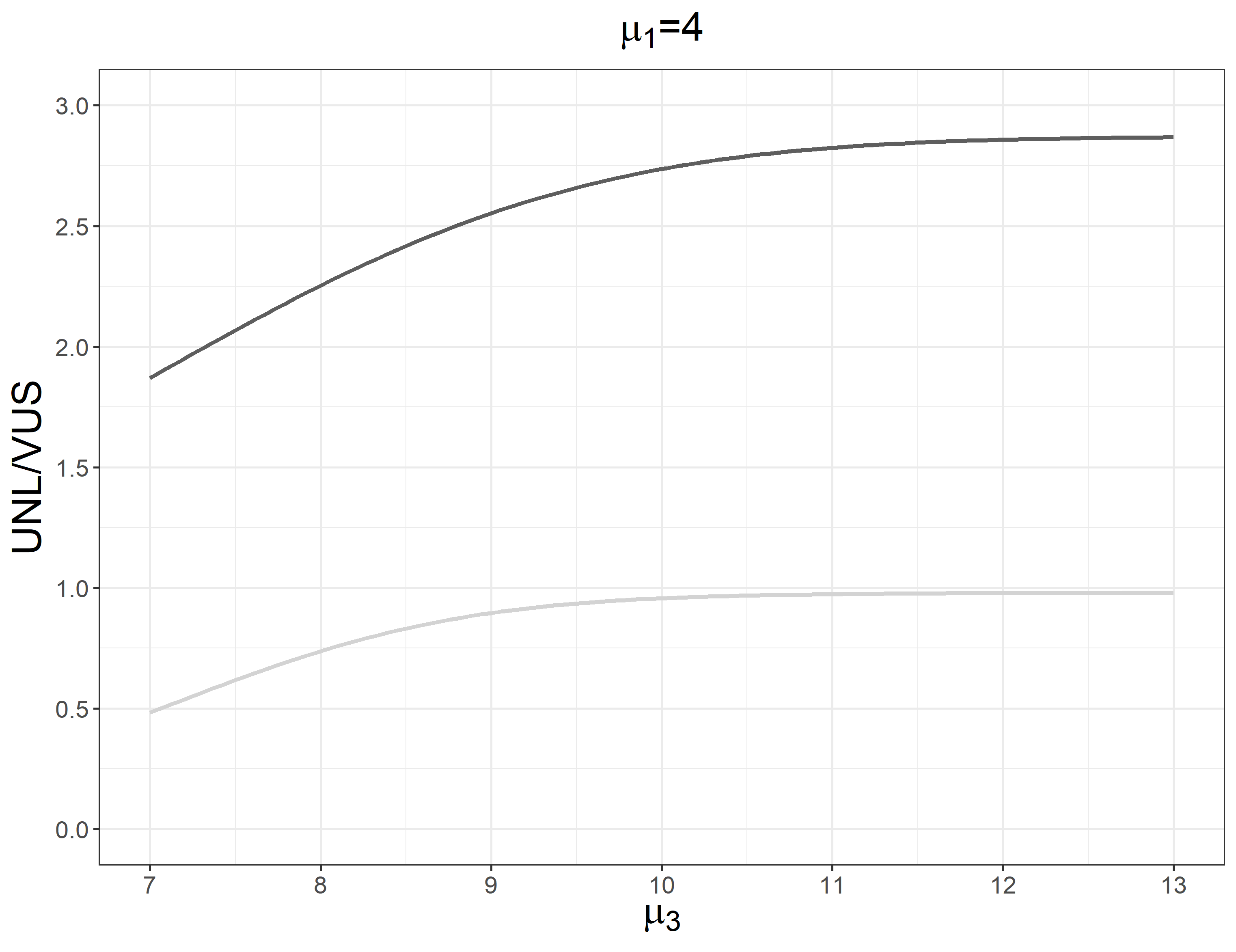}
	}
	\subfigure{
		\centering
		\includegraphics[width=0.3\textwidth]{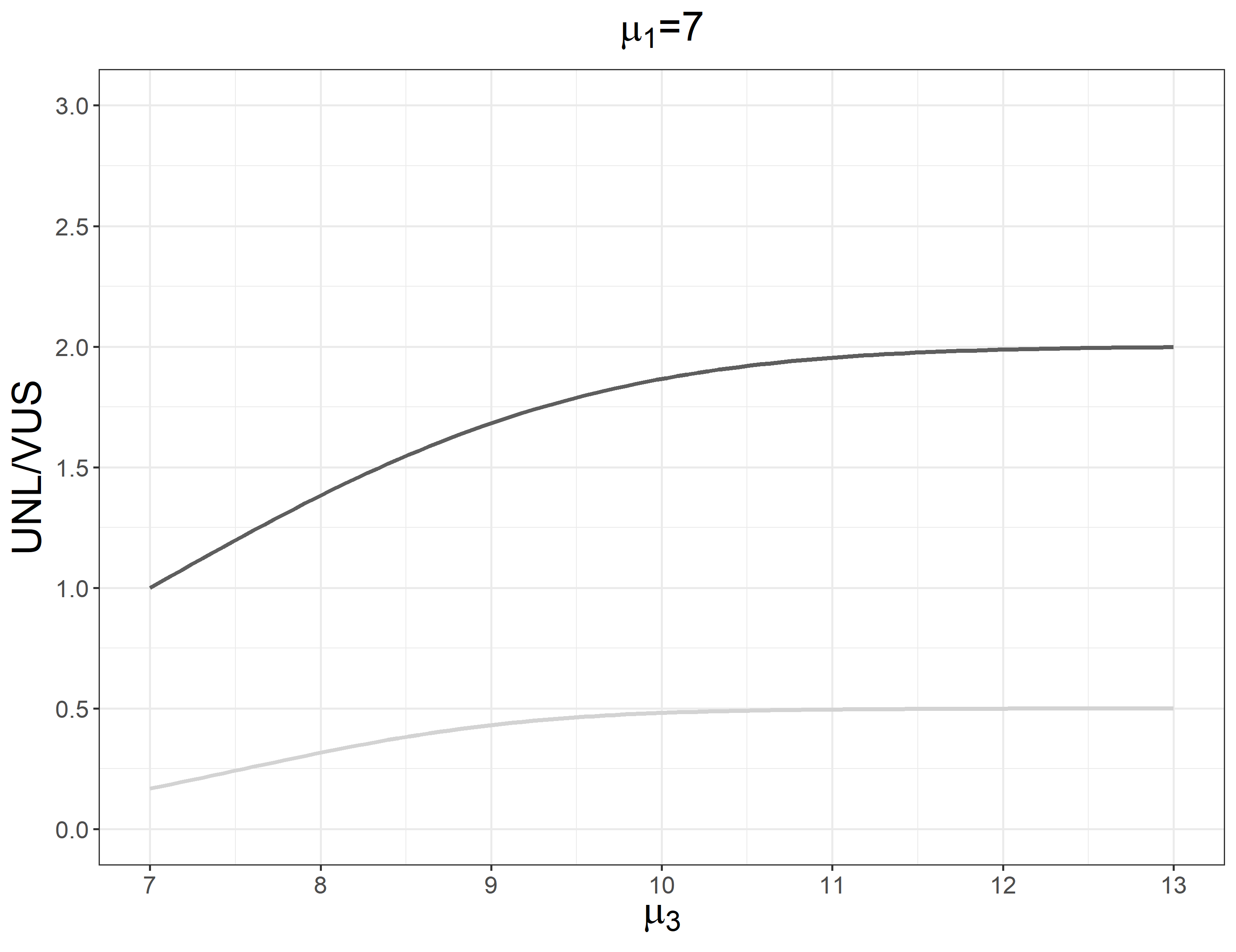}
	}
	\caption{VUS and UNL trace plots for the example in Section 2.4 of the main paper, where the densities of biomarker outcomes in all three groups follow a normal distribution with common variance. Here,  $\mu_2$ = 7. Dark grey line: UNL; light grey line: VUS.}
	\label{VUS_vs_UNL_2d_1}
\end{figure}
\clearpage

\subsection*{Posterior inference for (covariate-dependent) Dirichlet process mixtures of normals}
\subsubsection*{Gibbs sampler algorithm for the  Dirichlet process mixture of normals}
We describe the Gibbs sampler algorithm for the Dirichlet process mixture of normals model presented in Section 3.1. of the main text. Let $G_{1i}\in \{1,\ldots,L_1\}$ be a latent variable indicating the mixture component to which observation $i$ is assigned, with  $G_{1i}= l$ denoting that the $i$th observation is allocated to component $l$. The Gibbs sampling scheme proceeds by alternating through the following steps.

\begin{algorithm}
	\caption{Steps of the Gibbs sampler for the Dirichlet process mixture of normals model}\label{dpm_algorithm_table}
	\begin{algorithmic}
		\State \textbf{Step 1: Assign each observation $i=1,\ldots,n_1$ to a mixture component $l=1,\ldots,L_1$.}
		\For{$i = 1,\dots,n_1$}
		\State Sample $G_{1i}$ from $\{1,\dots,L_1\}$ with probabilities:
		\[
		\Pr(G_{1i}=l \mid \text{else})=\frac{\omega_{1l}\,\phi(y_{1i}\mid \mu_{1l},\sigma_{1l}^2)}
		{\sum_{l=1}^{L_1}\omega_{1l}\,\phi(y_{1i}\mid \mu_{1l},\sigma_{1l}^2)},\quad l=1,\ldots,L_1.
		\]
		\EndFor
		
		\State \textbf{Step 2: Update the inputs of the stick-breaking weights.}
		\For{$l = 1,\dots,L_1-1$}
		\State Sample 
		\[
		v_{1l} \mid \text{else} \sim \text{Beta}\Bigl(n_{1l}+1,\ \alpha_1+\sum_{m=l+1}^{L_1} n_{1m}\Bigr)
		\]
		\State where $n_{1l}=\sum_{i=1}^{n_1} I(G_{1i}=l)$.
		\EndFor
		
		\State \textbf{Step 3: Update the mean and variance parameters for each component.}
		\For{$l = 1,\dots,L_1$}
		\State Sample 
		\[
		\mu_{1l} \mid \text{else} \sim \text{N}\Biggl(
		\frac{a_{\mu_{1}}/b_{\mu_1}^2+\sum_{i:G_{1i}=l} y_{1i}/\sigma_{1l}^2}{1/b_{\mu_1}^2+n_{1l}/\sigma_{1l}^2},\ 
		\frac{1}{1/b_{\mu_1}^2+n_{1l}/\sigma_{1l}^2}
		\Biggr),
		\]
		\State Sample 
		\[
		\sigma_{1l}^2 \mid\text{else} \sim \text{IG}\Bigl(a_{\sigma^2_1}+\frac{n_{1l}}{2},\ b_{\sigma^2_1}+\frac{1}{2}\sum_{i:G_{1i}=l}(y_{1i}-\mu_{1l})^2\Bigr).
		\]
		\EndFor
		
	\end{algorithmic}
\end{algorithm}
\noindent An analogous scheme applies to groups 2 and 3.

\clearpage
\subsubsection*{Gibbs sampler algorithm for the covariate-dependent Dirichlet process mixture of normals}
We describe the Gibbs sampling scheme for the covariate-dependent Dirichlet process mixture, where the dependence of the stick-breaking weights on covariates is modelled through a logit stick-breaking process. For full details, we refer the reader to \cite{rigon2021tractable}. Our methods for conducting inference about the covariate-specific underlap coefficient are implemented using the \texttt{R} package \texttt{LSBP} developed by the authors.
Below, $\mathbf{Z}_{1l}$ and $\mathbf{U}_{1l}$ denote the $\bar{n}_{1l}\times Q_{1}^{v}$ and $n_{1l}\times Q_1^{\mu}$ predictor matrices with row entries $\mathbf{z}_{1i}^{\prime}$  and $\mathbf{u}_{1i}^{\prime}$, for observations $i$ such that $G_{1i} = l$ and $G_{1i}>l-1$, respectively. The Gibbs sampler cycles through the following steps.

\begin{algorithm}
	\caption{Steps of the Gibbs sampler for a covariate-specific mixture of normal distributions with a logit stick-breaking prior}\label{lsbp_conditional_algorithm_table}
	\begin{algorithmic}
		
		\State \textbf{Step 1: Assign each observation $i=1,\ldots,n_1$ to a mixture component $l=1,\ldots,L_1$.}
		\For{$i = 1,\dots,n_1$}
		\State Sample $G_{1i}$ from $\{1,\dots,L_1\}$ with probabilities:
		\begin{align*}
			&\Pr(G_{1i}=l\mid\text{else})=\frac{\omega_{1l}(\mathbf{x}_{1i})\phi(y_{1i}\mid \mu_1(\mathbf{x}_{1i},\boldsymbol{\beta}_{1l}),\sigma_{1l}^2)}{\sum_{l=1}^{L_1}\omega_{1l}(\mathbf{x}_{1i})\phi(y_{1i}\mid \mu_1(\mathbf{x}_{1i},\boldsymbol{\beta}_{1l}),\sigma_{1l}^2)},\quad l=1,\ldots, L_1.\\
			& \omega_{11}(\mathbf{x}_{1i}) = v_{11}(\mathbf{x}_{1i}),\quad \omega_{1l}(\mathbf{x}_{1i}) = v_{1l}(\mathbf{x}_{1i})\prod_{m=1}^{l-1}\{1-v_{1m}(\mathbf{x}_{1i})\},\quad \text{logit}(v_{1l}(\mathbf{x}_{1i})) = \mathbf{z}_{1i}^{\prime}\boldsymbol{\gamma}_{1l},\quad l\geq 1.
		\end{align*}
		\EndFor
		
		\State \textbf{Step 2: Update $\boldsymbol{\gamma}_{1l}$ by exploiting the reparametrisation of the weights as a set sequential logistic regressions and the  P\'olya-gamma data augmentation \citep{polson2013bayesian}.}
		\For{$l = 1,\dots,L_1-1$}
		\For{each $i$ with $G_{1i} > l-1$}
		\State Sample the P\'olya-gamma data $\zeta_{1il}$ from  $\zeta_{1il}\mid \text{else} \sim \text{PG}(1,\mathbf{z}_{1i}^{\prime}\boldsymbol{\gamma}_{1l})$.
		\EndFor
		\State Based on the P\'olya-gamma data, update $\boldsymbol{\gamma}_{1l}$ from $\boldsymbol{\gamma}_{1l}\mid\text{else} \sim\text{N}_{Q_1^{v}}(\mu_{\boldsymbol{\gamma}_{1l}}^{*},\Sigma_{\boldsymbol{\gamma}_{1l}}^{*})$, where
		\begin{align*}
			\mu_{\boldsymbol{\gamma}_{1l}}^{*} = \Sigma_{\boldsymbol{\gamma}_{1l}}^{*}\Bigl\{\mathbf{Z}_{1l}^{\prime}(g_{11l}-0.5,\ldots,g_{1\bar{n}_{1l}l}-0.5)^{\prime}+\boldsymbol{\Sigma}_{\boldsymbol{\gamma}_1}^{-1}\mathbf{\mu}_{\boldsymbol{\gamma}_1}\Bigr\},\quad\Sigma_{\boldsymbol{\gamma}_{1l}}^{*} = \Bigl\{ \mathbf{Z}_{1l}^{\prime}\operatorname{diag}(\zeta_{11l},\dots,\zeta_{1\bar{n}_{1l}l})\mathbf{Z}_{1l}+\boldsymbol{\Sigma}_{\boldsymbol{\gamma}_1}^{-1}\Bigr\}^{-1},
		\end{align*}
		with $g_{1il}=1$ if $G_{1i}=l$ and $g_{1il}=0$ if $G_{1i}>l$.
		\EndFor
		
		\State \textbf{Step 3: Update $\boldsymbol{\beta}_{1l}$.}
		\For{$l = 1,\dots,L_1$}
		\State Sample $\boldsymbol{\beta}_{1l}$ from the full conditional distribution $\boldsymbol{\beta}_{1l}\mid \text{else}\sim\text{N}_{Q_1^{u}}(\mu_{\boldsymbol{\beta}_{1l}}^{*},\Sigma_{\boldsymbol{\beta}_{1l}}^{*})$, where
		\begin{align*}
			\mu_{\boldsymbol{\beta}_{1l}}^{*} = \Sigma_{\boldsymbol{\beta}_{1l}}^{*}\Bigl\{\sigma_{1l}^{-2}\mathbf{U}_{1l}^{\prime} \mathbf{y}_{1l}+\boldsymbol{\Sigma}_{\boldsymbol{\beta}_1}^{-1} \mathbf{\mu}_{\boldsymbol{\beta}_1}\Bigr\}, \quad
			\Sigma_{\boldsymbol{\beta}_{1l}}^{*} = \Bigl\{\sigma_{1l}^{-2}\mathbf{U}_{1l}^{\prime}\mathbf{U}_{1l}+\boldsymbol{\Sigma}_{\boldsymbol{\beta}_1}^{-1}\Bigr\}^{-1},
		\end{align*}
		where $\mathbf{y}_{1l}$ is the $n_{1l}\times 1$ vector of biomarker outcomes for all subjects with $G_{1i}=l$.
		\EndFor
		
		\State \textbf{Step 4: Update $\sigma_{1l}^2$.}
		\For{$l = 1,\dots,L_1$}
		\State Sample $\sigma_{1l}^2$ from 
		\begin{equation*}
			\sigma_{1l}^2\mid\text{else}\sim
			\text{IG}\Bigl(a_{\sigma^2_1}+\frac{n_{1l}}{2},\,b_{\sigma^2_1}+\frac{1}{2}\sum_{i:G_{1i}=l}(y_{1i}-\mathbf{u}_{1i}^{\prime}\boldsymbol{\beta}_{1l})^2\Bigr),\quad n_{1l}=\sum_{i=1}^{n_1} I(G_{1i}=l).
		\end{equation*}    
		\EndFor
		\end{algorithmic}
\end{algorithm}
\noindent An analogous scheme applies to groups 2 and 3.

\clearpage

\subsection*{Simulation study: additional figures and tables}
\subsubsection*{Unconditional case}
\begin{figure}[htpb]
	\centering
	\subfigure{
		\centering
		\includegraphics[width=0.28\textwidth]{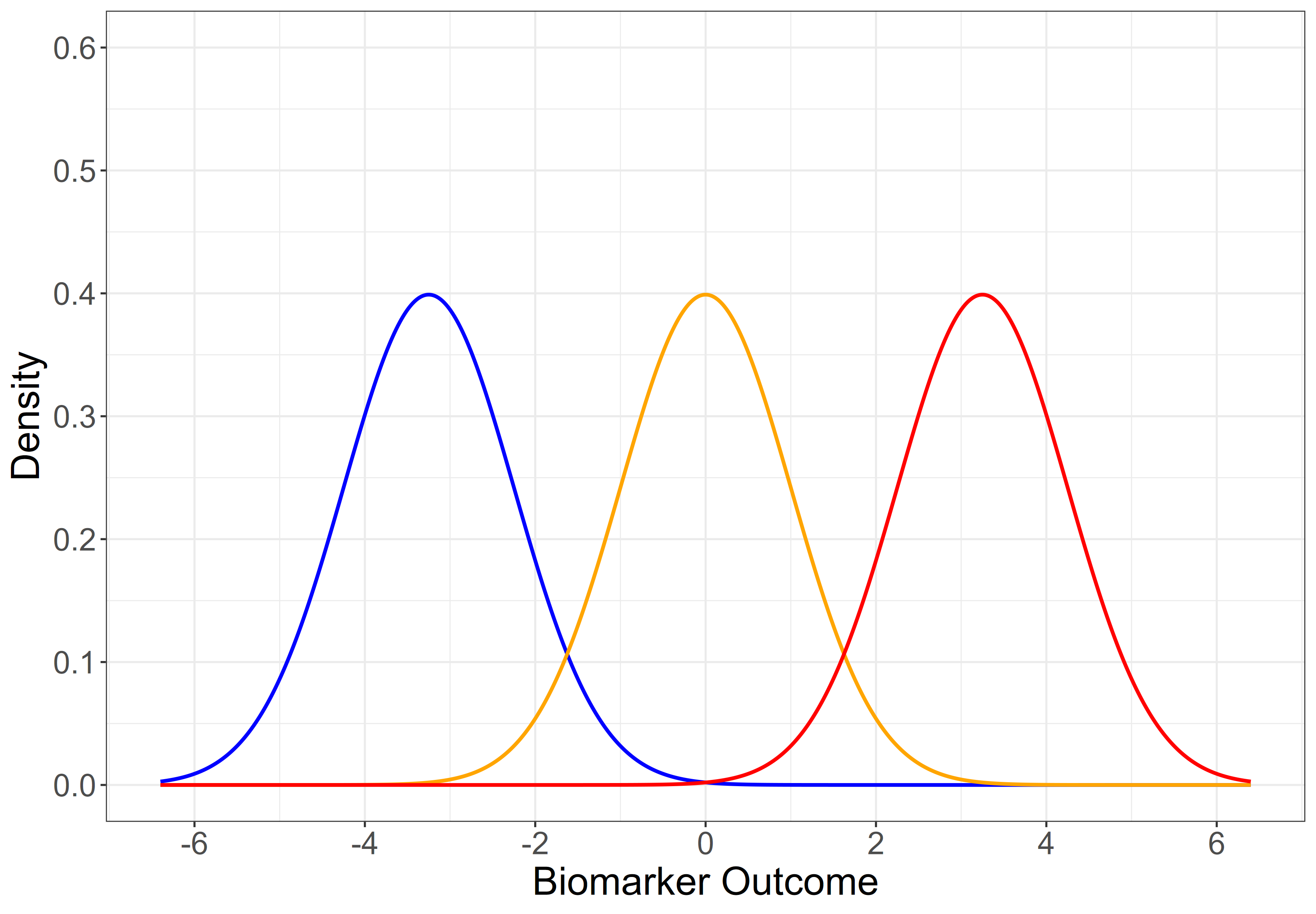}
	}
	\subfigure{
		\centering
		\includegraphics[width=0.28\textwidth]{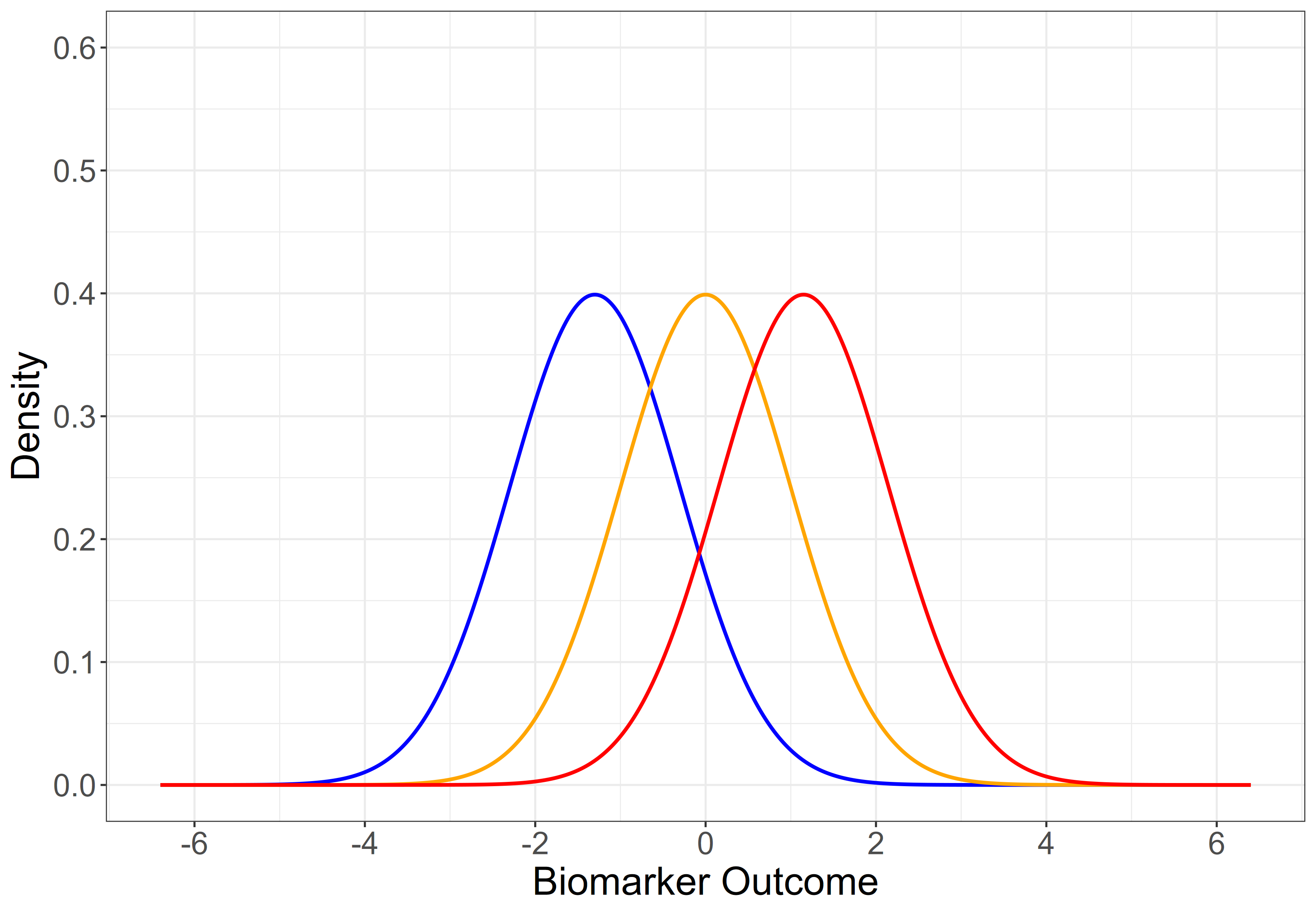}
	}
	\subfigure{
		\centering
		\includegraphics[width=0.28\textwidth]{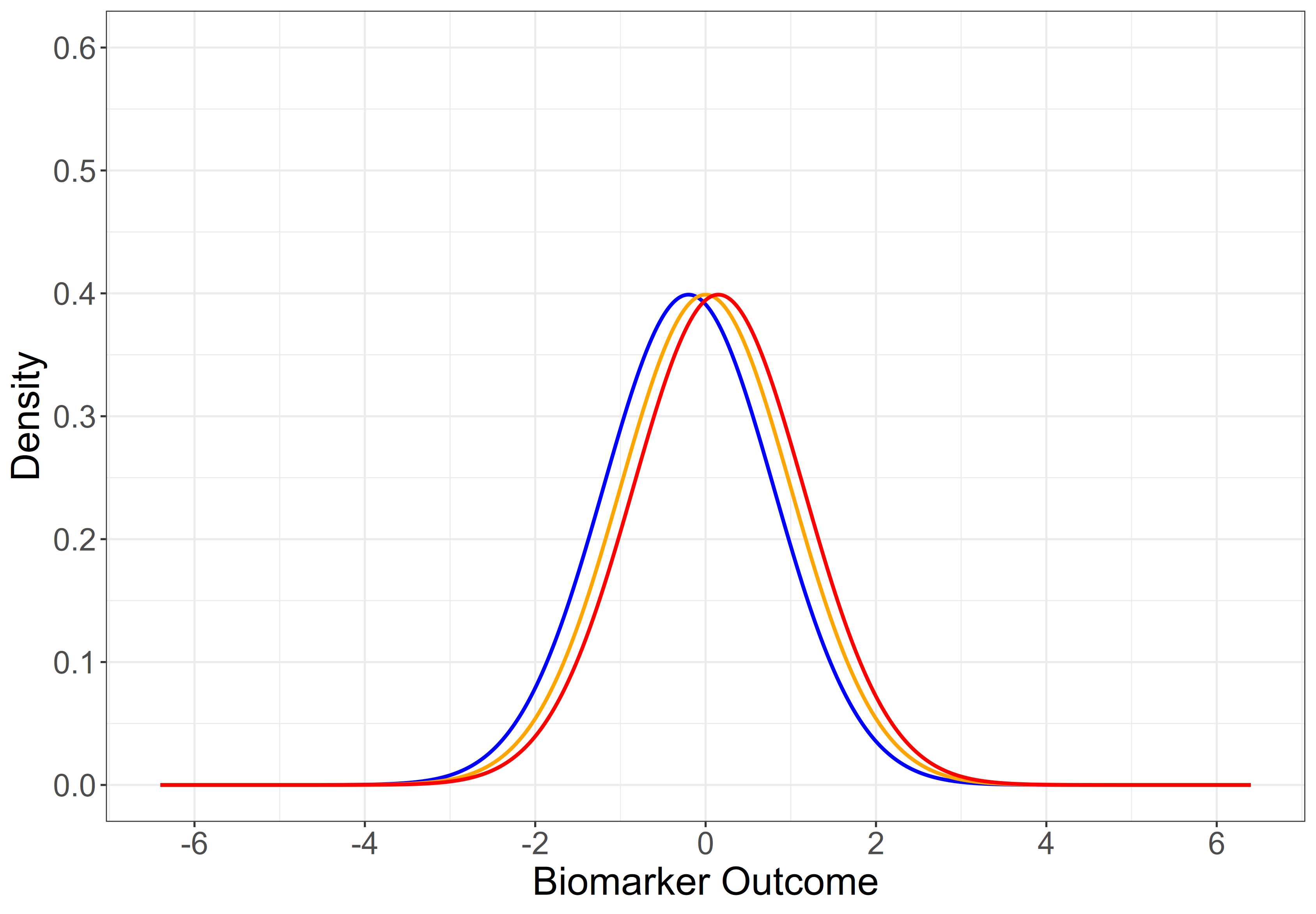}
	}
	\\
	\subfigure{
		\centering
		\includegraphics[width=0.28\textwidth]{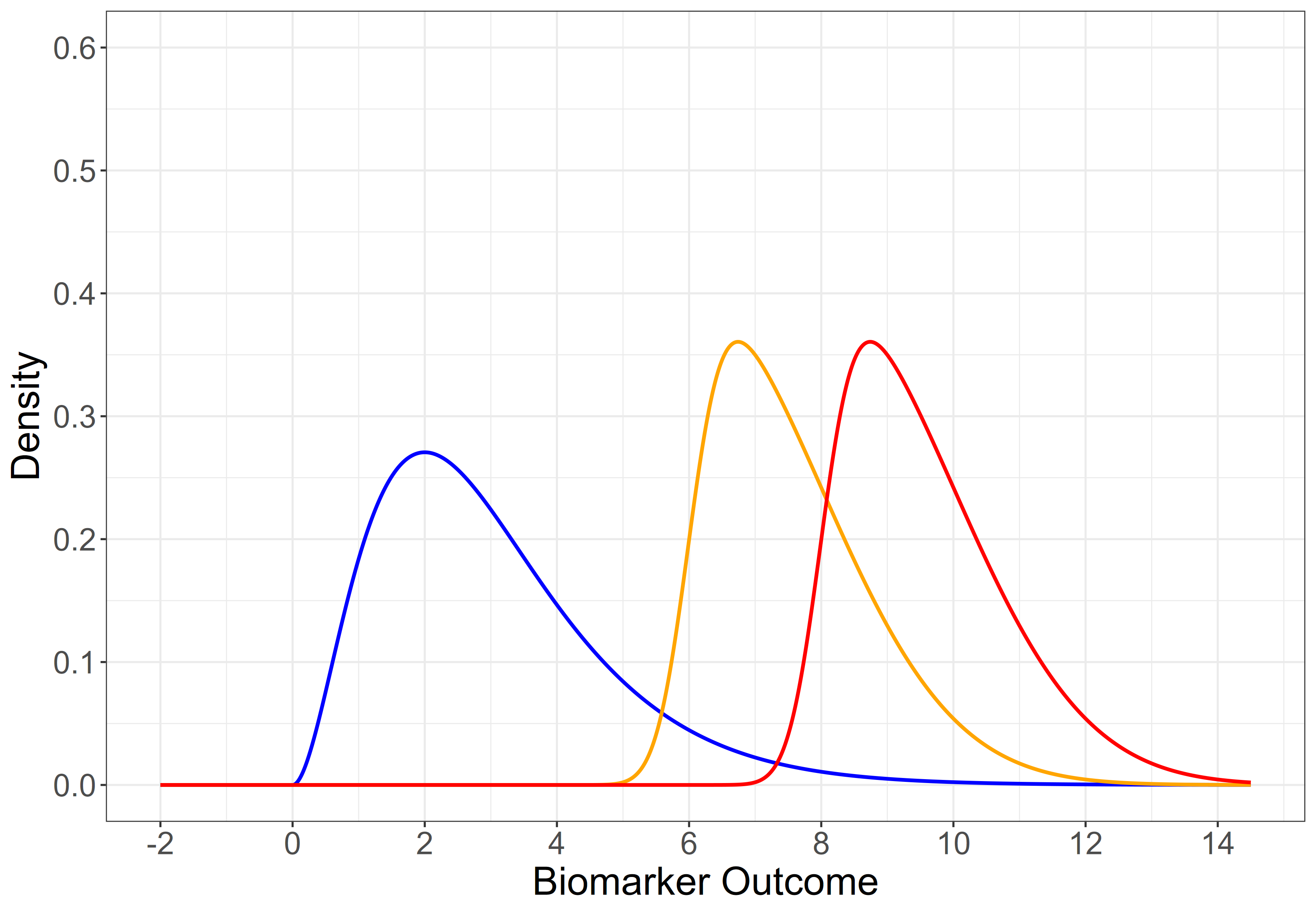}
	}
	\subfigure{
		\centering
		\includegraphics[width=0.28\textwidth]{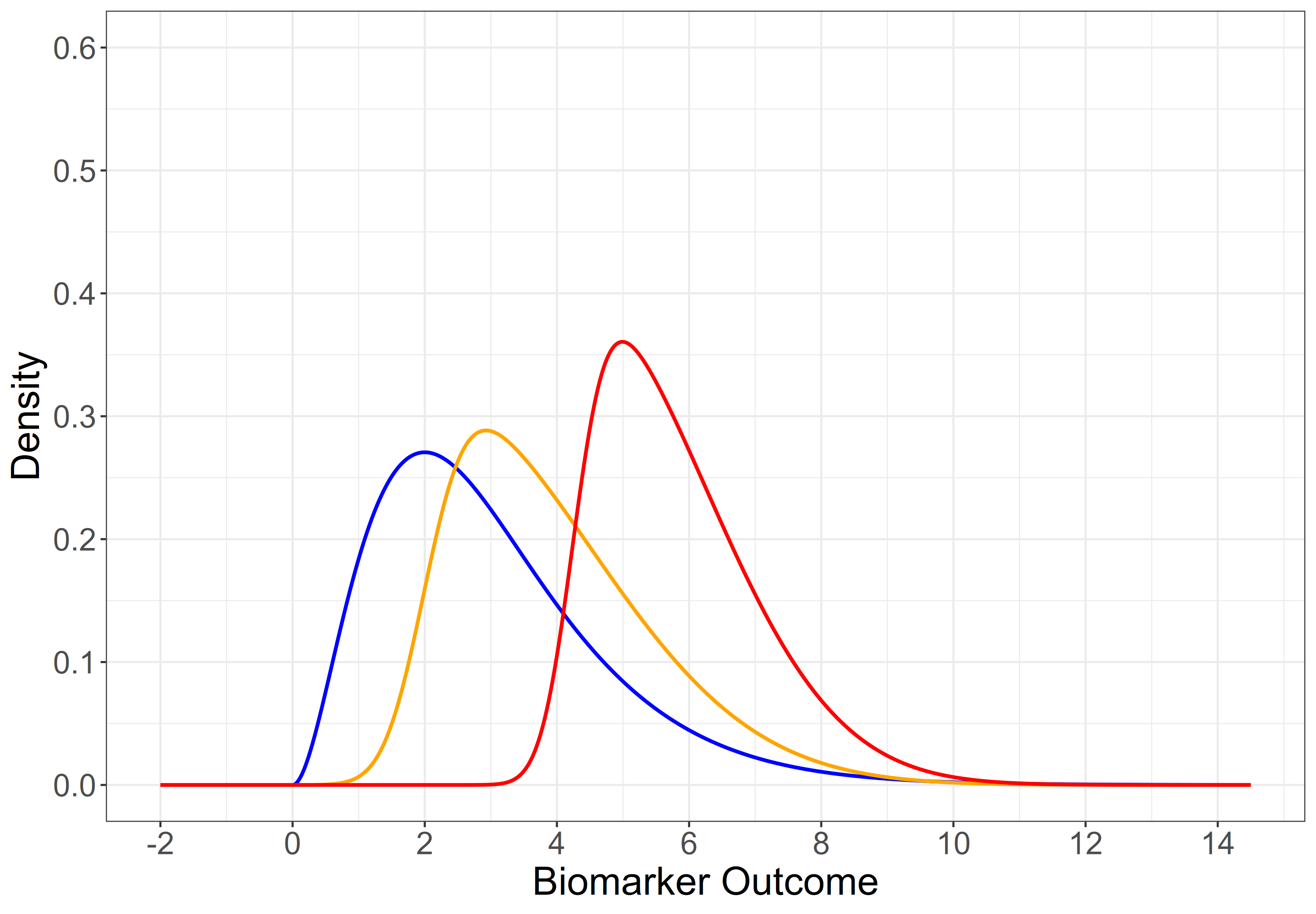}
	}
	\subfigure{
		\centering
		\includegraphics[width=0.28\textwidth]{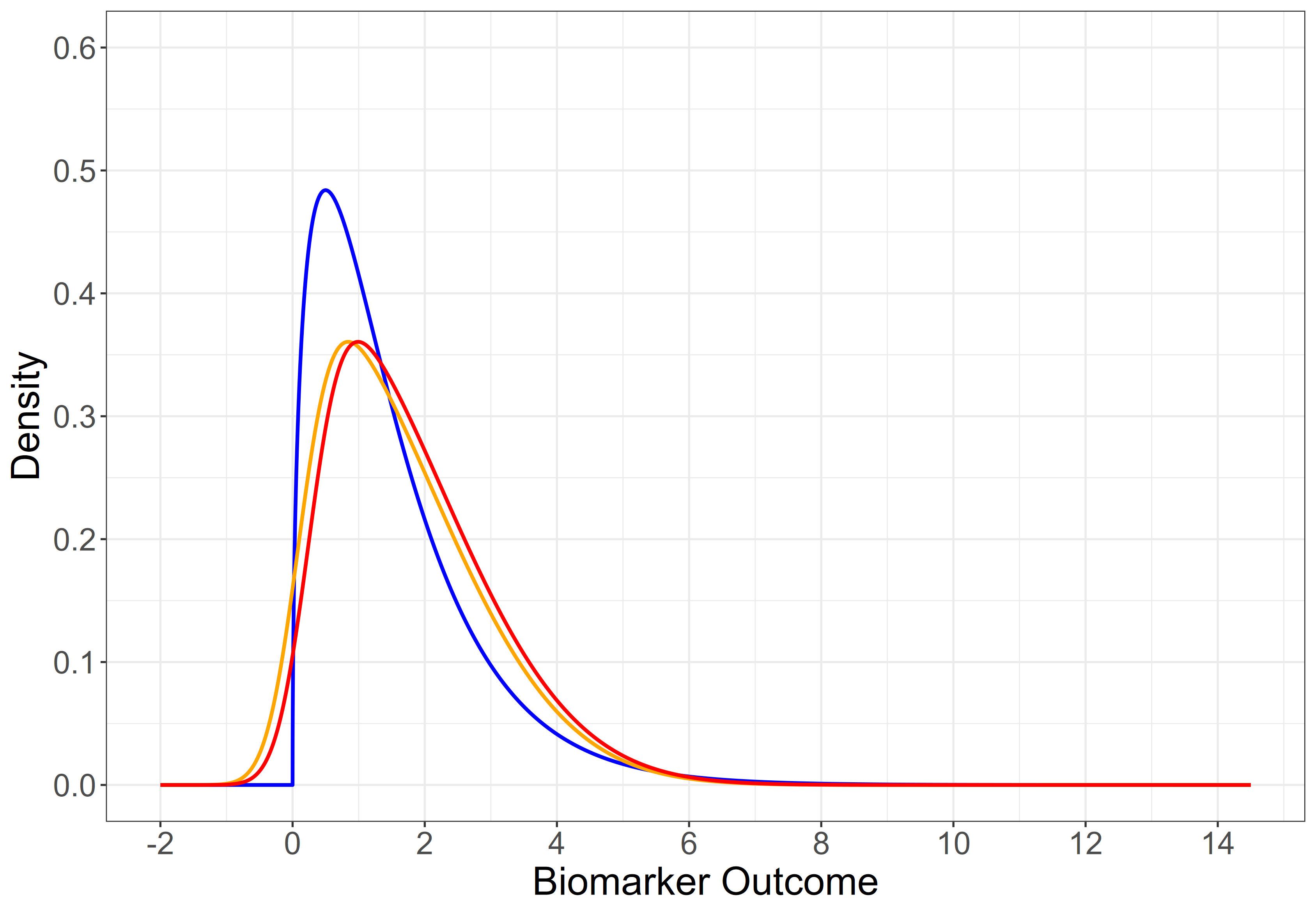}
	}
	\\
	\subfigure{
		\centering
		\includegraphics[width=0.28\textwidth]{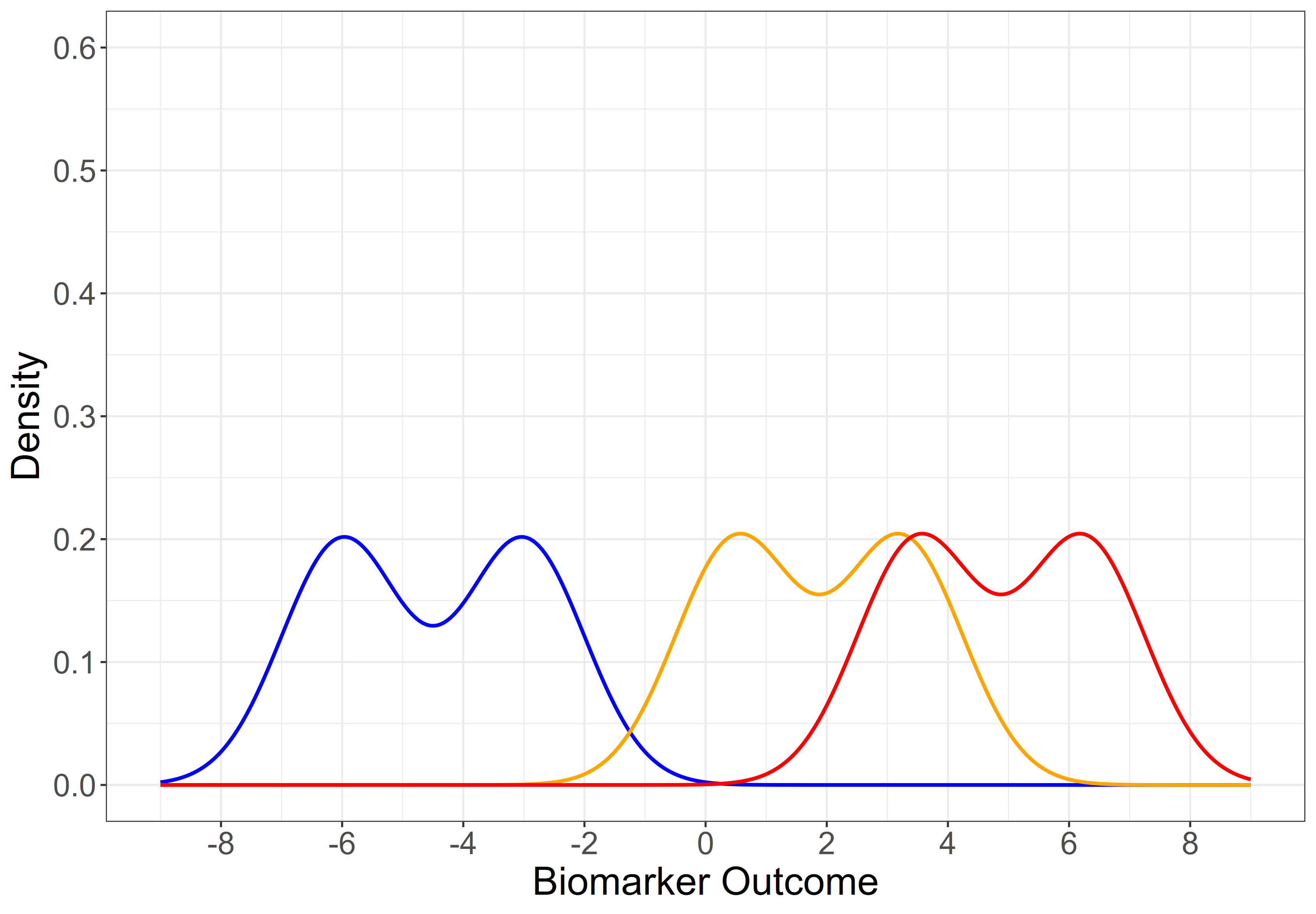}
	}
	\subfigure{
		\centering
		\includegraphics[width=0.28\textwidth]{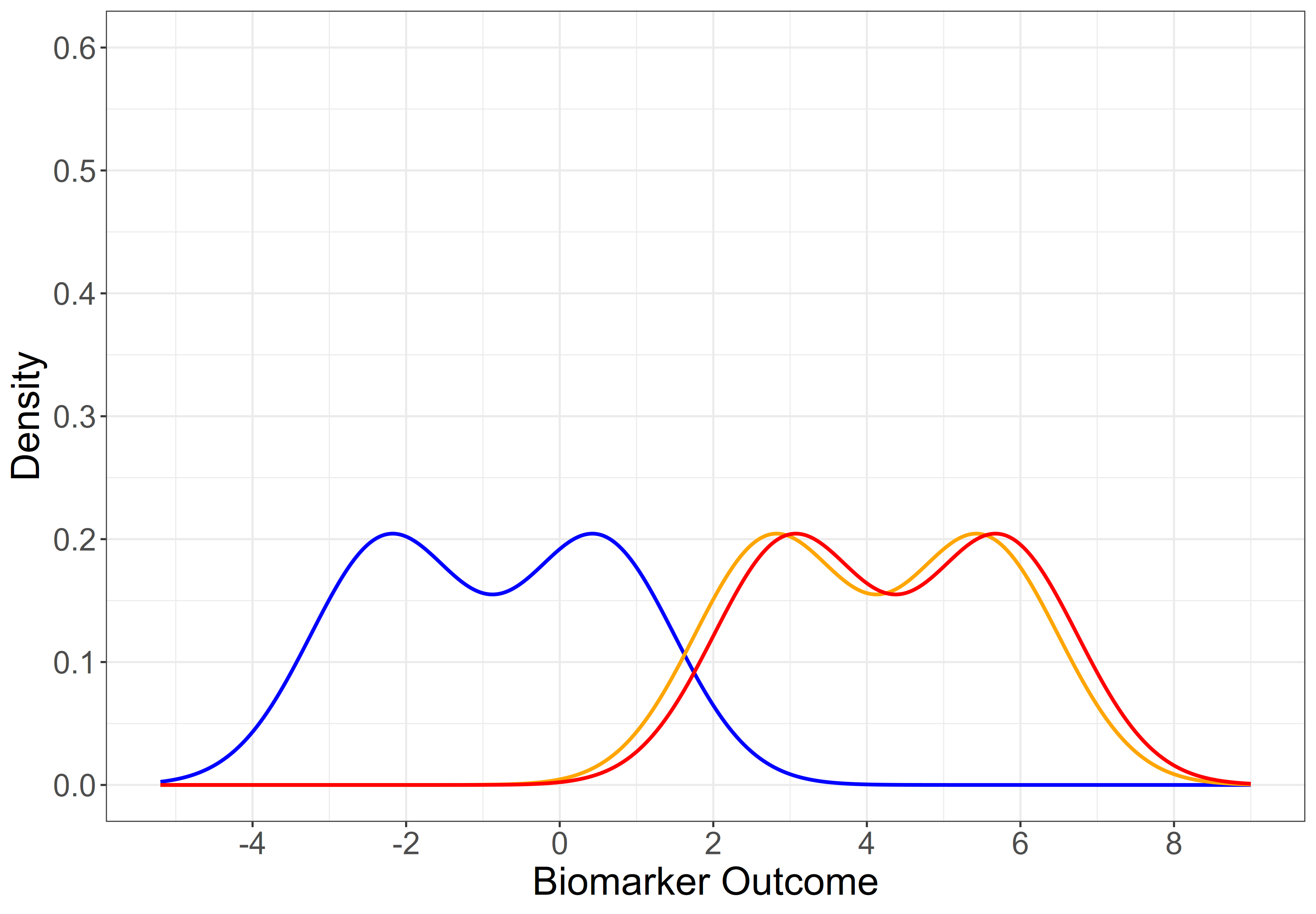}
	}
	\subfigure{
		\centering
		\includegraphics[width=0.28\textwidth]{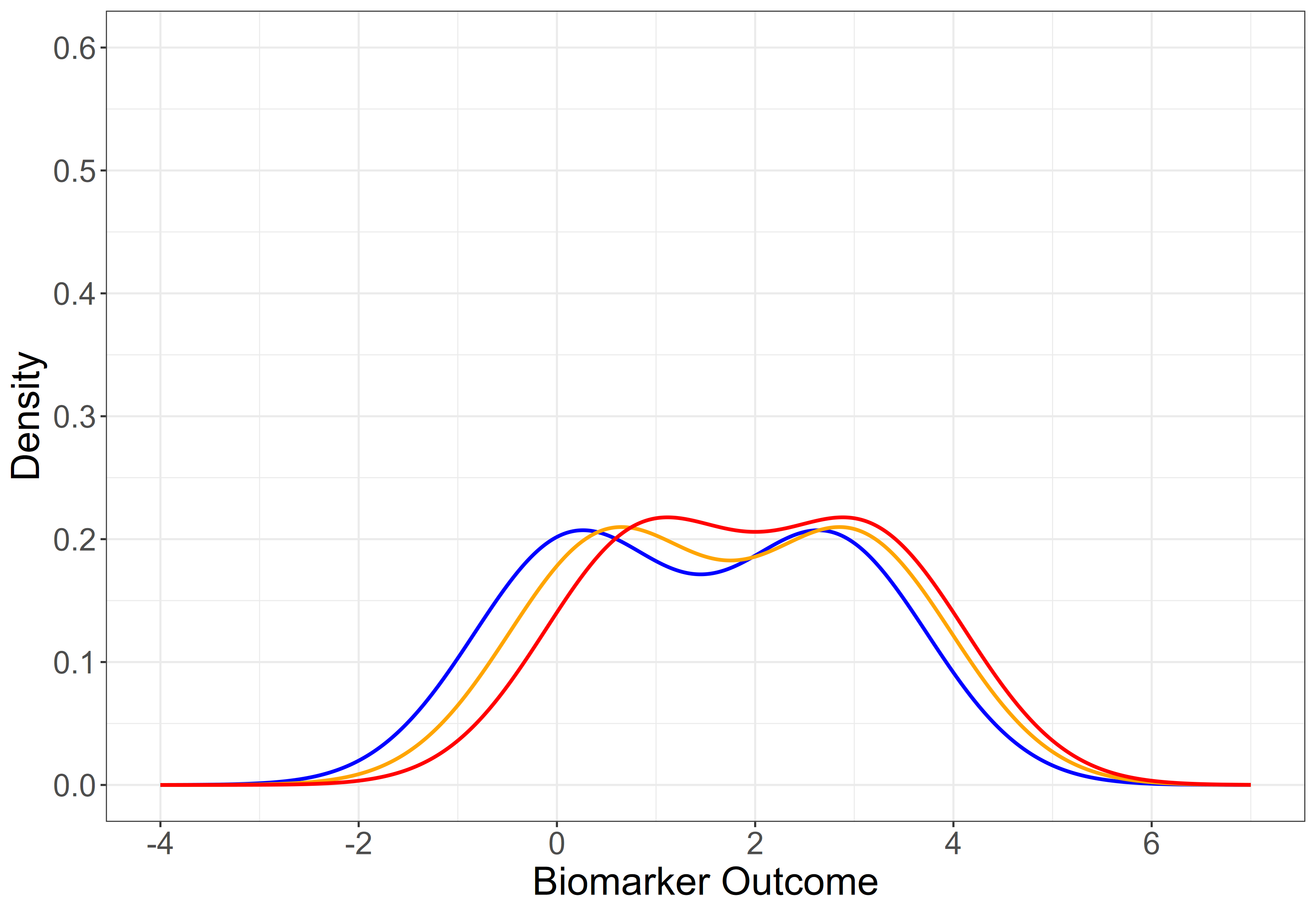}
	}
	\centering
	\caption{True densities of biomarker outcomes in groups 1 (blue lines), 2 (orange lines), and 3 (red lines), for simulation Scenarios I (top row), II (middle row), and III (bottom row).}
	\label{s123_diagrams}
\end{figure}

\begin{table}[H]
	\centering
	\begin{adjustbox}{width=0.8\textwidth,center}
		\begin{tabular}{c c c c c c c}
			\multicolumn{7}{c}{Sample size} \\
			\multicolumn{7}{c}{$(n_1,n_2,n_3)$} \\
			\rowcolor{gray!100}
			Scenario& UNL &$(100,100,100)$&$(200,200,200)$&$(500,500,500)$&
			$(1000,1000,1000)$&$(100,300,500)$\\
			
			I & 2.792& 0.95 &0.97 &0.96 & 0.99 & 1.00 \\ 
			& 1.919&0.93 & 0.99 & 0.98 & 0.97 & 0.99 \\ 
			& 1.139&0.62 & 0.80 & 0.92 & 0.91 & 0.90 \\ 
			\rowcolor{gray!50}
			II & 2.527&0.97 & 0.86 & 0.96 & 0.96 & 0.94 \\ 
			\rowcolor{gray!50}
			& 1.855&0.98 & 0.94 & 0.91 & 0.93 & 0.92 \\ 
			\rowcolor{gray!50}
			& 1.191&0.59 & 0.70 & 0.91 & 0.95 & 0.82 \\ 
			III & 2.508 &0.95 & 0.95 & 0.93 & 0.94 & 0.97 \\ 
			& 1.933&0.60 & 0.82 & 0.83 & 0.88 & 0.88 \\ 
			& 1.143&0.26 & 0.51 & 0.70 & 0.86 & 0.68 \\ 
			\hline
		\end{tabular}
	\end{adjustbox}
	\caption{Empirical frequentist coverage probabilities of the 95\% credible intervals for the underlap coefficient. Results are presented for each configuration of distribution parameters, sample size, and simulation scenario.}
	\label{simulation_uncon_coverage}
\end{table}

\clearpage
\begin{figure}[htpb]
	\centering
	\subfigure{
		\centering
		\includegraphics[width=0.3\textwidth]{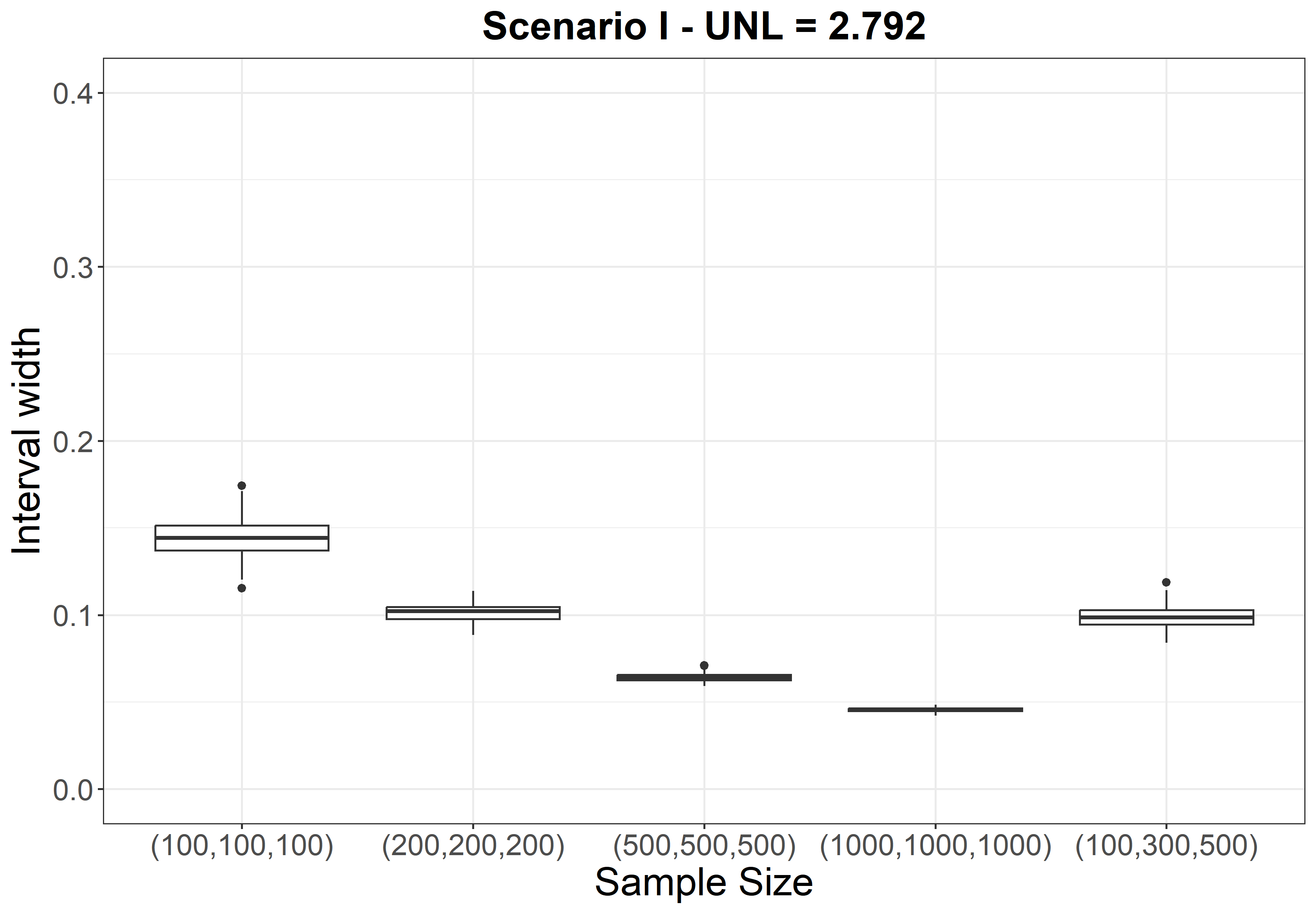}
	}
	\subfigure{
		\centering
		\includegraphics[width=0.3\textwidth]{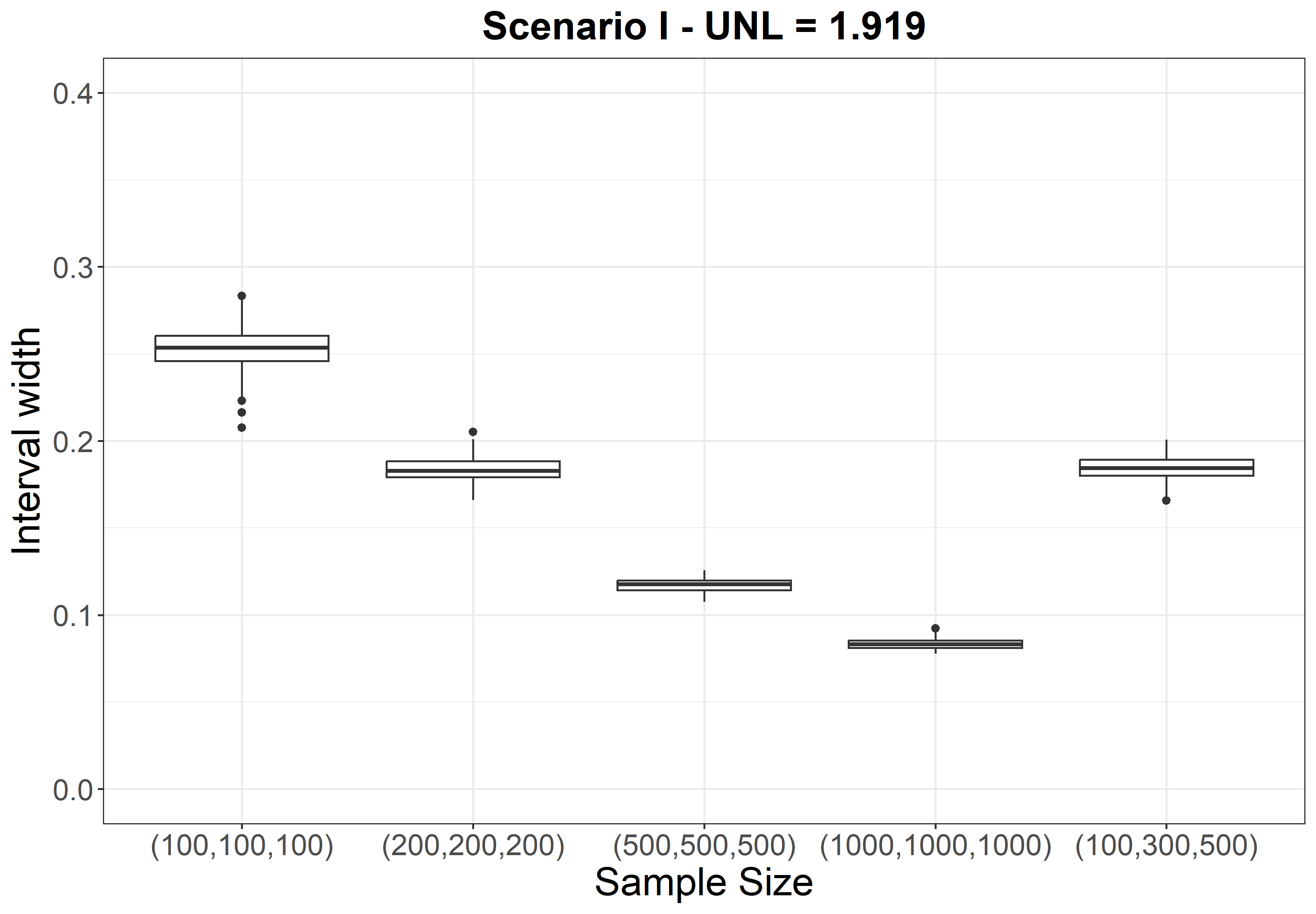}
	}
	\subfigure{
		\centering
		\includegraphics[width=0.3\textwidth]{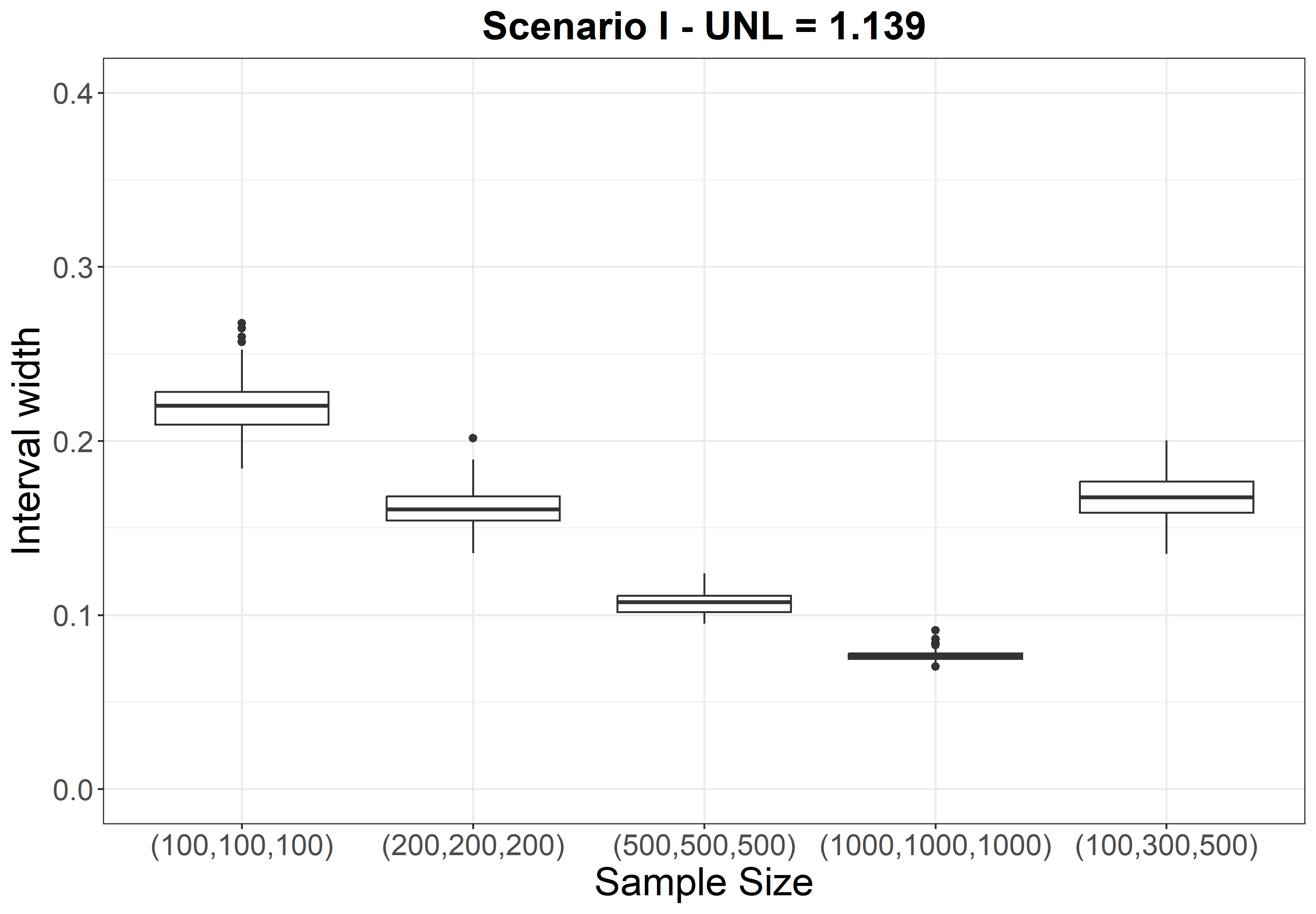}
	}
	\\
	\subfigure{
		\centering
		\includegraphics[width=0.3\textwidth]{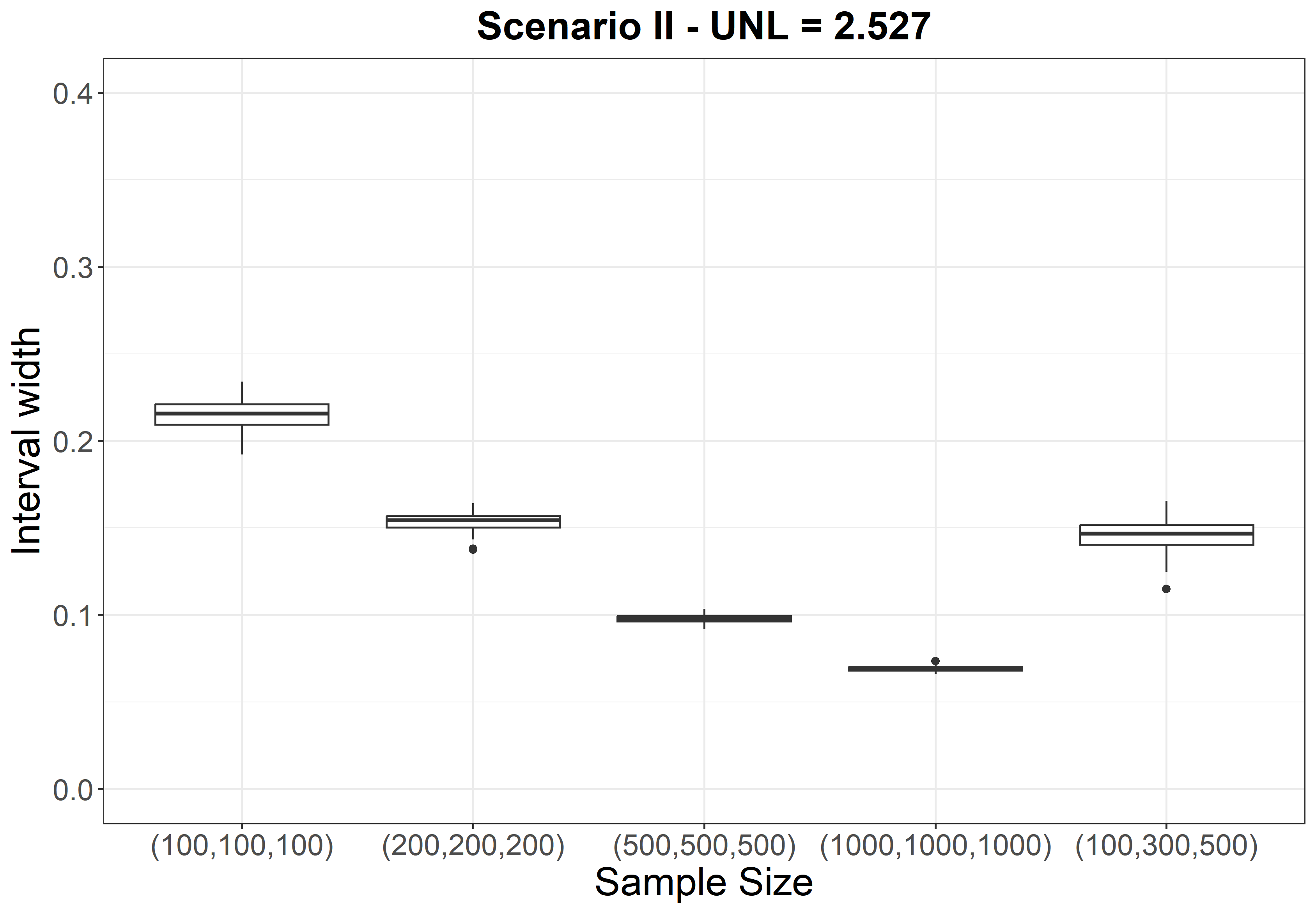}
	}
	\subfigure{
		\centering
		\includegraphics[width=0.3\textwidth]{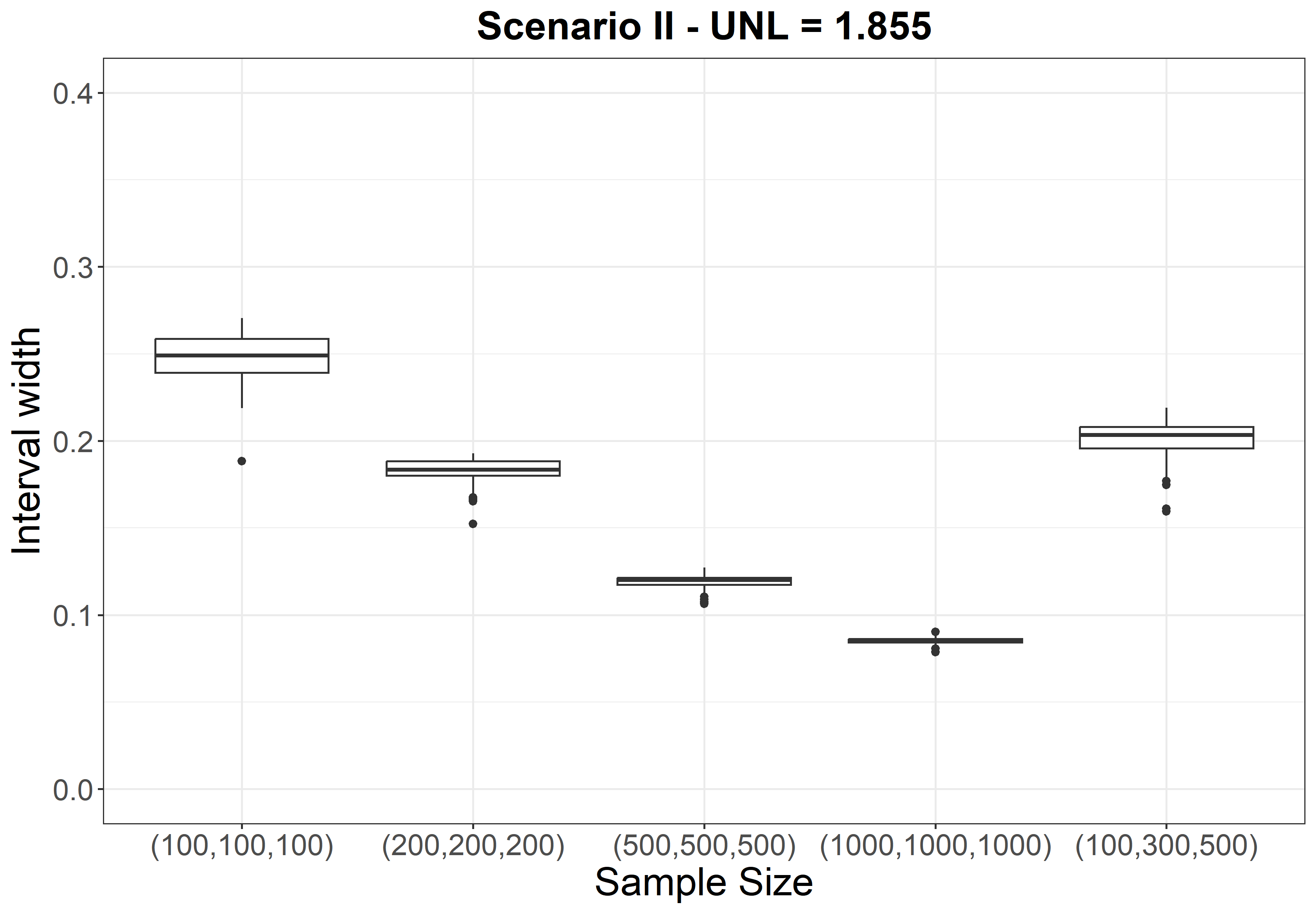}
	}
	\subfigure{
		\centering
		\includegraphics[width=0.3\textwidth]{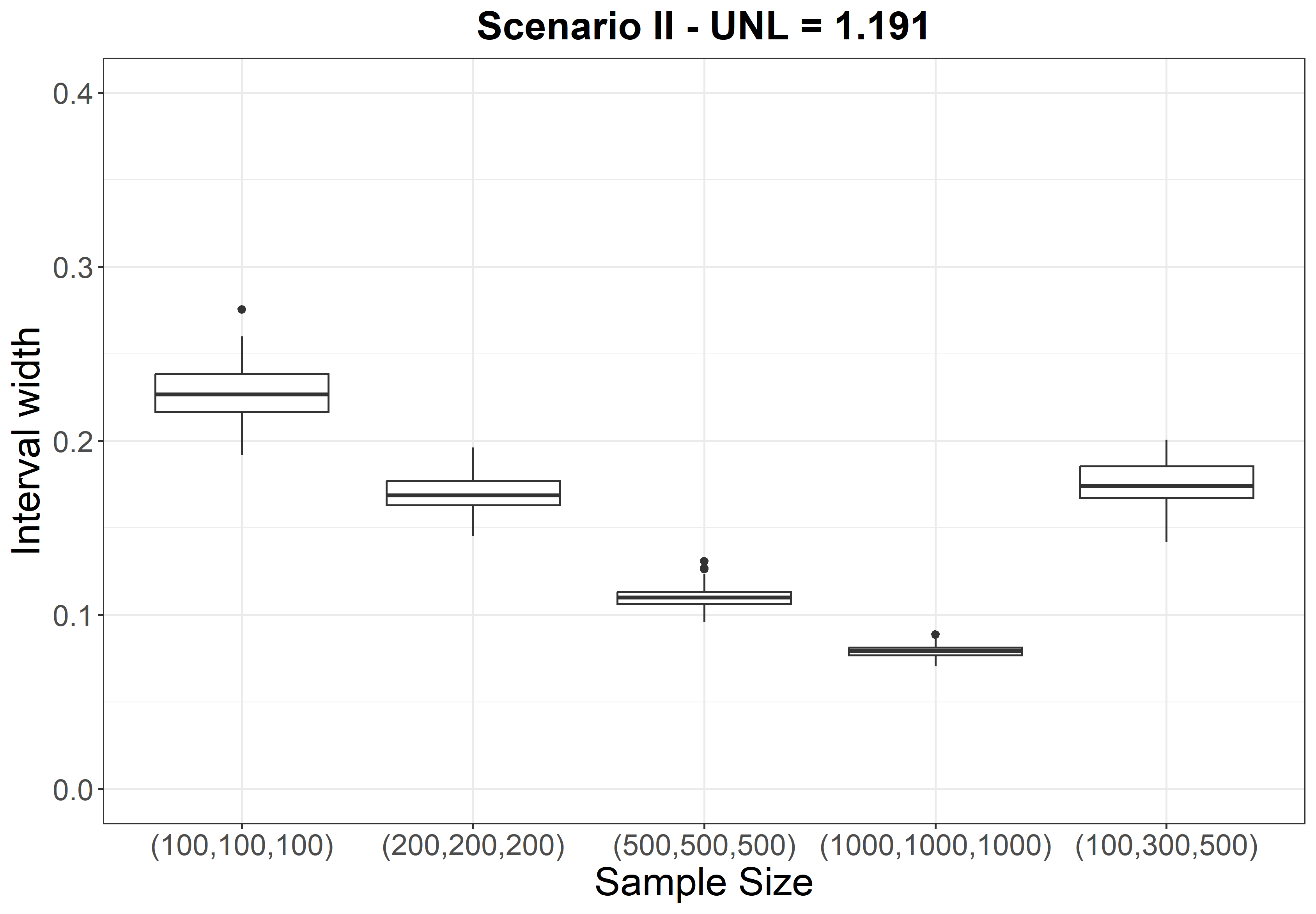}
	}
	\\
	\subfigure{
		\centering
		\includegraphics[width=0.3\textwidth]{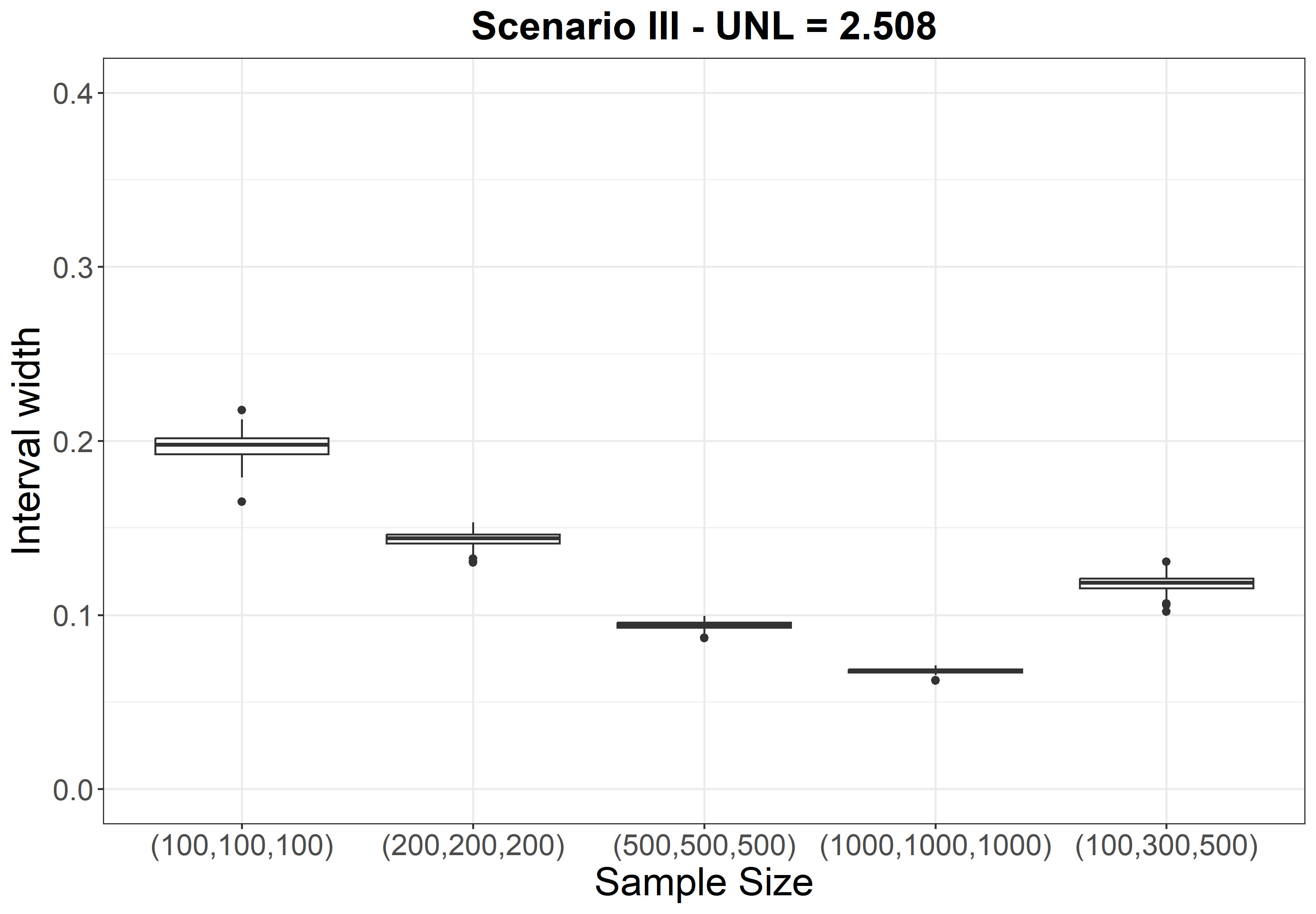}
	}
	\subfigure{
		\centering
		\includegraphics[width=0.3\textwidth]{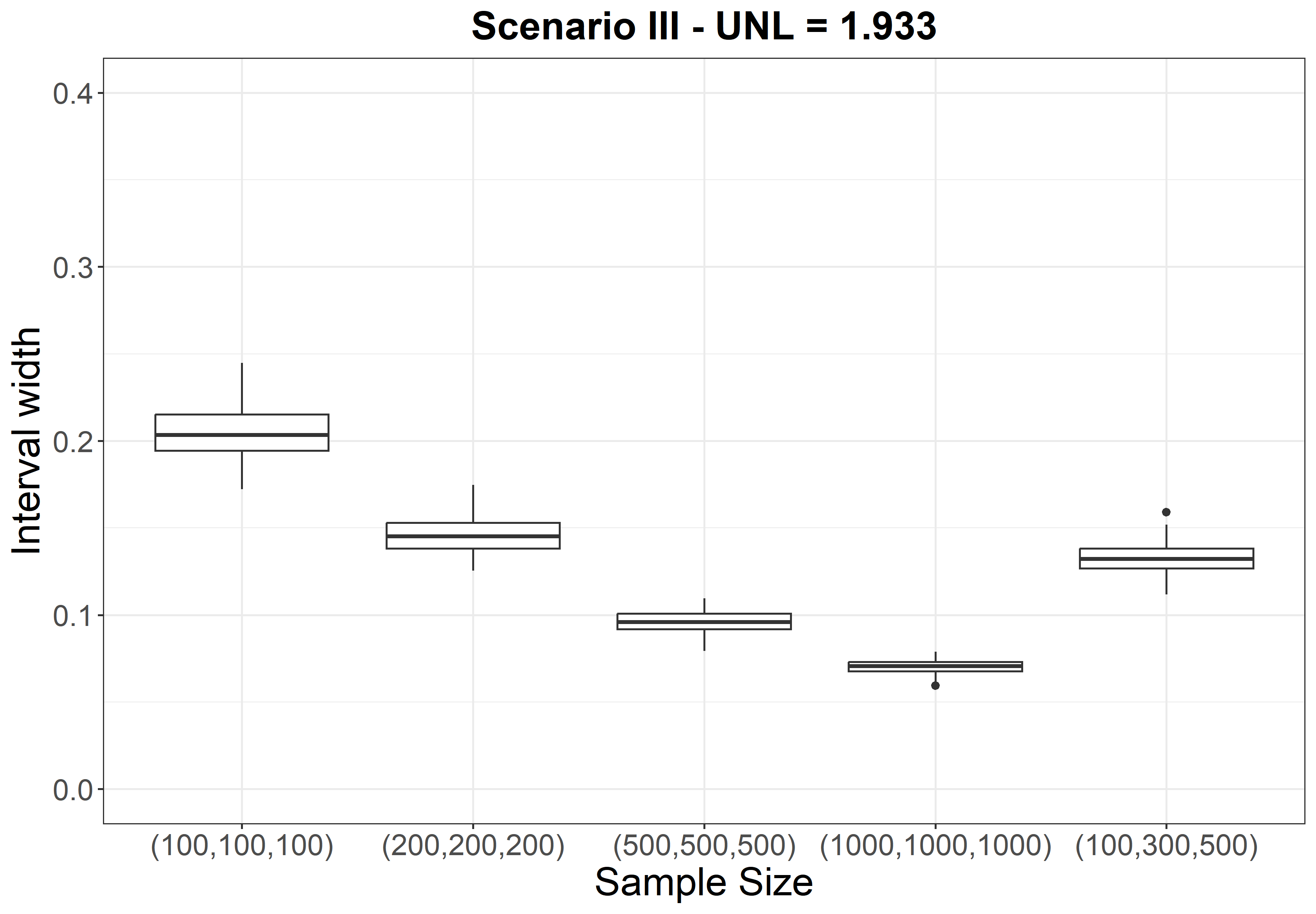}
	}
	\subfigure{
		\centering
		\includegraphics[width=0.3\textwidth]{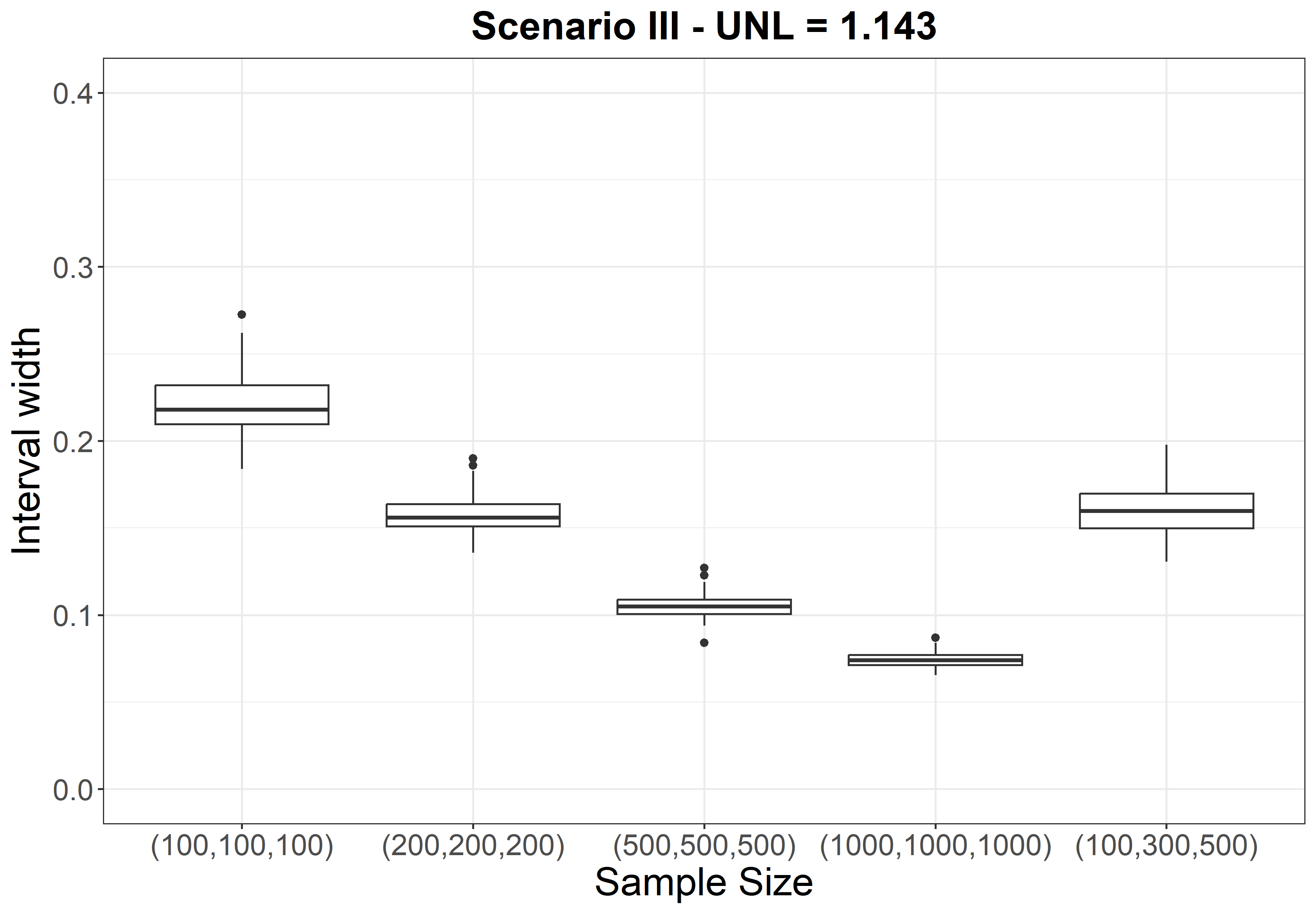}
	}
	\centering
	\caption{Boxplot of the width of the 95\% credible intervals for the underlap coefficient, across the 100 simulated datasets for each configuration of distribution parameters, sample size, and simulation scenario.}
	\label{width_box_uncons}
\end{figure}

\newpage
\subsubsection*{Covariate-specific case}

\begin{figure}[H]
	\centering
	\subfigure{
		\includegraphics[width=0.3\linewidth]{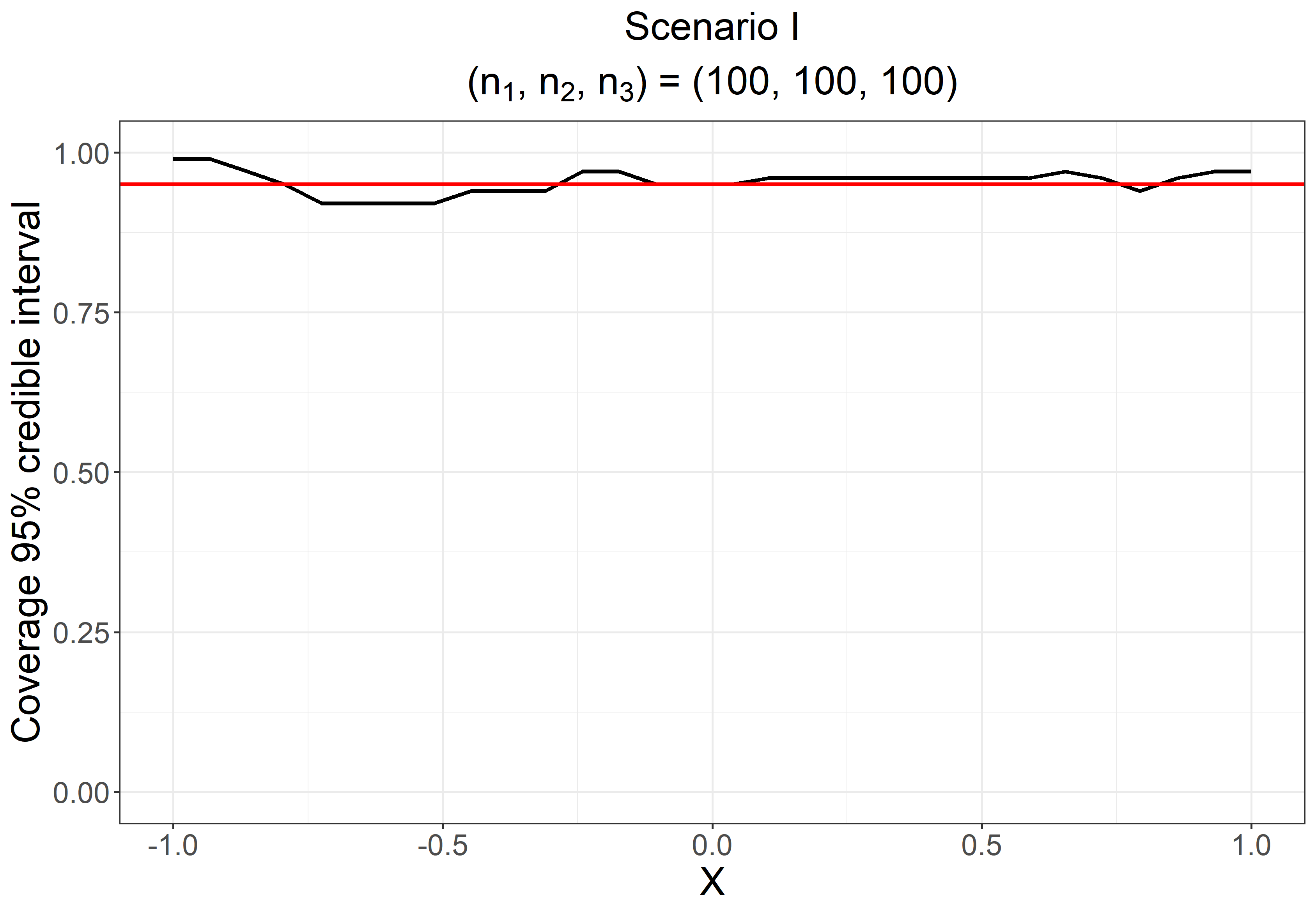}
	}
	\subfigure{
		\includegraphics[width=0.3\linewidth]{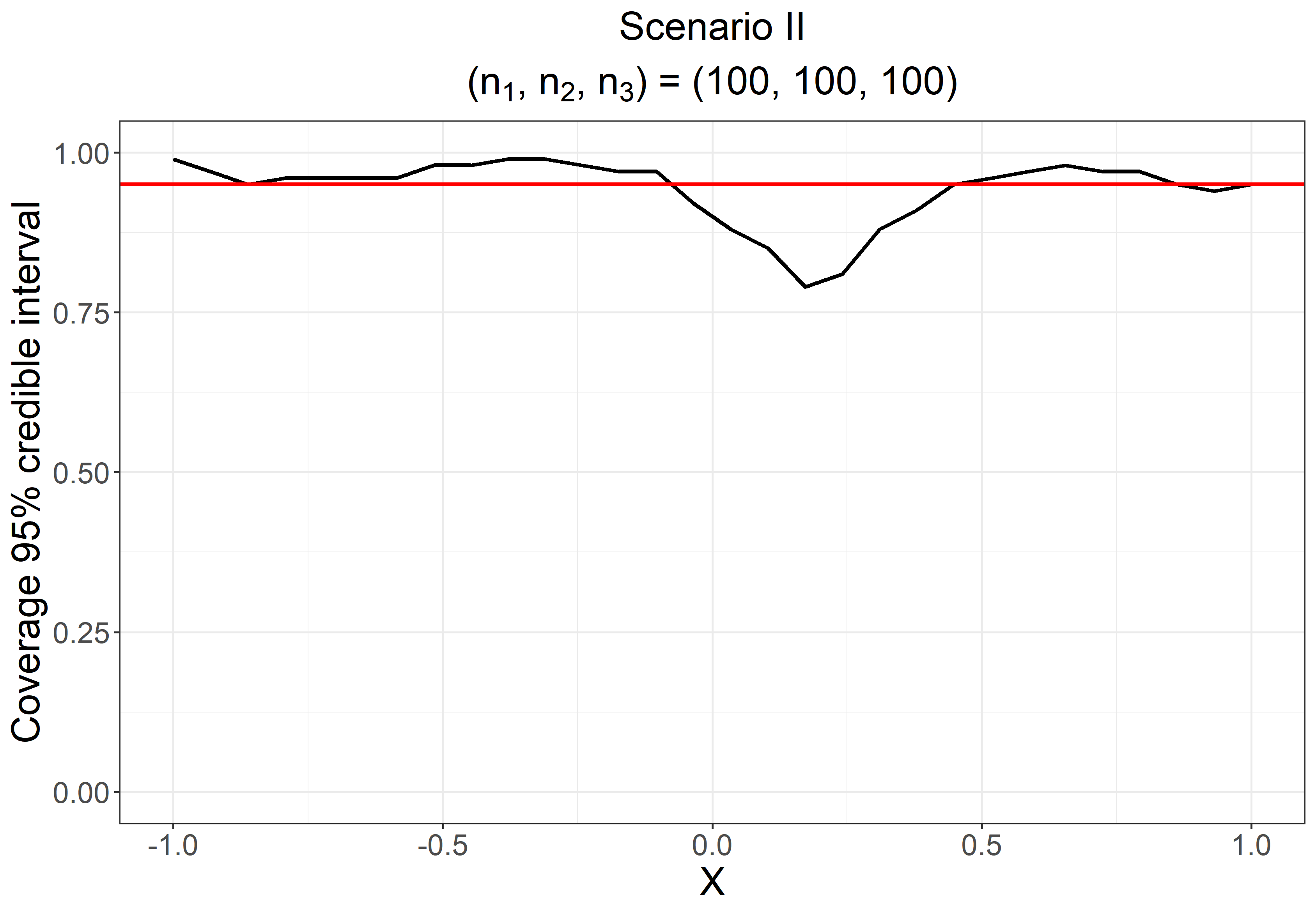}
	}
	\subfigure{
		\includegraphics[width=0.3\linewidth]{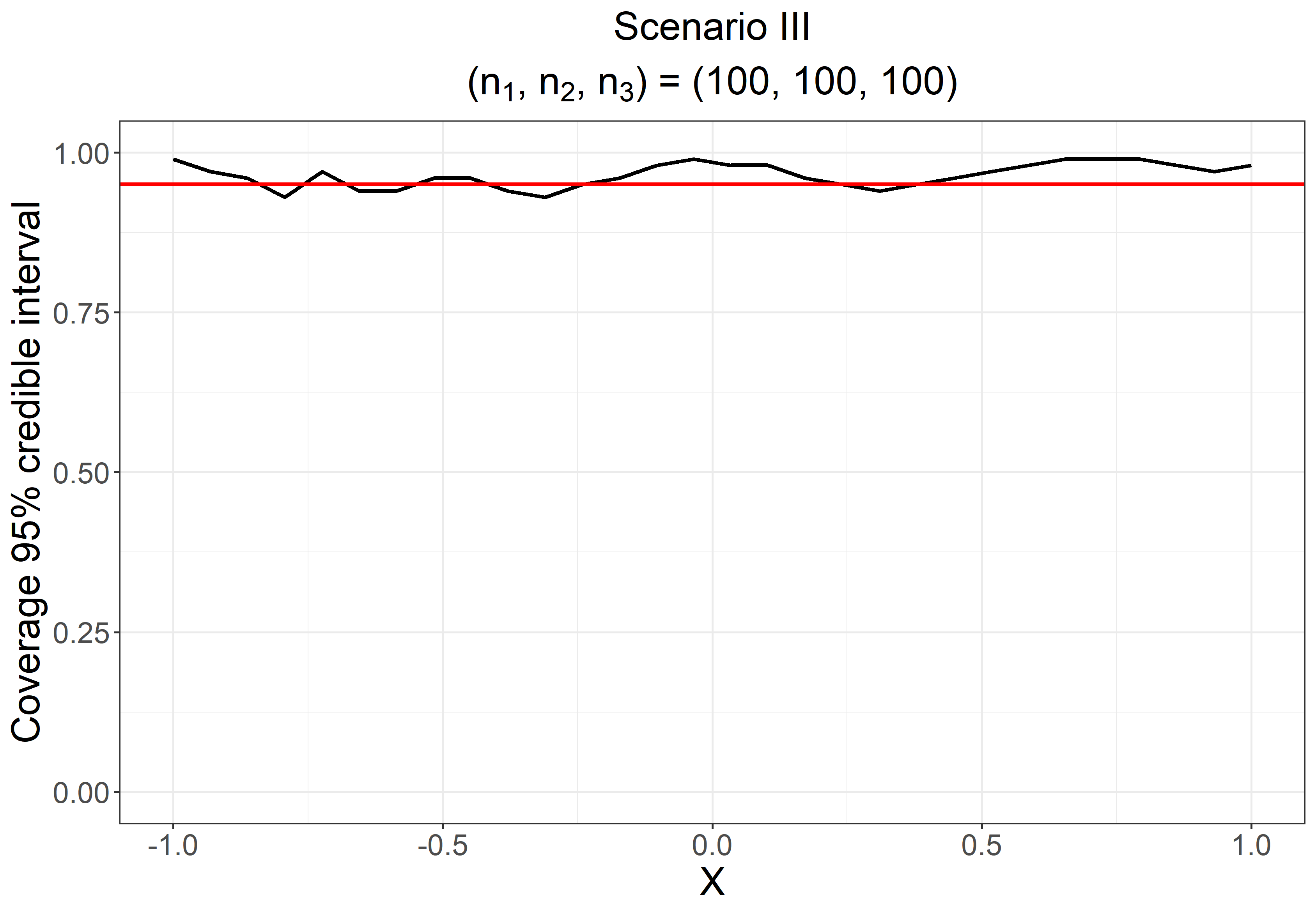}
	}
	\\
	\subfigure{
		\includegraphics[width=0.3\linewidth]{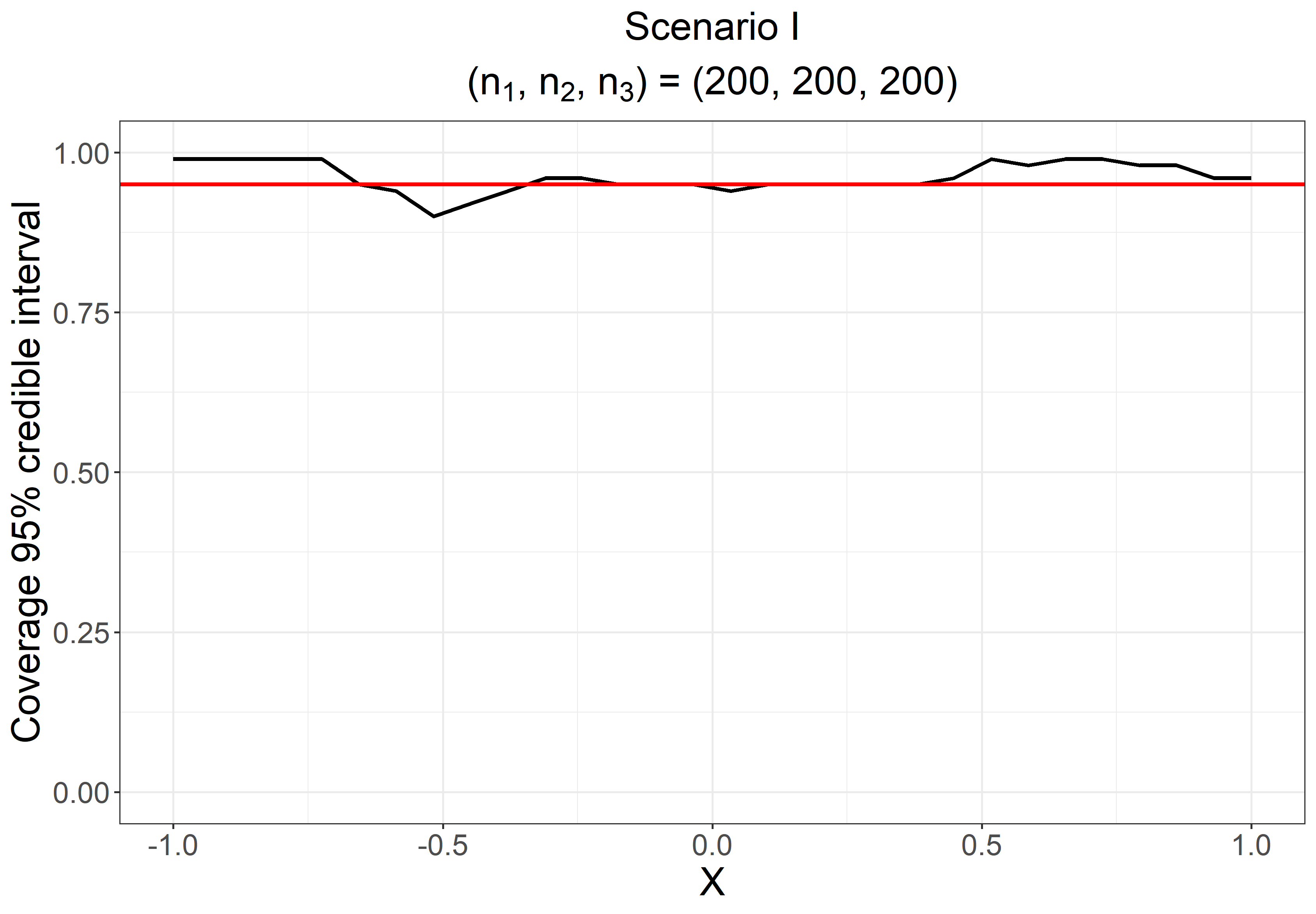}
	}
	\subfigure{
		\includegraphics[width=0.3\linewidth]{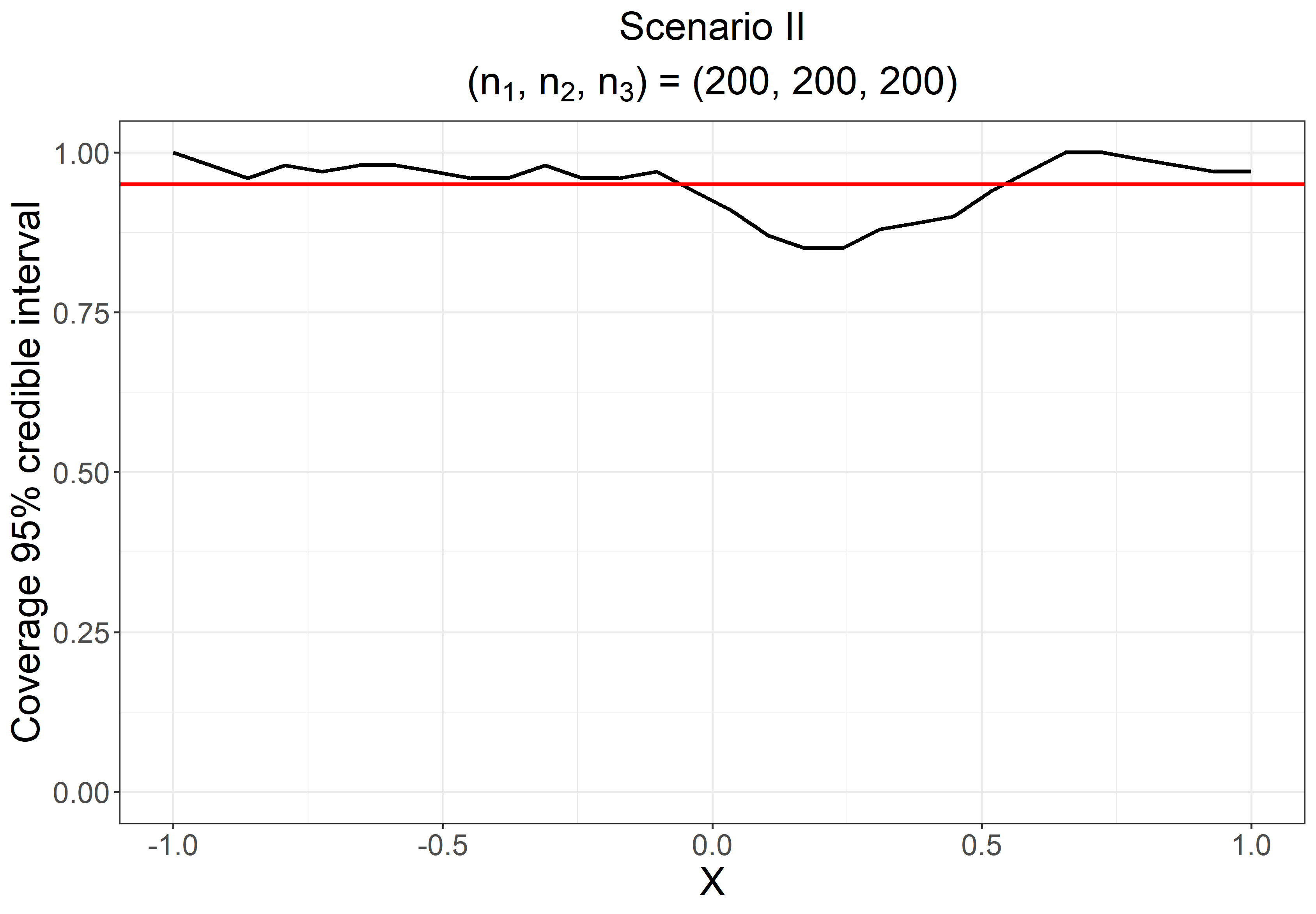}
	}
	\subfigure{
		\includegraphics[width=0.3\linewidth]{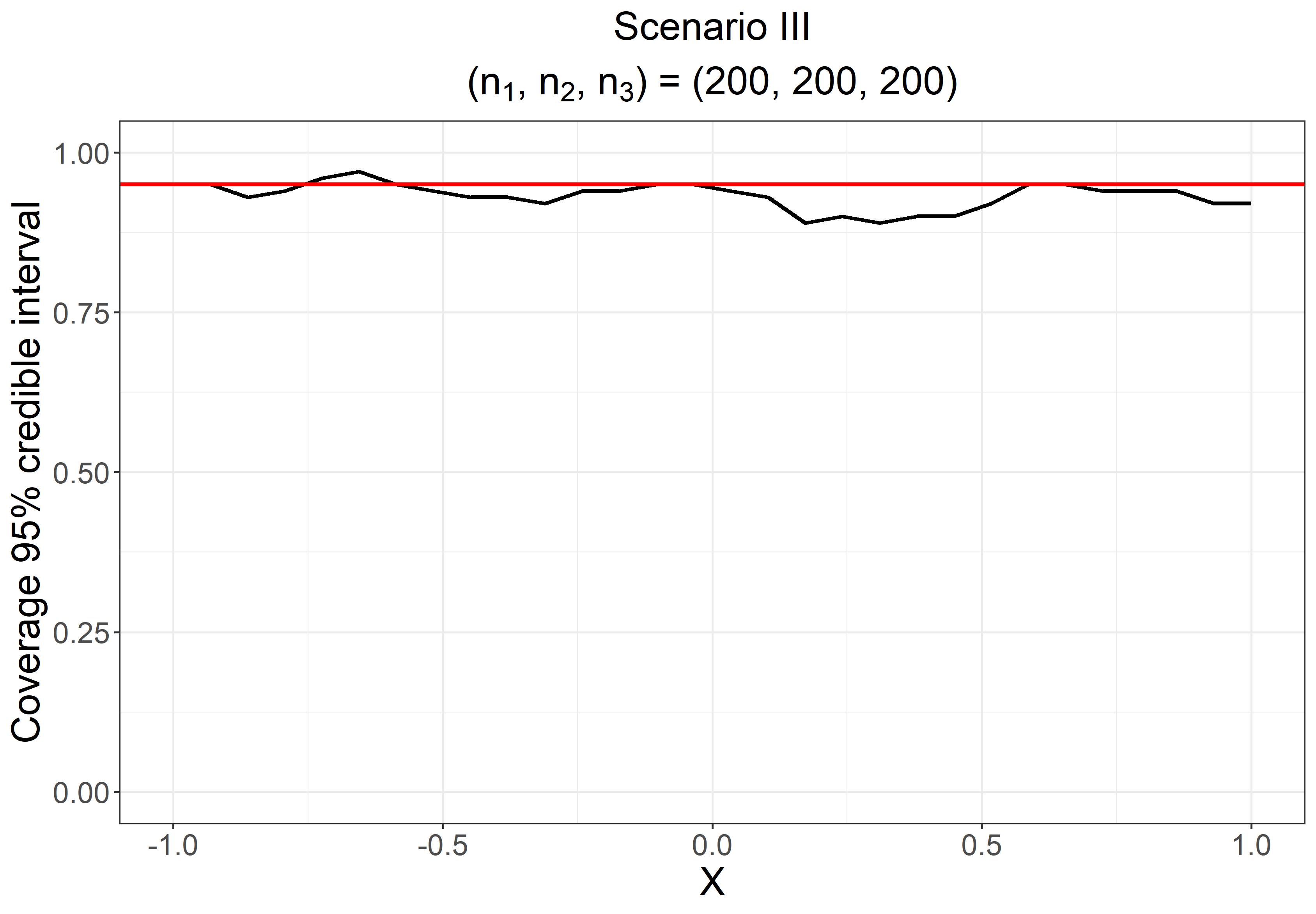}
	}
	\\
	\subfigure{
		\includegraphics[width=0.3\linewidth]{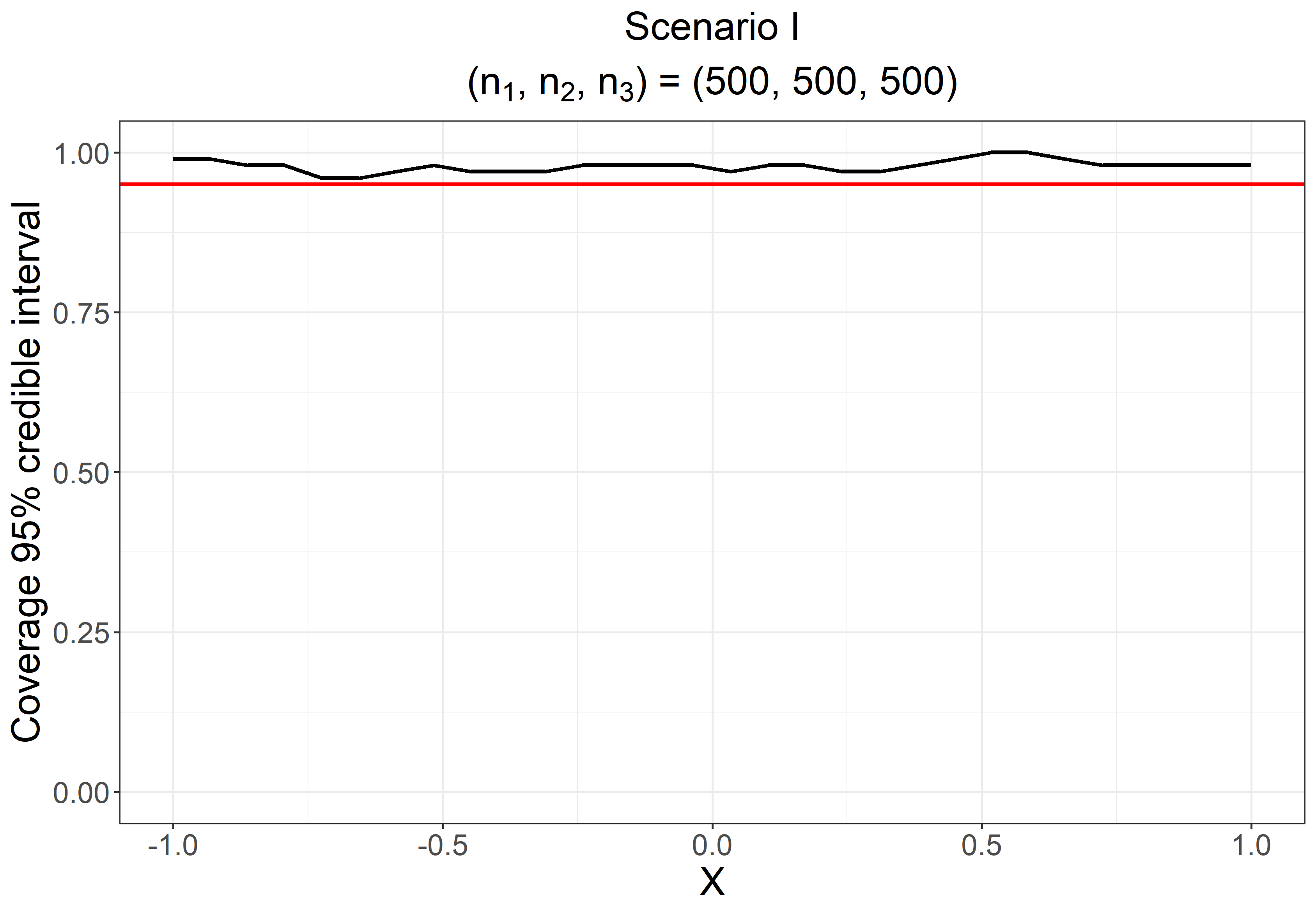}
	}
	\subfigure{
		\includegraphics[width=0.3\linewidth]{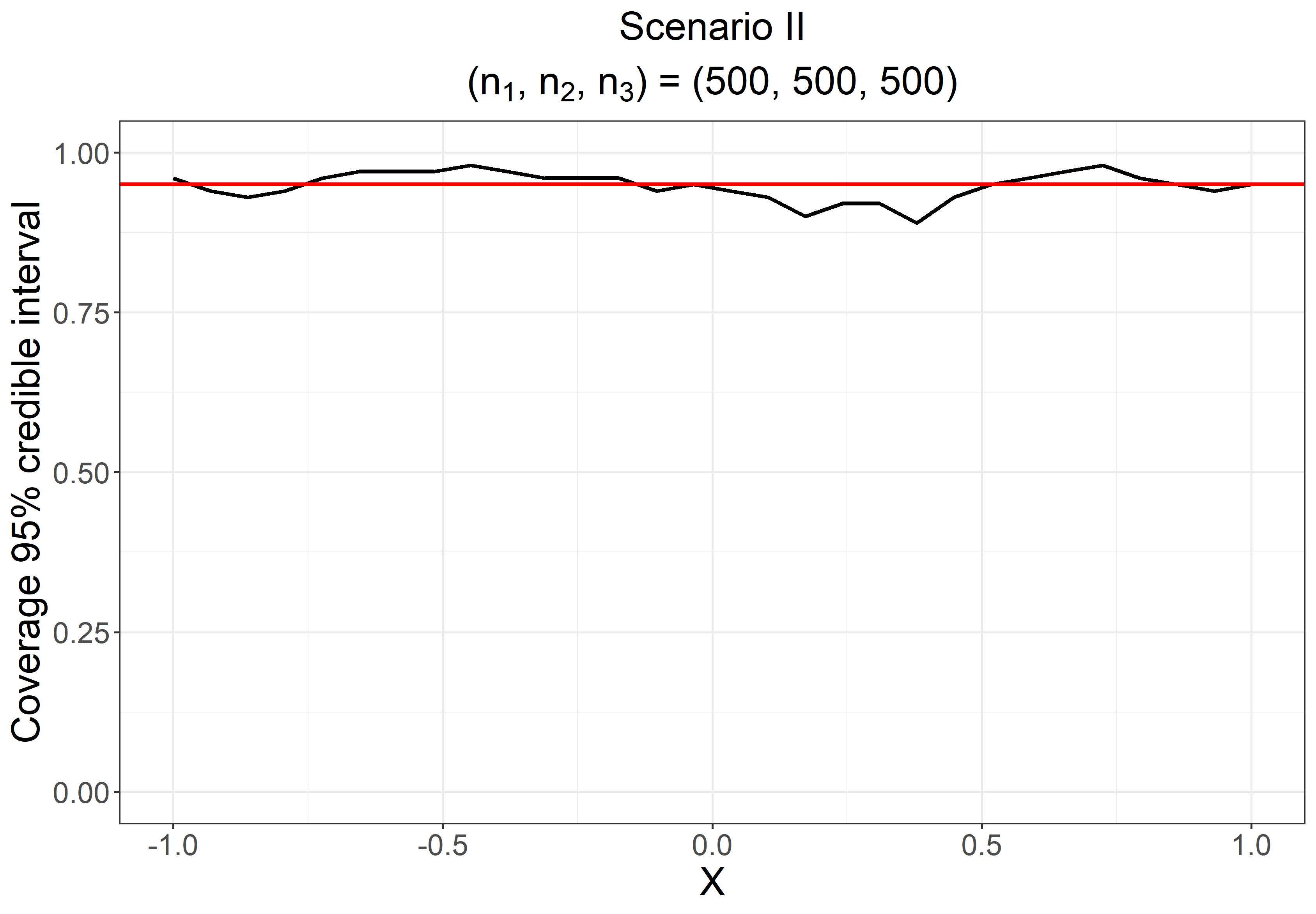}
	}
	\subfigure{
		\includegraphics[width=0.3\linewidth]{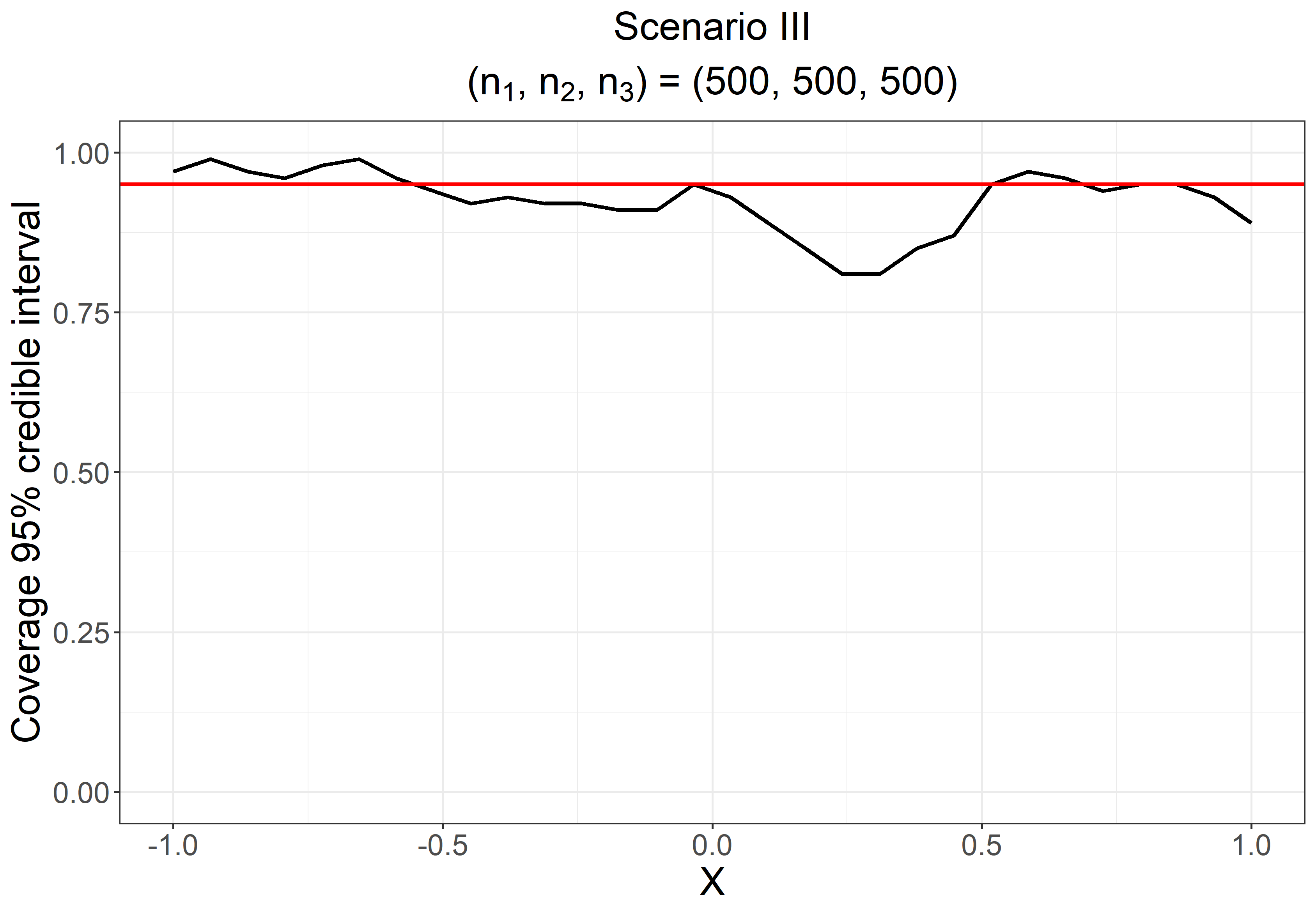}
	}
	\\
	\subfigure{
		\includegraphics[width=0.3\linewidth]{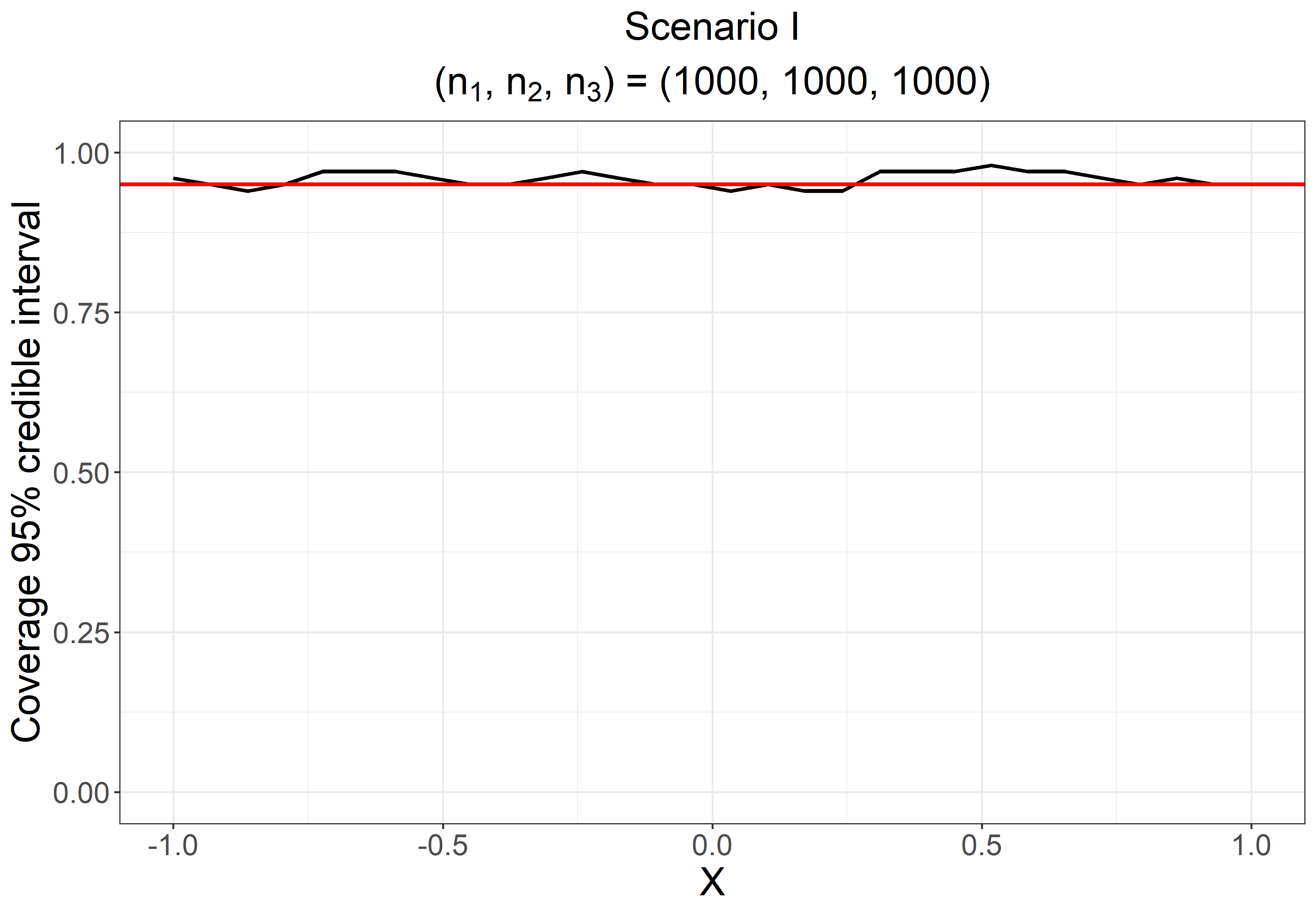}
	}
	\subfigure{
		\includegraphics[width=0.3\linewidth]{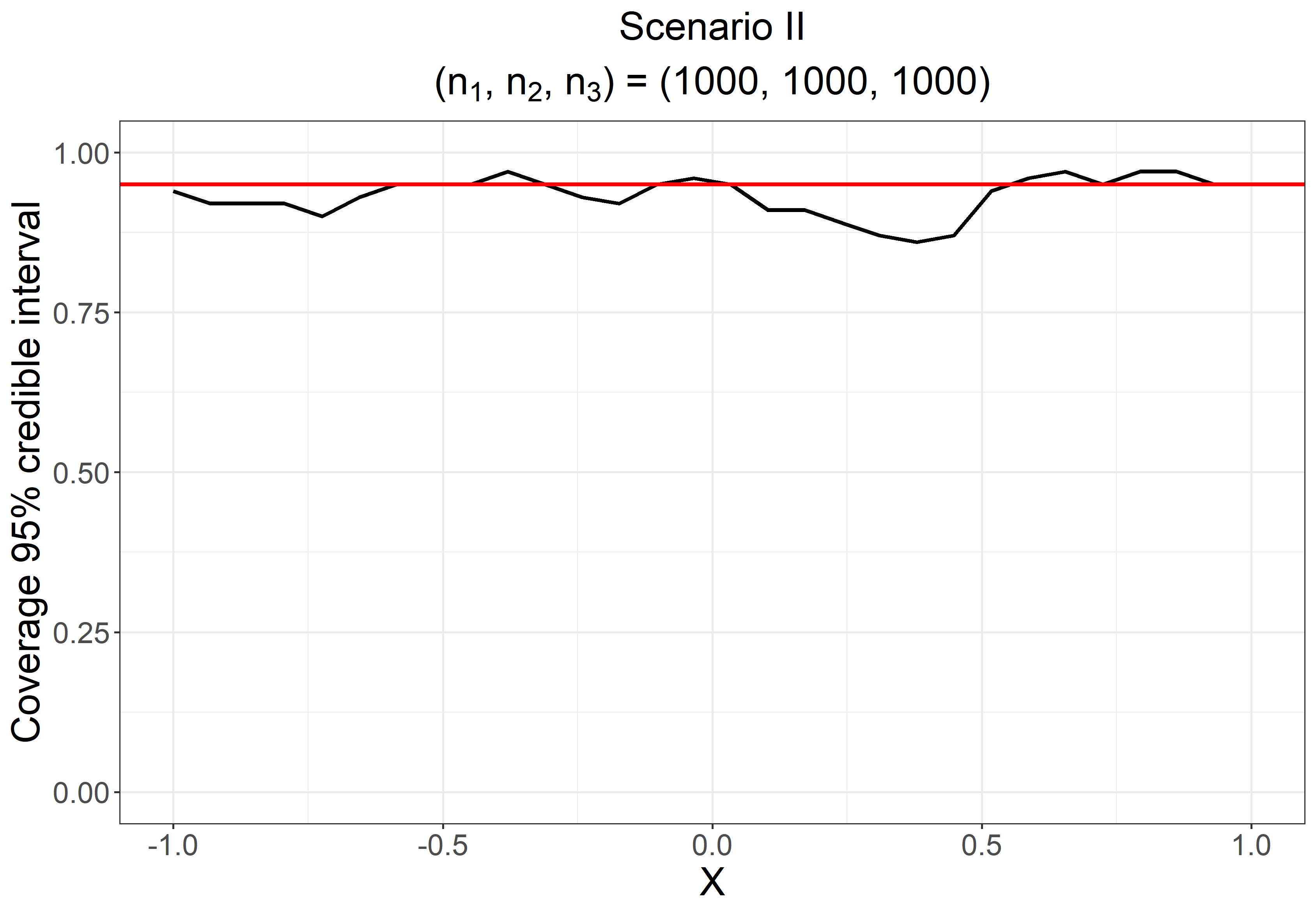}
	}
	\subfigure{
		\includegraphics[width=0.3\linewidth]{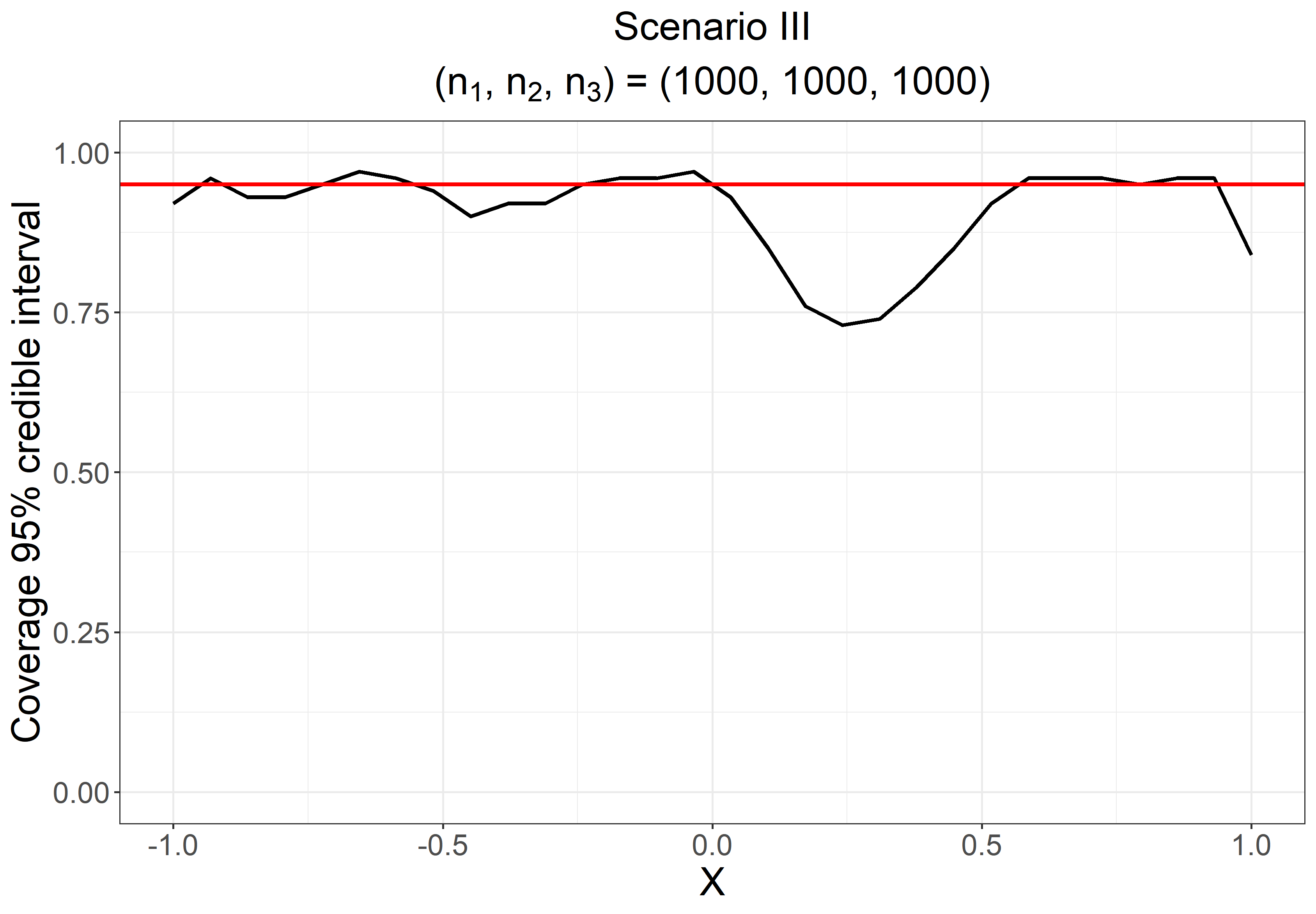}
	}
	\\
	\subfigure{
		\includegraphics[width=0.3\linewidth]{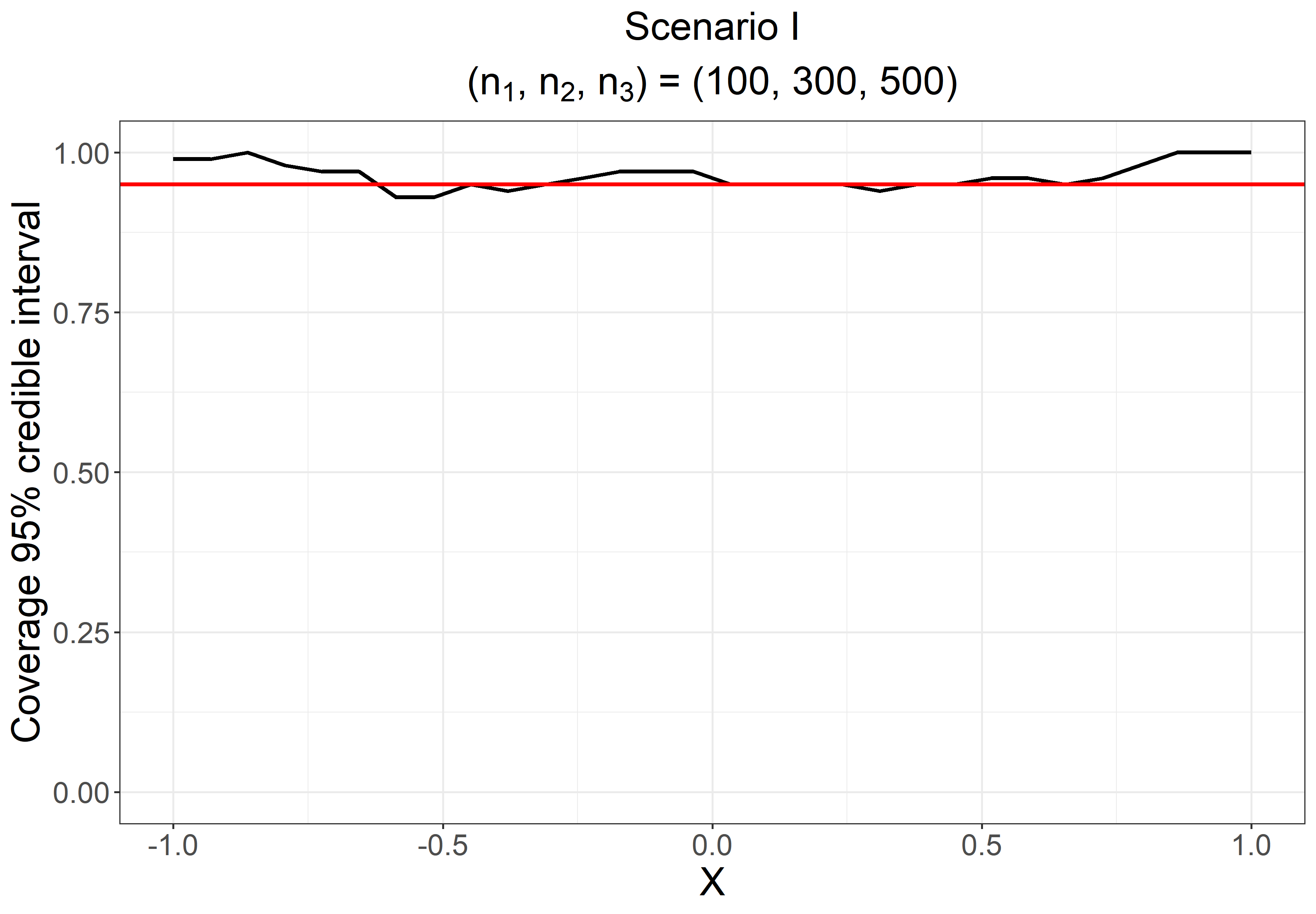}
	}
	\subfigure{
		\includegraphics[width=0.3\linewidth]{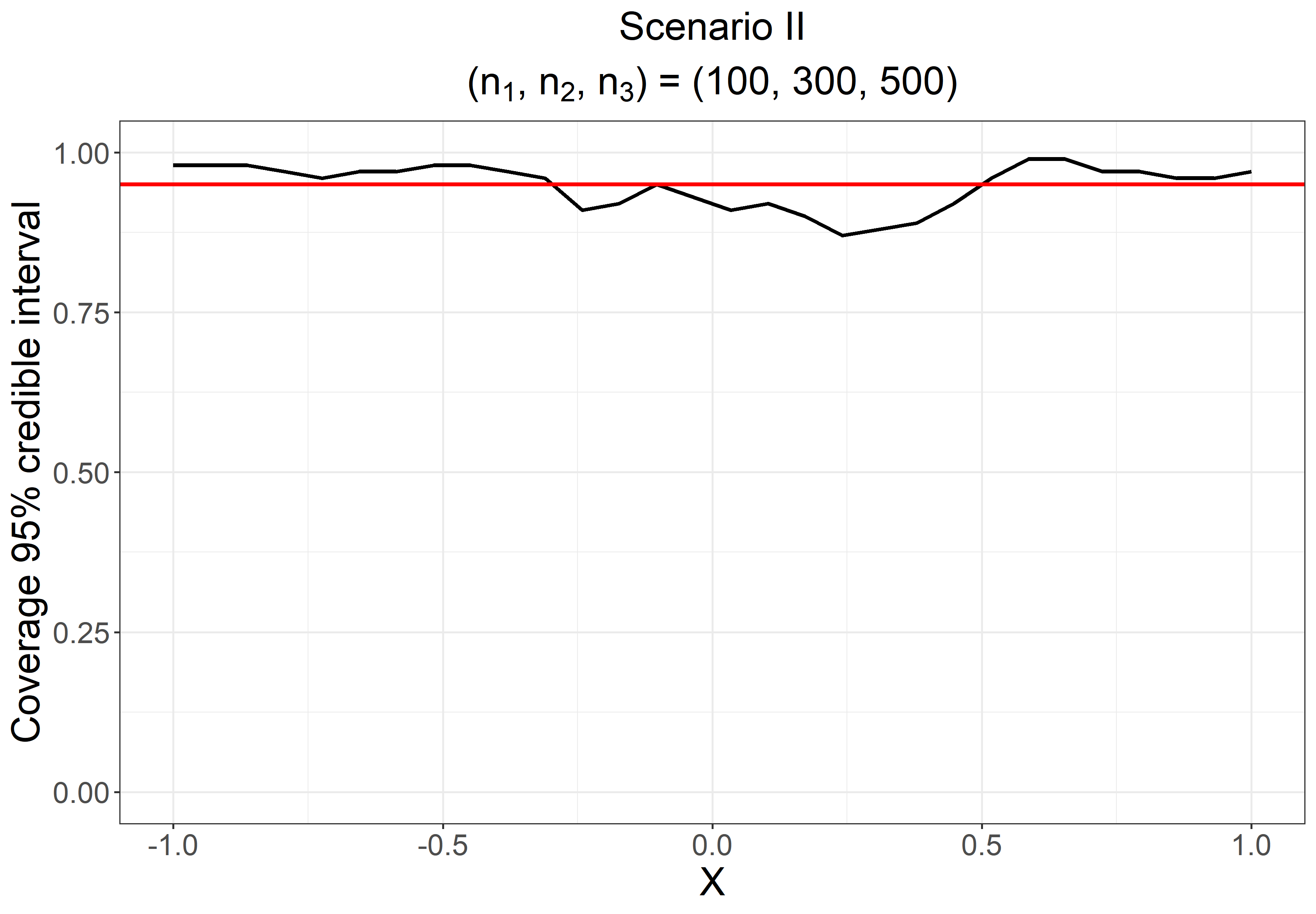}
	}
	\subfigure{
		\includegraphics[width=0.3\linewidth]{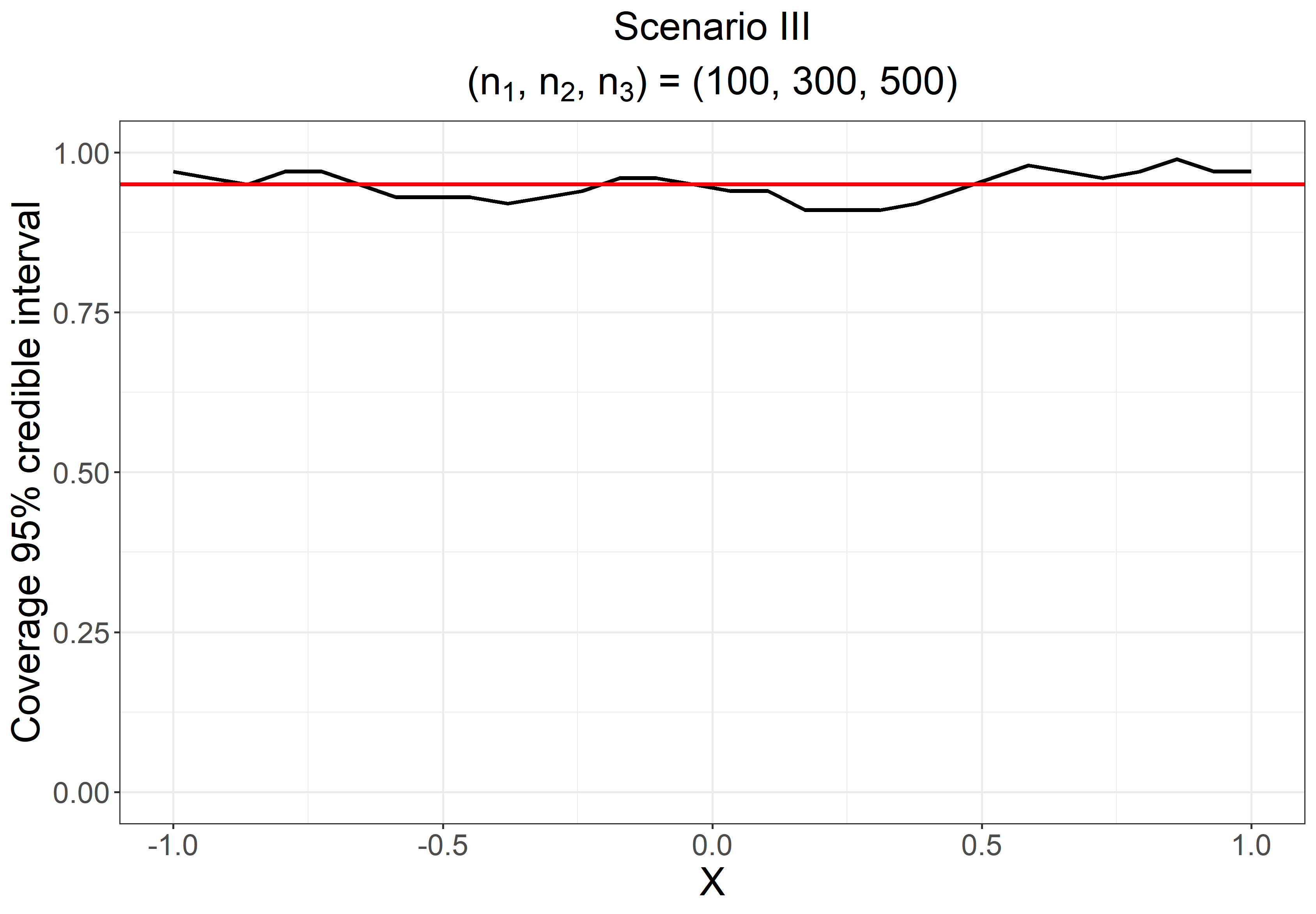}
	}
	\caption{Pointwise empirical frequentist coverage probabilities for the 95\% credible intervals associated to our Bayesian nonparametric estimator of the covariate-specific underlap coefficient. }
	\label{plot_con_coverage_kernel}
\end{figure}

\clearpage

\subsection*{Alzheimer's disease application: additional figures}
\begin{figure}[H]
	\vspace{-0.5ex}
	\centering
	\subfigure{
		\includegraphics[width=0.23\linewidth]{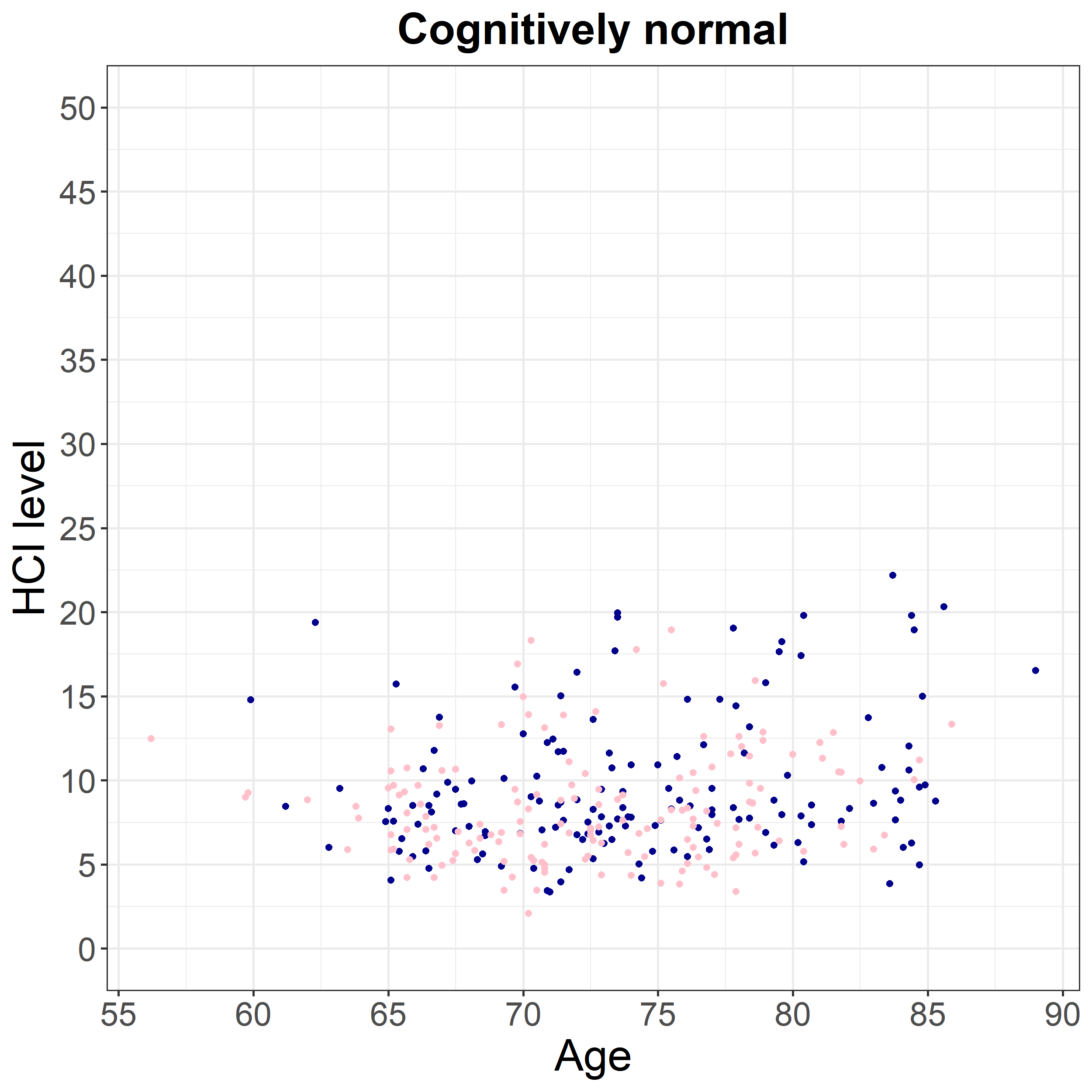}
	}
	\subfigure{
		\includegraphics[width=0.23\linewidth]{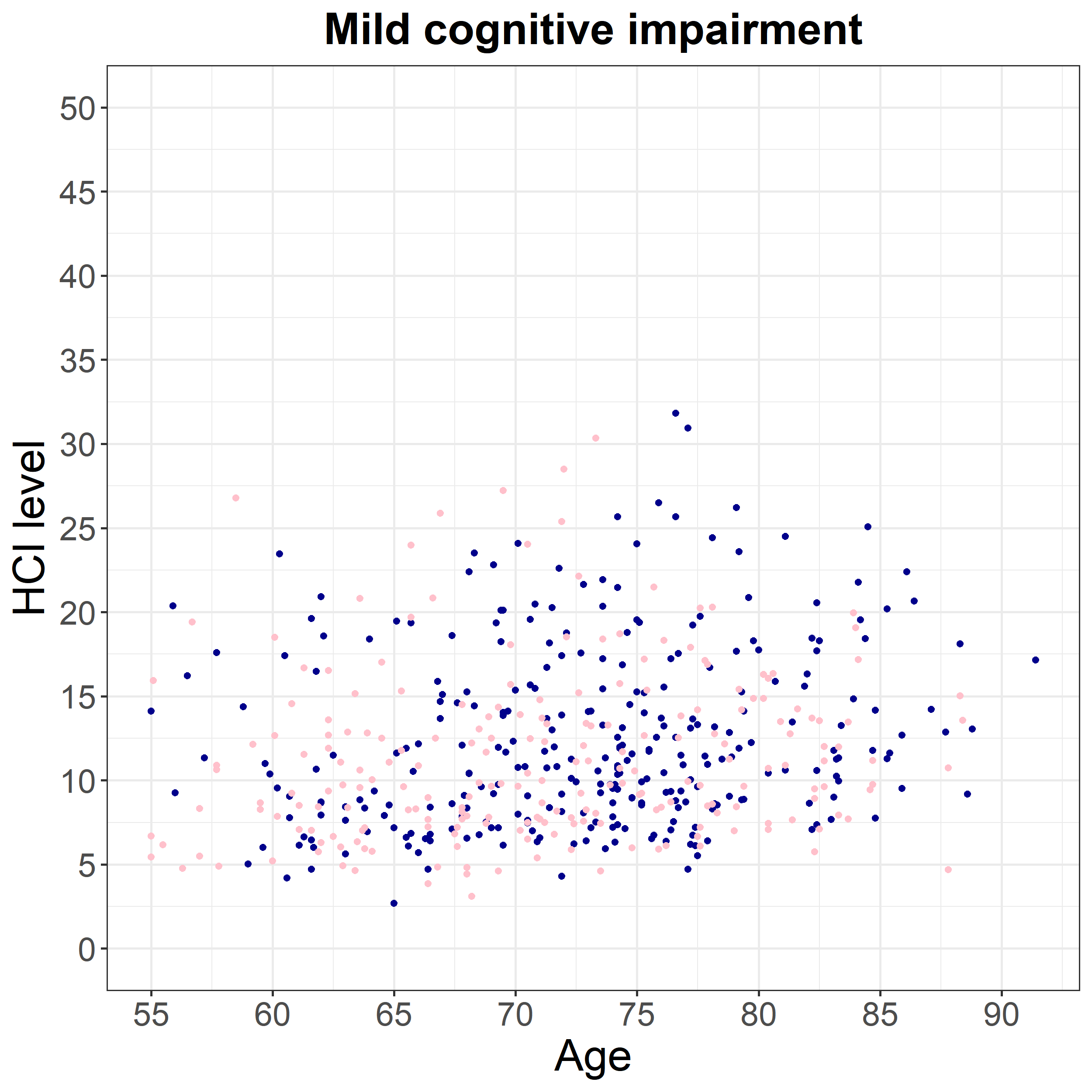}
	}
	\subfigure{
		\includegraphics[width=0.23\linewidth]{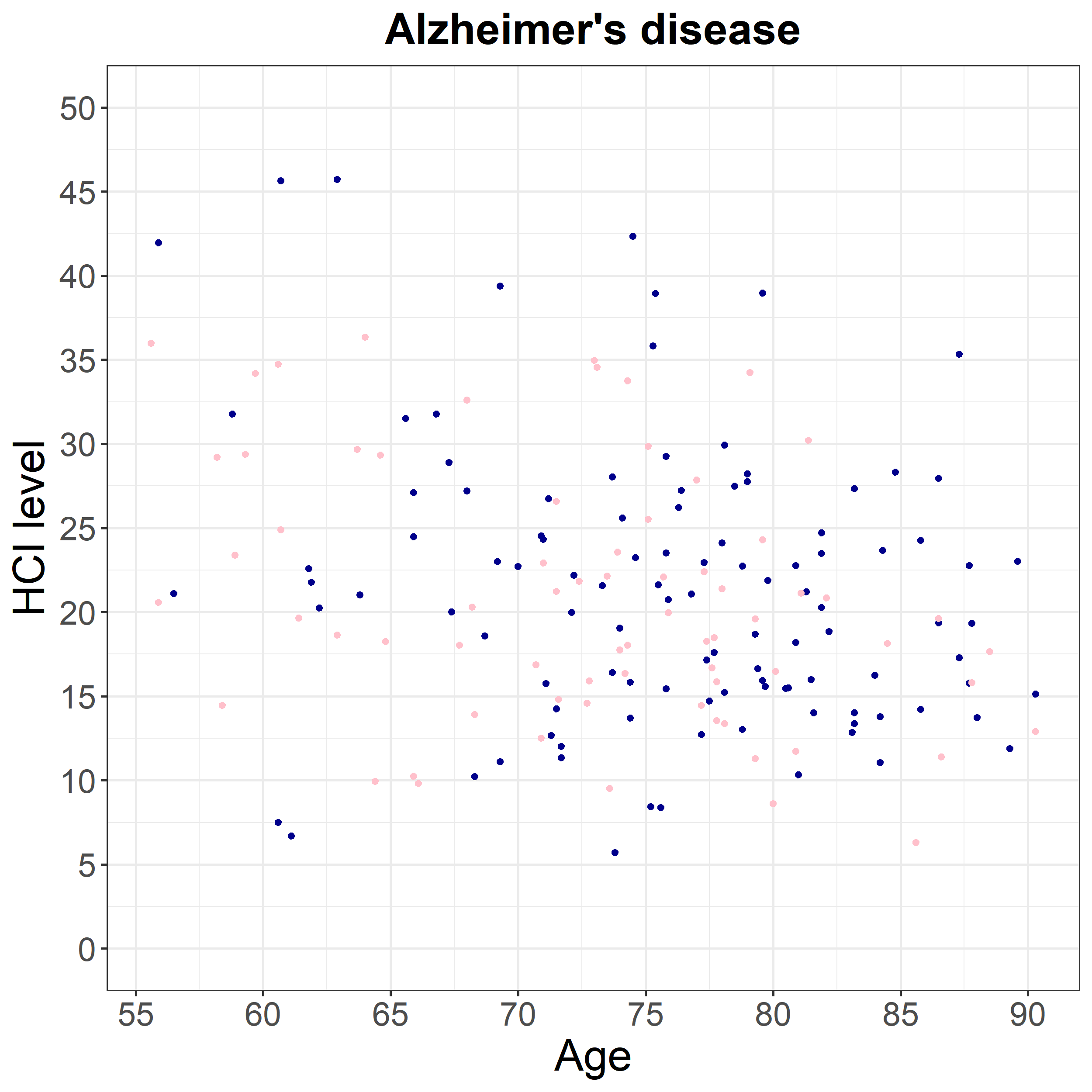}
	}
	\\
	\subfigure{
		\includegraphics[width=0.23\linewidth]{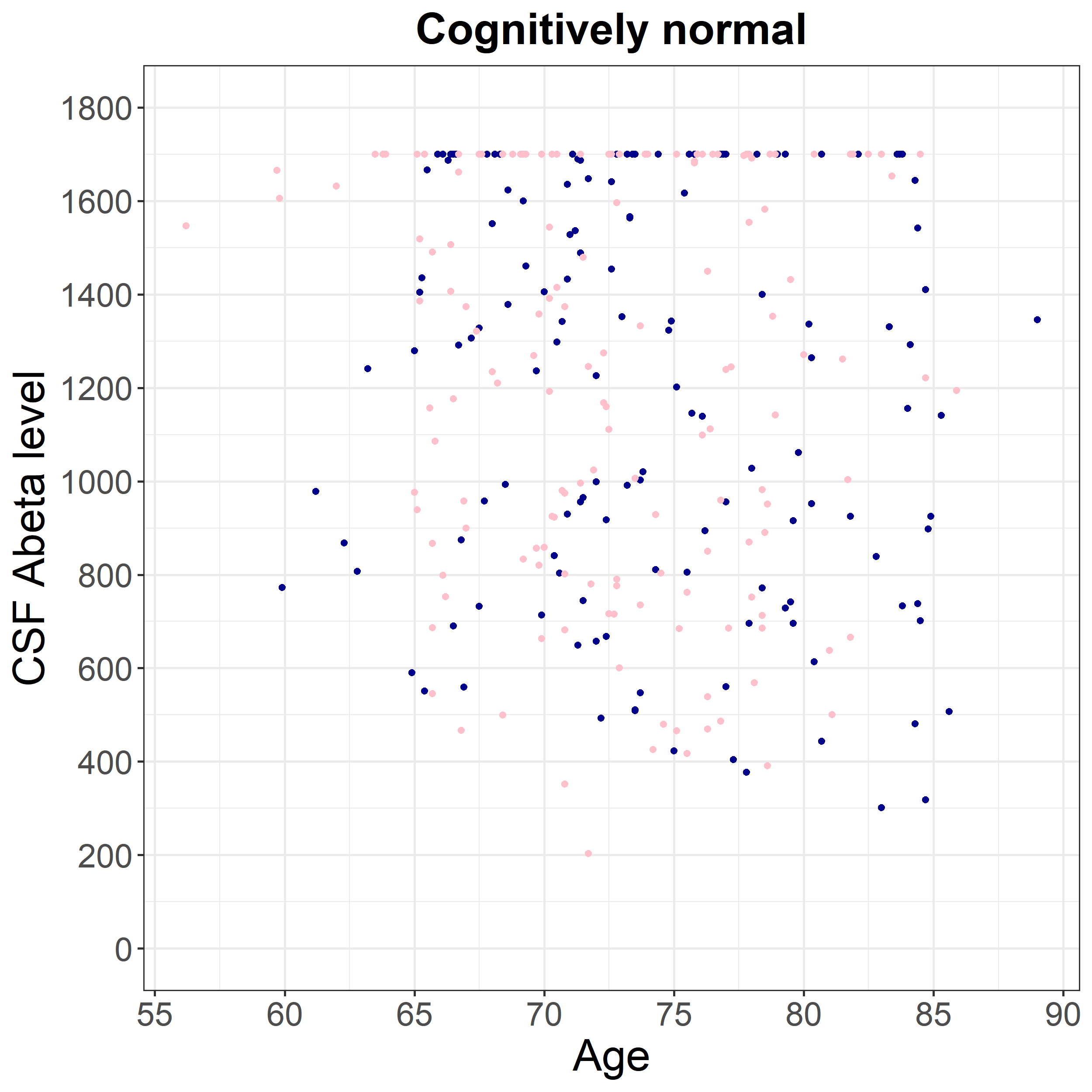}
	}
	\subfigure{
		\includegraphics[width=0.23\linewidth]{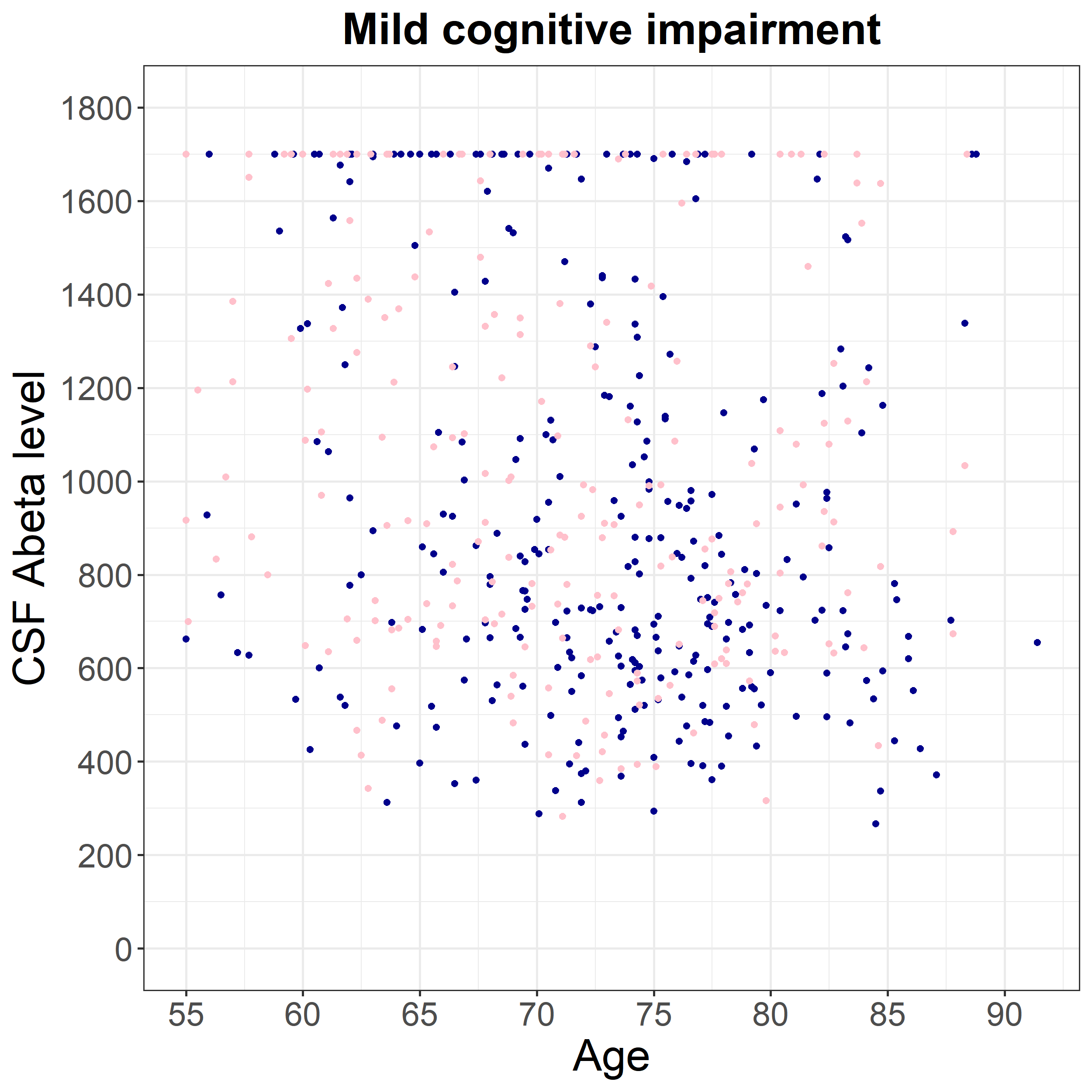}
	}
	\subfigure{
		\includegraphics[width=0.23\linewidth]{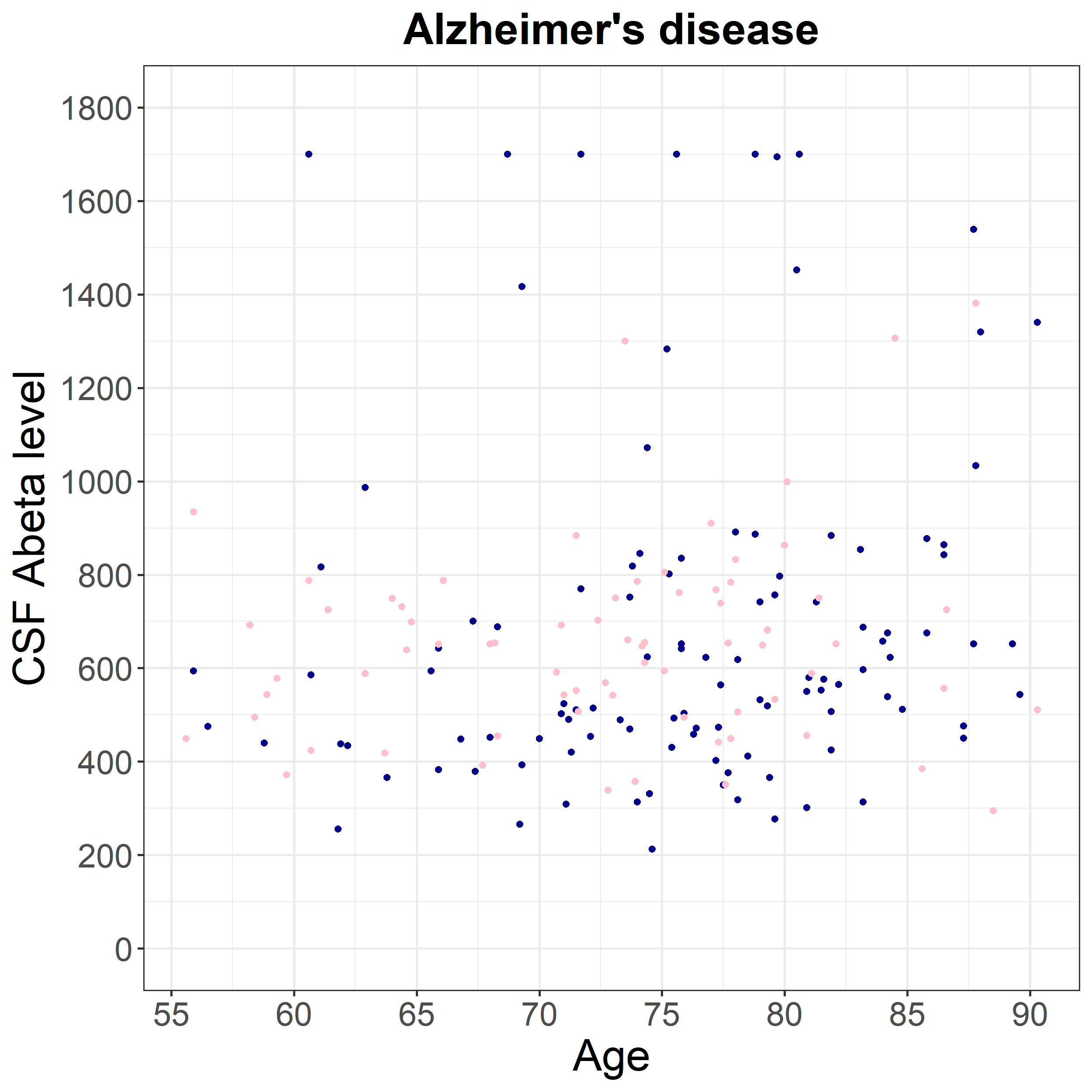}
	}
	\\
	\subfigure{
		\includegraphics[width=0.23\linewidth]{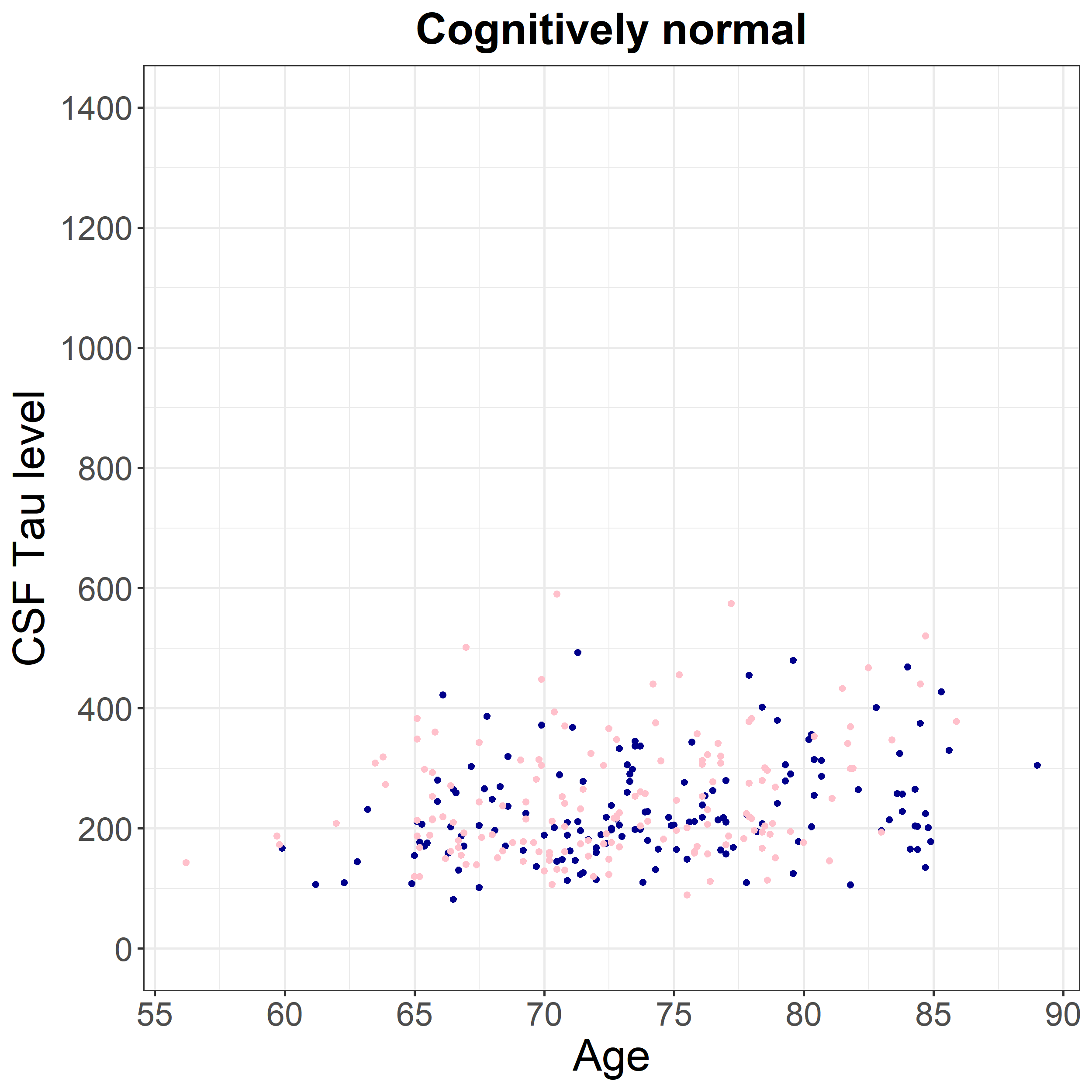}
	}
	\subfigure{
		\includegraphics[width=0.23\linewidth]{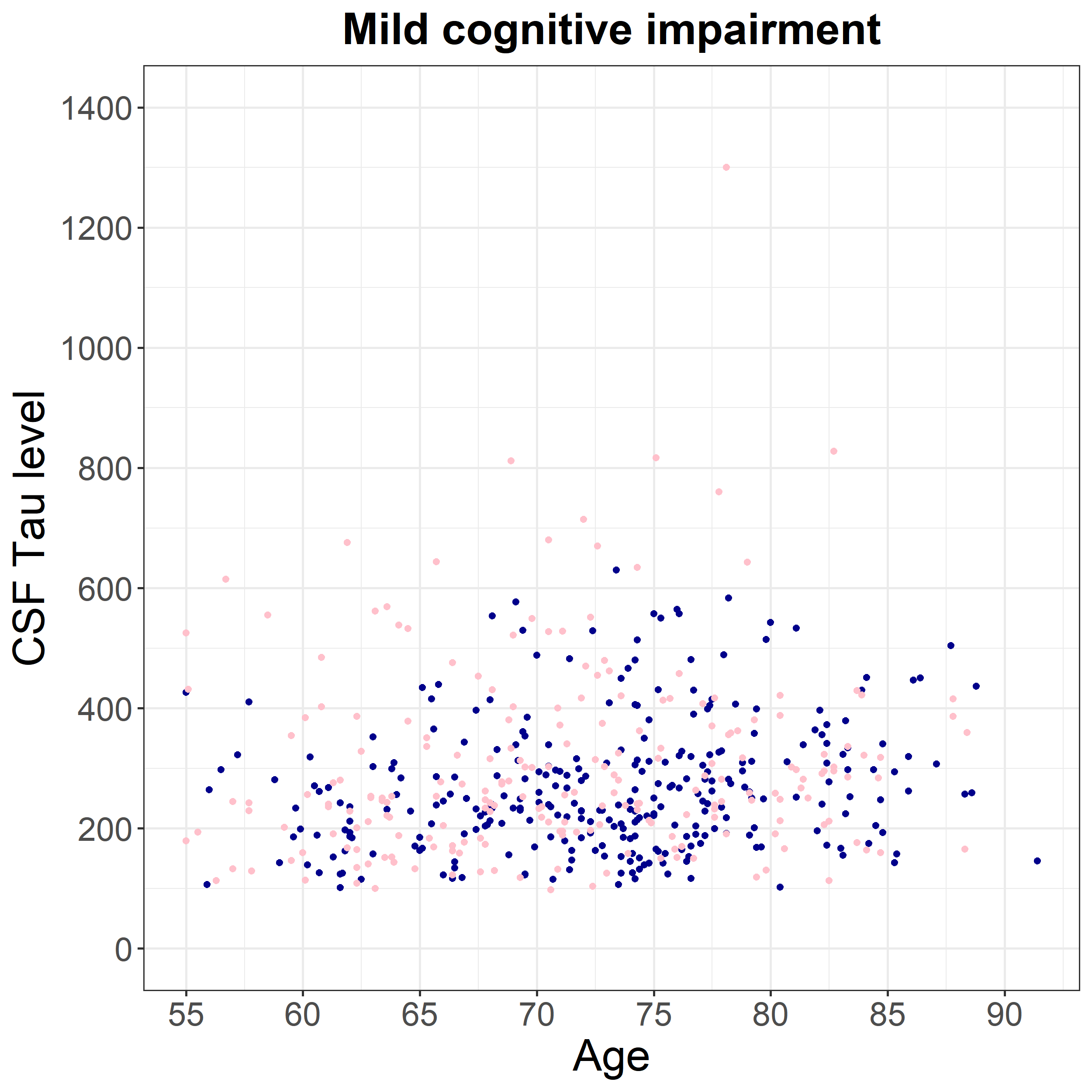}
	}
	\subfigure{
		\includegraphics[width=0.23\linewidth]{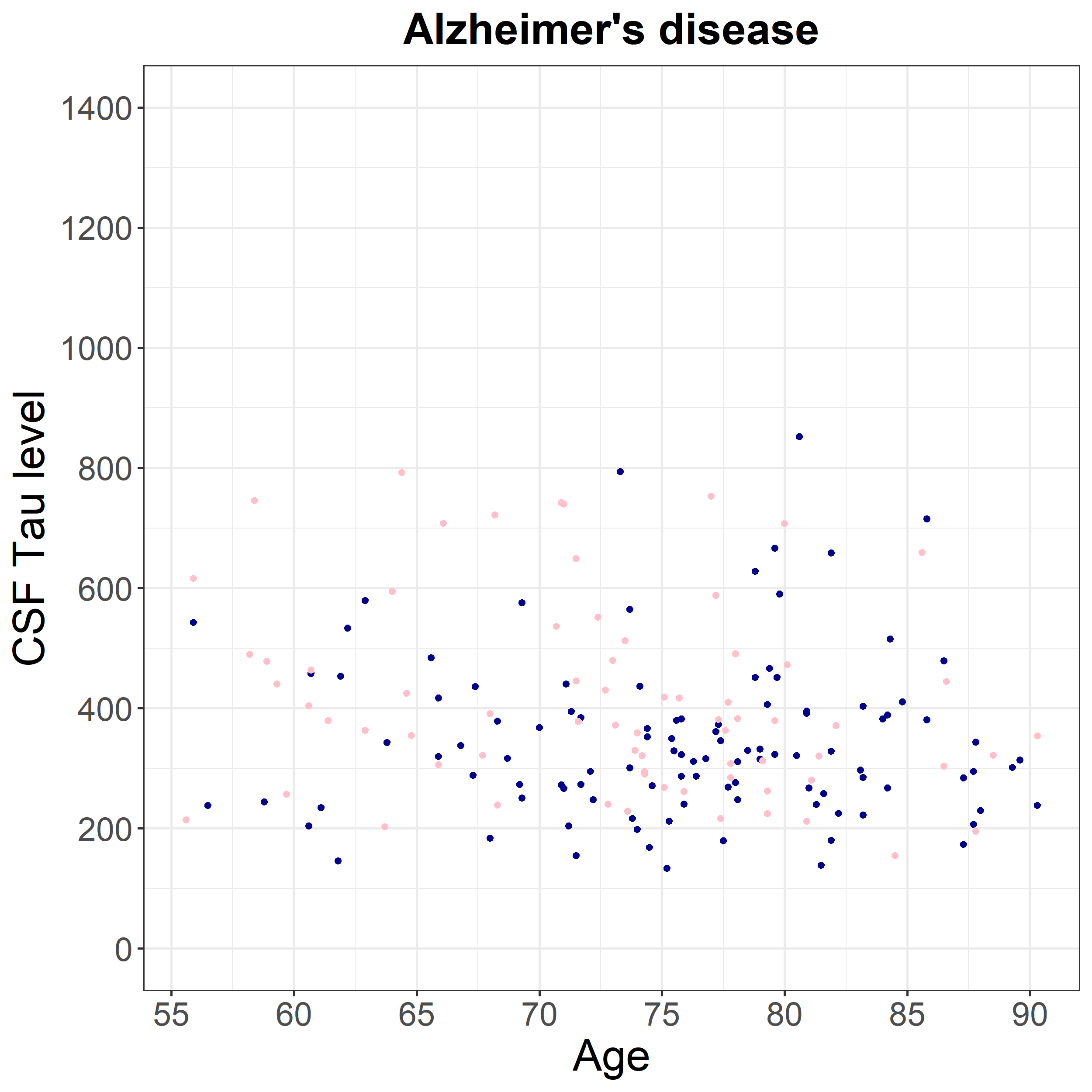}
	}
	\\
	\subfigure{
		\includegraphics[width=0.23\linewidth]{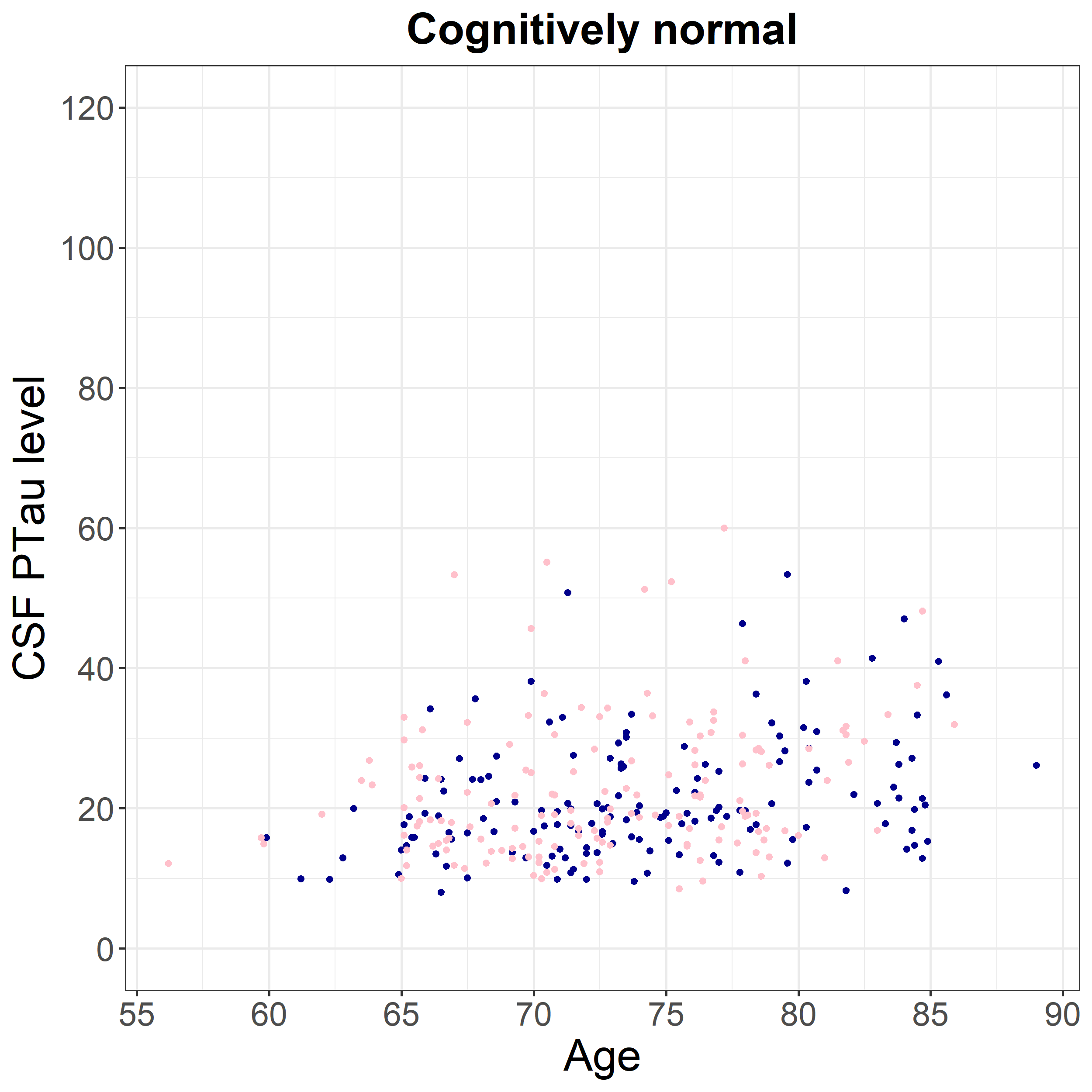}
	}
	\subfigure{
		\includegraphics[width=0.23\linewidth]{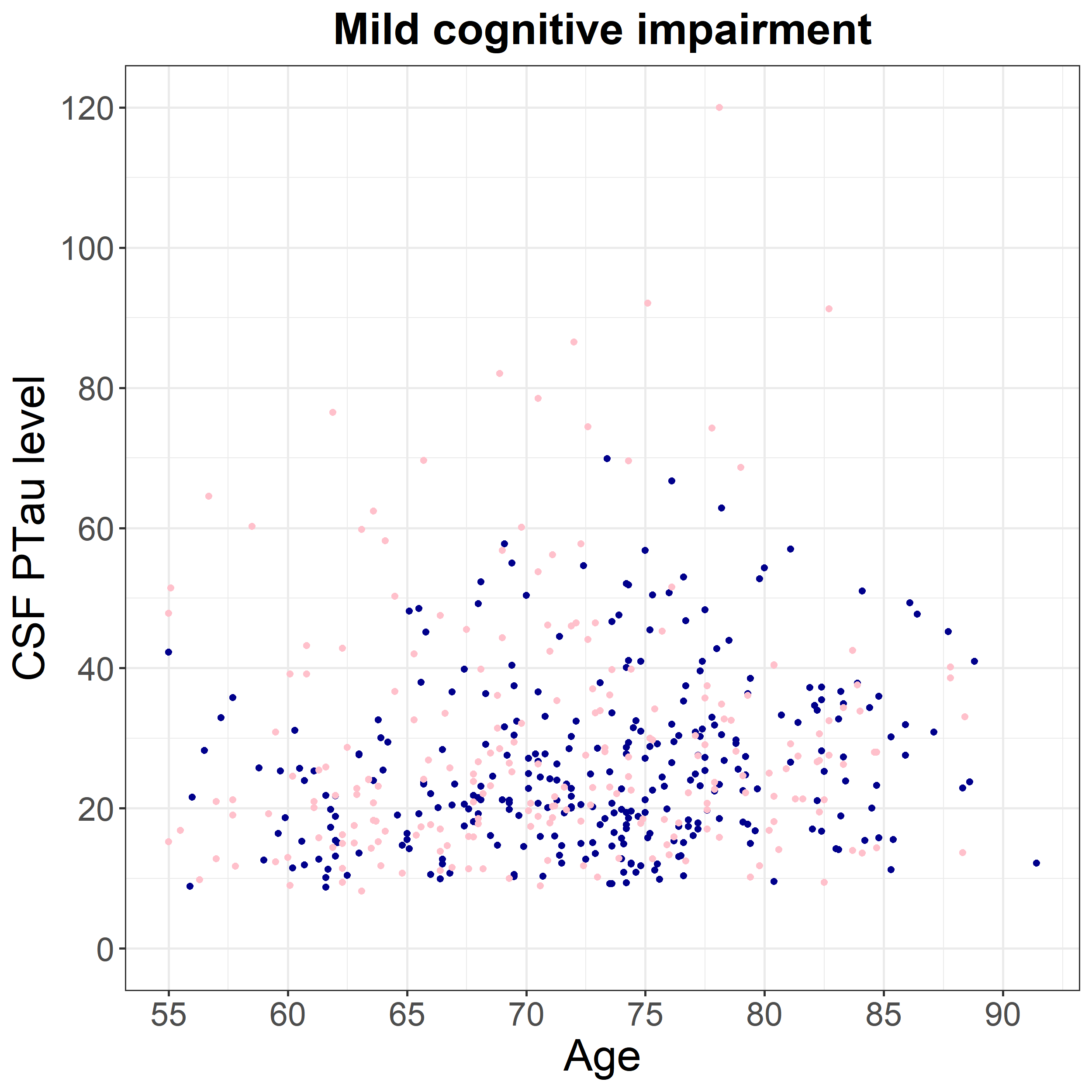}
	}
	\subfigure{
		\includegraphics[width=0.23\linewidth]{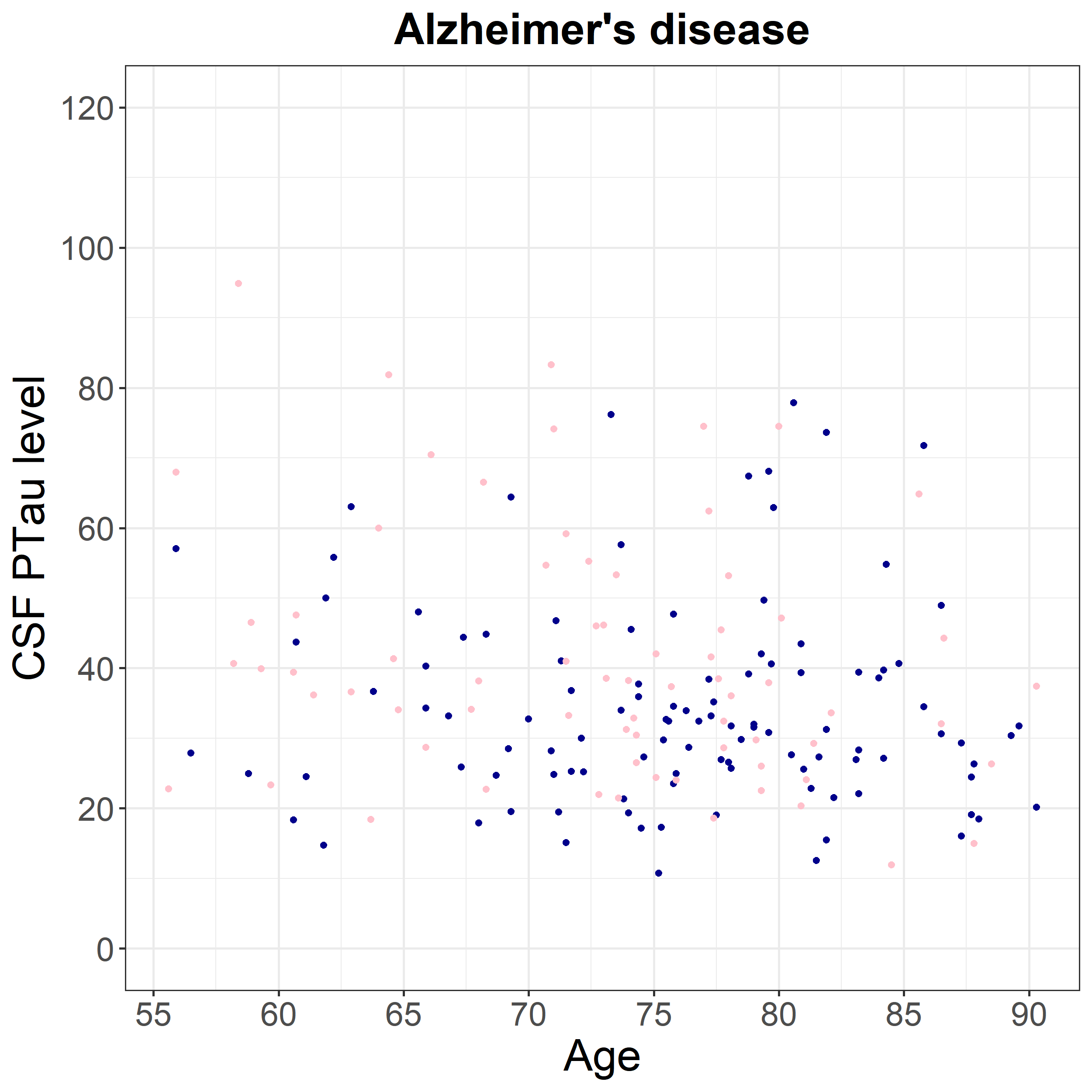}
	}
	\caption{Exploratory data analysis. Scatter plots of the HCI, CSF Abeta, CSF Tau, and CSF pTau biomarker outcomes against age in the three groups. Dark blue points represent males, while pink points represent females.}
	\label{scatter_adni}
	\vspace{-1ex}
\end{figure}

\clearpage

\begin{figure}[htpb]
	\centering
	\subfigure{
		\includegraphics[width=0.3\linewidth]{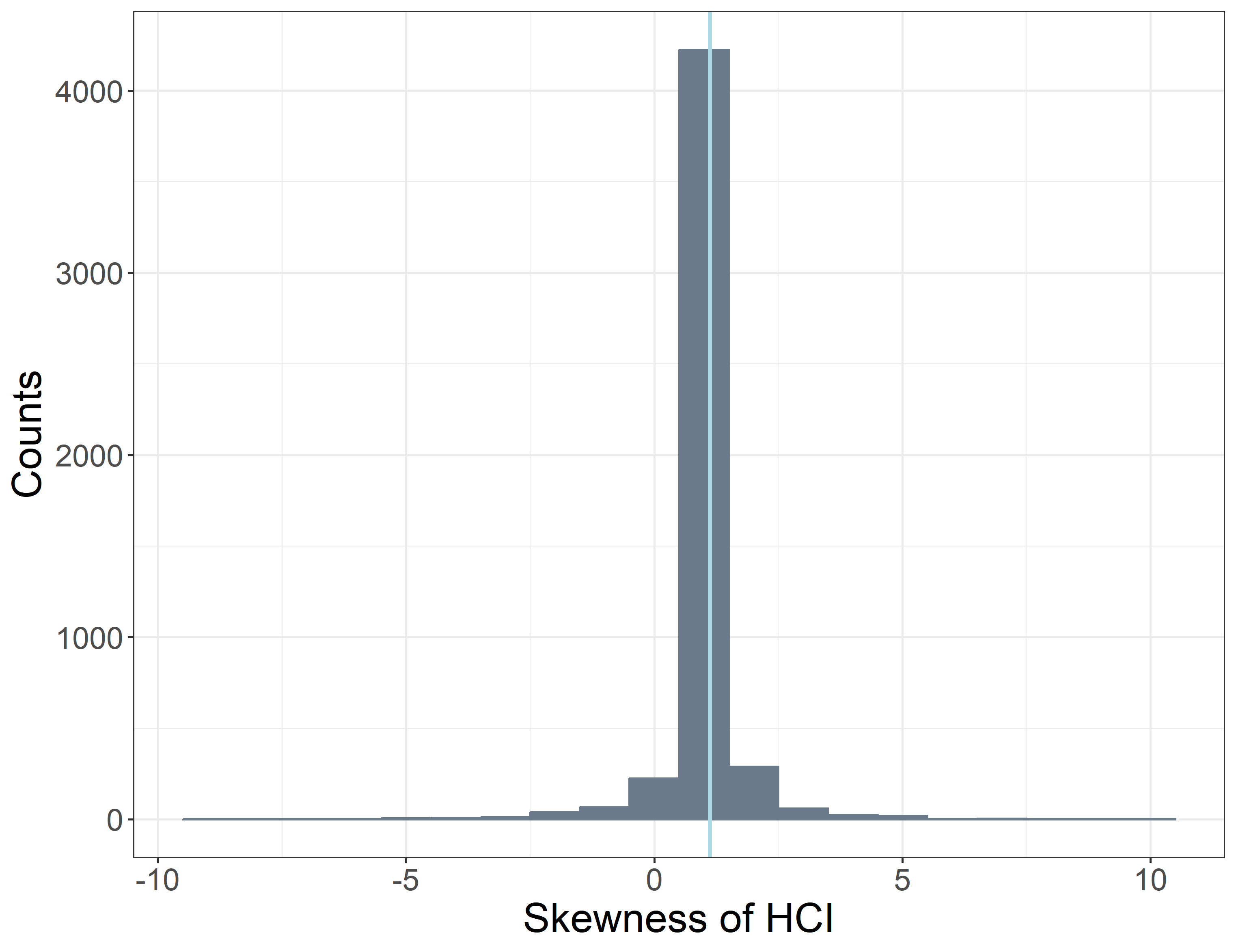}
	}
	\subfigure{
		\includegraphics[width=0.3\linewidth]{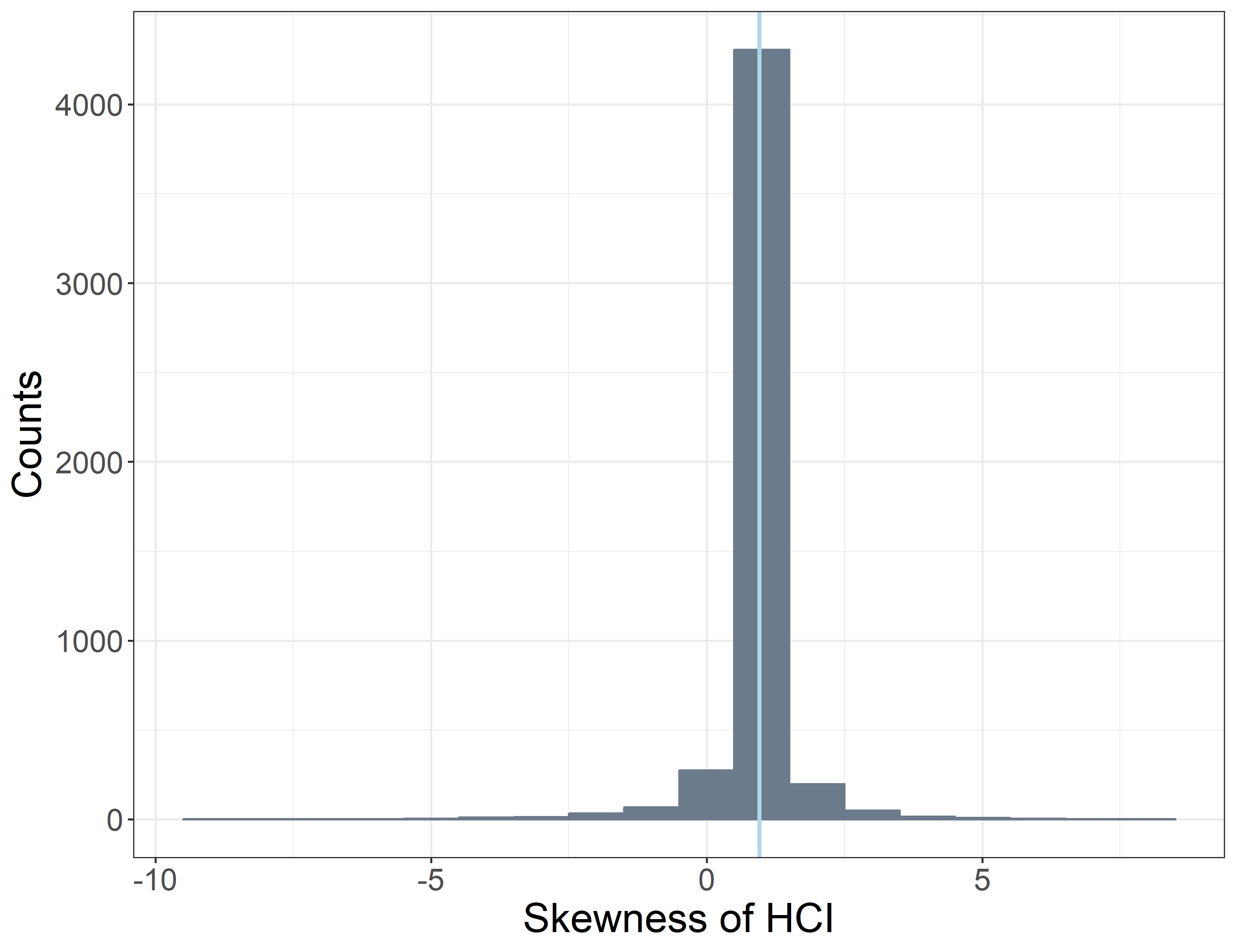}
	}
	\subfigure{
		\includegraphics[width=0.3\linewidth]{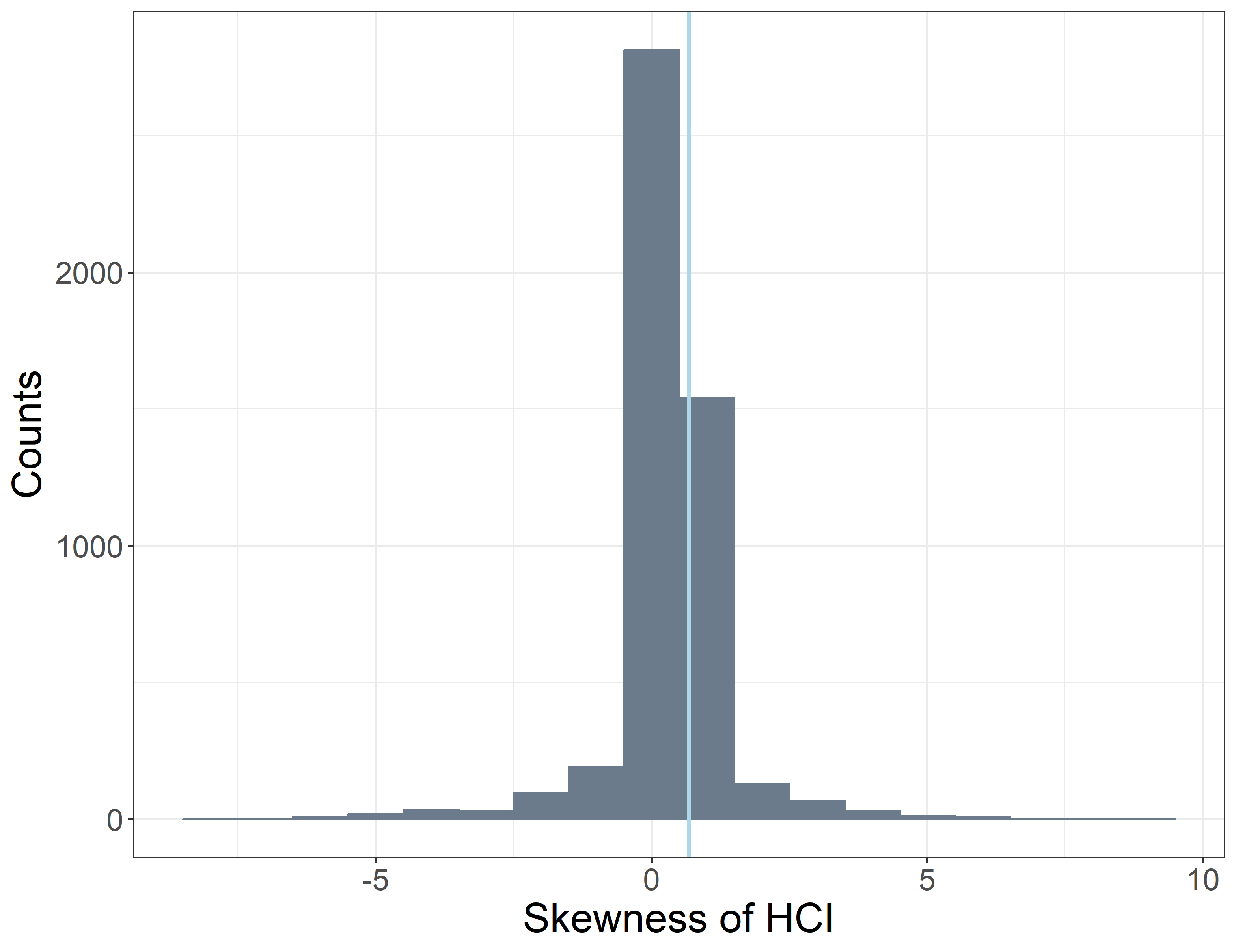}
	}
	\\
	\subfigure{
		\includegraphics[width=0.3\linewidth]{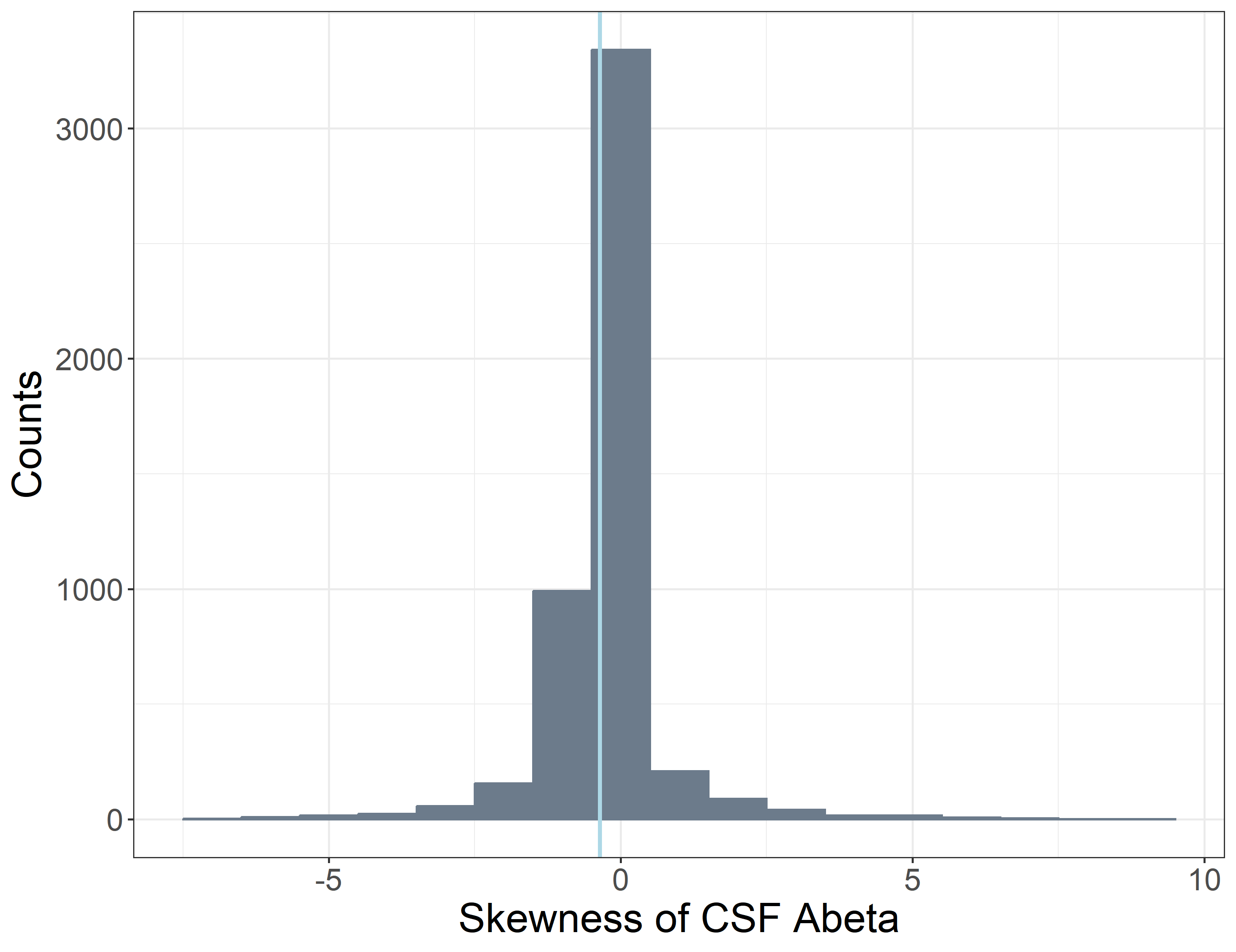}
	}
	\subfigure{
		\includegraphics[width=0.3\linewidth]{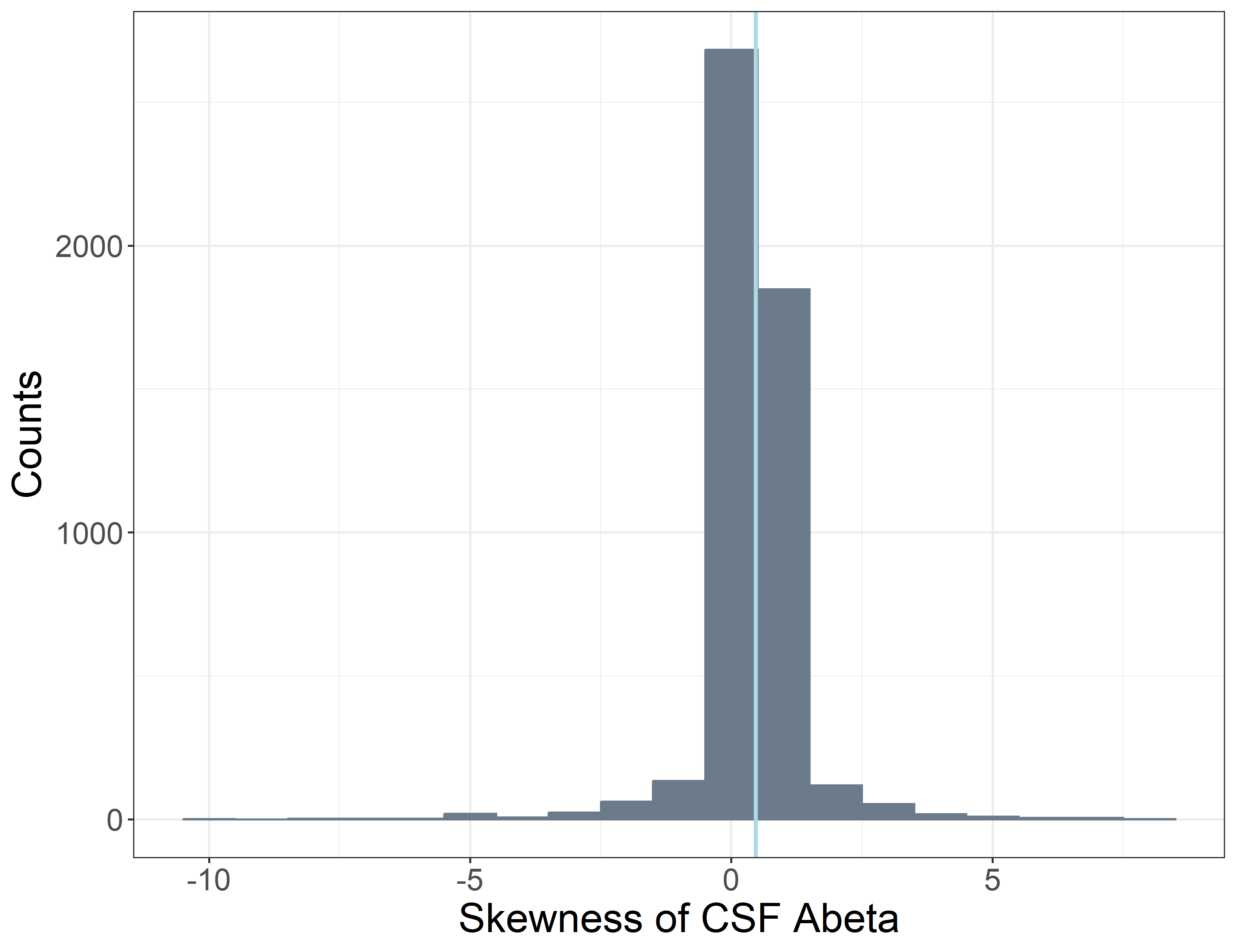}
	}
	\subfigure{
		\includegraphics[width=0.3\linewidth]{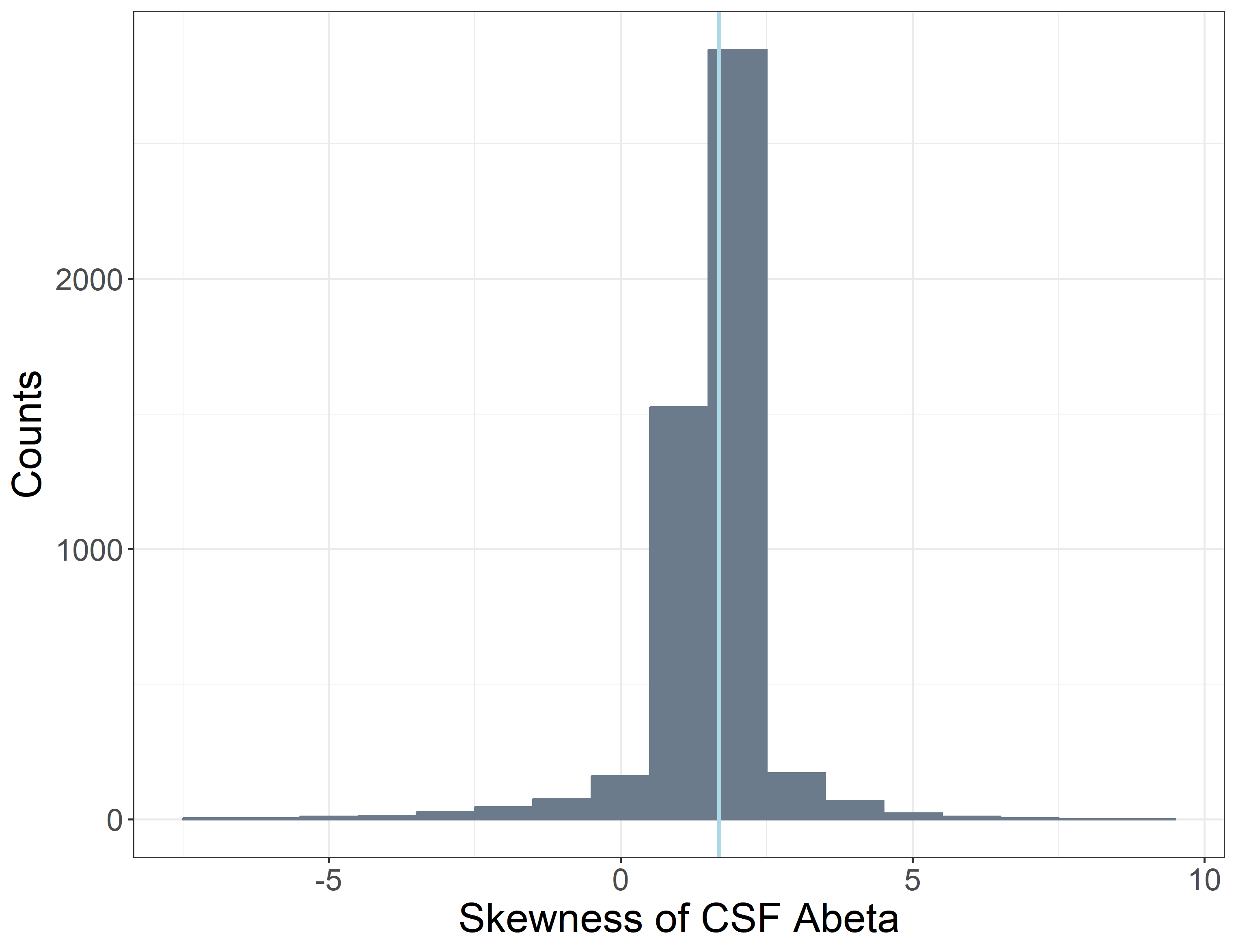}
	}
	\\
	\subfigure{
		\includegraphics[width=0.3\linewidth]{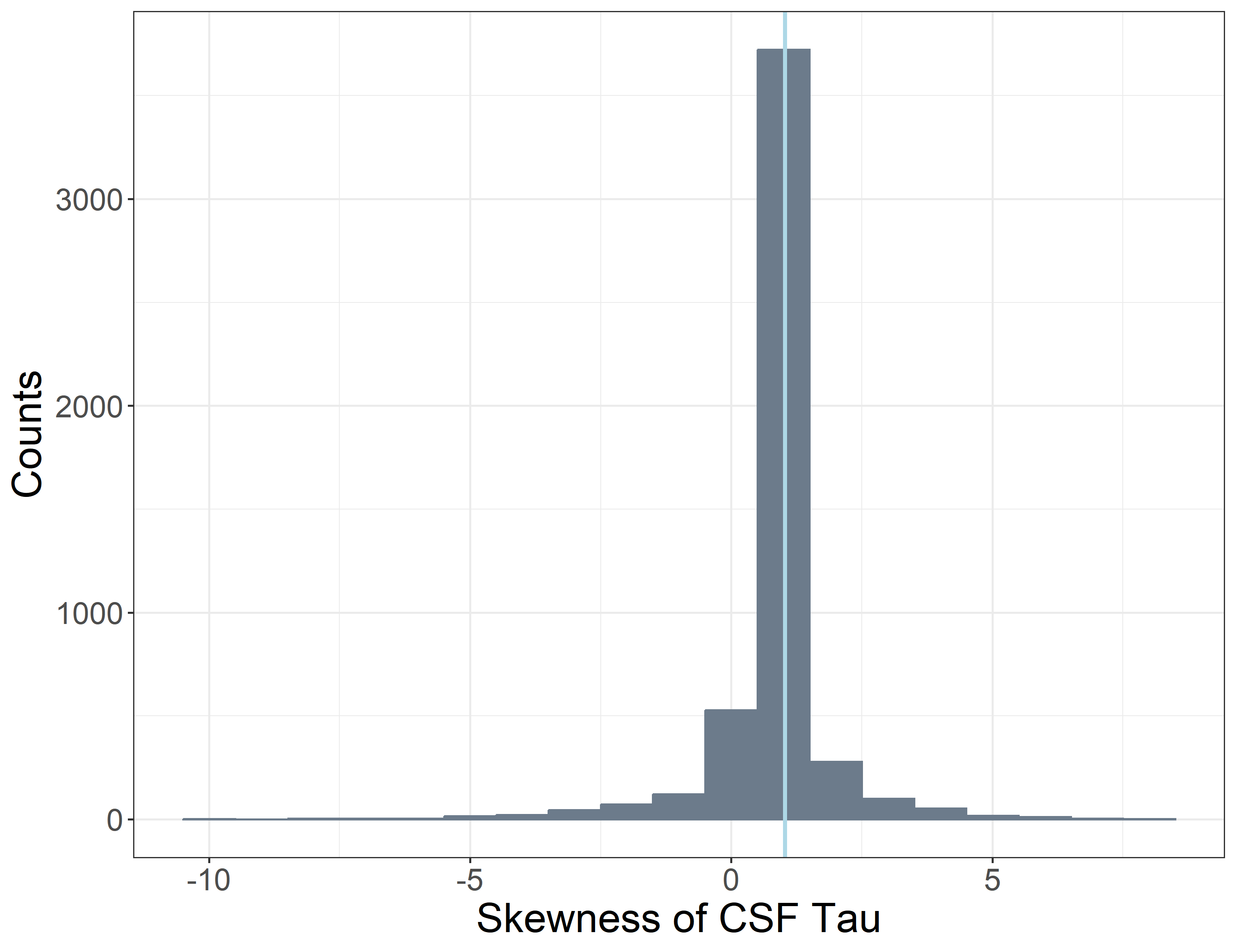}
	}
	\subfigure{
		\includegraphics[width=0.3\linewidth]{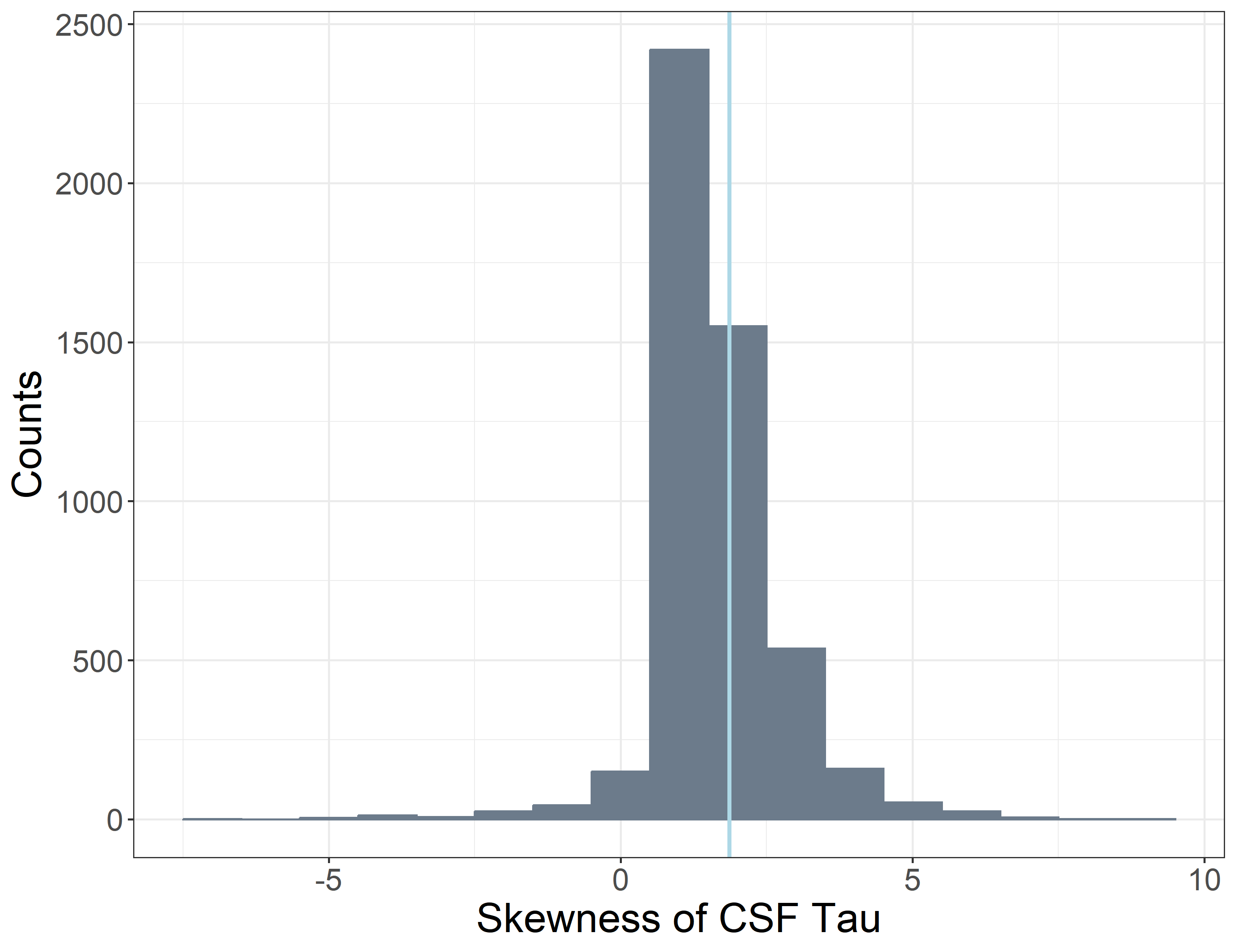}
	}
	\subfigure{
		\includegraphics[width=0.3\linewidth]{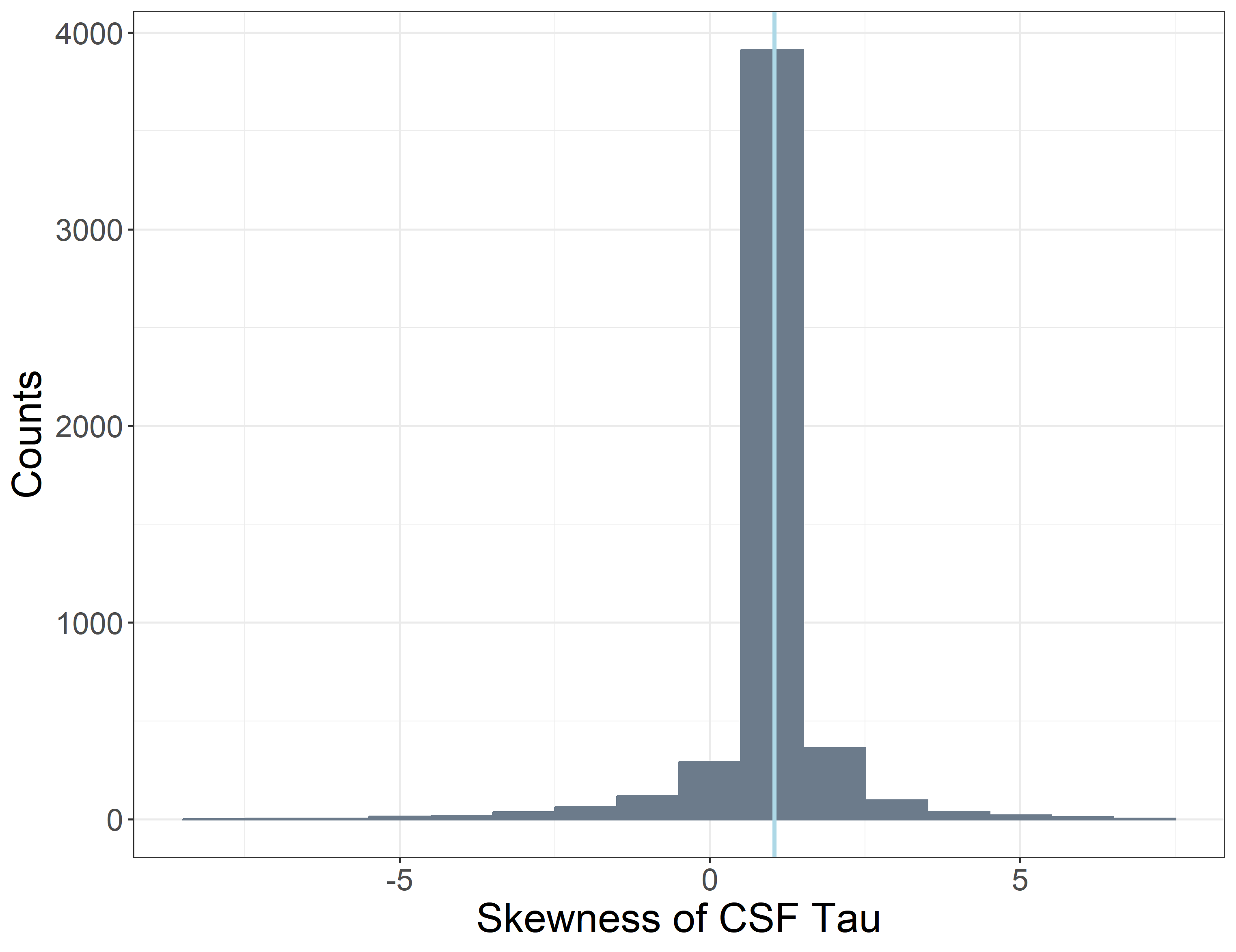}
	}
	\\
	\subfigure{
		\includegraphics[width=0.3\linewidth]{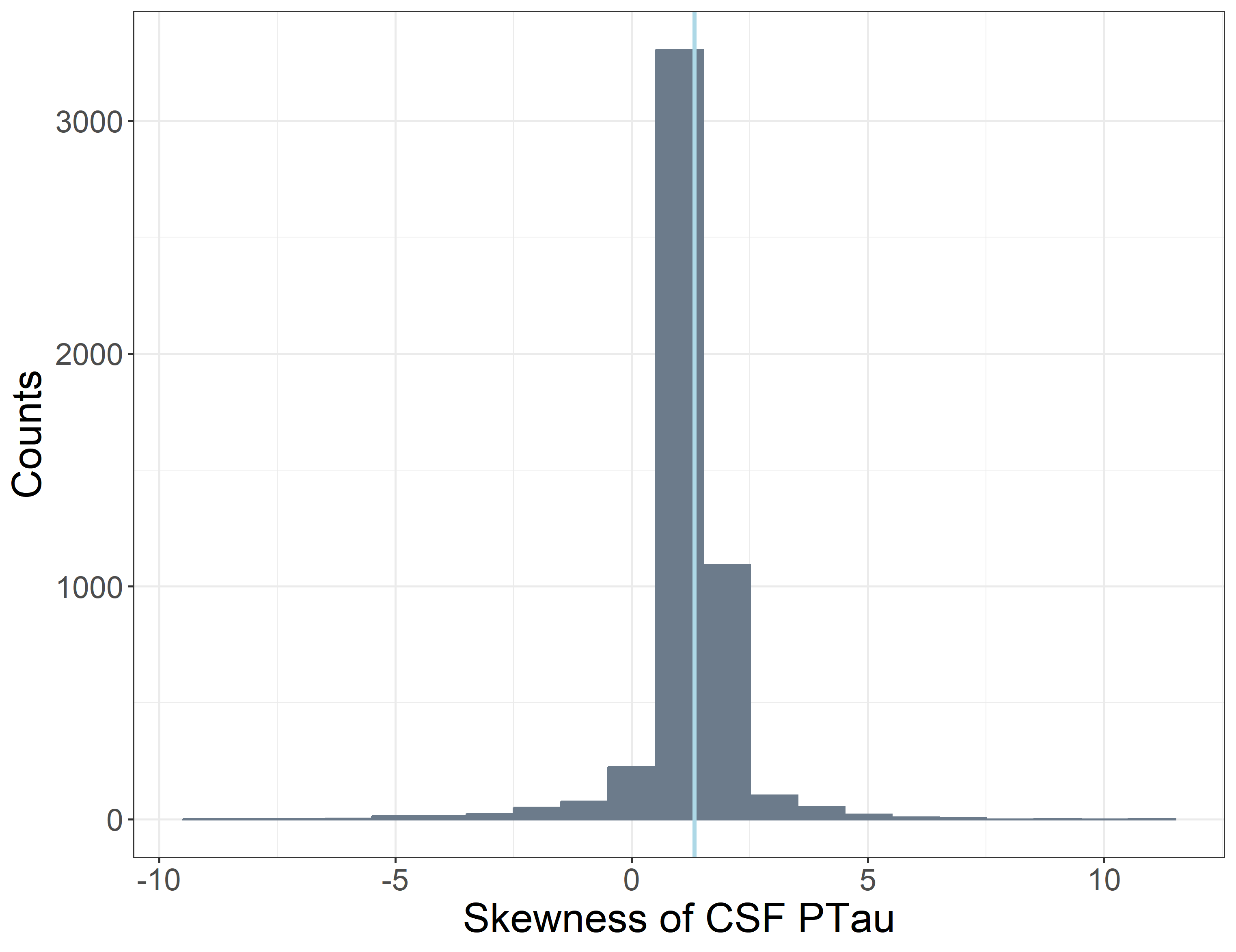}
	}
	\subfigure{
		\includegraphics[width=0.3\linewidth]{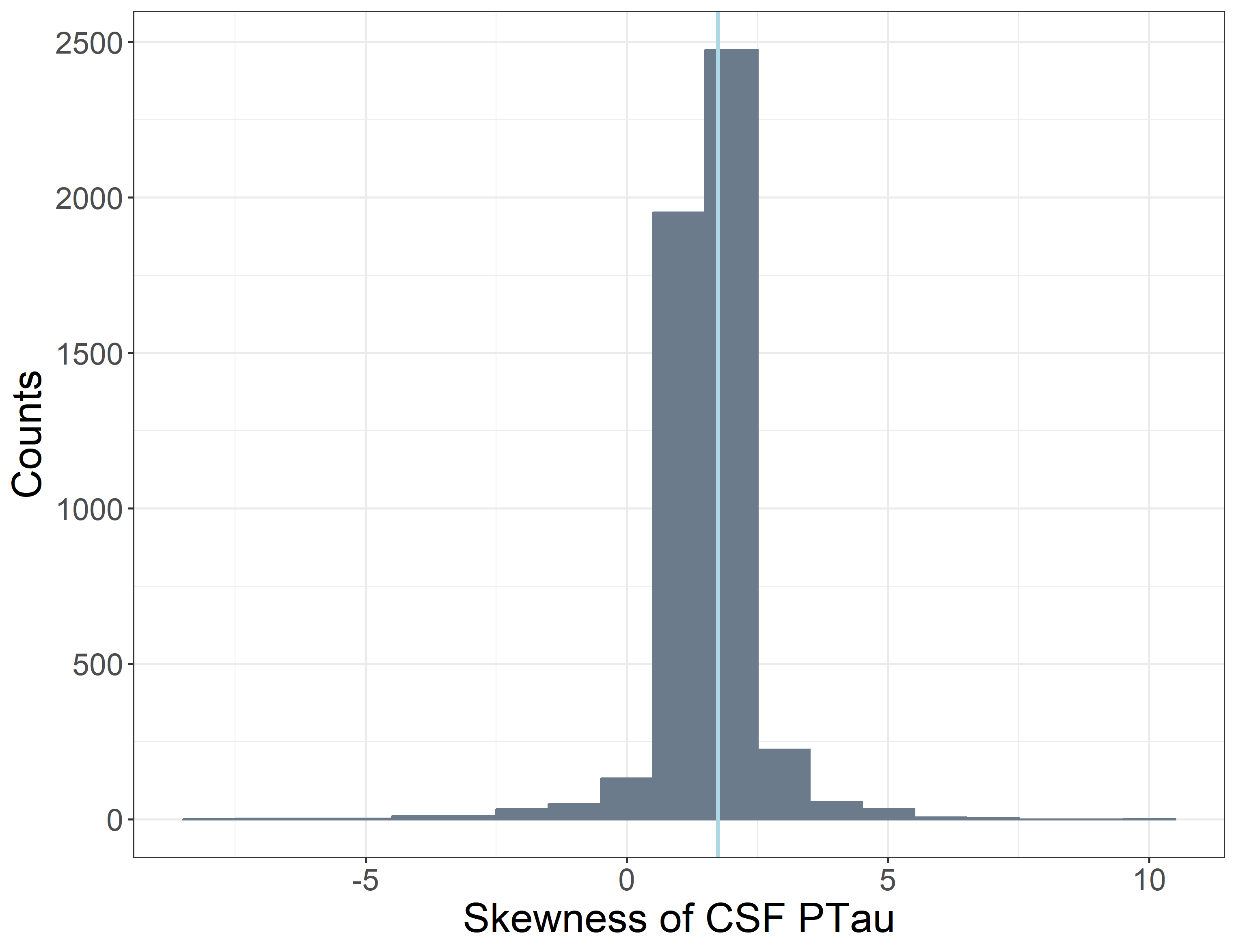}
	}
	\subfigure{
		\includegraphics[width=0.3\linewidth]{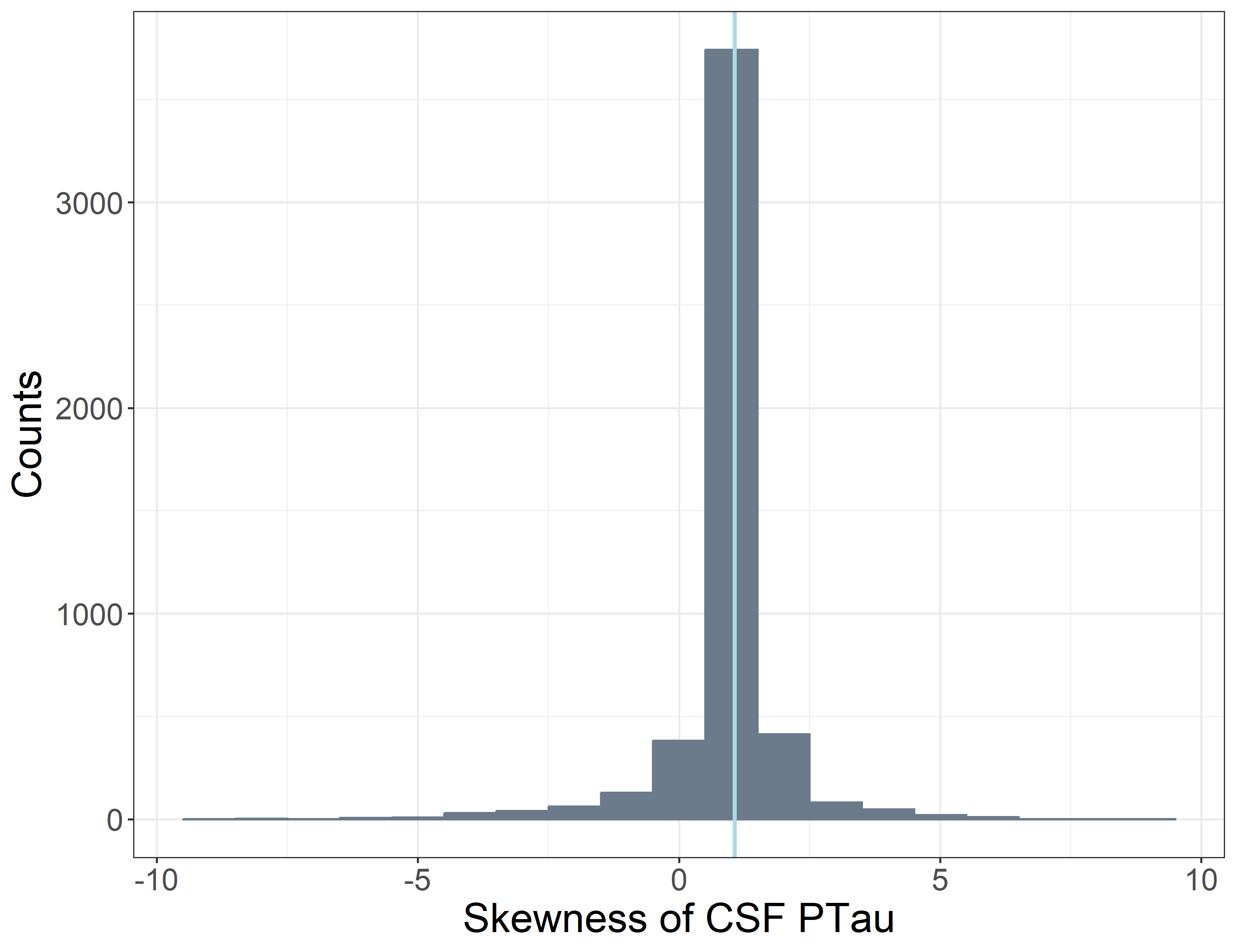}
	}
	\caption{Posterior predictive checks. Skewness estimate
		of the observed biomarker outcomes (light blue), shown alongside skewness estimates from the 5000 datasets drawn from the posterior predictive distribution (grey).}
	\label{posterior_predictive_skewness}
\end{figure}

\begin{figure}[htpb]
	\centering
	\subfigure{
		\includegraphics[width=0.3\linewidth]{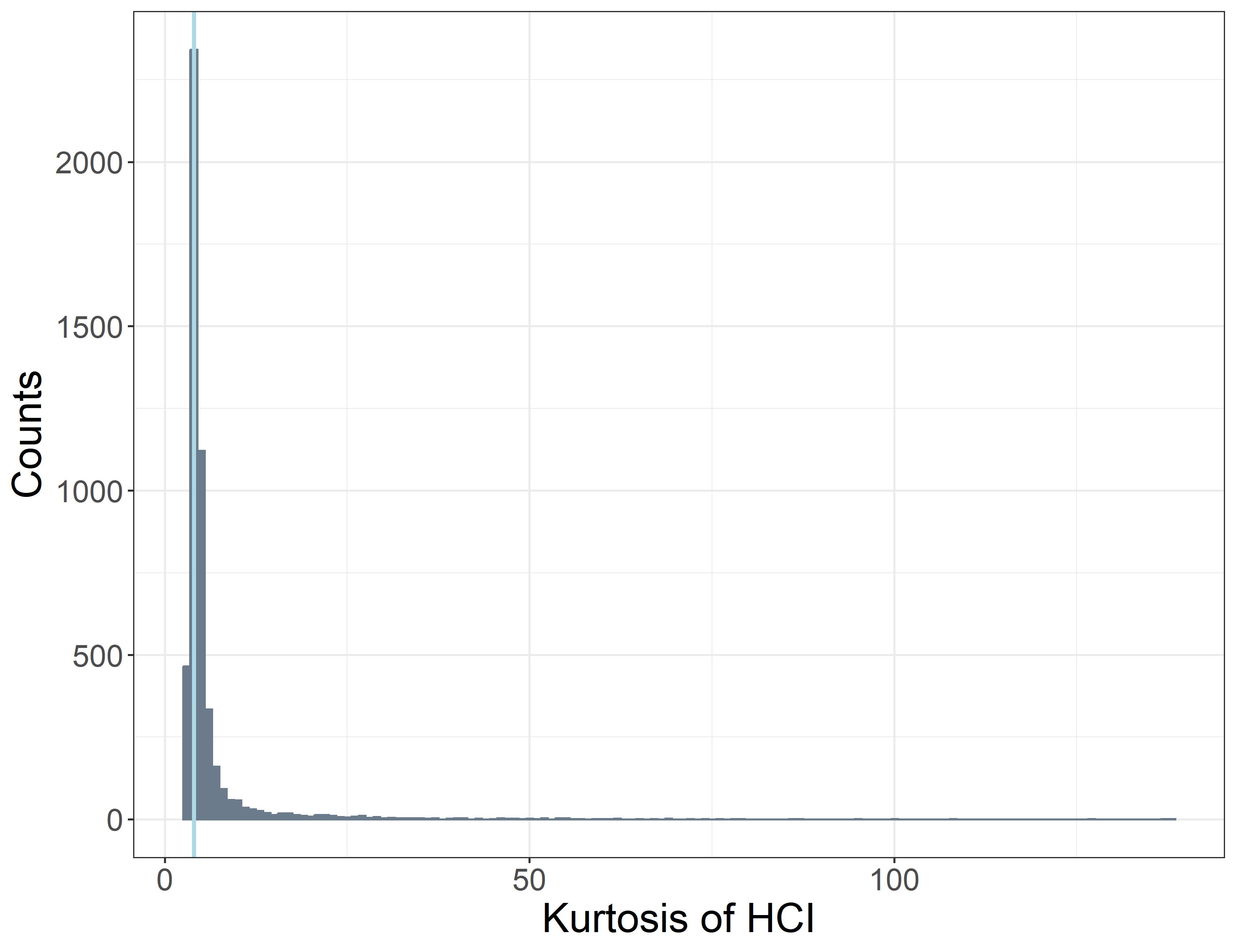}
	}
	\subfigure{
		\includegraphics[width=0.3\linewidth]{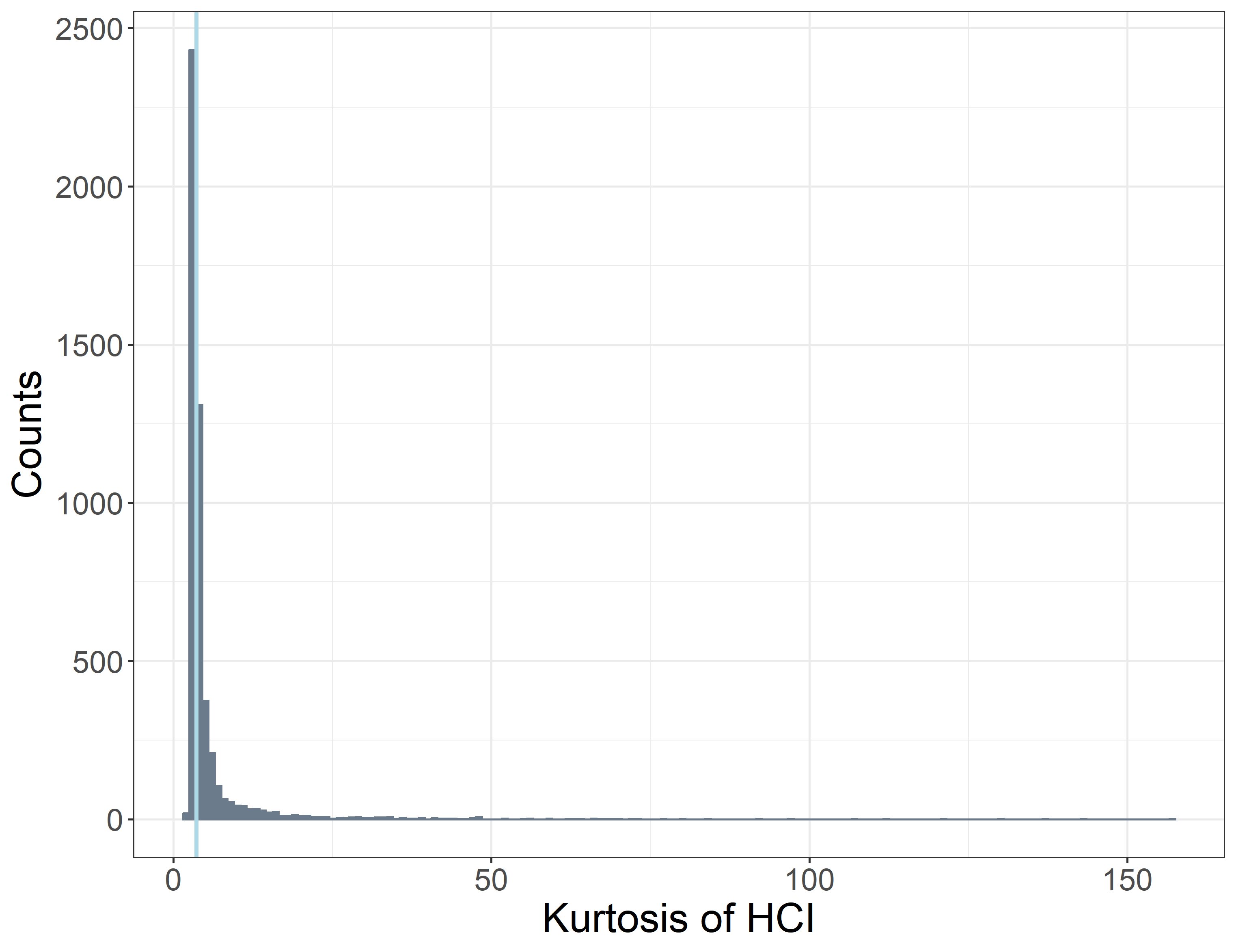}
	}
	\subfigure{
		\includegraphics[width=0.3\linewidth]{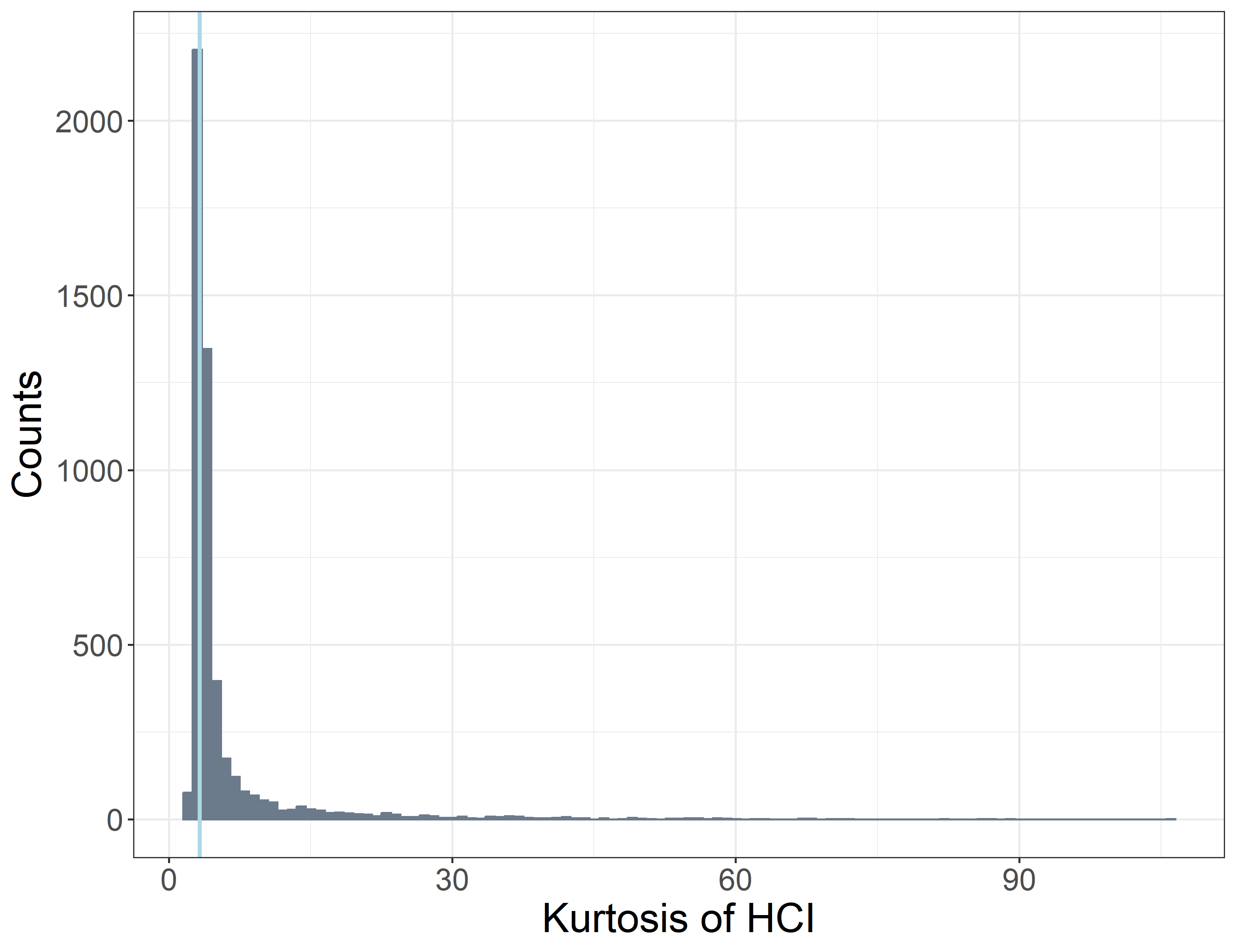}
	}
	\\
	\subfigure{
		\includegraphics[width=0.3\linewidth]{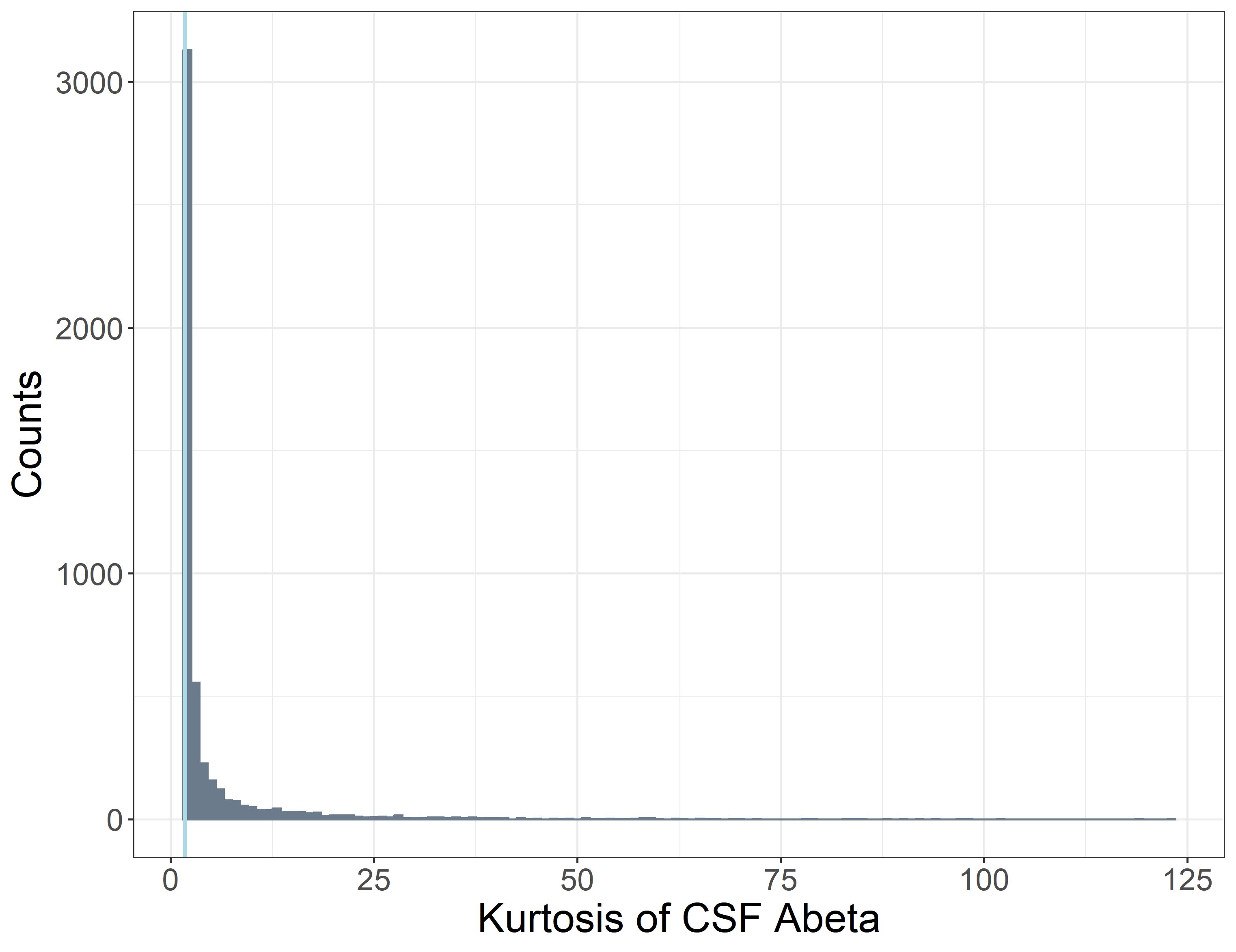}
	}
	\subfigure{
		\includegraphics[width=0.3\linewidth]{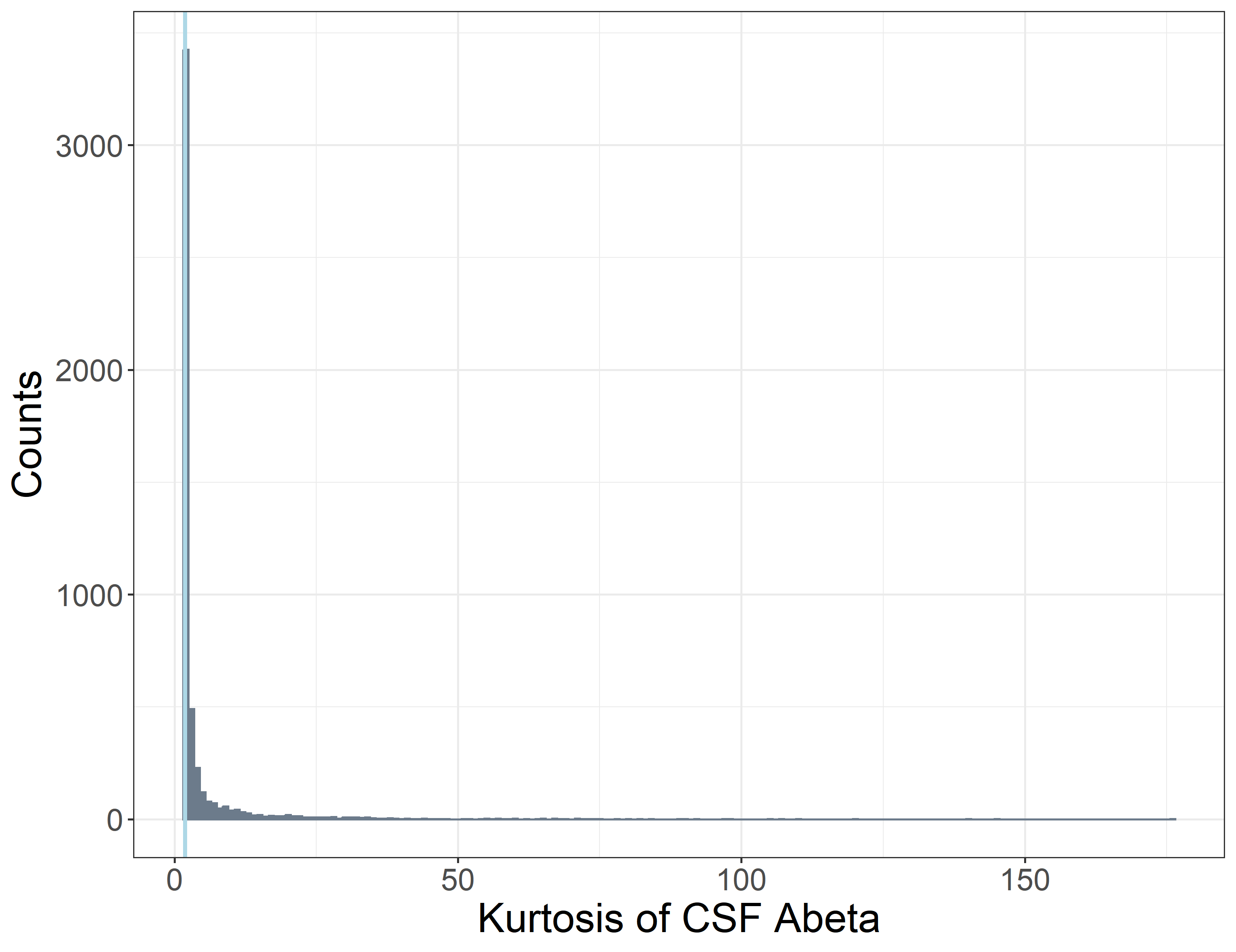}
	}
	\subfigure{
		\includegraphics[width=0.3\linewidth]{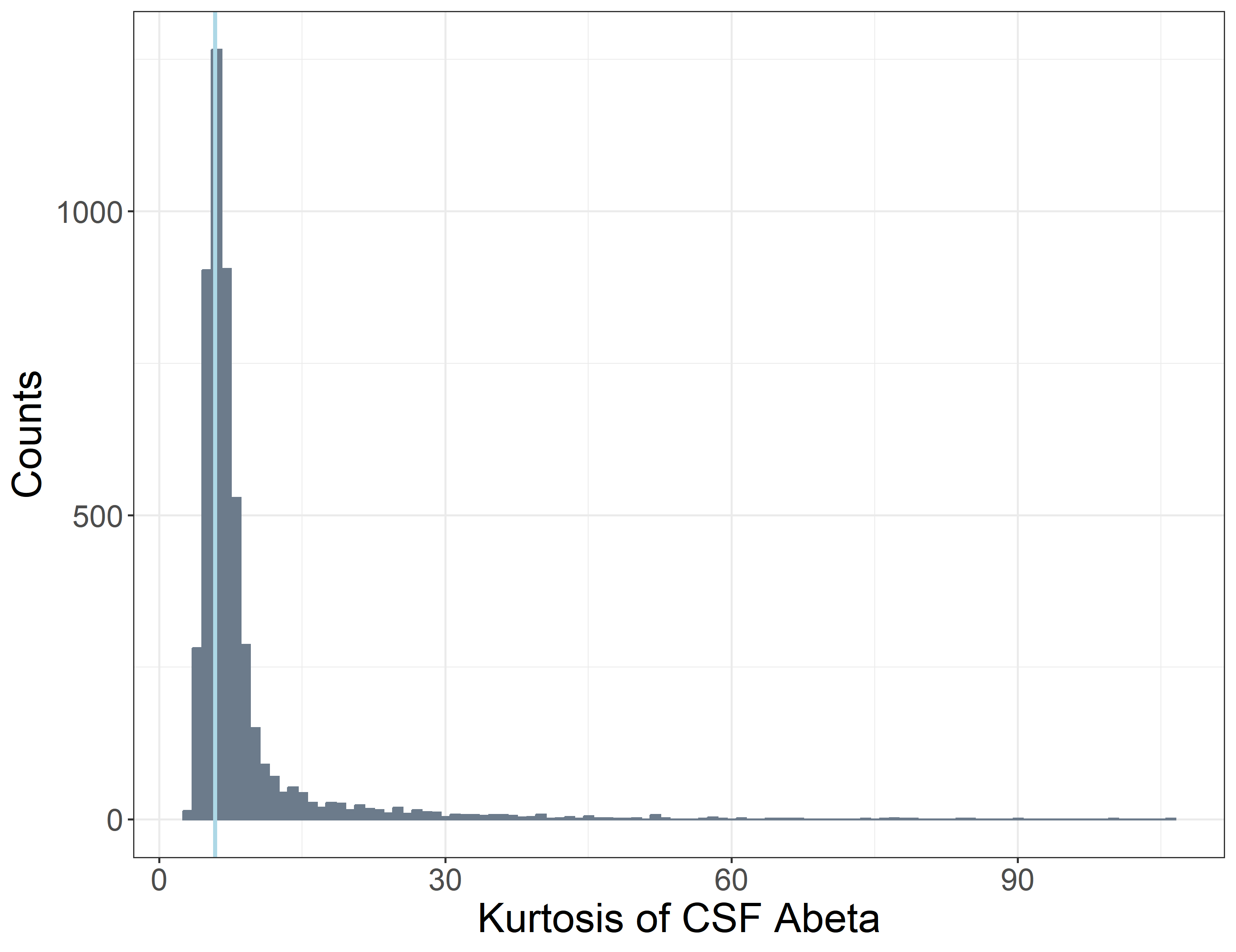}
	}
	\\
	\subfigure{
		\includegraphics[width=0.3\linewidth]{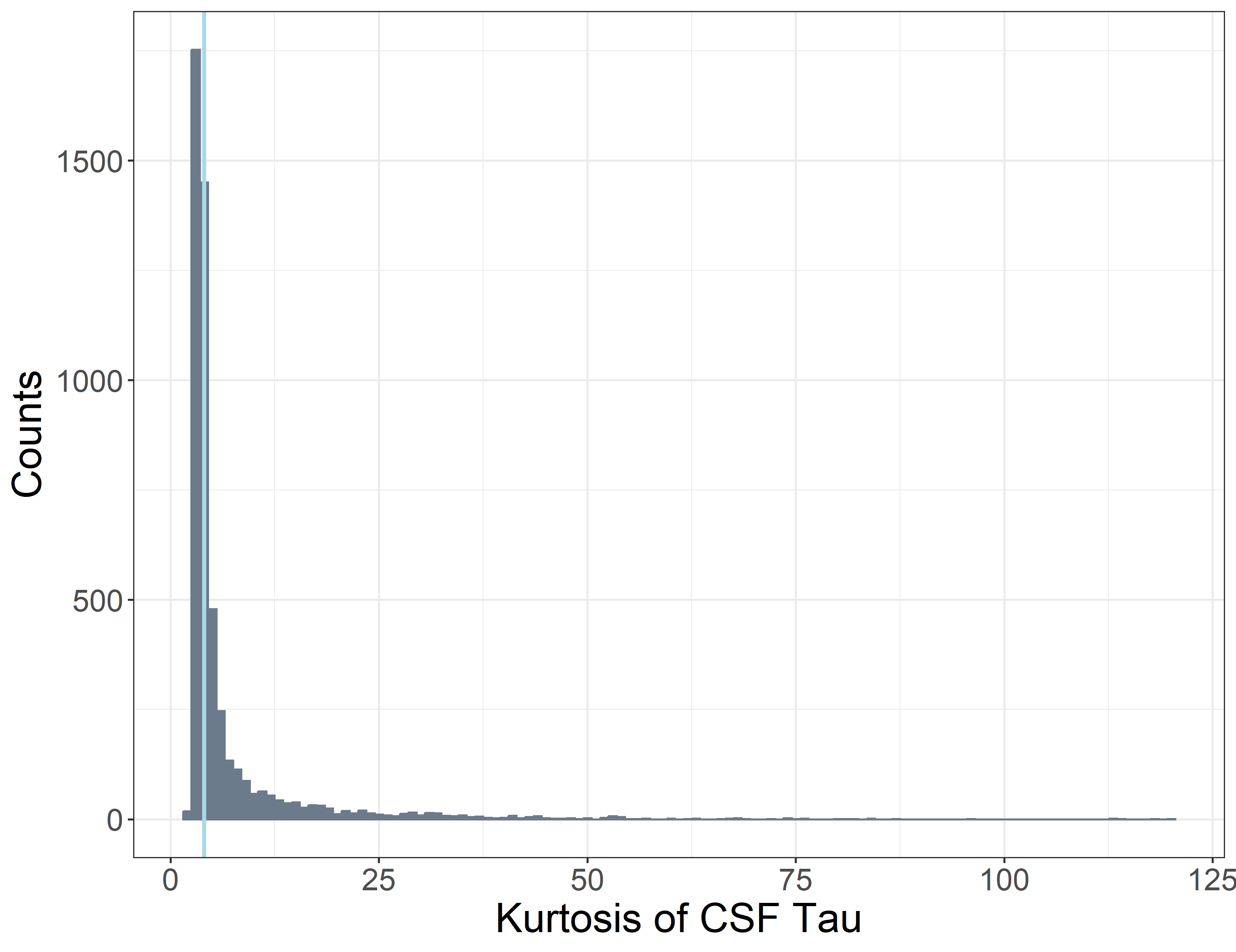}
	}
	\subfigure{
		\includegraphics[width=0.3\linewidth]{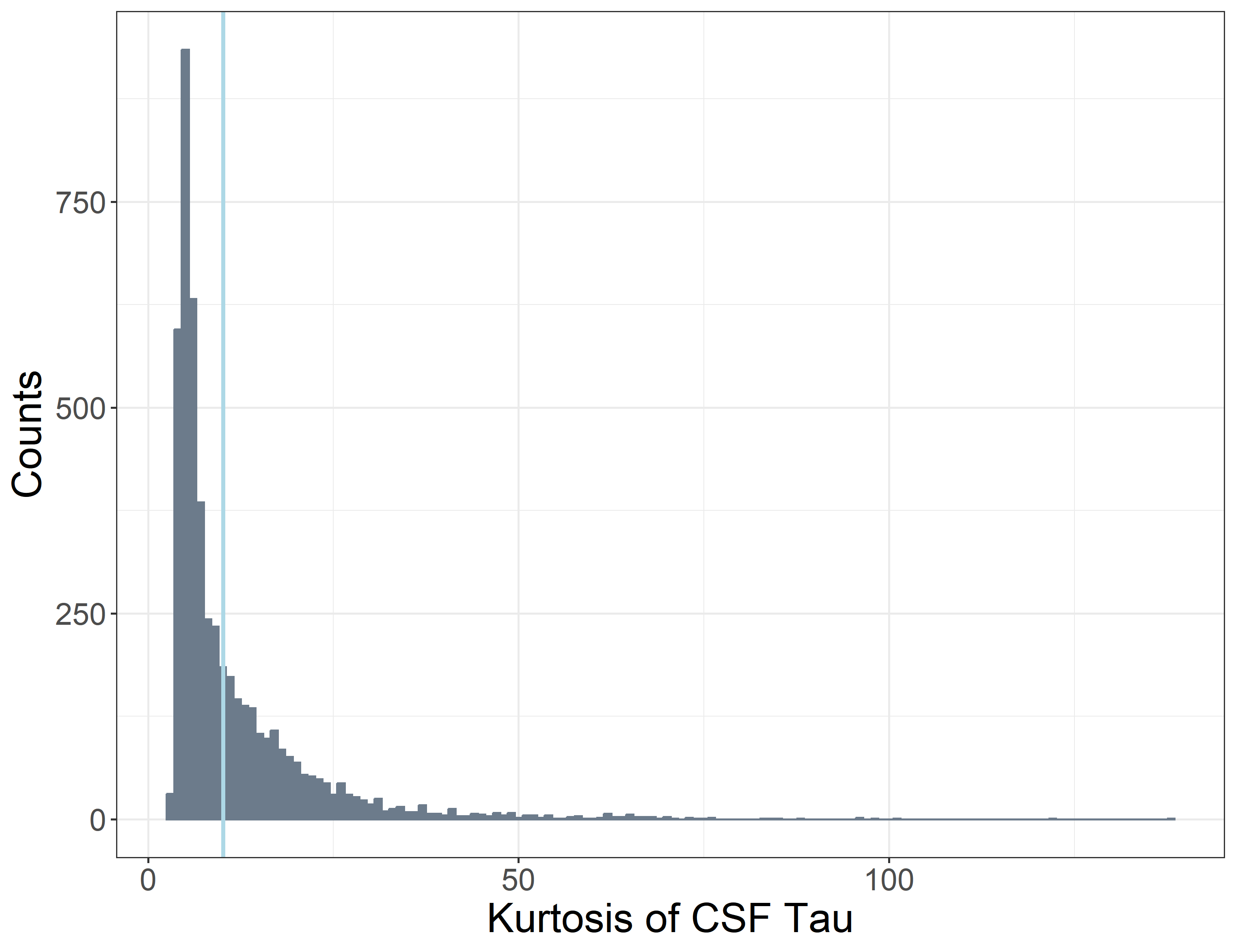}
	}
	\subfigure{
		\includegraphics[width=0.3\linewidth]{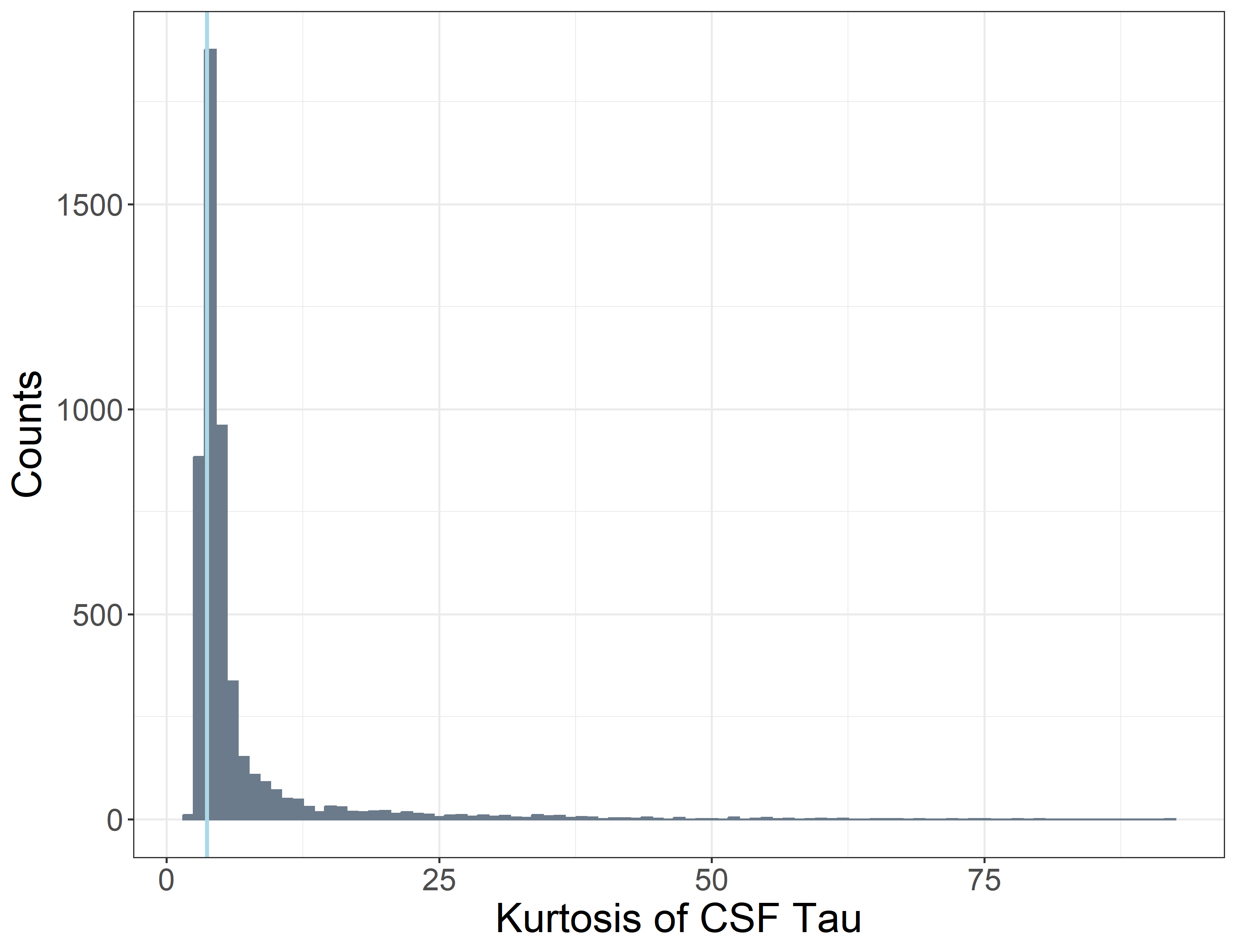}
	}
	\\
	\subfigure{
		\includegraphics[width=0.3\linewidth]{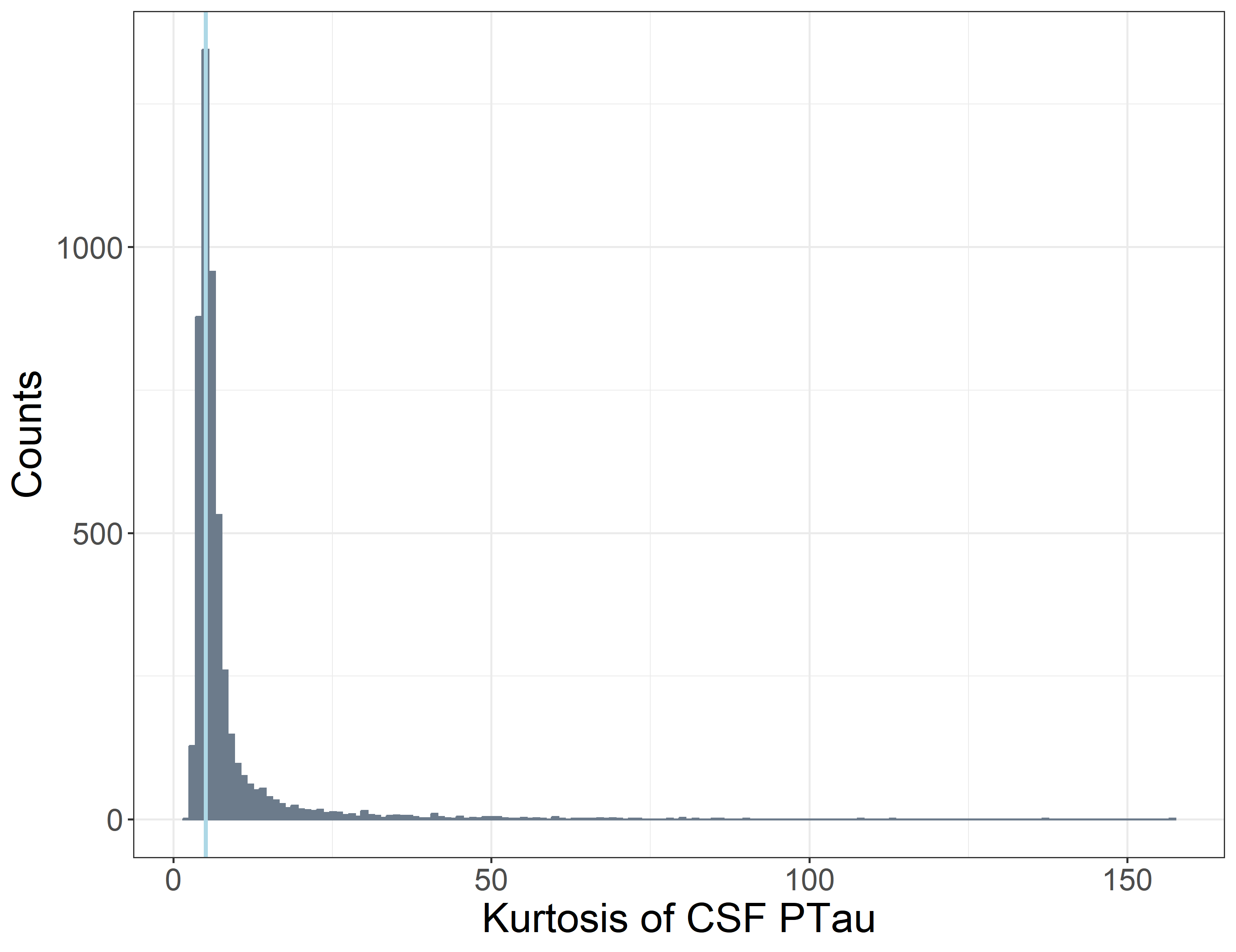}
	}
	\subfigure{
		\includegraphics[width=0.3\linewidth]{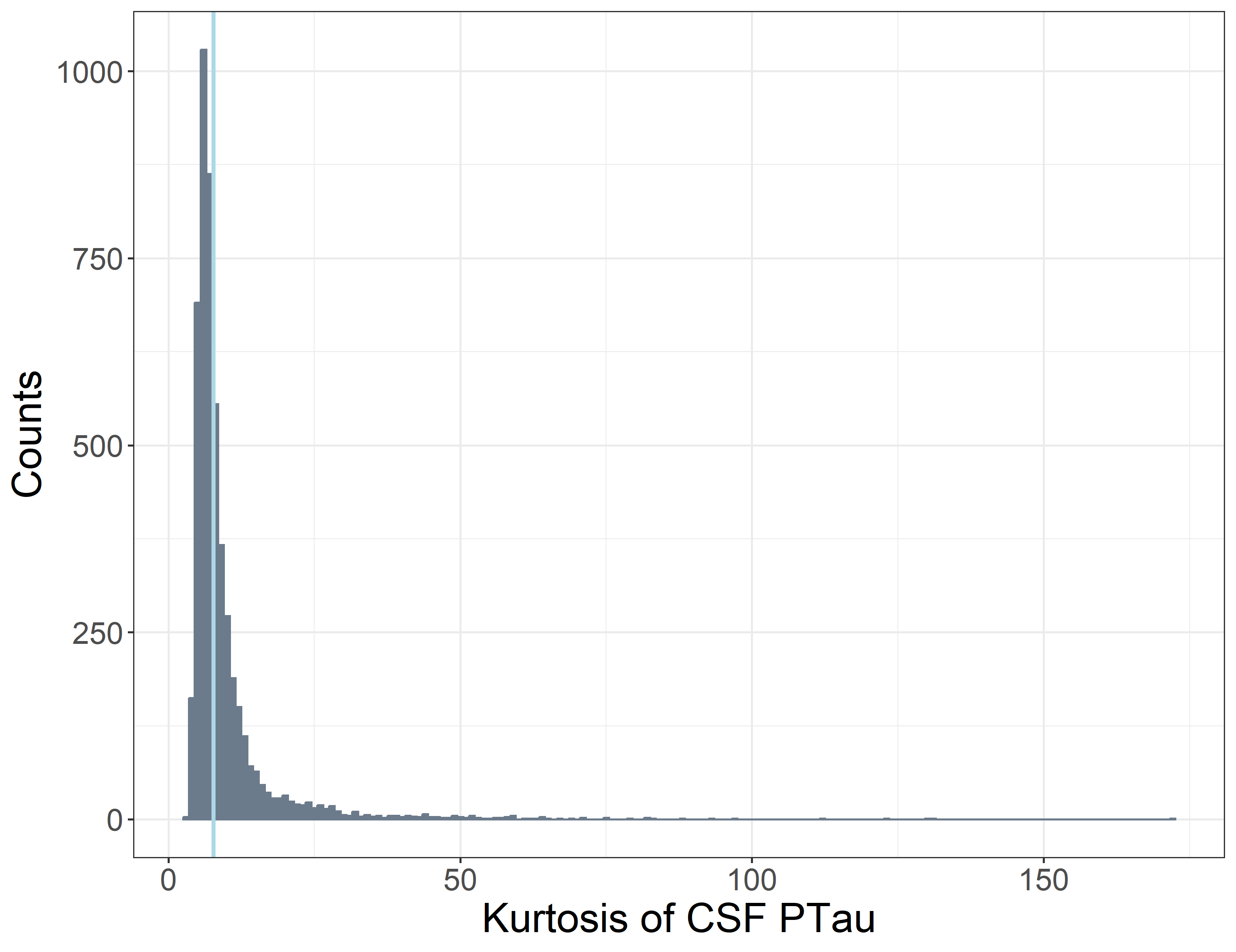}
	}
	\subfigure{
		\includegraphics[width=0.3\linewidth]{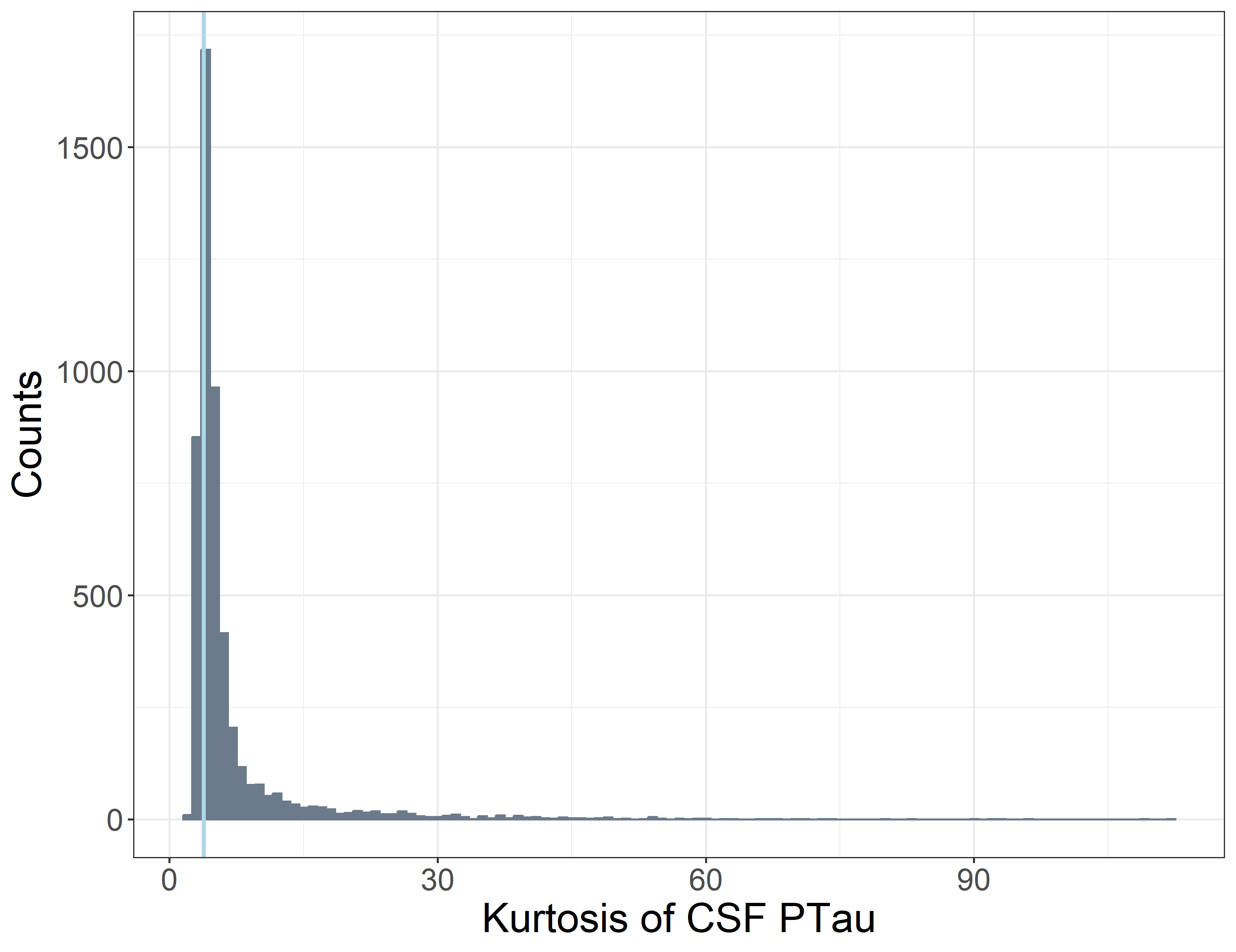}
	}
	\caption{Posterior predictive checks. Kurtosis estimate
		of the observed biomarker outcomes (light blue), shown alongside kurtosis estimates from the 5000 datasets drawn from the posterior predictive distribution (grey).}
	\label{posterior_predictive_kurtosis}
\end{figure}

\begin{figure}[htpb]
	\vspace{-2.5ex}
	\centering
	\subfigure{
		\includegraphics[width=0.3\linewidth]{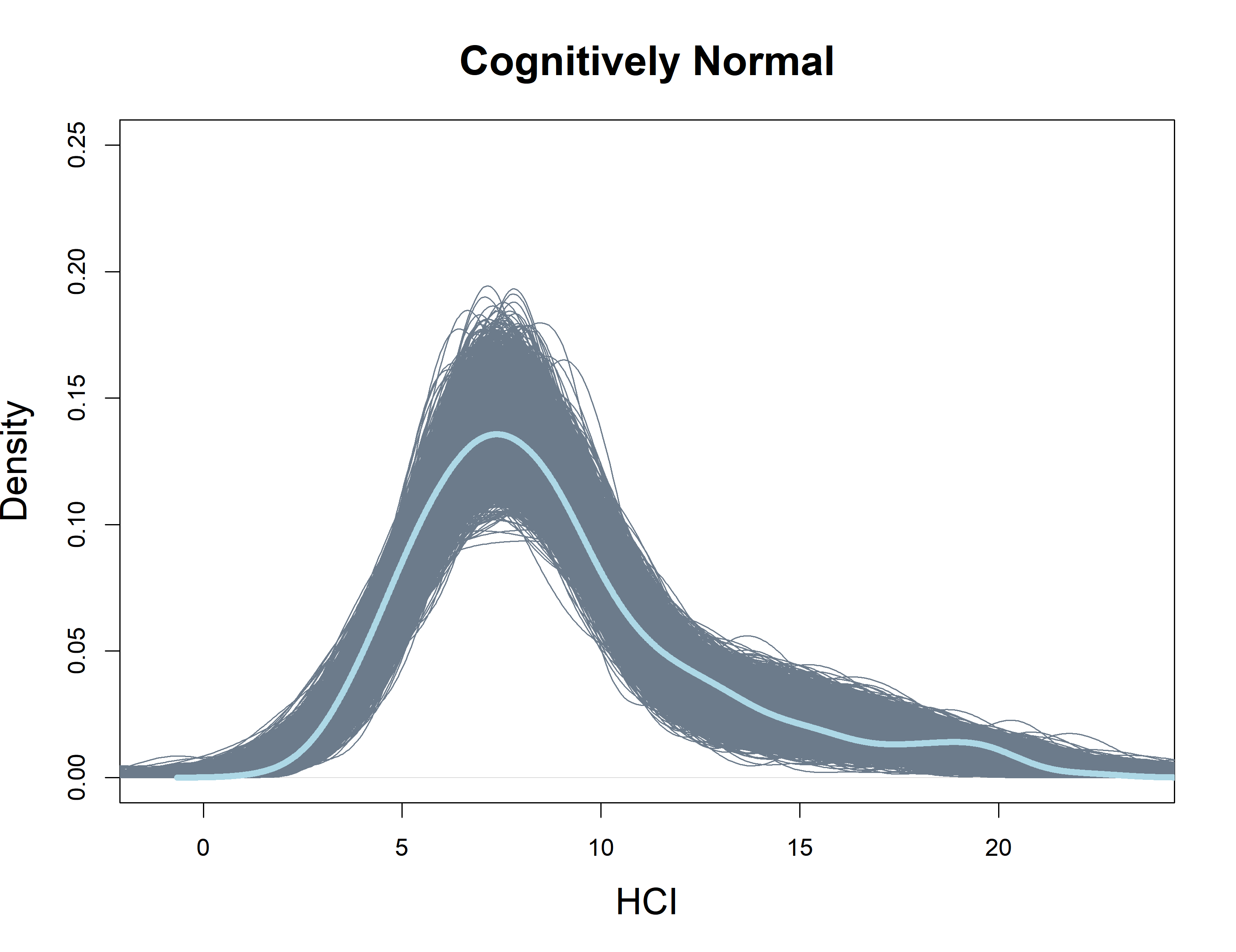}
	}
	\subfigure{
		\includegraphics[width=0.3\linewidth]{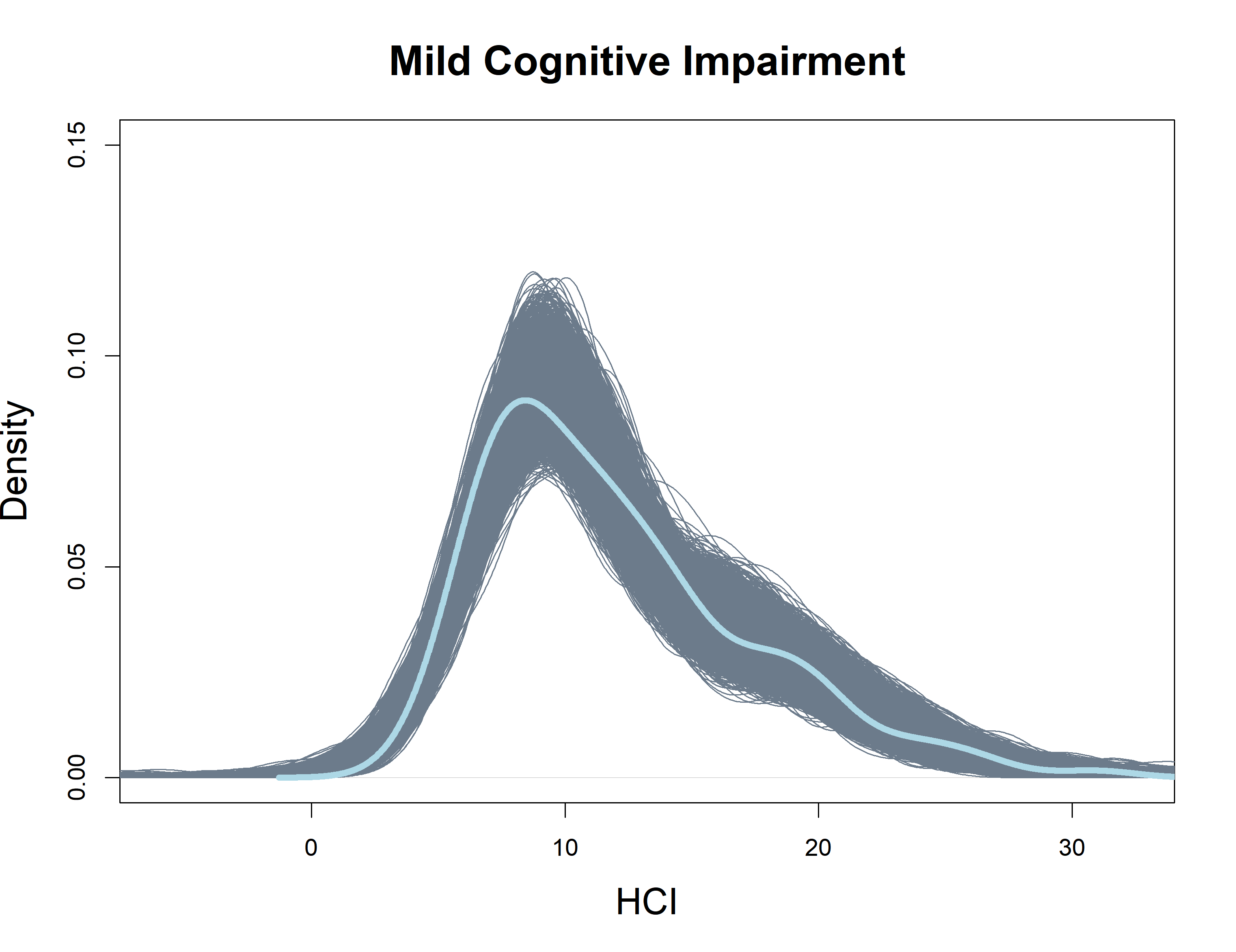}
	}
	\subfigure{
		\includegraphics[width=0.3\linewidth]{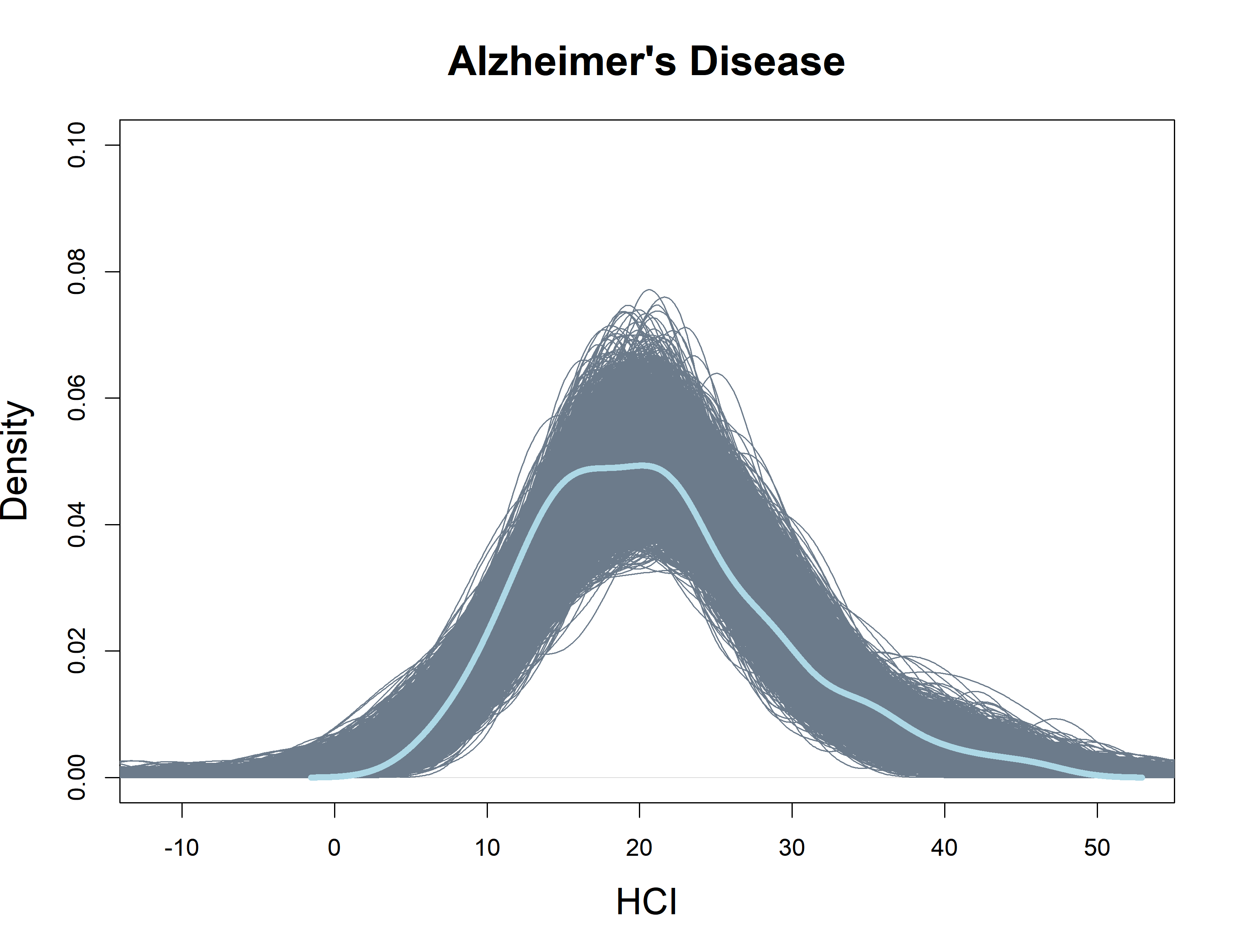}
	}
	\\
	\subfigure{
		\includegraphics[width=0.3\linewidth]{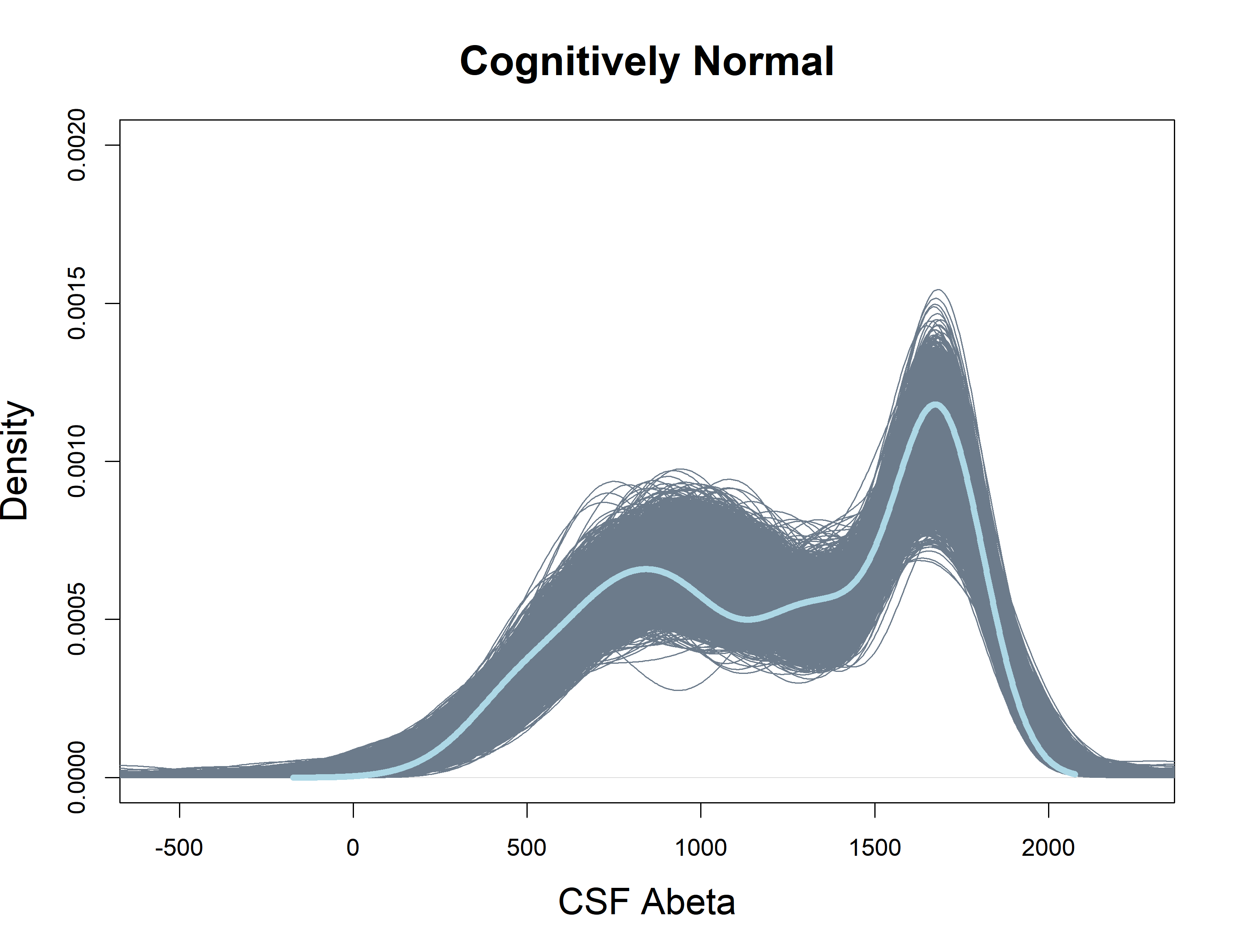}
	}
	\subfigure{
		\includegraphics[width=0.3\linewidth]{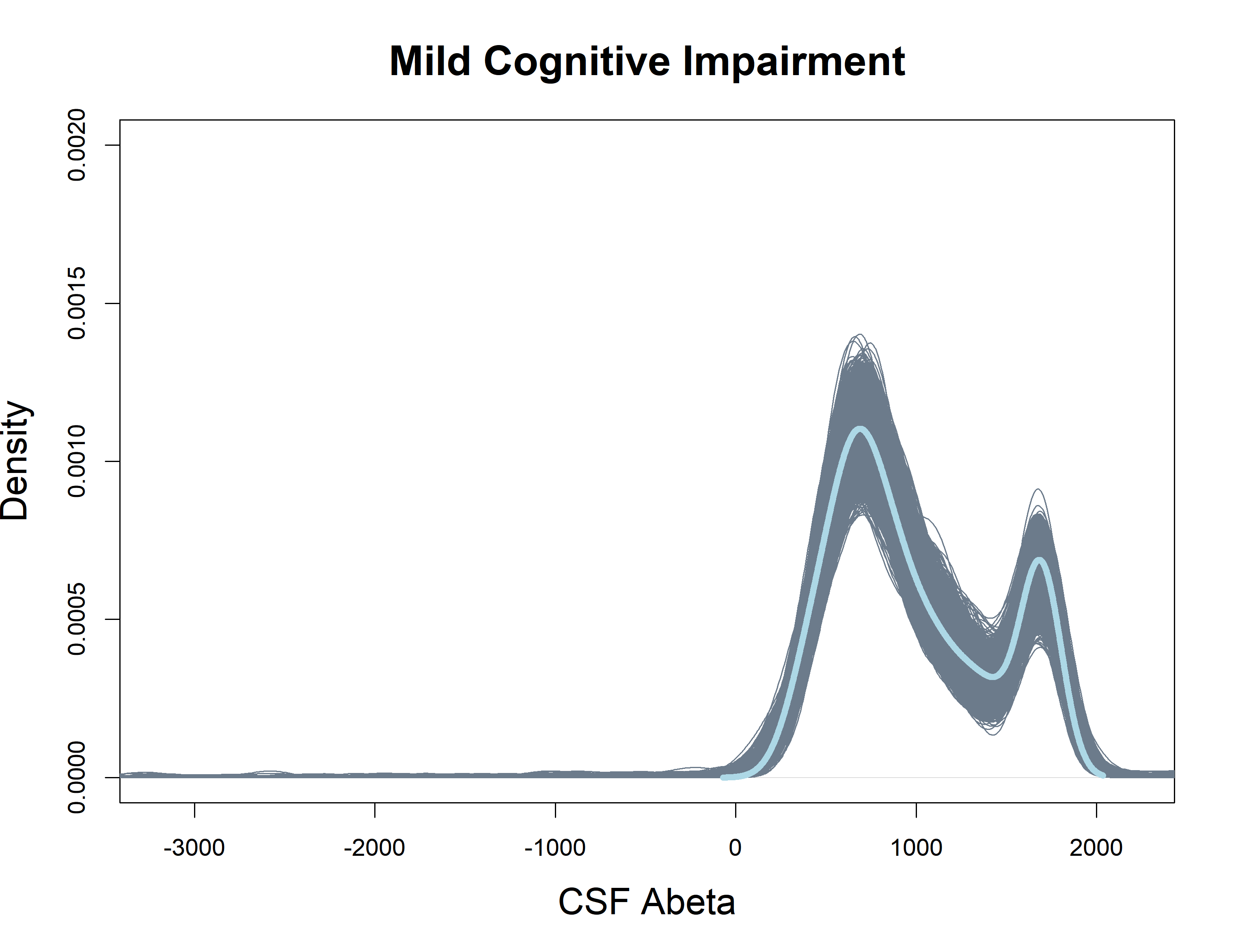}
	}
	\subfigure{
		\includegraphics[width=0.3\linewidth]{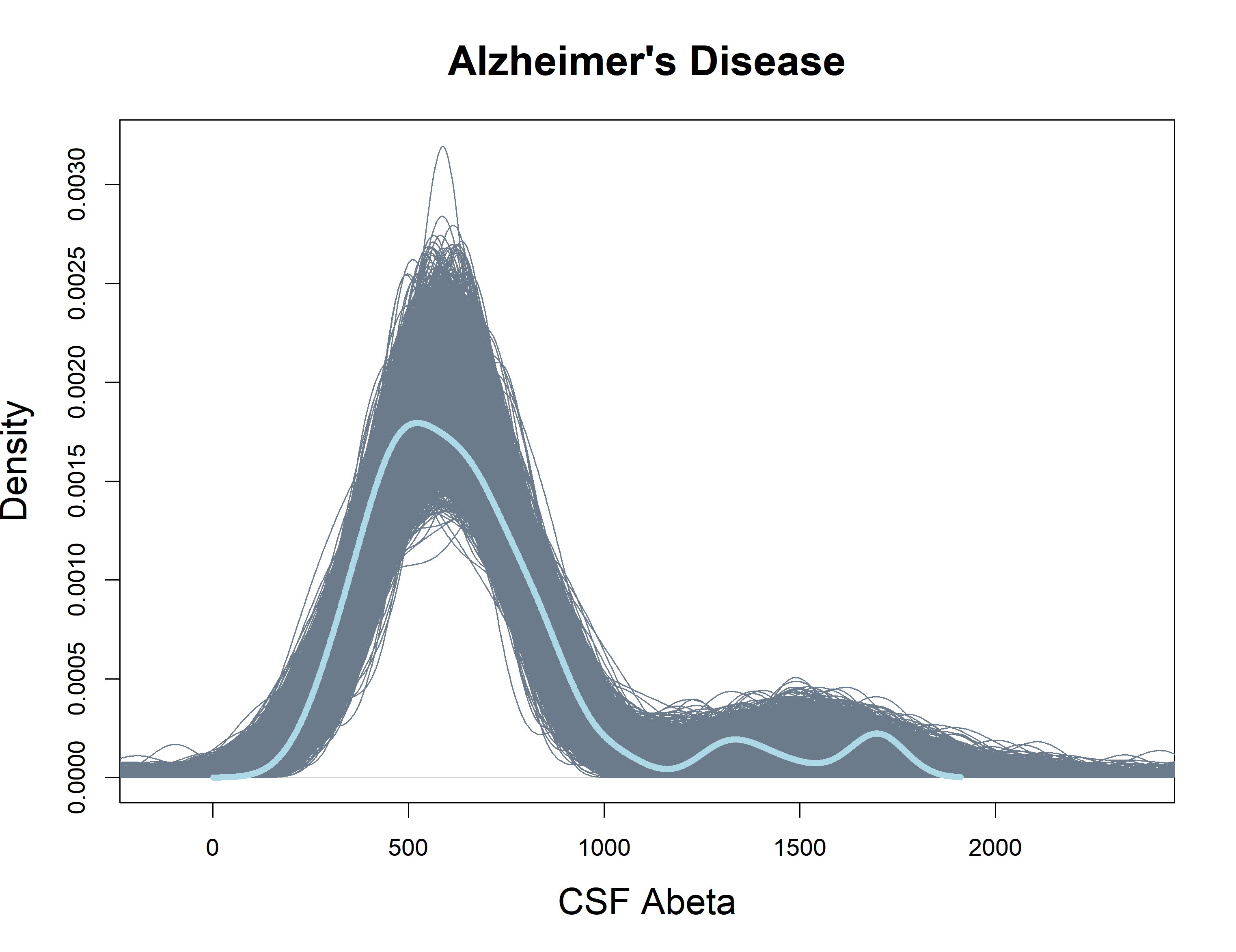}
	}
	\\
	\subfigure{
		\includegraphics[width=0.3\linewidth]{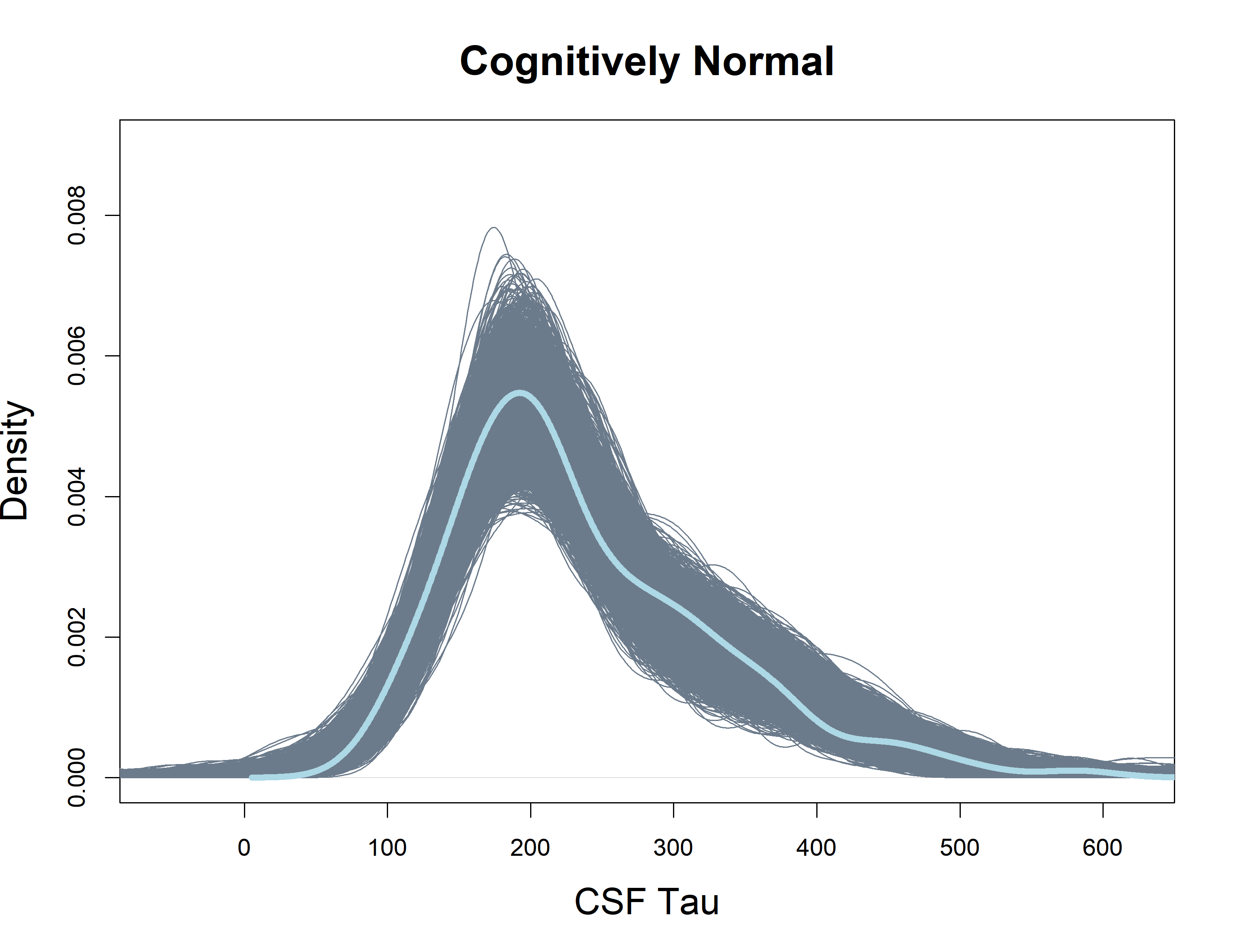}
	}
	\subfigure{
		\includegraphics[width=0.3\linewidth]{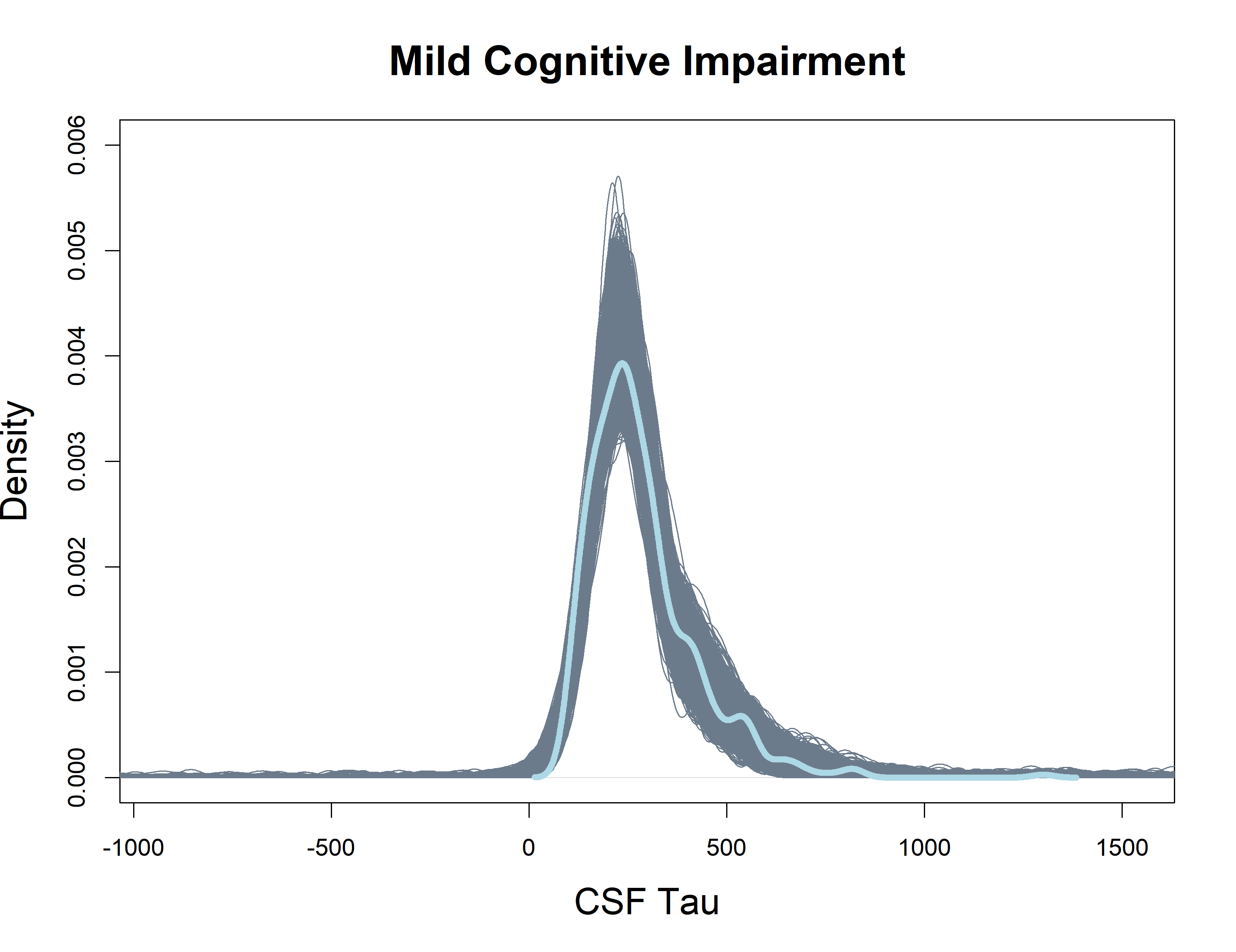}
	}
	\subfigure{
		\includegraphics[width=0.3\linewidth]{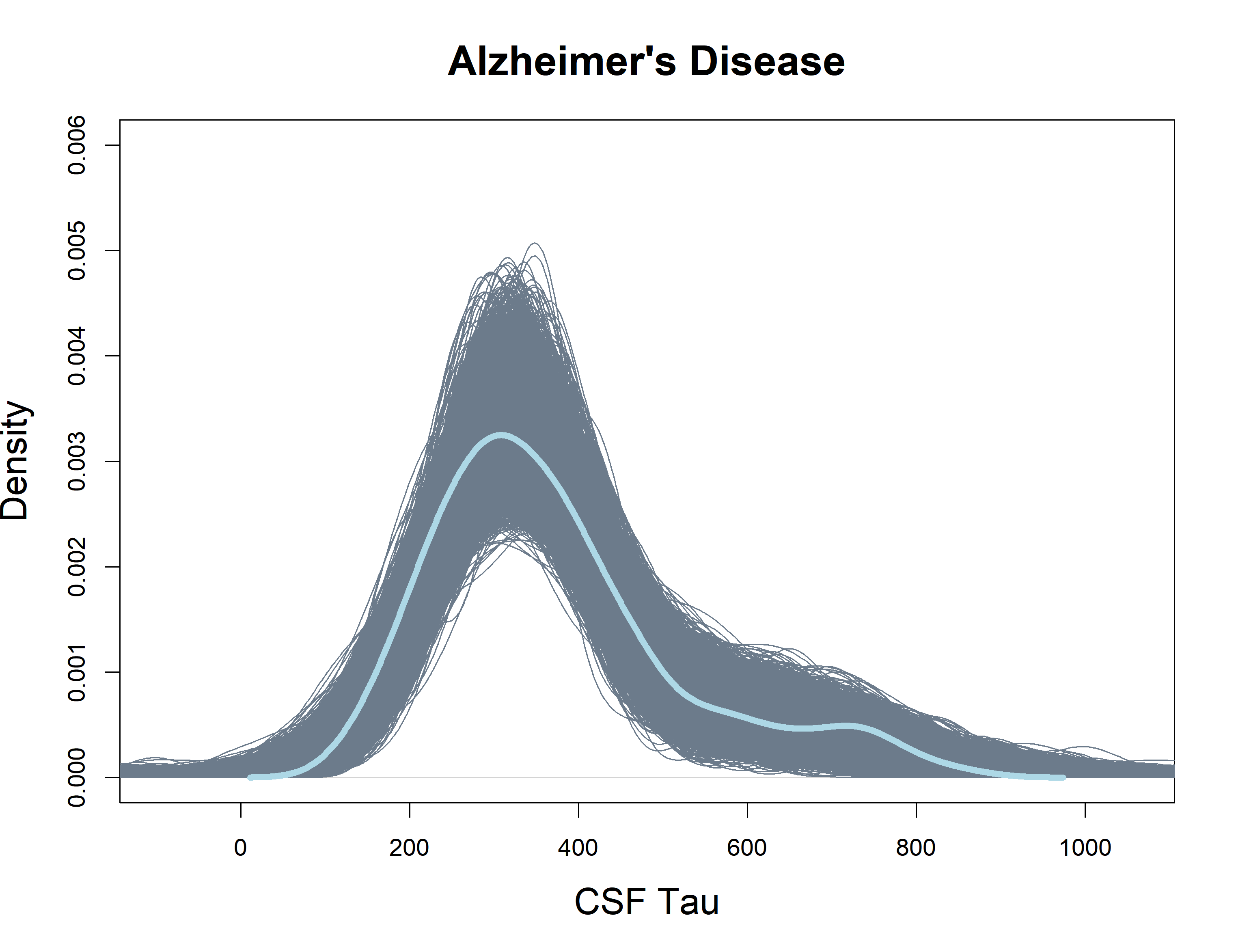}
	}
	\\
	\subfigure{
		\includegraphics[width=0.3\linewidth]{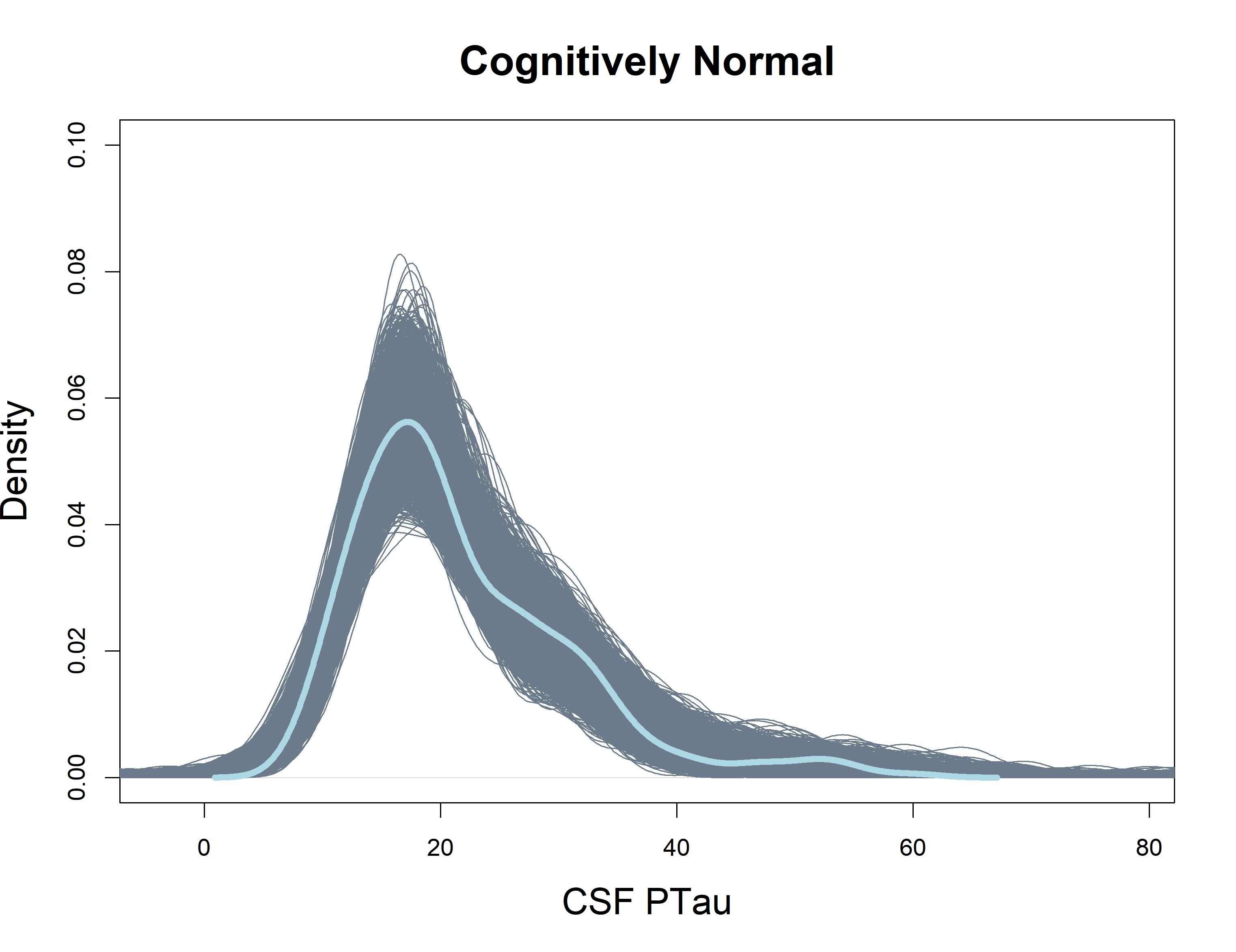}
	}
	\subfigure{
		\includegraphics[width=0.3\linewidth]{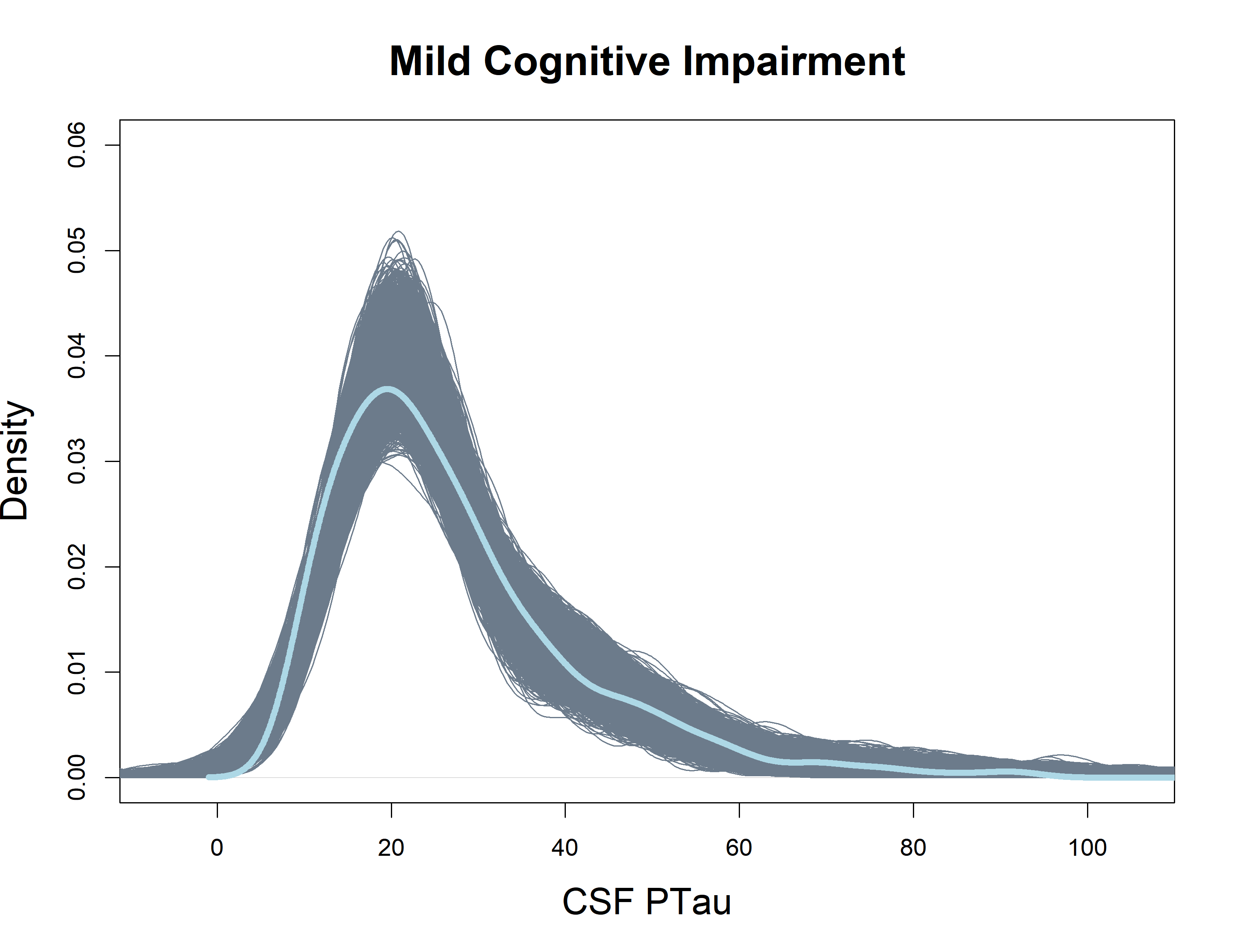}
	}
	\subfigure{
		\includegraphics[width=0.3\linewidth]{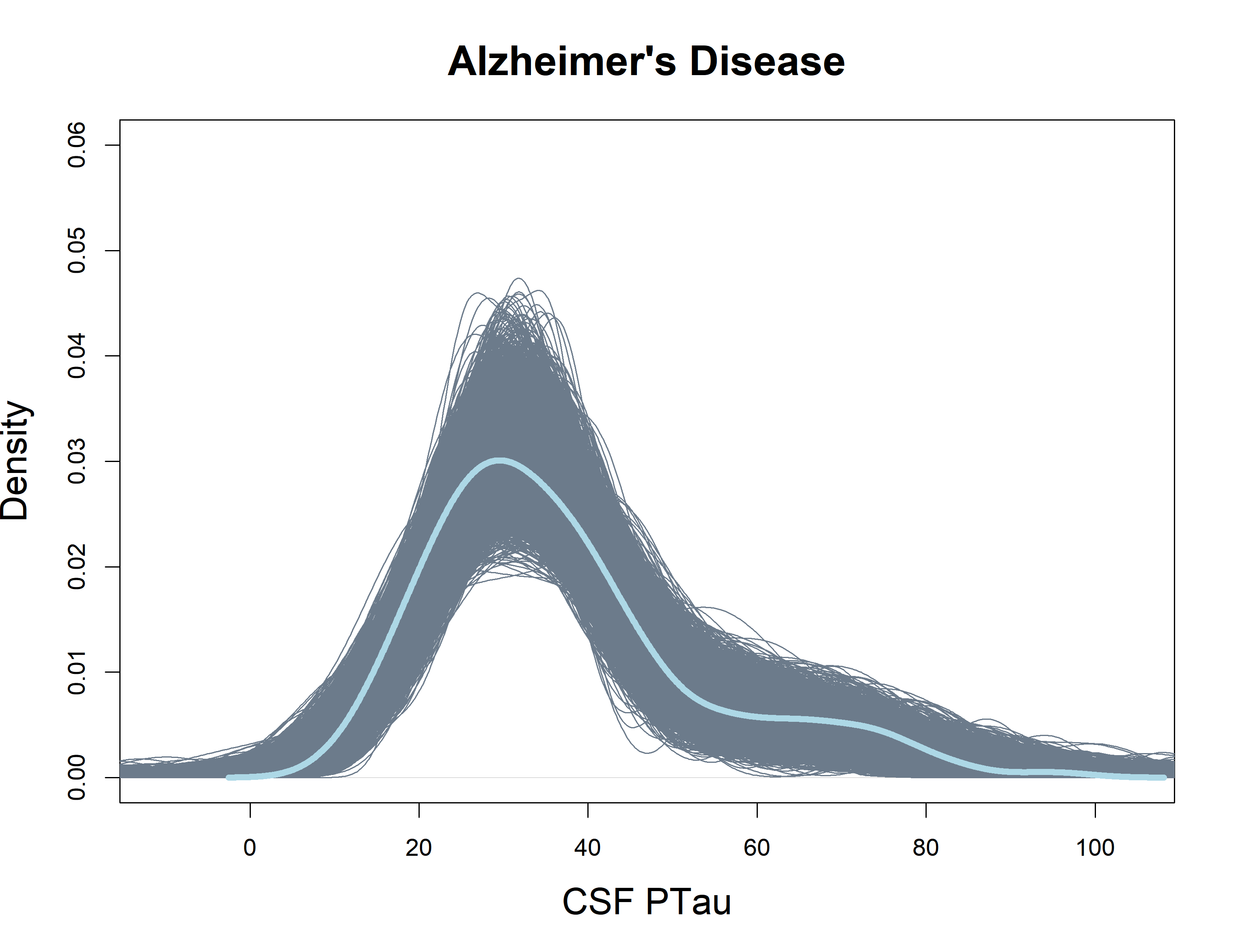}
	}
	\caption{Posterior predictive checks. Kernel density estimate of the observed biomarker outcomes (light blue), shown alongside kernel density estimates from the 5000 datasets drawn from the posterior predictive distribution (grey).}
	\label{posterior_predictive_density}
\end{figure}

\begin{figure}[htpb]
	\centering
	\subfigure{
		\includegraphics[width=0.3\linewidth]{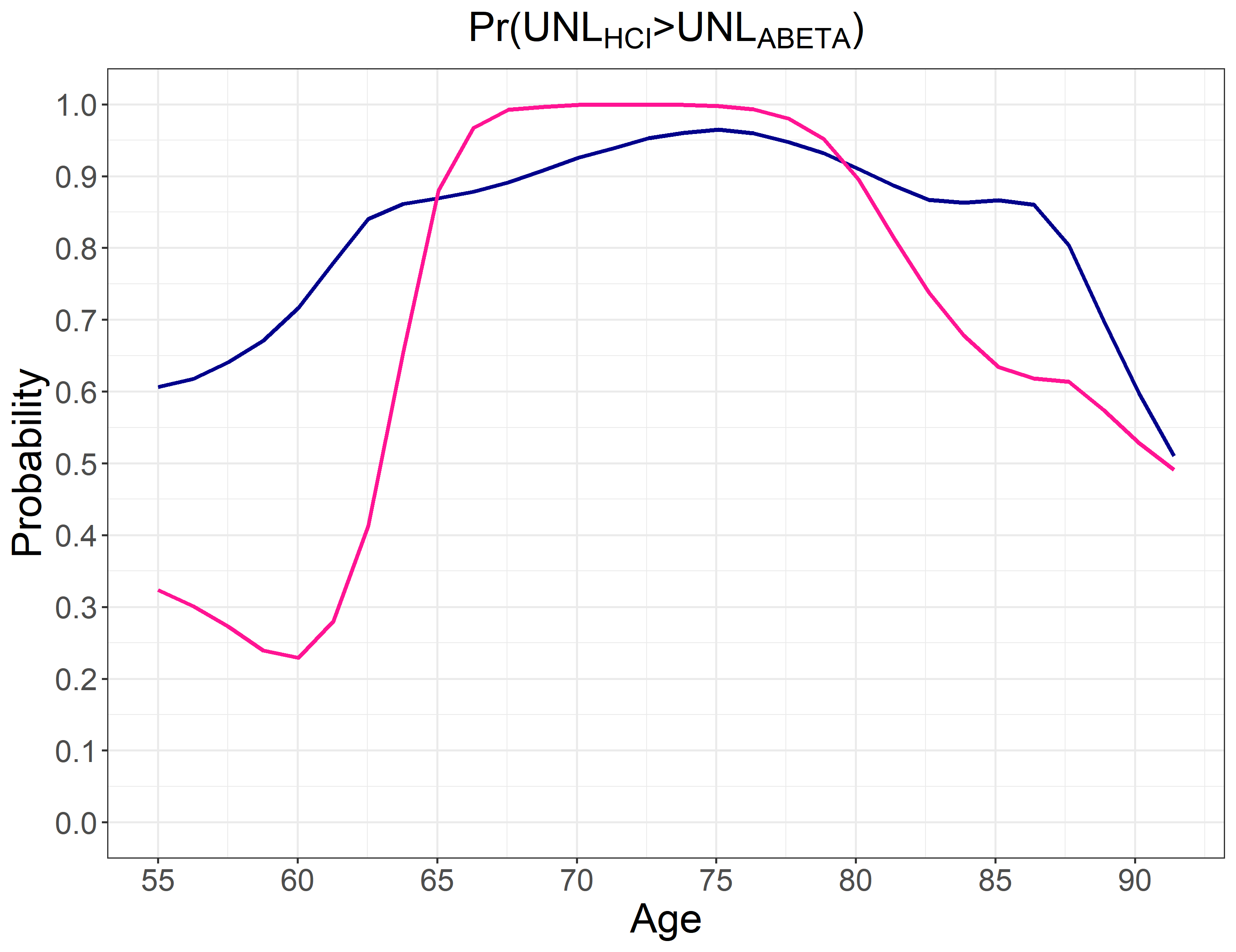}
	}
	\subfigure{
		\includegraphics[width=0.3\linewidth]{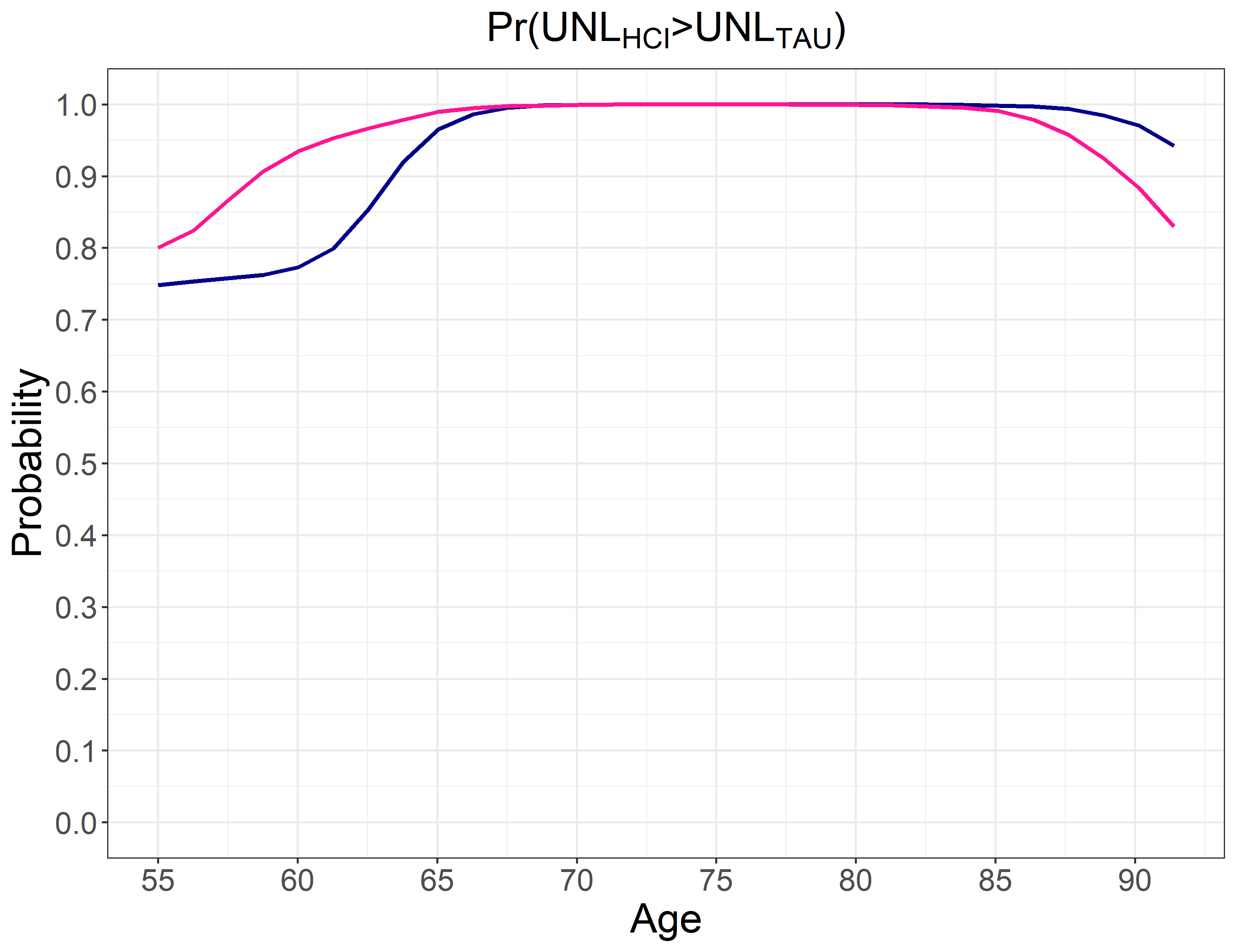}
	}
	\subfigure{
		\includegraphics[width=0.3\linewidth]{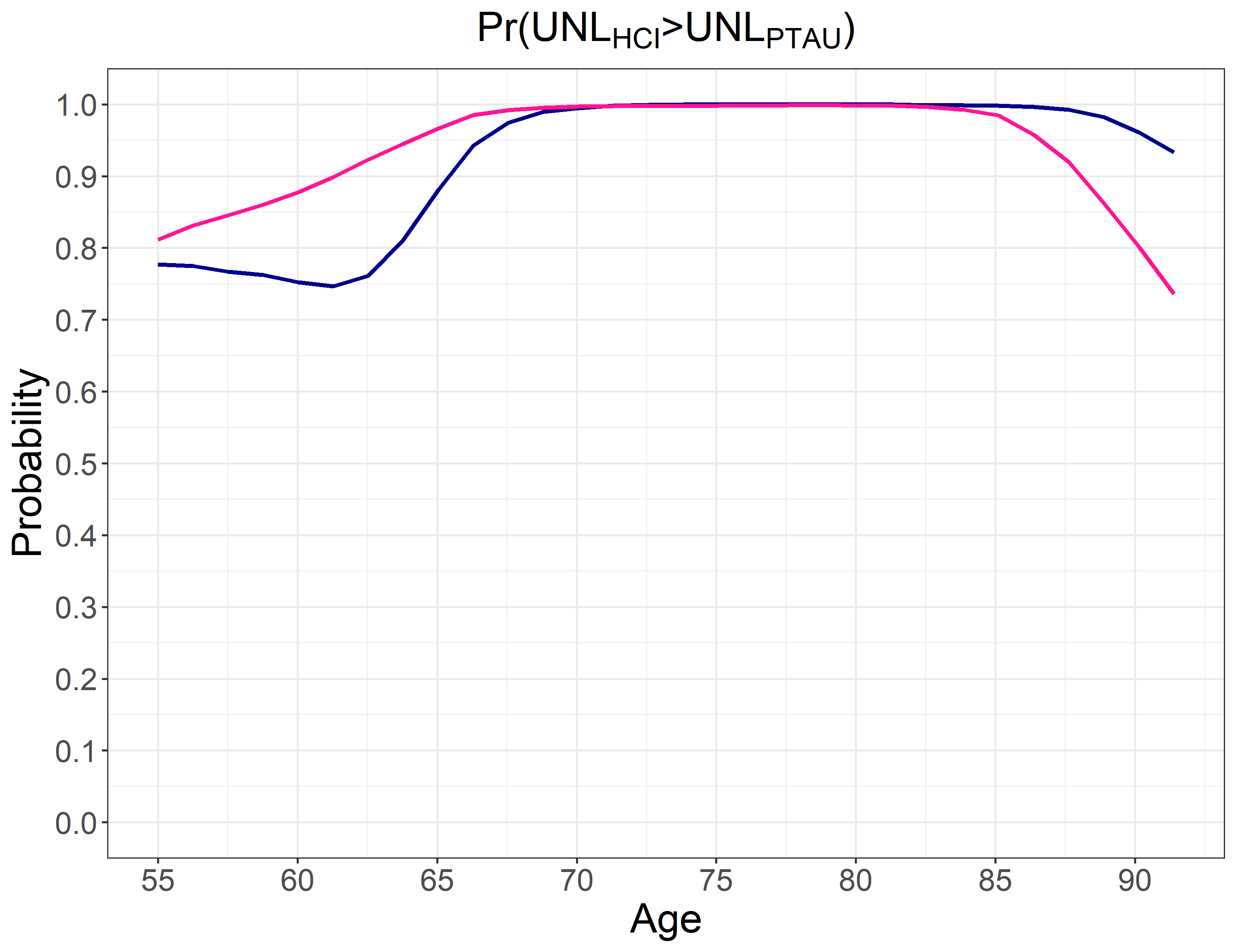}
	}
	\caption{Probability that the underlap coefficient of the HCI is higher than that of the other three biomarkers. The blue line represents the male group, while the pink line represents the female group. }
	\label{HCI_VS_others}
\end{figure}

\begin{figure}[htpb]
	\centering
	\includegraphics[width=0.8\textwidth]{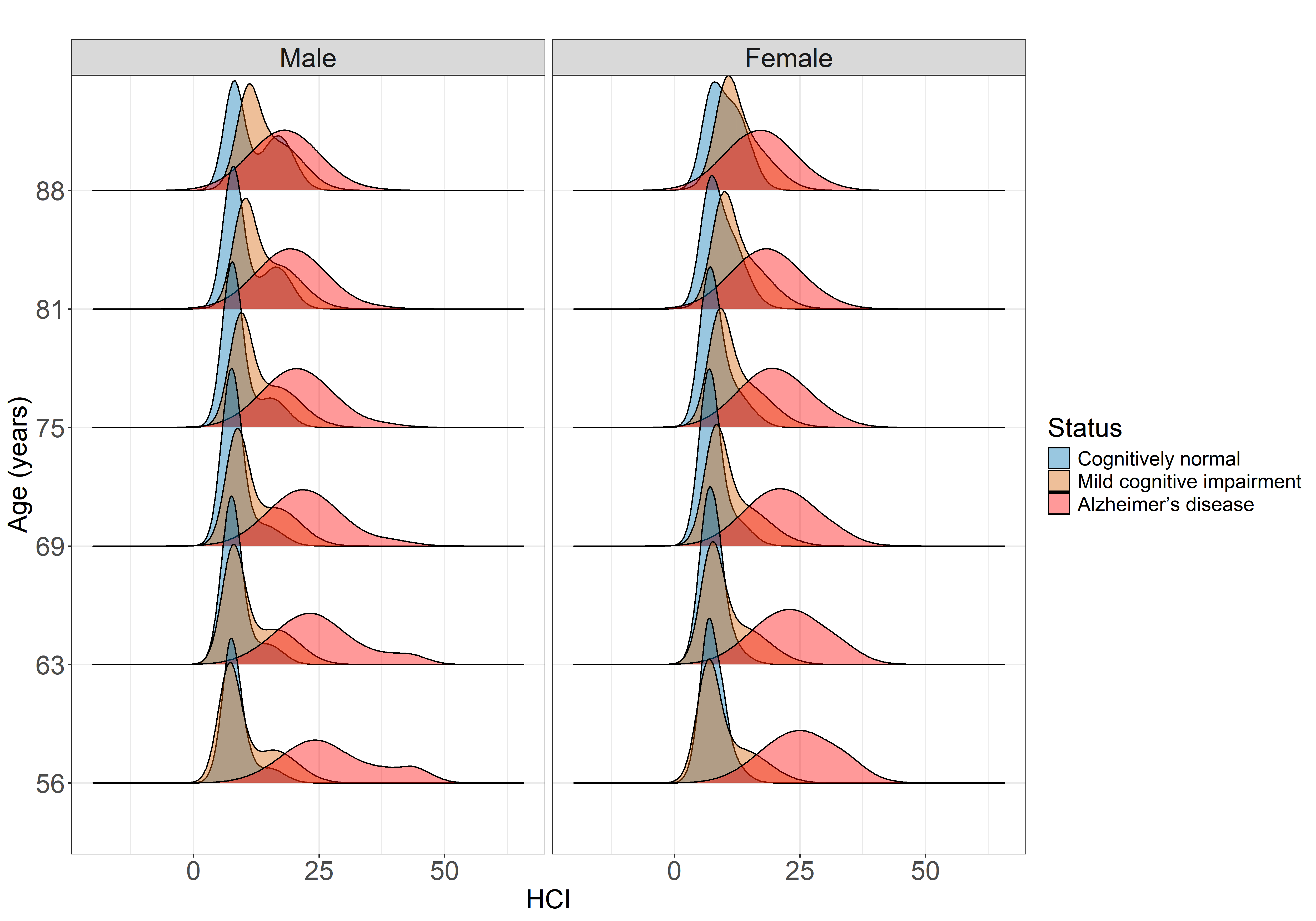}
	\caption{Estimated (posterior median) density functions of the HCI biomarker in the
		cognitively normal (blue), mild cognitive impairment (orange), and Alzheimer’s disease (red) groups, conditional on gender and age.}
	\label{HCI_density_plot}
\end{figure}

\begin{figure}[htpb]
	\centering
	\includegraphics[width=0.8\textwidth]{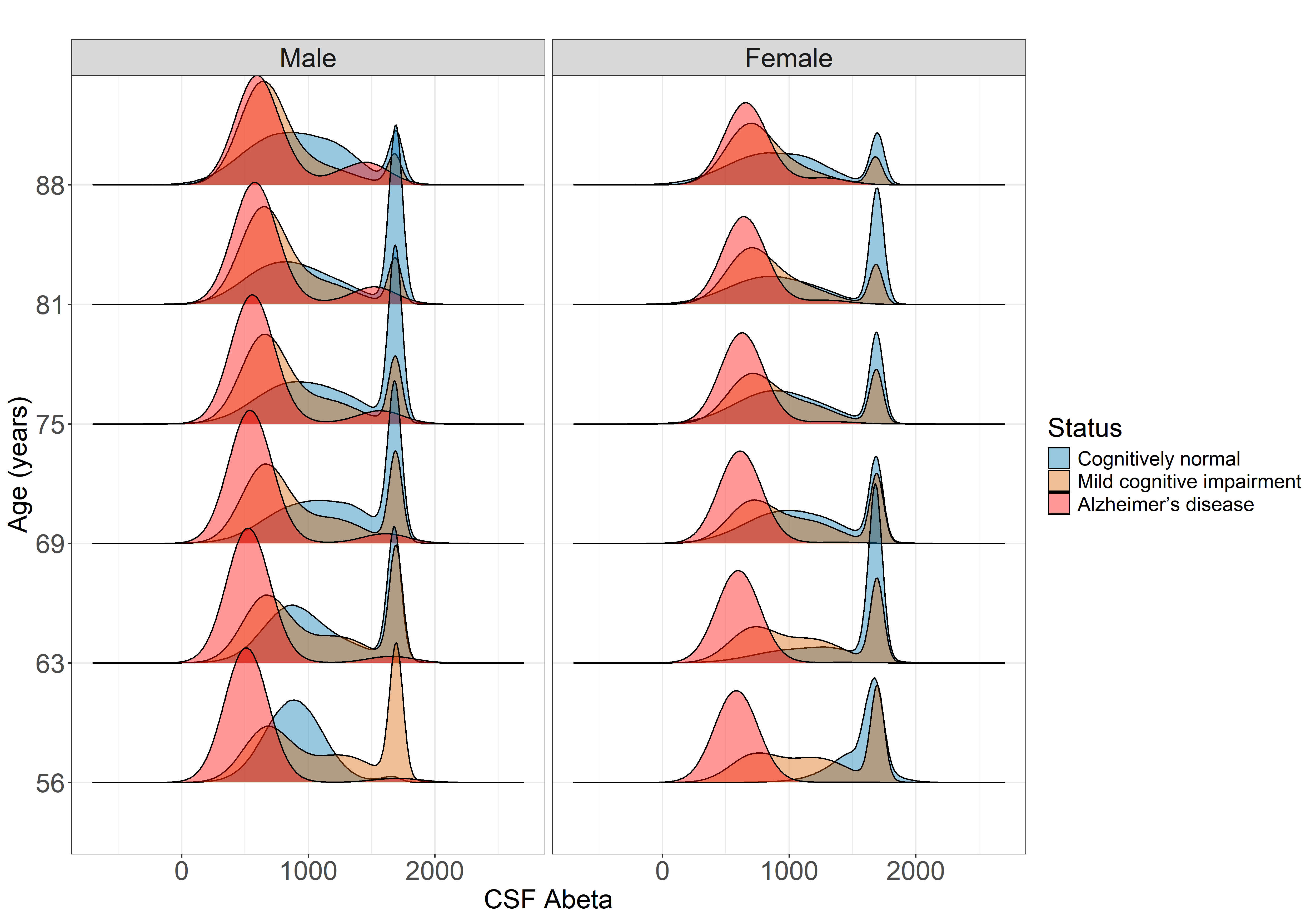}
	\caption{Estimated (posterior median) density functions of the CSF Abeta biomarker in the
		cognitively normal (blue), mild cognitive impairment (orange), and Alzheimer’s disease (red) groups, conditional on gender and age.}
	\label{abeta_density_plot}
\end{figure}

\begin{figure}[htpb]
	\centering
	\includegraphics[width=0.8\textwidth]{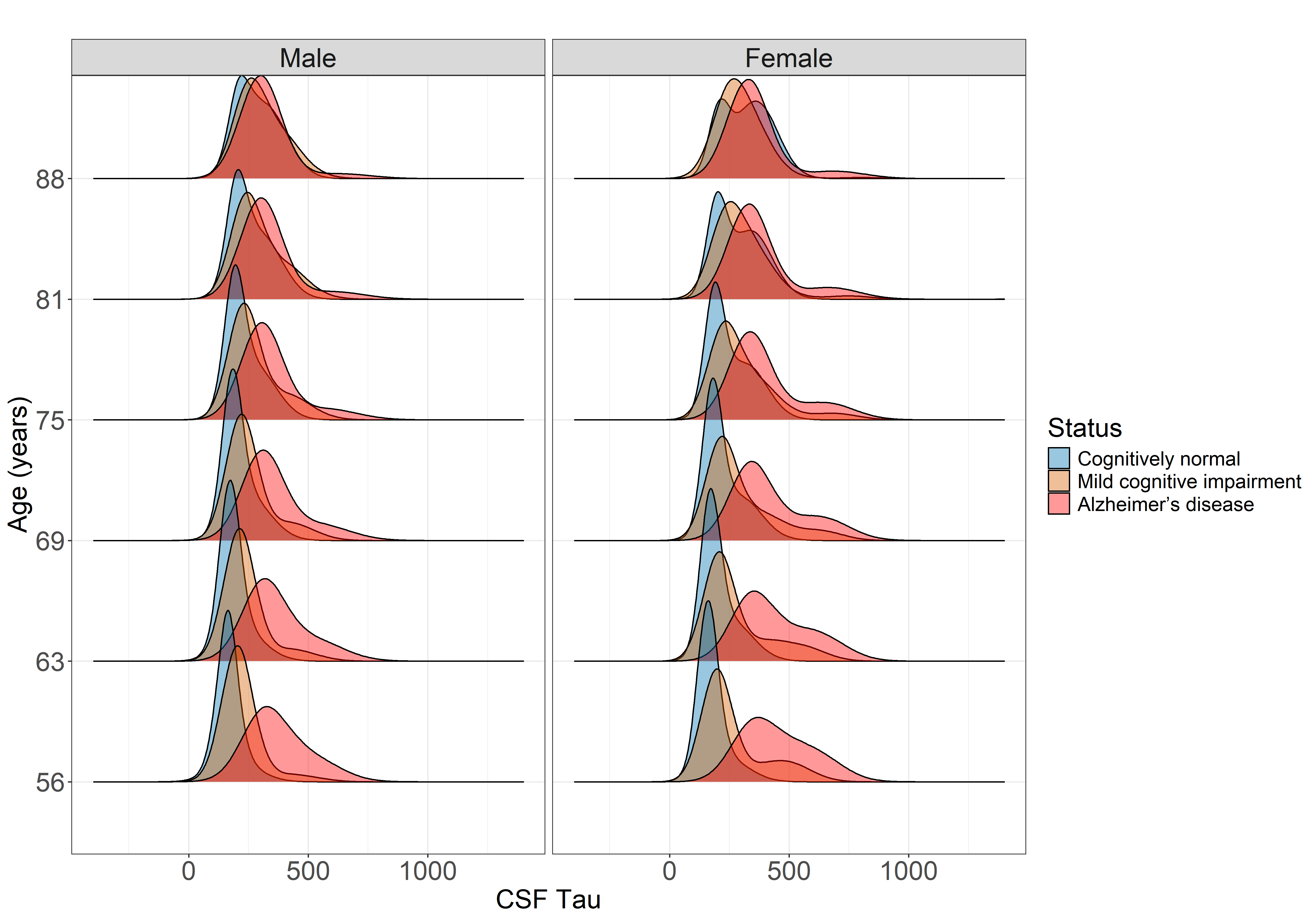}
	\caption{Estimated (posterior median) density functions of the CSF Tau biomarker in the
		cognitively normal (blue), mild cognitive impairment (orange), and Alzheimer’s disease (red) groups, conditional on gender and age.}
	\label{tau_density_plot}
\end{figure}

\begin{figure}[htpb]
	\centering
	\includegraphics[width=0.8\textwidth]{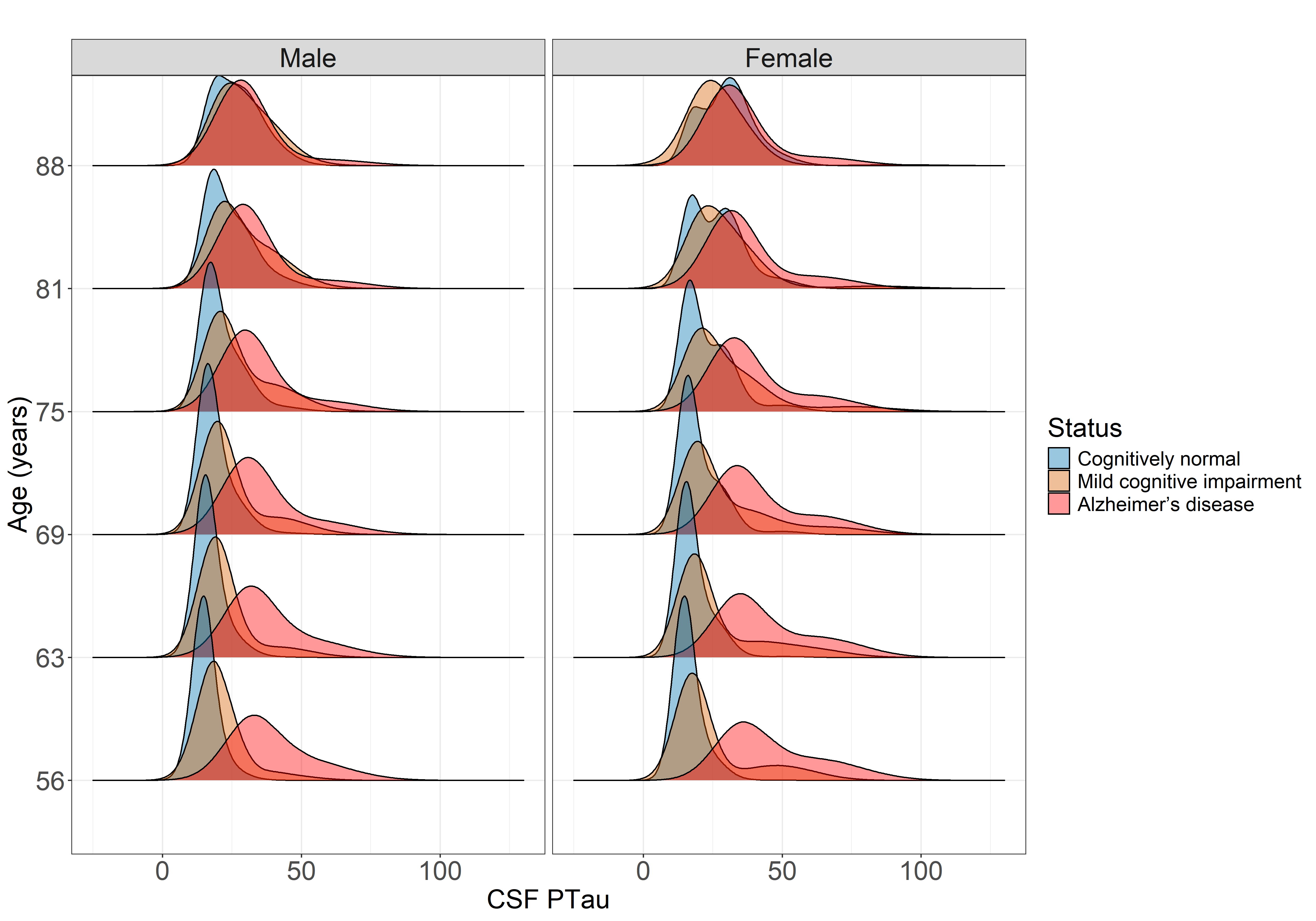}
	\caption{Estimated (posterior median) density functions of the CSF pTAU biomarker in the
		cognitively normal (blue), mild cognitive impairment (orange), and Alzheimer’s disease (red) groups, conditional on gender and age.}
	\label{ptau_density_plot}
\end{figure}

\begin{figure}[htpb]
	\centering
	\subfigure{
		\centering
		\includegraphics[width=0.3\textwidth]{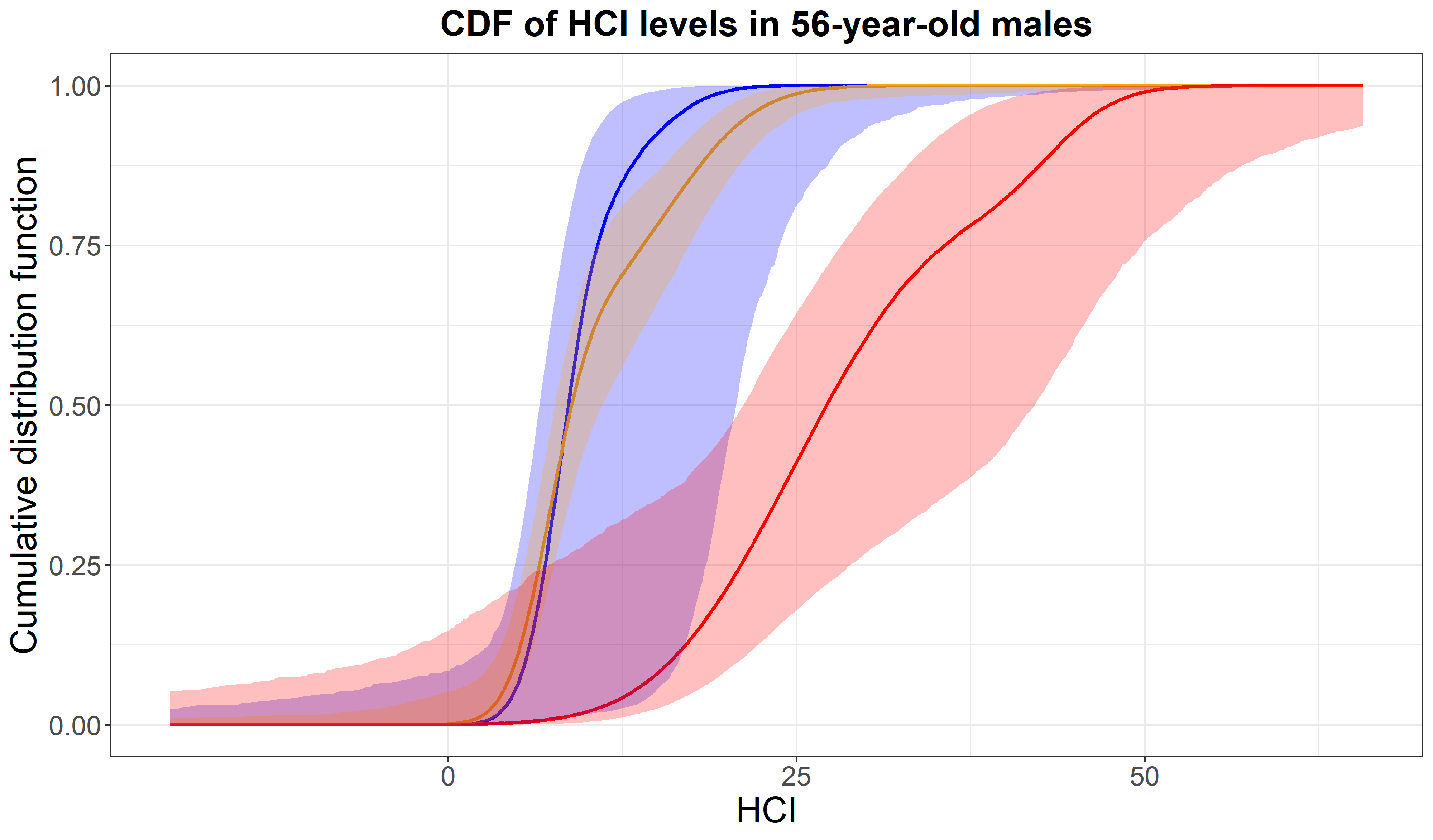}
	}
	\subfigure{
		\centering
		\includegraphics[width=0.3\textwidth]{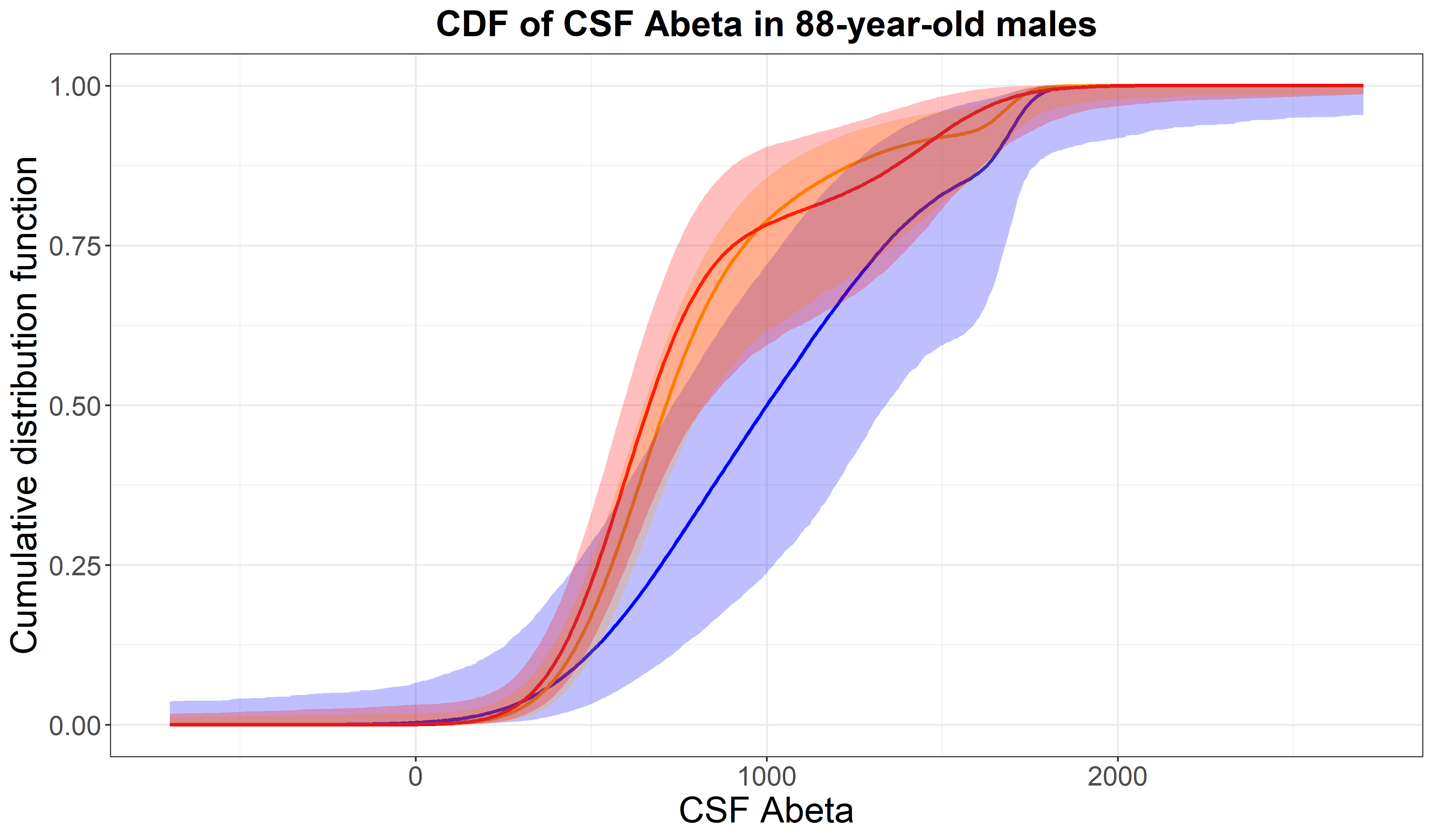}
	}
	\subfigure{
		\centering
		\includegraphics[width=0.3\textwidth]{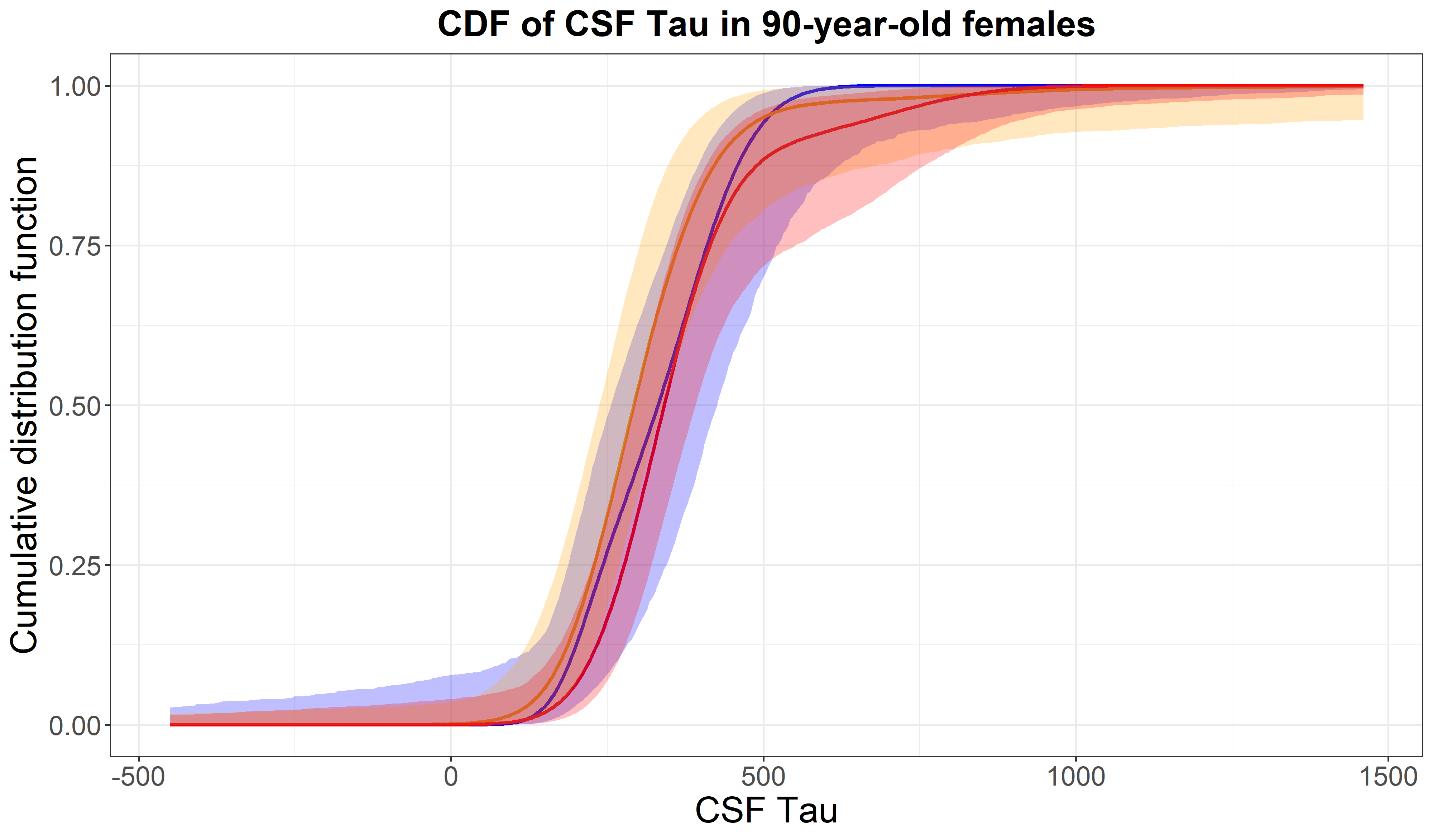}
	}
	\centering
	\caption{Examples illustrating  violations of the stochastic order assumption for three biomarkers. Left panel: Estimated (posterior median) cumulative distribution function of HCI in 56-year-old males, with pointwise 95\% credible bands. Middle panel: Estimated (posterior median) cumulative distribution function of CSF Abeta in 88-year-old males, with pointwise 95\% credible bands. Right panel: Estimated (posterior median) cumulative distribution function of CSF Tau in 90-year-old females, with pointwise 95\% credible bands. Blue lines and ribbons: cognitively normal group. Orange lines and ribbons: mild cognitive impairment group. Red lines and ribbons: Alzheimer’s disease group. It should be noted that, because we model the biomarker outcome densities as mixtures of normal distributions, the estimated densities may take nonzero values for negative outcomes. To ensure accurate estimation of the distribution functions, we perform estimation over a grid that includes negative values.}
	\label{adni_reverse_order_examples}
\end{figure}

\begin{figure}[htpb]
	\centering
	\subfigure{
		\includegraphics[width=0.48\linewidth]{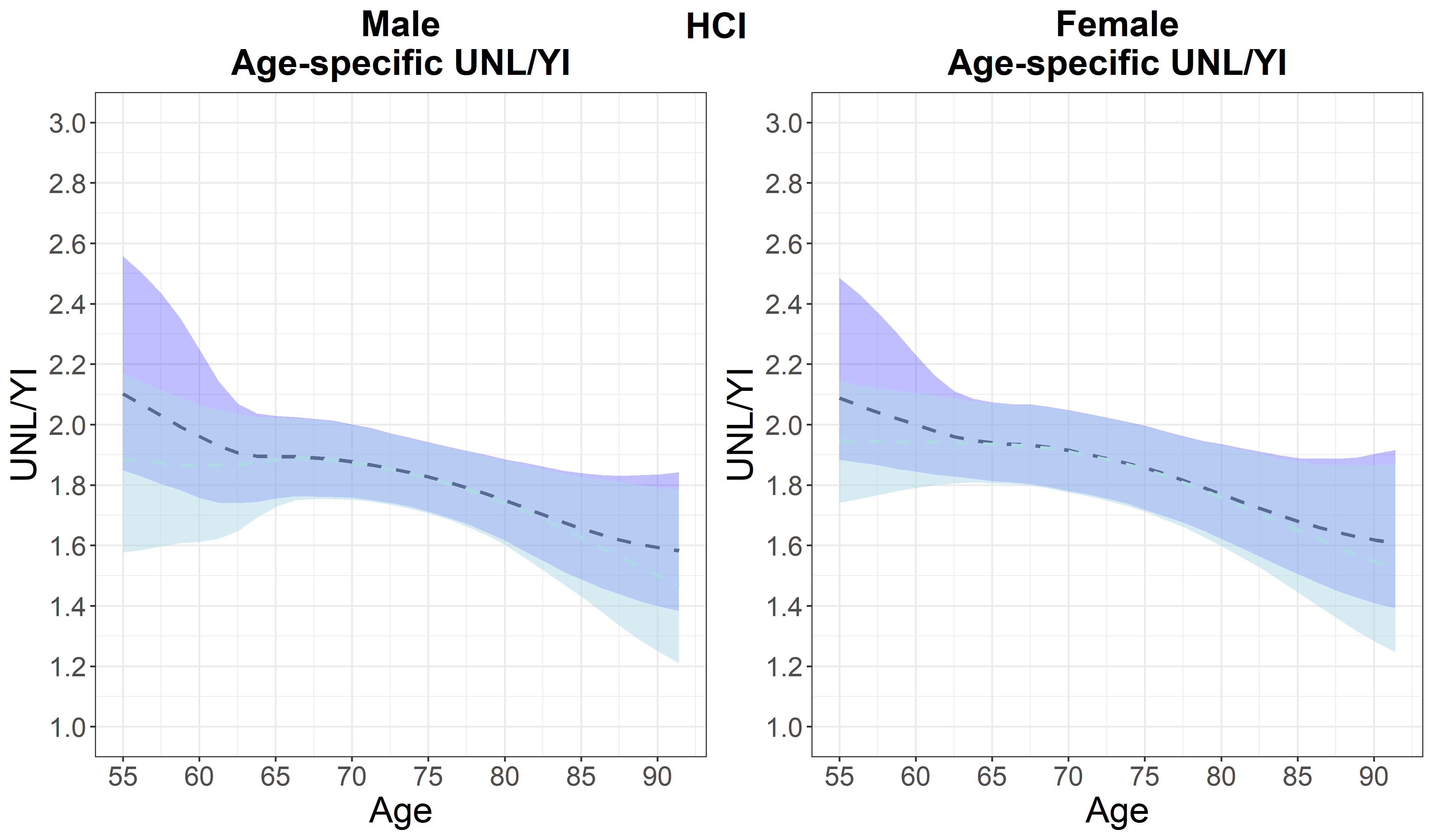}
	}
	\subfigure{
		\includegraphics[width=0.48\linewidth]{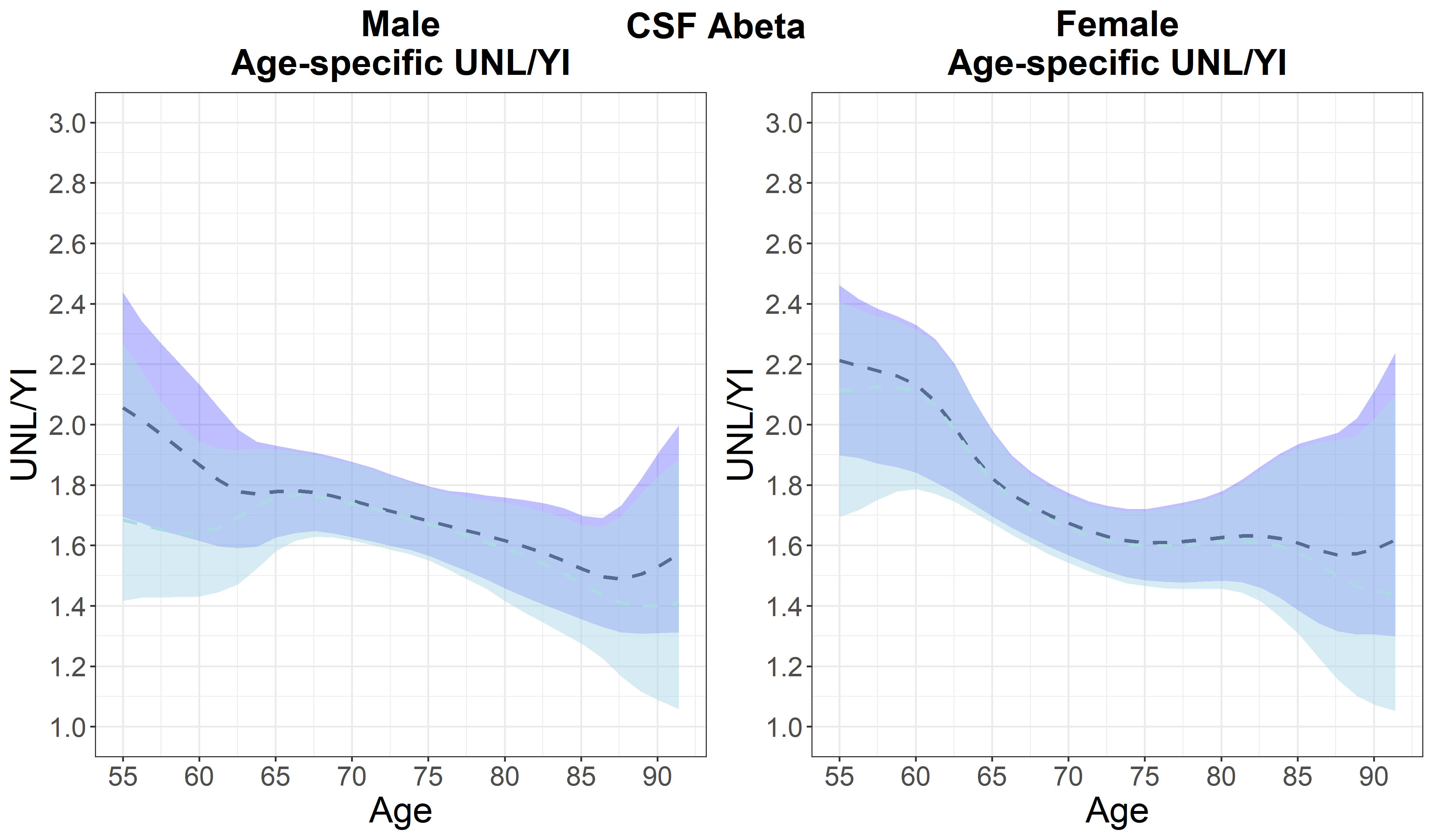}
	}
	\\
	\subfigure{
		\includegraphics[width=0.48\linewidth]{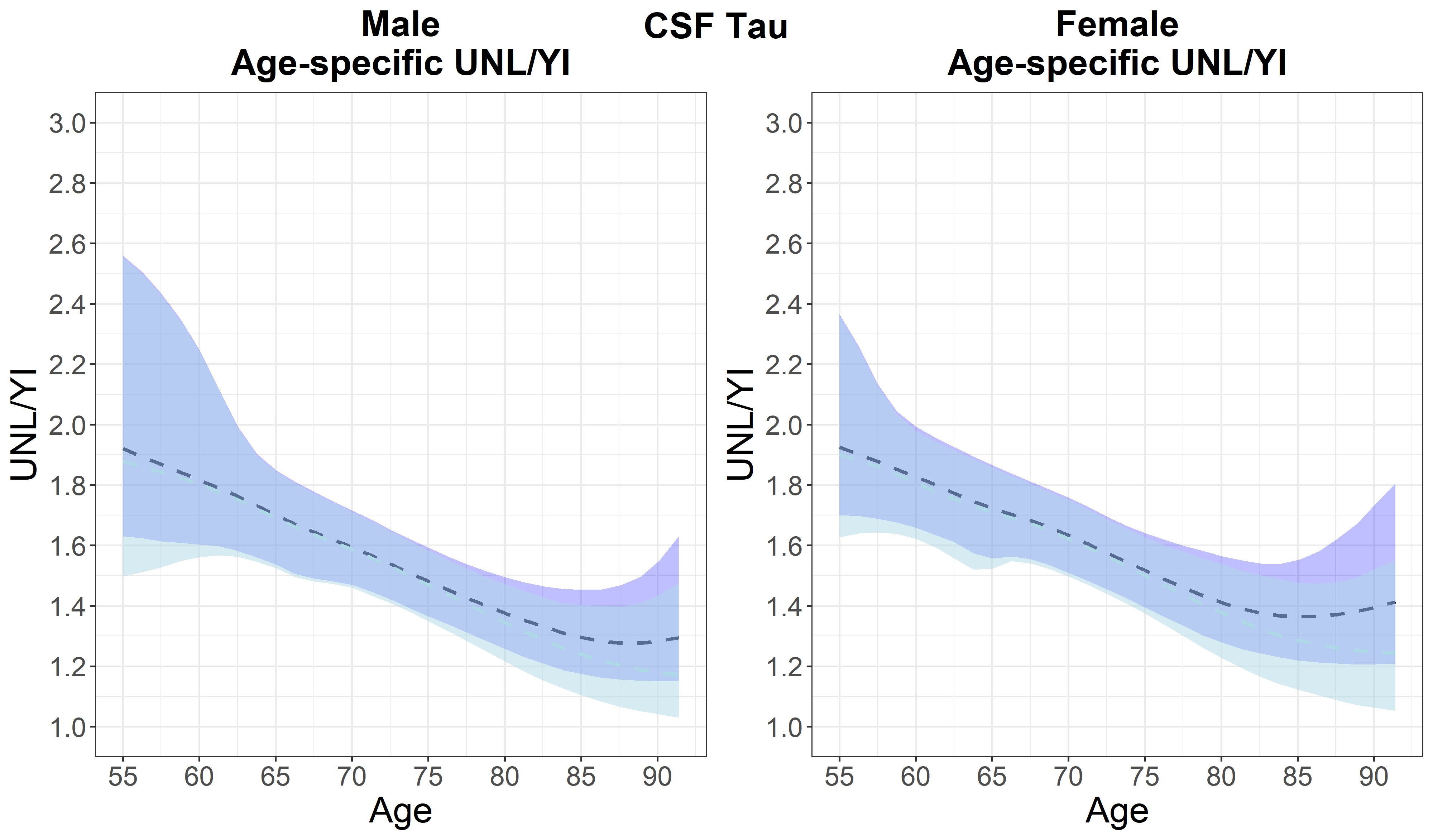}
	}
	\subfigure{
		\includegraphics[width=0.48\linewidth]{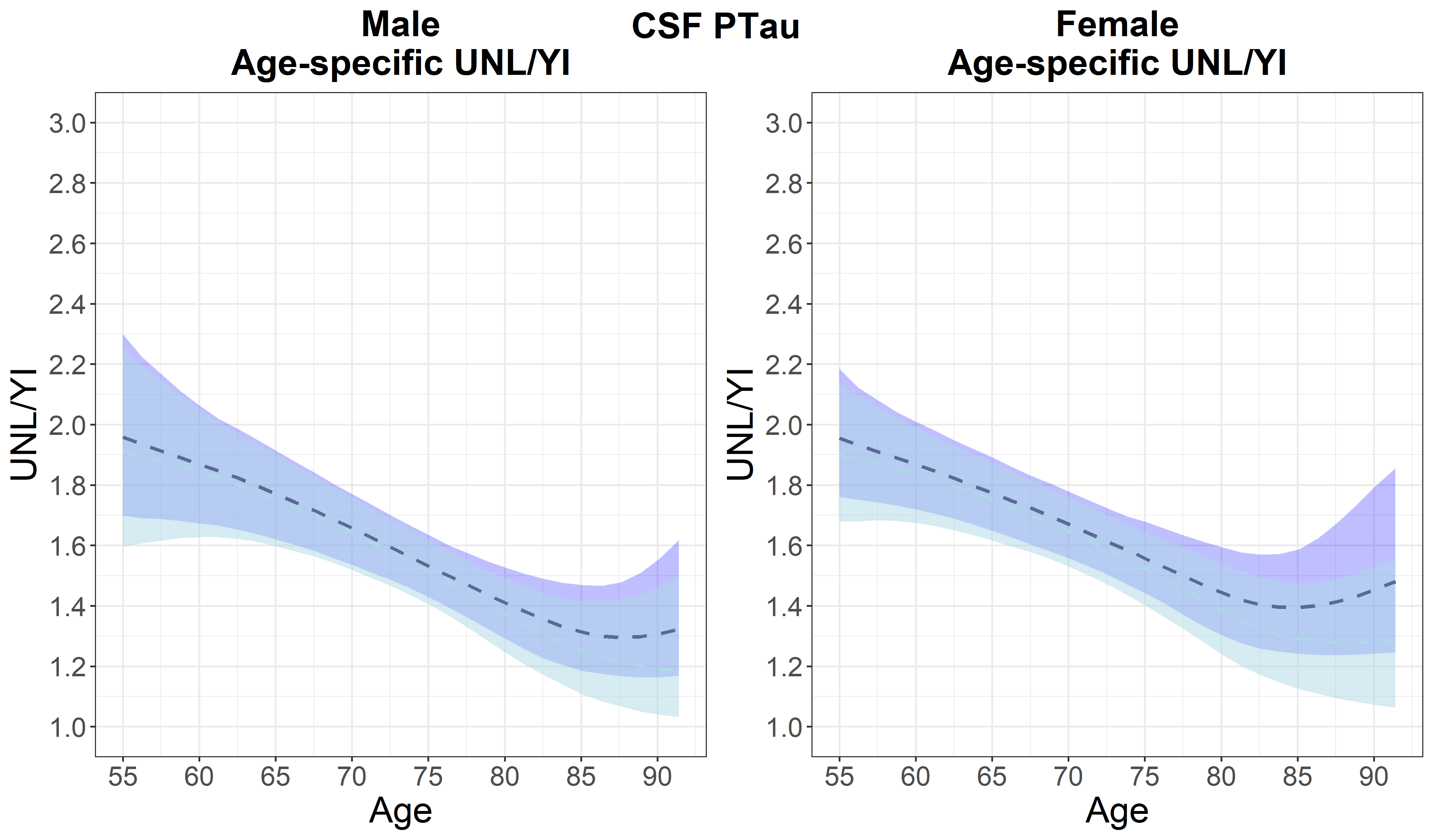}
	}
	\caption{Black dashed line and dark blue ribbon: posterior median and pointwise 95\% credible bands for the age-and-gender-specific underlap coefficient for the four biomarkers. 
		Light blue dashed line and ribbon: posterior median and pointwise 95\% credible bands for the age-and-gender-specific three-class Youden index.}
	\label{unl_YI_age_gender_specific}
\end{figure}

\begin{figure}[htpb]
	\centering
	\includegraphics[width=0.6\textwidth]{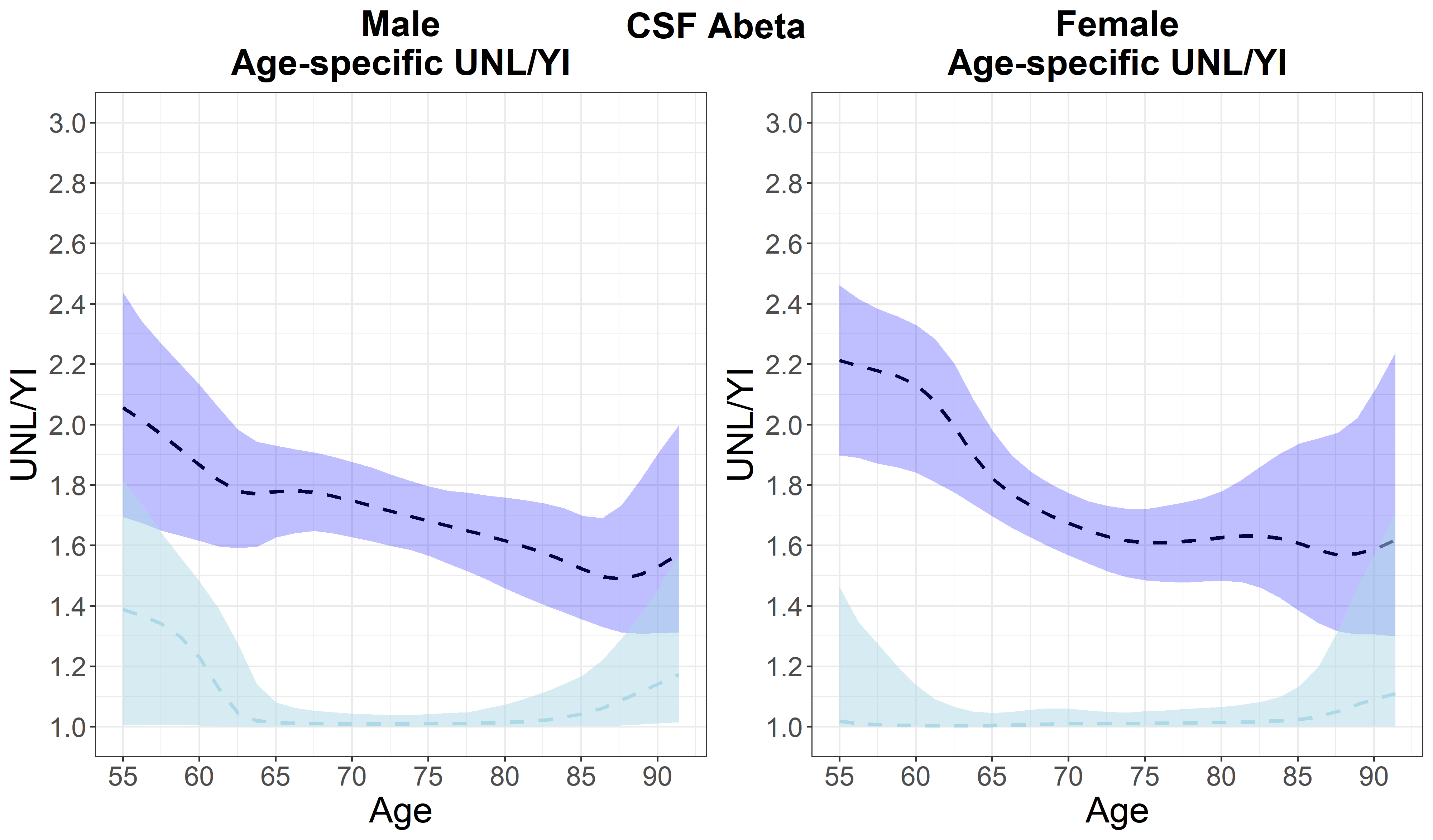}
	\caption{Black dashed line and dark blue ribbon: posterior median and pointwise 95\% credible bands for the age-and-gender-specific underlap coefficient for the CSF Abeta biomarker. Light blue dashed line and ribbon: posterior median and  pointwise 95\% credible bands for the age-and-gender-specific three-class Youden index for CSF Abeta using the original order.}
	\label{abeta_unl_YI_orderx}
\end{figure}

\end{document}